\documentclass{article}
\usepackage{amsmath}
\usepackage{amssymb}

\input{tcilatex}
\begin{document}

\title{Why must we work in the phase space?}

\author{J. J. S\l awianowski $^{1}$, F. E. Schroeck, Jr.$^{2}$, A. Martens $^{3}$\\
$^{1,3}$ Institute of Fundamental Technological Research,\\
Polish Academy of Sciences,\\
 Pawi\'{n}skiego 5B, 02-106 Warszawa, Poland\\
$^{2}$ Institute of Mathematics, Denver University,\\
Colorado, USA\\
e-mails: jslawian@ippt.pan.pl$^{1}$, Franklin.Schroeck@du.edu$^{2}$,\\ amartens@ippt.pan.pl$^{3}$\\}

\maketitle
\begin{abstract}

We are going to prove that the phase-space description is fundamental
both in the classical and quantum physics. It is shown that many problems
in statistical mechanics, quantum mechanics, quasi-classical theory and
in the theory of integrable systems may be well-formulated only in the
phase-space language.

\end{abstract}

\section{Some philosophy. Ancient and middle ages comprehensions of the phase space}

By abuse of language one can claim that our modern phase-space concepts have some roots in elementary human thinking, in elementary experience and in ancient philosophy.

In a sense, the phase space ideas go back to the ancient Greeks, namely they seem to have their origin in the famous Zeno's Paradox of Aporia. In spite of a manifold of mathematical and philosophical arguments aimed at justifying the reality of motion, all of them seem to be somehow superficial and unconvincing (hypocritical, one could say by abuse of language). It seems that the intuition of Greek sophists, the first critical positivists in European thought, was on the right track in spite of their deliberately provocative mode of expression.

One cannot resist the feeling that "common sense" was more reliable here. There is something wrong in considering motion "in vitro", killing it and trying to recover it again as a sequence of inanimate rest states. As an example, take a sequence of pictures which simulate some motion, as in a movie picture. You have a stage coach which seems to be going in some direction. But then you look at its wheels which appear to be rolling in the opposite direction! The elementary human intuition conceives motion "in vivo" as something primary and non-reducible to the sequence of instantaneous positions. According to the animalistic and naive but also non-corrupted pre-scientific human perception, it is not so that motion is a change of position; unlike this, change is a result of motion, the latter being perceived as something primary. And no modern mathematical distinctions between zero-measure and finite-measure subsets are able to overwhelm our doubts and bad feelings. One feels intuitively that the mentioned formally-logical shields offer a rather weak protection against at least some arguments of ancient sophists. Motion does not resurrect from instantaneous positions.

When trying to solve the "problem" of motion, one is faced immediately with another one. Namely, if it is not the instantaneous position but rather the instantaneous state of motion that is to be a primary concept, then our description must be shifted from the physical space $M$ in which a particle moves to some other manifold the points of which are labelled, roughly speaking, by instantaneous positions in $M$ and velocities. More generally, if one considers a multiparticle system, e.g., one consisting of $N$ material points, and subject it perhaps to some constraints, then $M$ is replaced by some submanifold $Q\subset M^{N}$, the configuration space of the system. And when analyzing motion we must lift the description from $Q$ to some byproduct $TQ$ the points of which contain the information about instantaneous configurations $q\in Q$ and systems of instantaneous velocities when $q$ is passed. This manifold $TQ$ of mechanical states in the sense of Newton is what mathematicians call the tangent bundle over $Q$. We quote a more explicit definition below. Points of $TQ$ are labelled by generalized coordinates $q^{i}$ taken from $Q$ and generalized velocities $v^{i}$. In $Q$ motion is described by time-parametrized curves $\varrho: \mathbb{R}\rightarrow Q$ analytically represented by $q^{i}(t)$-the time dependence of $q^{i}$. Then $\varrho$ is lifted to $\dot{\varrho}: \mathbb{R} \rightarrow TQ$, analytically represented by the system of quantities $q^{i}(t)$, $v^{i}(t)$, where 
\[
v^{i}(t)=\dot{q}^{i}(t)=\frac{d q^{i}}{dt}(t).
\]

But then $TQ$ may be used as a "new $Q$" and the Zeno paradox reappears on the level of $TQ$. By induction, we construct higher-floor spaces, i.e., iterated tangent bundles $T^{n}Q:=T(T^{n-1} Q)$ and the iterated lifted curves in these spaces, $\varrho^{(n)}: \mathbb{R}\rightarrow T^{n}Q$ and apparently something like the infinite regression of Zeno paradoxes; although, let us notice that once $\varrho$ is fixed as a curve in $Q$, then all lifts $\varrho^{(n)}$ are automatically defined as byproducts of $\varrho$. One feels intuitively that the solution is somewhere in dynamics, and it was suggested by some medieval XIV-century "positivists" like Wilhelm Ockham, Jean Buridan (rector of the Paris University) and Peter Olvi. The latter two invented the idea of "impetus", i.e., amount of motion. Roughly speaking, this was to be a kind of "charge" proportional both to the velocity and to the "amount of matter" in the body. This "amount of matter", mass in our language, was not  rigorously defined; they remained on the heuristic level. But it was clear for them that besides the velocity itself, there existed some other characteristics of the body, internal ones this time, which together with velocity (but unlike the velocity alone) led to some conserved quantity, balanced during collisions. It is difficult to understand what a kind of intellectual speculation had led those people to rejecting the Aristotlean mechanics with its rest-state as a natural situation. Their invention preceded Galileo a few centuries. In any case they were probably the first to decide that rectilinear uniform motion is a natural situation which needed a sufficient reason to be changed. This may be a naive speculation but perhaps the catastrophic defeat of the French at Poitiers (1356) due to the efficiency of English archers was a convincing argument. The arrows moved, had some internal ability to act at the impact (mass) and once put in motion just needed some external factor to be stopped. The terrible invention of the arbalest later on was even more convincing.

Let us go back to modern language. Obviously, "impetus" is linear momentum. Without external influences it is a constant of motion. And it is something other than velocity. It is measured by its impact (and damage) on other bodies when colliding with them; velocity is measured kinematically by "tachometers". They are logically different concepts. From the modern point of view we are aware of a subtle relationship between them. The proper manifold of states is what physicists call the phase space and mathematicians the symplectic cotangent bundle $T^{\ast}Q$. The manifold $T^{*}Q$ is dual to $TQ$ in the sense that it consists of linear functions of velocities. Analytically speaking, its points are labelled by quantities $(q^{i}, p_{i})$, $i=1, \cdots , {\rm dim} Q$, where the $q^{i}$ are again generalized coordinates on $Q$ and $p_{i}$ are their conjugate canonical momenta. They are covariant vectors in the sense that under the change of coordinates from some "old" ones $q^{i}$ to "new" ones $q^{i\prime}(q^{j})$, the components of canonical momenta $p_{i}$ transform to
\[
p_{i\prime}=p_{j}\frac{\partial q^{j}}{\partial q ^{i\prime}},
\]
thus, contragradiently to the transformation rule for velocities as contravariant vectors tangent to curves (motions),
\[
v^{i\prime}=\frac{\partial q^{i\prime}}{\partial q ^{j}}v^{j}.
\]

Because of this $p_{i}v^{i}=p_{i\prime}v^{i\prime}$ and this number is a well-defined (independent on the choice of coordinates) evaluation of canonical momenta $p$ on virtual velocities $v$. If $Q$ has no additional structure, there is no canonical diffeomorphism of $TQ$ onto $T^{*}Q$. Just like there is no distinguished linear isomorphism between linear spaces $V$ and their $V^{*}$ (spaces of linear functions on $V$), in the finite-dimensional case  $V$ and $V^{*}$ are isomorphic in an infinity of ways. And this is just the proper hint we need. We mentioned above that the proper key is hidden in dynamics, i.e., roughly speaking, in a particular relationship between $TQ$ and $T^{*}Q$. But assuming $T^{*}Q$ as something primary is just in the spirit of medieval reasoning of Buridan and Olvi, in the spirit of the idea that besides the instantaneous position there is something additional, conceptually independent of position and its change (velocity), but fundamental for the true definition of "mechanical state". At the same time, the non-existence of natural identifications between  $TQ$ and $T^{*}Q$ removes the mentioned "problem" of the infinite regression of spaces $T^{n}Q$ and curves $\varrho^{(n)}$. 

The above philosophical and historical remarks with a small addition of roughly introduced mathematical concepts might seem a little pretentious, however we have the feeling they shed some light onto the problem of the fundamental role of the phase space in mechanics. And now, after all this "literature" we should pass to rigorous mathematical concepts. First of all we must remind ourselves of the concepts of the tangent and cotangent bundle.

We consider a mechanical system, the configuration space of which is a finite-dimensional differential manifold. A manifold is defined by an atlas of open sets $U_{q}$ such that $q\in U_{q}\subset Q$ and each $U_{q}$ is isomorphic to an open subset of $\mathbb{R}^{n}$. Its dimension $n$ is what one calls the number of degrees of freedom \cite{28}. At any point $q\in Q$ of this manifold there exists its tangent space $T_{q}Q$. If $Q$ is a general manifold, then for different points $q$ the corresponding $T_{q}Q$ are different, completely unrelated linear spaces of dimension $n$.

Velocity vectors, being tangent to curves describing motion, are always attached at some points of $Q$. So there is nothing like simply velocity and the Newtonian state given by a pair: configuration and velocity. If the time axis is identified with $\mathbb{R}$ (the origin and time unit fixed) and motions are described by curves $\varrho:\mathbb{R} \rightarrow Q $, then $\varrho(t)\in Q$ is an instantaneous position at the time instant $t\in \mathbb{R}$ and its tangent vector $\dot{\varrho}(t)\in T_{\varrho(t)}Q$ is an instantaneous velocity at the time instant $t$, when the object passes the configuration $q=\varrho(t)$. And strictly speaking, the Newtonian state at the time instant $t$ is given by a single object $\dot{\varrho}(t)$ and the information about $\varrho(t)$ is automatically contained in $\dot{\varrho}(t)$; by its very definition $\dot{\varrho}(t)$ is attached just at the fixed $\varrho(t)$. The set of all possible mechanical states (in the sense of Newton) is given by the set-theoretical union of all possible spaces $T_{q}Q$,
\[
TQ=\bigcup_{q\in Q} T_{q}Q.
\]
This is the tangent bundle over $Q$ \cite{1}, \cite{2}, \cite{30}, \cite{Schf}, \cite{JJSa}, \cite{Synge}-\cite{55}. As every $v\in TQ$ belongs to some (and only one) $T_{q}Q$, there exists a natural projection $\tau_{Q}:TQ\rightarrow Q$ such that $q=\tau_{Q}(v)$ ($v\in T_{q}Q\subset TQ$). It is shown in differential geometry that $TQ$ has a natural structure of a differential manifold induced from that of $Q$. This structure is given by an atlas induced by the complete atlas in $Q$. In more easy, analytical terms: suppose we are given local coordinates $q^{i}$, $i=1, \cdots , n$ working in some neighbourhood in $Q$. These coordinates give rise to the system of basic vectorfields defined on their domain and identified with differential operators $e_{i}=\partial / \partial q^{i}$. Now, any vector $v\in T_{q}Q$ has some components $v^{i}$ with respect to coordinates $q^{i}$, explicitly,
\[
v=v^{i}e_{iq}=v^{i}\frac{\partial}{\partial q^{i}}|_{q}.
\]

In this way the elements $v$ of $TQ$ are labelled by the system of $2n$ parameters $q^{i}, v^{i}$,  $i=1, \cdots , n$. By taking an atlas in $Q$ we obtain some atlas in $TQ$; one shows that all axioms of a differential manifold are then satisfied in $TQ$. The resulting differential structure in $TQ$ is of the same class of smoothness as the original one in $Q$. The local coordinates in $TQ$ adapted to $q^{i}$ will be denoted by ($q^{i}, v^{i}$), although strictly speaking, one should make a distinction between $q^{i}$ as functions on $Q$ and their pull-backs $\bar{q^{i}}=q^{i} \circ \tau_{Q}$ as functions on $TQ$. However, to avoid a multitude of symbols, usually one does not distinguish graphically between $q^{i}$ and $\bar{q^{i}}$ and the adapted coordinates are denoted simply by ($q^{i}, v^{i}$). Similarly, if there is no danger of confusion one writes simply $\tau$ instead $\tau_{Q}$. Once the differential structure is defined in $TQ$, one can use there any other system of coordinates, not necessarily the $Q$-adapted ones. The only restriction is the systems' compatibility with the atlas in $TQ$ induced by one in $Q$. It is important that every curve in $Q$, $\varrho:\mathbb{R} \rightarrow Q$, may be uniquely lifted to one in $TQ$, $\dot{\varrho}:\mathbb{R} \rightarrow TQ$, just by the differentiation procedure. 

If $Q$ is a general manifold, then in $TQ$ only the vertical fibres $T_{q}Q=\tau_{Q}^{-1}(q)$ are well-defined. There are no well-defined horizontal fibres in the objective sense independent of the choice of coordinates. Because of this, the elements of $TQ$ are not pairs (position-velocity) but just velocities, because any $v \in TQ$ belongs to exactly one $T_{q}Q$ and the $q$-information is automatically contained in $v$. (We are inclined to think otherwise because of our habits motivated by work in affine spaces and, in particular, Euclidean ones). If $Q$ is an affine space, there exists some fixed linear space of translations $V$ and the mapping from $Q \times Q$ onto $V$ assigning to any pair of points $a, b \in Q$ some vector $\vec{ab}\in V$. This assignment satisfies the well-known axioms of affine geometry. Let us remind ourselves of them:
\begin{itemize}
\item[$(i)$] $\vec{ab}+\vec{bc}+\vec{ca}=0$  for any $a, b, c\in Q$,\\ 
\item[$(ii)$] for any fixed $b\in Q$, the assignment $Q\ni a \rightarrow \vec{ba}\in V$ is a bijection of $Q$ onto $V$.
\end{itemize}
Obviously, the first condition implies immediately that  $\vec{aa}=0$, and $\vec{ab}=-\vec{ba}$ for any $a, b \in Q$.

It is obvious and well-known that there exists a canonical diffeomorphism of $TQ$ onto $Q \times V$; thus, horizontal directions are also well-defined and Newtonian states are pairs $(q, v)\in Q \times V$, (position, velocity). Positions and velocities may be independently manipulated. There is nothing like this in a general manifold $Q$. 

For conceptual purity, one thing should be stressed here. Namely, $V$ is the translation space in $Q$, so, roughly speaking, it has to do with quantities measured in centimeters. The use of $V$ as the space of velocities measured in ${\rm cmsec}^{-1}$ is justified only if the time axis is simply identified with $\mathbb{R}$. But it is more correct to define the time axis as a one-dimensional affine space of the linear space of translations $\Lambda$. Additional structures like the time unit (metric) in $\Lambda$ and the arrow of time (orientation in $\Lambda$) are used but this does not change the essence of the problem; in any case one should mention that such a problem does exist. To avoid it one must introduce as the space of velocities the tensor product $V_{\Lambda}:=V \otimes \Lambda^{*}$; obviously this space is also $n$-dimensional because $\Lambda$ is one-dimensional. More generally, when $Q$ is a manifold, then the manifold of Newtonian states should be defined as the bundle $TQ \otimes \Lambda^{*}$ rather than $TQ$ itself. By the last tensor product we mean the bundle with fibres $T_{q}Q \otimes  \Lambda^{*}$. However, in the sequel we do not go into such subtle points and simply put $\Lambda=\mathbb{R}$. It is sufficient for our purposes here.

Now let us remind ourselves of the cotangent bundle concept. If the configuration space $Q$ was an affine space with the linear space of translations $V$, then the phase space would be the Cartesian product $Q \times V^{*}$- the structure dual to the space of Newtonian states $Q \times V$. Its elements are pairs $(q, p)$ - configuration and canonical momentum. To be precise, if we do not identify the time axis $T$ with the numerical field $\mathbb{R}$, we would have used rather $V_{\Lambda}^{*}=V^{*}\otimes \Lambda$, the dual of $V_{\Lambda}=V \otimes \Lambda^{*}$. But, as was said above, we shall not deal with such subtle distinctions.

If $Q$ is a general manifold, then dually to the picture of Newtonian states as elements of $TQ$, we shall use the manifold $T^{*}Q$ consisting of linear functions on all possible tangent spaces $T_{q}Q$. So, for any $q$ we take $T_{q}^{*}Q$ - the dual space of $T_{q}Q$, and then let $q$ run over the manifold $Q$. In other words \cite{1}, \cite{2}, \cite{GuiSte}, \cite{20}, \cite{Schf}, \cite{JJSa}, \cite{Synge}-\cite{55}, 
\[
T^{*}Q=\bigcup_{q\in Q} T_{q}^{*}Q.
\]
Just following the pattern of $TQ$, we introduce in $T^{*}Q$ the natural differential structure starting from some atlas on $Q$. Then, if $q^{i}$ are local coordinates in some neighbourhood in $Q$, we introduce the field of basic covectors $e^{i}$ on the domain of coordinates $q^{i}$, namely as differentials $e^{i}=dq^{i}$. Then any covector $p\in T_{q}^{*}Q$ may be expanded with respect to basic covectors at $q$,
\[
p=p_{i}e^{i}_{q}=p_{i}dq^{i}_{q}.
\]

Using more common expressions: when some coordinates $q^{i}$ are fixed in some domain, then any tensor attached at some point $q$ of this domain is analytically represented by the system of its components, depending obviously on coordinates $q^{i}$ and transforming under their change according to the rule specific for this particular kind of tensor object. The same concerns all geometric objects, not only tensors (thus, e,g., tensor densities). 

The natural projection of $T^{*}Q$ onto $Q$ will be denoted by $\tau_{Q}^{*}$, or simply by $\tau^{*}$ if there is no danger of confusion. By definition $\tau^{*}$:$T^{*}Q \rightarrow Q$ is given by:
\[
\tau^{*-1}(q)=T_{q}^{*}Q,
\]
i.e., for any $p\in T_{q}^{*}Q$ we have $\tau_{Q}^{*}(p)=q$. In a complete analogy to the tangent bundle, if $Q$ has no additional structure like e.g., an affine space, or more generally - a group manifold, there is no natural isomorphism between different fibres $T_{q}^{*}Q$. So, there exist well-defined vertical fibres, but there are no objective, coordinate-independent transversal fibres.

If  $p\in T_{q}^{*}Q \subset T^{*}Q$, then it is labelled by the system of $2n$ numbers $q^{i}$, $p_{i}, i=1, \cdots, n$. Those may be considered as coordinates in the domain $\tau_{Q}^{*-1}(U)$, where $U$ is the domain of coordinates $q^{i}$. One can easily show that the so defined atlas in $T^{*}Q$ satisfies all necessary axioms and $T^{*}Q$ is a differential manifold of the same class of smoothness as $Q$ itself. This $2n$-dimensional manifold is just the phase space of a system moving in $Q$; mathematically this is the cotangent bundle over $Q$.

Let us observe an important structural difference between $TQ$ and $T^{*}Q$. Any curve in $Q$, $\varrho: \mathbb{R} \rightarrow Q$ may be lifted to $TQ$ by the intrinsic prescription resulting in the curve $\dot{\varrho}: \mathbb{R} \rightarrow TQ$; just $\dot{\varrho}(t) \in T_{\varrho(t)}Q$ is the tangent vector at $\varrho(t)$, i.e., for the parameter (time) value $t \in \mathbb{R}$. Unlike this, without some additional structure in $Q$ such a prescription does not exist in $T^{*}Q$ at all. And conversely, in $T^{*}Q$ every (sufficiently smooth) curve is a priori acceptable as a description of some kinematically possible motion in $T^{*}Q$ as a manifold of states. But it is not so in $TQ$, where only curves of the form $\dot{\varrho}$ are acceptable. Analytically speaking: the time dependence of coordinates, $q^{i}(t)$ is a priori arbitrary (up to the smoothness class) on the level of kinematics, but the time dependence of $v^{i}(t)$ is then rigidly soldered to  $q^{i}(t)$, namely by the differentiation procedure, $v^{i}(t)=dq^{i}(t) /dt$. In the phase space manifold $T^{*}Q$, a priori, on the level of kinematics, i.e., before formulating equations of motion, any time dependence $q^{i}(t)$, $p_{i}(t)$ of all $2n$ state variables $q^{i}$, $p_{i}$ is acceptable (if sufficiently smooth).

Of course in $TQ$ one is dealing with quantities which are operationally interpretable. It is not so in $T^{*}Q$, where as yet the canonical momentum is something mysterious. As mentioned, its interpretation is based on fixing some particular dynamical model.

In Lagrange-Hamilton mechanical theory the dynamics is encoded in a Ha\-miltonian function $H:T^{\ast}Q\rightarrow \mathbb{R}$ or, in an equivalent sense, in the corresponding Lagrange function $L:TQ\rightarrow \mathbb{R}$. The relationship is given by the Legendre transformation
\[
\mathcal{L}:TQ\rightarrow T^{\ast}Q,
\]
where, for any $v\in T_{q}Q\subset TQ$,
\[
\mathcal{L}(v)=p=D_{v}\left(L|T_{q}Q\right),
\]
$D_{v}$ denoting the Frechet derivative of $L|T_{q}Q$ at $v$. Analytically, in coordinates induced canonically from $Q$,
\begin{equation}\label{eq3}
p_{i}=\frac{\partial L\left(q^{a},v^{b}\right)}{\partial v^{i}}.
\end{equation}
Conversely, for any $p\in T^{\ast}_{q}Q$
\[
\mathcal{L}^{-1}(p)=v=\delta_{q}\cdot D_{p}\left(H|T^{\ast}_{q}Q\right),
\]
where $\delta_{q}:T^{\ast\ast}_{q}Q\rightarrow T_{q}Q$ denotes the canonical isomorphism of the second dual $T^{\ast\ast}_{q}Q$ onto the original space $T_{q}Q$ (canonical due to the finite dimension of linear spaces, of course). Analytically
\[
v^{i}=\frac{\partial H\left(q^{a},p_{b}\right)}{\partial p_{i}}.
\]
Obviously,
\[
E=H\circ\mathcal{L},\qquad H=E\circ\mathcal{L}^{-1},
\]
where $E:TQ\rightarrow \mathbb{R}$ is the energy function,
\[
E(v)=\left\langle \mathcal{L}(v),v\right\rangle-L(v),
\]
i.e., analytically,
\[
E\left(q^{a},v^{b}\right)=v^{i}\frac{\partial L\left(q^{a},v^{b}\right)}{\partial v^{i}}-L\left(q^{a},v^{b}\right)=H\left(q^{a},p_{b}\right),
\]
provided that (\ref{eq3}) holds. It is well-known that in standard mechanical theory $\mathcal{L}$ is invertible ($\mathcal{L}^{-1}$ exists); therefore the following Hessians do not vanish:
\[
\det\left[\frac{\partial^{2}L}{\partial v^{i}\partial v^{j}}\right]\neq 0,\qquad \det\left[\frac{\partial^{2}H}{\partial p_{i}\partial p_{j}}\right]\neq 0.
\]
(The inverse statements hold only locally.)

In the regular case of invertible $\mathcal{L}$ the variational principles
\[
\delta\int L\left(q^{a},\dot{q}^{b}\right)dt=0,\qquad \delta\int\left(p_{a}\frac{dq^{a}}{dt}-H\left(q^{b},p_{c}\right)\right)dt=0
\]
are equivalent to the equations of motion:
\begin{itemize}
    \item Lagrange-second kind:
\[
\frac{D}{Dt}\frac{\partial L\left(q,\dot{q}\right)}{\partial \dot{q}^{a}}-\frac{\partial L\left(q,\dot{q}\right)}{\partial q^{a}}=0,
\]

    \item Hamilton:
\[
\frac{dq^{i}}{dt}=\frac{\partial H\left(q,p\right)}{\partial p_{i}},\qquad \frac{dp_{i}}{dt}=-\frac{\partial H\left(q,p\right)}{\partial q^{i}}.
\]
\end{itemize}
The situation becomes complicated when $\mathcal{L}$ is not invertible. Then according to the Dirac procedure \cite{Dirac_wiezy}-\cite{15}, \cite{21}, \cite{JJSa}, a technically difficult problem of dynamical constraints appears.

But even in the regular case the "impetus" approach based on $T^{\ast}Q$ is much more adequate, just as Buridan and Olvi expected. The main point is that both the position and momentum are on equal footing as the components of the physical situation in mechanics. The Zeno paradox vanishes in a sense; moreover, the infinite regression of Zeno paradoxes based on the hierarchy of bundles $T^{n}Q$ disappears as well. The reason is that on a purely kinematical basis one is unable to lift the curves $\varrho:\mathbb{R}\rightarrow Q$ to the phase space $T^{\ast}Q$. Such a lifting may be done only in the tangent bundle $TQ$ and its higher-order levels $T^{n}Q$. In $T^{\ast}Q$, where the "impetus" is a primary quantity, one can achieve this on a dynamical basis by performing the Legendre transformation.

\section{Phase space geometry and Poisson manifolds}

Once founded and developed by the giant geniuses of J. L. Lagrange and R. W. Hamilton, originally on the basis of purely mathematical and aesthetic speculation and without any practical necessity, phase space geometry became the apriori knowledge underlying everything essential in statistical and quantum mechanics. The modern sophisticated language of differential geometry provides a formal synthesis and opens some new perspectives.

Classical statistical mechanics as developed by J. W. Gibbs \cite{G} is based very deeply on the geometric a priori of phase spaces, i.e., on the symplectic structure to be defined below. This is a striking example of how strongly a properly chosen geometry underlies and implies the shape of physical theories.

Obviously, statistical distributions and their time evolutions, i.e., stochastic processes, may be defined in quite general probabilistic spaces, including ones with very poor structures, even without topology and differentiability. But the Gibbs statistical theory is something fundamentally different than the usual statistical models used in technical problems and applied physics, even if the latter models appear often as its distant consequences or byproducts.

The main peculiarities of Gibbs' approach have to do with the energy concept, symplectic structure, canonical transformations, the existence and conservation of the phase space canonical volumes, and therefore, the existence of entropy and the resulting special probability distributions like microcanonical, macrocanonical and other physically distinguished ensembles.

Using modern language we would say that Gibbs theory was conceptually based on the cotangent bundle $T^{\ast}Q$ and it was just here from which its generality and efficiency emerged. Let us stress that it is also impossible to overestimate the contribution of Boltzmann \cite{B}. But, by abuse of lan
guage, one can say that his way of thinking was inspired by the tangent bundle $TQ$-geometry, i.e., as one says sometimes, by the "$\mu$-phase space" (often with the special stress on the one-particle six-dimensional $\mu$-phase space) \cite{KH63}. Is there some contradiction here? In principle the contradiction does not exist just because Boltzmann and his followers were dealing with dynamical models where the Legendre transformation $\mathcal{L}$ established a very simple identification of $TQ$ and $T^{\ast}Q$. Nevertheless, the full generality and efficiency of statistical methods was attained only due to the consequent use of the $T^{\ast}Q$-phase-space description following the simplest pattern of the $6N$-dimensional phase space of the system of $N$ identical structureless material points. This includes quite unexpected areas like quantum statistics. Incidentally, it is difficult to imagine even the very rise of quantum mechanics without formulating the classical theory in symplectic phase-space terms. But this is a different story for now.

It is standard geometric knowledge that any cotangent bundle 
\[
T^{\ast}Q=\bigcup_{q\in Q} T^{\ast}_{q}Q 
\]
is endowed with the intrinsic symplectic structure given by the two-form
\[
\gamma=d\theta,
\]
where $\theta$ denotes the intrinsic Cartan one-form to be discussed next \cite{1}, \cite{2}, \cite{JJS}. In adapted local coordinates $\left(q^{i},p_{i}\right)$ induced from $Q$: 
\begin{equation}\label{eq14}
\theta=p_{i}dq^{i},\qquad \gamma=dp_{i}\wedge dq^{i};
\end{equation}
both expressions do not depend on the particular choice of $q^{i\prime}$-s in $Q$. 

Let us remind ourselves of the intrinsic definition of $\theta$. Denoting the natural projection onto the base $Q$ by
\[
\tau^{\ast}_{Q}:T^{\ast}Q\rightarrow Q
\]
($\tau^{\ast}_{Q}\left(T^{\ast}_{q}Q\right)=\{q\}$), we have for any $p\in T^{\ast}_{q}Q$ and for any vector $X\in T_{p}\left(T^{\ast}_{q}Q\right)$ the following evaluation:
\[
\left\langle\theta_{p},X\right\rangle:=\left\langle p,T\tau^{\ast}_{Q}\cdot X\right\rangle.
\]
In other words: the contraction of $\theta_{p}$ with $X$ equals by definition the contraction of $p$ with the projection of $X$ onto $T_{q}Q$.

Now, being a differential, $\gamma$ is automatically closed:
\[
d\gamma=0.
\]
Furthermore it is seen that $\gamma$ is non-singular, because in adapted coordinates
\[
\left[\gamma_{ab}\right]=\left[
\begin{array}{cc}
O & -I\\
I & O
\end{array}
\right],
\]
where $O$, $I$ are respectively the $n\times n$ null and identity matrices. Therefore, one immediately obtains:
\[
\det\left[\gamma_{ab}\right]\neq 0;
\]
a fact independent of a choice of coordinates.

This structure, canonical in $T^{\ast}Q$, motivated the general definition of symplectic geometry as given by a pair $(P,\gamma)$, $P$ denoting a differential manifold and $\gamma$ being a two-form on $P$ (skew-symmetric twice covariant tensor field) subject to two restrictions:
\begin{itemize}
\item[$(i)$] $\gamma$ \ is non-degenerate, \ $\det\left[\gamma_{ab}\right]\neq 0$, \ i.e., \ $X=0$ \ if \ for any other vector \newline field $Y$
\begin{equation}\label{eq21}
\gamma(X,Y)=\gamma_{ab}X^{a}Y^{b}=0.
\end{equation}
(This implies that $P$ is even-dimensional, $\dim P=2n$.)

\item[$(ii)$] $\gamma$ is closed, $d\gamma=0$; i.e., analytically, $\gamma$ satisfies the "source-free group of Maxwell equations". Let $\partial_{a}=\partial / \partial q^{a}$. Then
\begin{equation}\label{eq23}
\partial_{a}\gamma_{bc}+\partial_{b}\gamma_{ca}+\partial_{c}\gamma_{ab}=0.
\end{equation}  
\end{itemize}
Such more general symplectic structures appear in mechanics and are crucial for many problems.

Seemingly, such a phase-space structure resembles Riemannian geometry, because $\gamma$ defines something like the scalar product of vectors (\ref{eq21}). But the antisymmetry of $\gamma$,
\[
\gamma_{ab}=-\gamma_{ba},
\]
implies that
\[
\gamma(X,Y)=-\gamma(Y,X).
\]
In particular, for any vector $X$ we have
\[
\gamma(X,X)=0;
\]
so the concept of the length of a vector does not exist if $Q$ is a structureless manifold. And besides, the $\gamma$-orthogonality ($\gamma$-duality, more precisely) of vectors,
\[
\gamma(X,Y)=0
\]
is completely exotic in comparison with the Riemannian or pseudo-Riemannian orthogonality (duality). It happens quite easily that non-vanishing vectors $\gamma$-orthogonal ($\gamma$-dual) to a submanifold $M\subset P$ are at the same time tangent to $M$. Moreover, this is just a tool of classification of submanifolds, crucial for a variety of problems, including Dirac's analysis of singular Lagrangians and dynamical constraints. There is nothing like this in Riemannian geometry. In pseudo-Riemannian manifolds there exist isotropic surfaces ("light-fronts"), but their nature and other classification problems are structurally quite different than in symplectic manifolds.

An important common feature with Riemannian (and pseudo-Riemannian) geometry is, however, the existence of the natural isomorphisms between $T_{p}P$ and $T^{\ast}_{p}P$, i.e., between vectors and covectors. One must fix only the ordering of contracted and non-contracted indices (antisymmetry of $\gamma$). Thus, using the obvious shorthands, we have
\[
u_{a}=u^{b}\gamma_{ba}
\]
for the isomorphism of $T_{p}P$ onto $T^{\ast}_{p}P$, and conversely,
\[
u^{a}=u_{b}\gamma^{ba},
\]
where
\[
\gamma^{ac}\gamma_{cb}=\delta^{a}{}_{b}.
\]
In virtue of (\ref{eq23}), the reciprocal bivector field (field of reciprocal skew-symmetric twice contravariant tensors) satisfies:
\begin{equation}\label{eq31}
\gamma^{ai}\partial_{i}\gamma^{bc}+\gamma^{bi}\partial_{i}\gamma^{ca}+
\gamma^{ci}\partial_{i}\gamma^{ab}=0.
\end{equation}
An important remark: when $\gamma$ is non-singular, the biform and bivector approaches are equivalent. However, in many problems of mechanics and in one form of quantization it is convenient to start from the bivector field $\gamma^{ij}$ as something primary and subject it only to (\ref{eq31}), but not necessarily to the non-singularity condition. These are so-called Poisson structures \cite{24}, \cite{KarMas}, \cite{26}, \cite{27}, \cite{34}, \cite{36}, \cite{37}. It is well known that they split into a family of leaves which carry natural symplectic geometries.

Poisson manifolds are sufficient for this type of quantization analysis, because they have enough structure for constructing Poisson brackets:
\begin{equation}\label{eq32}
\{F,G\}:=\gamma^{ab}\partial_{a}F\partial_{b}G.
\end{equation}
Here $F$, $G$ are differentiable functions on $P$ and indices $a$, $b$, etc. label some, in general arbitrary, coordinates in $P$. This bracket has all the properties known from analytical mechanics:
\begin{itemize}
\item[$(i)$] It is bilinear over constants:
\begin{eqnarray}
\{F,aG+bH\}&=&a\{F,G\}+b\{F,H\},\label{eq33a}\\ 
\{aF+bG,H\}&=&a\{F,H\}+b\{G,H\},\label{eq33b}
\end{eqnarray}
for arbitrary functions $F$, $G$, $H$ and arbitrary constants $a$, $b$. 

\item[$(ii)$] It is skew-symmetric,
\begin{equation}\label{eq34}
\{F,G\}=-\{G,F\},
\end{equation}
and in particular,
\begin{equation}\label{eq35}
\{F,F\}=0.
\end{equation}

\item[$(iii)$] It satisfies the Jacobi identity:
\begin{equation}\label{eq36}
\{\{F,G\},H\}+\{\{G,H\},F\}+\{\{H,F\},G\}=0,
\end{equation}
which is just the direct consequence of (\ref{eq31}).

\item[$(iv)$] It satisfies also:
\begin{equation}\label{eq37}
\left\{F\left(f_{1},\ldots,f_{m}\right),G\right\}=
\sum^{m}_{k=1}F_{,k}\left(f_{1},\ldots,f_{m}\right)\{f_{k},G\},
\end{equation}
and analogously with respect to the second argument. Obviously, for any $F:\mathbb{R}^{m} \rightarrow \mathbb{R}$, $F,_{k}$ is an abbreviation for the partial derivative of $F$ with respect to the $k$-th variable.
\end{itemize}
In symplectic manifolds, when $\det\left[\gamma^{ab}\right]\neq 0$ and adapted coordinates 
\[
\left(\ldots,z^{a},\ldots\right)=
\left(\ldots,q^{i},\ldots;\ldots,p_{i},\ldots\right)
\]
are used, one obtains the usual formula:
\[
\{F,G\}=\frac{\partial F}{\partial q^{i}}\frac{\partial G}{\partial p_{i}}-
\frac{\partial F}{\partial p_{i}}\frac{\partial G}{\partial q^{i}}.
\]

The link between symplectic manifolds and Poisson manifolds is very intimate and very important both in classical and quantum mechanics. Poisson manifolds belong to the very widely understood field of phase space geometry \cite{24}, \cite{KarMas}, \cite{26}, \cite{27}. Let us remind ourselves of some minimum of important facts. Let $(M, \Gamma)$ be a Poisson manifold; $\Gamma$ is a bivector field on the differential manifold $M$, satisfying the above demands (\ref{eq31}) with $\Gamma$ substituted for $\gamma$. The Poisson bracket is defined by (\ref{eq32}), also with $\gamma$ replaced by $\Gamma$. It has all the properties (\ref{eq33a})-(\ref{eq37}). The Jacobi identity (\ref{eq36}) follows from the fact that $\Gamma$ satisfies (\ref{eq31}). As usual the $\Gamma^{ab}$ denote the components of $\Gamma$ with respect to some system of local coordinates $z^{a}$ in $M$. In general it is not assumed that the matrix $[\Gamma^{ab}]$ has a constant rank all over $M$. On the contrary, in certain problems the points of $M$ where this rank suffers some jump are very important for understanding the structure of $(M, \Gamma)$.

Hamiltonian vector fields with generators $H$ are defined by the usual formula known from symplectic geometry; in terms of local coordinates it reads
\[
X_{H}^{a}=\Gamma^{ab}\partial_{b}H=\Gamma^{ab}\frac{\partial H}{\partial z^{b}}.
\]
One can also consider locally Hamiltonian vector fields, where instead of the differential $dH$ some closed one-form is substituted, not necessarily a differential of a function. $\Gamma$ produces vectors from linear forms (covectors); i.e., it gives rise to the field of linear mappings acting from the tangent spaces $T_{z}M$ to cotangent ones $T_{z}^{*}M$. If the rank of $\Gamma$ at $z$ is smaller than $n$, those mappings are not invertible. The vector in $T_{z}M$ obtained from the form $x \in T_{z}^{*}M$ is denoted by $\widetilde{x} \in T_{z}M$; analytically it is given by 
\[
\widetilde{x}^{a}=\Gamma^{ab}x_{b}.
\]

At any $z\in M$ we are given the linear subspace $V_{z}\in T_{z}M$ consisting of all such vectors,
\[
V_{z}:=\left\{\widetilde{x}: x\in T_{z}^{*}M\right\}.
\]

We obtain an assignment
\[
M \ni z \mapsto V_{z} \subset T_{z}M \subset TM.
\]

In open regions where the rank of $\Gamma$ is constant, this assignment is a distribution. Its smoothness class is inherited from that of $\Gamma$. One can show that this distribution is integrable; i.e., vector fields tangent to it (fields $X$ such that at any $z\in M$, $X_{z}\in V_{z}$) form a Lie algebra under the usual Lie bracket of vector fields. Then the mentioned regions (constancy of ${{\rm Rank}} \ \Gamma$) may be foliated be the family of integral leaves. It is an important fact \cite{26,27} that those leaves carry natural structures of symplectic manifolds. Namely, take two vectors $u, v\in V_{z}\subset T_{z}M$. Represent them as $\widetilde{U}$, $\widetilde{V}$, where $U, V\in T_{z}^{*}M$. $\widetilde{U}$, $\widetilde{V}$ are not unique if ${\rm Rank} \ \Gamma_{z}< {\rm dim}M$. Define a two-form $\gamma_{z}$ on $V_{z}$, $\gamma_{z}\in V_{z}^{*} \wedge  V_{z}^{*}$ by the formula 
\[
\gamma_{z}(u, v):=\Gamma_{z}(U, V)=\Gamma_{z}^{ab}U_{a}V_{b}.
\]

It is obvious that $\gamma_{z}(u, v)$ does not depend on the choice of representatives $U, V$ $\Gamma$ - projecting onto $u, v$. So indeed this number is assigned to vectors $u, v\in V_{z}$. If $z$ runs over some integral leaf $N\subset M$ of the assignment $M\ni z \mapsto V_{z} \subset T_{z}M$, we obtain in this way a differential two-form $\gamma(N)$ on $N$ as a differential manifold in itself (the surrounding $M$ is forgotten). One proves easily that $\gamma(N)$ is non-degenerate and closed
\[
d\gamma(N)=0.
\]
The last fact is a consequence of the differential identity satisfied by $\Gamma$. In this way $(M, \Gamma)$ is foliated by some family of symplectic leaves. If the rank of $\Gamma$ jumps, then so does the dimension of the leaves.

This is an interesting prescription for constructing symplectic manifolds, different from cotangent bundles or certain of their submanifolds or some quotients of submanifolds. This type of symplectic manifold is important in quantization problems and theory of group representations. A particular role is played by canonical Poisson structures on Lie co-algebras. Before going any further, we remind ourselves of their construction.

Let $G$ be a Lie group and $G'\in T_{e}G$ - its Lie algebra, identified with the tangent space at the identity element $e\in G$. Take the dual space $G'^{*}$, i.e., the space of linear functions on  $G'$. It turns out that $G'^{*}$ has a canonical Poisson structure. Take two differentiable functions $f, g:G'^{*} \rightarrow  \mathbb{R}$ on the Lie co-algebra and fix some point $z\in G'^{*}$. Differentials of $f, g$ at $z, df_{z}, dg_{z}$ are linear functions on $T_{z}G'^{*}\simeq G'^{*}$. The last equality is meant as the canonical isomorphism between the linear space and all its tangent spaces. If $G'$ is a finite-dimensional linear space, it is canonically isomorphic with its second dual, i.e., the dual space of $G'^{*}$. Therefore, $df_{z}, dg_{z}$ are canonically identical with some vectors in $T_{z}G'\simeq G'$. Let us denote those vectors by $\widetilde{df_{z}}, \widetilde{dg_{z}}$. As they are elements of a Lie algebra, we can take their Lie bracket $[\widetilde{df_{z}}, \widetilde{dg_{z}}]\in G'^{*}$. And finally we evaluate the linear form $z\in G'^{*}$ on the vector $[\widetilde{df_{z}}, \widetilde{dg_{z}}]$, and quote:
\[
\left\{f, g \right\}(z):=\langle z, [\widetilde{df_{z}}, \widetilde{dg_{z}}] \rangle.
\]
When $z$ runs over $G'^{*}$, one obtains in this way the prescription for some function $\left\{f, g \right\}:G'^{*} \rightarrow \mathbb{R}$. One proves that this prescription satisfies all the properties of the Poisson bracket, therefore, $(G'^{*}, \left\{\cdot, \cdot \right\})$ is a Poisson manifold. Let $z_{a}$ denote some linear coordinates on $G'^{*}$. Being linear functions on $G'^{*}$, they may be identified with some elements of $G'$, more precisely, with elements of some basis in $G'$. In terms of these coordinates the above Poisson bracket may be analytically written down as follows:
\[
\left\{f, g \right\}=z_{a}C^{a}{}_{bd}\frac{\partial f}{\partial z_{b}}\frac{\partial g}{\partial z_{d}},
\]
where obviously, $C^{i}{}_{jk}$ are structure constants,
\[
[z_{a}, z_{b}]=z_{d}C^{d}{}_{ab}.
\]
In particular, we have the following Poisson brackets:
\[
\left\{z_{a}, z_{b}\right\}=z_{d}C^{d}{}_{ab},
\]
just analogously to the previous formula.

It turns out that symplectic leaves are identical with co-adjoint orbits, i.e., orbits of the co-adjoint action of $G$ on $G'^{*}$. Let us describe this briefly. In $G$ we are given two natural groups, namely left and right translations:
\[
L_{g}: G \rightarrow G, \qquad L_{g}(h)=gh,
\]
\[
R_{g}: G \rightarrow G, \qquad R_{g}(h)=hg.
\]

The corresponding tangent mappings of $TG$ onto $TG$, denoted respectively as $L_{g *}$, $R_{g *}$, or $TL_{g}$, $TR_{g}$, operate between tangent spaces as follows:
\[
L_{g*}: T_{h}G \rightarrow T_{gh}G, \qquad R_{g*}: T_{h}G  \rightarrow T_{hg}G.
\]
Matrices of the above linear mappings are obviously Jacobi matrices of $L_{g}$, $R_{g}$, calculated at the appropriate point $h$.

The inner automorphisms induced by $g\in G$ are given by
\[
L_{g}\circ R_{g^{-1}}=R_{g^{-1}} \circ L_{g}.
\]

Obviously, any $L_{g}\circ R_{g^{-1}}$ does preserve the identity.

Therefore, its tangent mapping $L_{g} \circ R_{g^{-1}}=T(L_{g}\circ R_{g})$ does preserve $G'=T_{e}G$. The resulting linear mapping will be denoted by ${ \rm Ad}_{g}:G' \rightarrow G'$. It is an isomorphism of the Lie algebra in the sense of preserving brackets:
\[
{ \rm Ad}_{g}[\xi, \eta]={ \rm Ad}_{g}\xi,{ \rm Ad}_{g}\eta].
\]

Linear mappings give rise to their adjoint mappings acting in the opposite direction. So, one obtains the transformation ${ \rm Ad}_{}{g}^{*}:T^{*}G \rightarrow T^{*}G$ of the cotangent bundle, operating between cotangent spaces as follows:
\[
L_{g}^{*}: T^{*}_{gh}G \rightarrow T^{*}_{h}G,  \qquad R_{g}^{*}: T^{*}_{hg}G \rightarrow T^{*}_{h}G.
\]
Obviously, we have the following representation rules:
\[
{ \rm Ad}_{gh}={ \rm Ad}_{g}{ \rm Ad}_{h},  \qquad { \rm Ad}^{*}_{gh}={ \rm Ad}^{*}_{h}{ \rm Ad}^{*}_{g}.
\]

It is clear that the groups ${ \rm Ad}_{G}$, ${ \rm Ad}^{*}_{G}$ do not act transitively; for example, the null element is an orbit in itself. It turns out that the symplectic leaves of $G'^{*}$ as a Poisson manifold, coincide with the orbits of the co-adjoint action. Those orbits may have different dimensions, for example, as mentioned, $\left\{ 0\right\}$ is a single-element, thus, a null-dimensional, orbit.

It turns out that orbits are common value-surfaces of the systems of Casimir invariants in the enveloping associative algebra of $G'^{*}$.

Canonical transformations of a Poisson manifold $(M, \Gamma)$, in particular of the Lie co-algebra $G'^{*}$ endowed with the Poisson bracket as above, are diffeomorphisms of $M$ preserving $\Gamma$. In particular, for every differentiable function $F:M \rightarrow \mathbb{R}$, the corresponding Hamiltonian vector field $X_{F}$ on $M$ is an infinitesimal canonical transformation. By this one means that its one-parameter group consists of canonical transformations. In particular, if $H:M \rightarrow \mathbb{R}$ has the physical meaning of the Hamiltonian, then $X_{H}$ generates the (local) one-parameter group of canonical transformations describing motion. In $C^{\infty}(M)$ this group is generated by the differential operator
\[
C^{\infty}(M)\ni F \mapsto \left\{F, H\right\} \mapsto C^{\infty}(M).
\]

Hamiltonian equations of motion may be expressed in the form 
\[
\frac{dF}{dt}=\left\{F, H\right\},
\]
where $F$ runs over some maximal family of functionally independent functions on $M$.

Canonical transformations preserve symplectic leaves, therefore, the above system splits into a family of the usual Hamilton equations on those leaves. If $N\subset M$ is such a leaf, then the corresponding reduced equations of motion are generated by the restricted Hamiltonian $H|N$. 

Above we mentioned similarities and differences between symplectic and Riemann structures. One can ask by analogy to Poisson manifolds what would be geometries based on the contravariant symmetric tensors, not necessarilly non-singular ones. One of them is the four-dimensional description of Galilean physics and Newtonian gravitation. The corresponding contravariant metrics are once degenerate.

Let us go back to phase spaces. Their automorphisms, i.e., canonical transformations, are defined as diffeomorphisms preserving the symplectic two-form $\gamma$. 

Staying with one phase space with its symplectic two-form $\gamma$, with its automorphisms (canonical transformations, symmetries) defined as diffeomorphisms preserving $\gamma$, we ask " in what sense is this analogous to Riemannian isometries". Canonical transformations are structurally quite different. Namely, for phase spaces their symplectic group is infinite-dimensional; its elements are labelled by arbitrary functions. Unlike this, Riemannian and pseudo-Riemannian isometry groups are always finite-dimensional; the highest possible dimension equals $(1/2)\dim M(\dim M +1)$ and is attained in constant-curvature (pseudo-)Rieman\-nian manifolds $(M,g)$. Infinitesimal canonical transformations are given by vector fields $X$ on $P$ such that the Lie derivative of $\gamma$ vanishes
\[
\pounds_{X}\gamma=0;
\]
they generate one-parameter groups of canonical transformations. It is well known that the $\gamma$-related covector field $\widetilde{X}=X\rfloor\gamma$, or analytically,
\[
(\widetilde{X}_{a})=X^{b}\gamma_{ba}
\]
is a closed differential one-form,
\[
d\widetilde{X}=0,\quad {\rm i.e.},\quad \partial_{a}X_{b}-\partial_{b}X_{a}=0.
\]
Therefore, in simply connected domains they are also exact,
\[
\widetilde{X}=dF,\qquad X_{a}=\partial_{a}F,
\]
where $a$ labels local coordinates in $P$. If $\widetilde{X}$ is globally exact in $P$, we say that it is a Hamiltonian vector field generated by $F$; the function $F$ itself is referred to as a Hamiltonian of $X$. To indicate the relationship between $X$ and $F$ one uses the symbol $X_{F}$, thus,
\[
X_{F}=\frac{\partial F}{\partial p_{i}}\frac{\partial}{\partial q^{i}}- \frac{\partial F}{\partial q^{i}}\frac{\partial}{\partial p_{i}},
\]
i.e.,
\[
X^{i}_{F}=\frac{\partial F}{\partial p_{i}},\qquad X^{n+i}_{F}=-\frac{\partial F}{\partial q^{i}},\qquad i=1,\ldots,n.
\]
Obviously, $F$ is defined up to an additive constant. The functional label $F$ just indicates explicitly that the group of canonical transformations is infinite-dimensional.

In particular, the time evolution is given by the one-parameter group of the vector field $X_{H}$. One says: "motion is a canonical transformation".

It may be instructive here to mention the relationship with the evolution space structure. Let us mention here various concepts used in the book by J. L. Synge "Classical Dynamics" \cite{Synge}, cf also J. J. S\l awianowski "Geometry of Phase Spaces" \cite{JJS}. Synge used the terms like "$(Q,P)$-space, $(Q,T,P)$-space, $(Q,T,P, \newline H)$-space". Let us concentrate now on the $(Q,T,P)$-description. So, if $\dim Q=n$, then $\dim (Q,T,P)=2n+1$; independent coordinates being generalized coordinates $q^{i}$, the time variable $t$ and canonical momenta $p_{i}$ conjugate to $q^{i}$. Then instead of the Cartan form on $T^{\ast}Q$ we are dealing with the Cartan form on $T^{\ast}Q\times \mathbb{R}$, or more precisely, on something which locally might be interpreted as $T^{\ast}Q\times \mathbb{R}$. Unlike the usual Cartan form $\omega$, the form $\omega_{H}$ depends on the dynamical structure; locally,
\[
\omega_{H}=p_{i}dq^{i}-H(q,p,t)dt.
\]
This is a differential form in the $(2n+1)$-dimensional manifold (evolution space) $T^{\ast}Q\times \mathbb{R}$. Its exterior differential
\[
\gamma_{H}=d\omega_{H}=dp_{i}\wedge dq^{i}-dH\wedge dt
\]
has the maximal possible order $2n$, so, in a sense, $\left(T^{\ast}Q\times \mathbb{R},\omega_{H}\right)$ is a contact manifold. It has a one-dimensional family of singular directions. If the corresponding vector field is "normalised" with respect to the parameter $t$, then it has the form:
\[
\widetilde{X}_{H}=\frac{\partial}{\partial t}+\frac{\partial H}{\partial p_{i}}\frac{\partial}{\partial q^{i}}-\frac{\partial H}{\partial q^{i}}\frac{\partial}{\partial p_{i}},
\]
so that
\begin{equation}\label{eq15a4}
\left\langle dt,\widetilde{X}_{H}\right\rangle=\widetilde{X}_{H}t=1.
\end{equation}
In this way, for integral curves one obtains the usual Hamilton canonical equations, assuming that the time variable $t$ coincides with the curve parameter. This choice, although not necessary, is most natural and convenient in analytical procedures. And obviously,
\[
\widetilde{X}_{H}\rfloor d\omega_{H}=0,
\]
which, together with the normalisation condition (\ref{eq15a4}) defines $\widetilde{X}_{H}$ uniquely.

This "contact" mode of description is exactly what J. L. Synge \cite{Synge} called the $(Q,T,P)$-space approach. He also used the "$(Q,T,P,H)$-approach", when time and energy were a priori assumed to be canonically conjugate variables. The system with $n$ degrees of freedom is then described in terms of the $2(n+1)$-dimensional "over-phase space", e.g., the cotangent bundle $T^{\ast}X$ over the space-time manifold. Then dynamics is defined by fixing some $(2n+1)$-dimensional hypersurface $M\subset T^{\ast}X$. Being odd-dimensional, it is endowed with a degenerate field of the restricted two-form $\gamma\|M:=i_{M}^{\ast}\cdot \gamma$, $i_{M}$ denoting the natural immersion of $M$ into $T^{\ast}X$. And as it is a hypersurface (co-dimension one), its field of degenerate directions is just some line field; i.e., the space of singular vectors tangent to $M$ at any point $x\in M$ is one-dimensional. Therefore, $M$ is foliated by a congruence of the corresponding integral curves, or more precisely, one-dimensional submanifolds. Projecting them from $M$ onto $T^{\ast}X$ (we assume that $T^{\ast}X$ projects onto the whole $X$), one obtains the family of dynamically admissible world-lines in space-time manifolds $X$. This is the so-called homogeneous representation of dynamics. Usually $M$ is described analytically by what Synge called the "energy equation", denoted by
\begin{equation}\label{eq15c}
\Omega\left(x^{\mu},p_{\mu}\right)=0,
\end{equation}
or more precisely
\[
M=\left\{p\in T^{\ast}X:\Omega(p)=0\right\}.
\]
The corresponding singular lines of $M$ satisfy the Hamilton-type equations with $\Omega$ as a "Hamiltonian":
\[
\frac{dx^{\mu}}{d\tau}=\frac{\partial\Omega}{\partial p_{\mu}},\qquad \frac{dp_{\mu}}{d\tau}=-\frac{\partial\Omega}{\partial x^{\mu}},
\]
where $\tau$ is some parameter, not necessarily anything like the "absolute time" or "proper time". The "Hamiltonian" $\Omega$ is obviously the "constant of motion", i.e., the first integral of the system. Only the integral curves placed on $M$, i.e., satisfying (\ref{eq15c}) are physically interpretable as realistic motions. Obviously, $\Omega$ in (\ref{eq15c}) may be chosen in an infinity of ways; any particular choice corresponds to some choice of the parameter $\tau$. Let us give two particularly extreme examples.
\begin{enumerate}
\item  Non-relativistic Hamiltonian dynamics formulated in space-time terms. Space-time coordinates: $t$ --- time, $q^{i}$, $i=1,\ldots,n$ --- generalized coordinates. Various notations:
\begin{eqnarray}\label{eq15c2}
\left(\ldots,x^{\mu},\ldots\right)&=&\left(x^{0},\ldots,x^{i},\ldots\right)=
\left(t,\ldots,q^{i},\ldots\right),\\
\left(\ldots,x^{\mu},\ldots\right)&=&\left(x', \ldots,x^{i},\ldots,x^{n+1}\right)=
\left(q', \ldots,q^{i},\ldots,t\right).
\end{eqnarray}
In a completely non-relativistic language: there is nothing like the standard of velocity like "$c$"; so, there is no way to use some "$x^{0}$" of the length dimension. Then we put
\begin{equation}\label{eq15d2}
\Omega=p_{0}+H\left(q^{i},p_{i},t\right)
\end{equation}
and after the usual manipulations we obtain simply 
\begin{eqnarray}
\frac{dx^{0}}{d\tau}=\frac{\partial \Omega}{\partial p_{0}}=1,&\quad&
\frac{dq^{i}}{d\tau}=\frac{\partial \Omega}{\partial p_{i}}=\frac{\partial H}{\partial p_{i}},\label{eq15e2}\\
\frac{dp_{0}}{d\tau}=-\frac{\partial \Omega}{\partial q^{0}}=-\frac{\partial H}{\partial t},&\quad&
\frac{dp_{i}}{d\tau}=-\frac{\partial \Omega}{\partial q^{i}}=-\frac{\partial H}{\partial q^{i}}.\label{eq15f}
\end{eqnarray}
After substituting (\ref{eq15e2}) into (\ref{eq15f}), i.e., using $x^{0}=t+{\rm const}$, we finally obtain
\[
\frac{dq^{i}}{dt}=\frac{\partial H}{\partial p_{i}},\qquad \frac{dp_{i}}{dt}=-\frac{\partial H}{\partial q^{i}};
\]
i.e., the usual (in general explicitly time-dependent) Hamilton equations.
\item Relativistic Hamiltonian dynamics. When dealing with a relativistic particle of mass $m$ and electric charge $e$, and taking natural units when $c=1$, we have the following natural model of $\Omega$:
\begin{equation}\label{eq15h}
\Omega\left(x^{\nu},p_{\nu}\right)=\frac{1}{2m}g^{\mu\nu}\left(
p_{\mu}-eA_{\mu}\right)\left(p_{\nu}-eA_{\nu}\right)-\frac{m}{2},
\end{equation}
where $g$ denotes the metric tensor of the physical space-time $X$. It need not be flat; so, the simultaneous influence of electromagnetic and gravitational fields may be taken into account. The choice of (\ref{eq15h}) as the left-hand side of equations for $M$ corresponds exactly to the choice of the proper time $s$ (the natural parameter) as a parameterization of world-lines. By analogy to  (\ref{eq15d2}) this corresponds to using the absolute time as a parameter.
\end{enumerate}

\section{Symplectic versus Riemannian geometry. Classical statistical mechanics and microcanonical ensembles}

It was mentioned that there exist both some formal similarities but also drastic differences between symplectic and Riemannian geometry. The problem of "measuring" in symplectic geometry is very essential for the basic concepts of Gibbs' theory. But there are many confusions and ill-defined concepts in the literature.

The first confusion is one concerning the distance concept in mechanical phase spaces. As mentioned, the symplectic "metric" $\gamma$ is in this respect completely useless. But usually (not always!) our phase spaces are cotangent bundles $T^{\ast}Q$, and configuration spaces $Q$ are endowed with some metric tensors $g$, e.g., ones underlying the kinetic energy forms,
\[
T=\frac{1}{2}g_{ij}(q)\frac{dq^{i}}{dt}\frac{dq^{j}}{dt}.
\]
In a sense $g$ induces almost canonically some metric tensors, i.e., Riemann structures both in $TQ$ and $T^{\ast}Q$. The idea is as follows: the metric tensor $g$ gives rise to the Levi-Civita affine connection $\Gamma^{i}{}_{jk}$ equal to 
\[
\Gamma^{i}{}_{jk}=\frac{1}{2}g^{im}\left(g_{mj,k}+g_{mk,j}-g_{jk,m}\right).
\]

PROPOSITION 1. {\it The above ideas enables one to construct in a canonical way $n$-dimensional ($\dim Q=n$) "horizontal" linear subspaces $H_{v}\subset T_{v}TQ$, $H[p]\subset T_{p}T^{\ast}Q$ at any points $v$, $p$ of the tangent and cotangent bundles. Analytically, in terms of adapted coordinates $\left(q^{i},v^{i}\right)$, $\left(q^{i},p_{i}\right)$, the basic vector fields of horizontal distributions are given by
\begin{eqnarray}
H_{a}\left(q^{r},v^{r}\right)&=&\frac{\partial}{\partial q^{a}}-
v^{k}\Gamma^{i}{}_{ka}\left(q^{r}\right)\frac{\partial}{\partial v^{i}},\label{eq46}\\
H_{a}\left[q^{r},p_{r}\right]&=&\frac{\partial}{\partial q^{a}}+
p_{k}\Gamma^{k}{}_{ia}\left(q^{r}\right)\frac{\partial}{\partial p_{i}},\label{eq47}
\end{eqnarray}
where, obviously, $a=1,\ldots,n$}.

PROPOSITION 2. {\it The complementary "vertical" distributions do not depend on the connection $\Gamma$ (do not depend on $g$) and are simply given by linear spaces tangent to the fibres of $TQ$ and $T^{\ast}Q$
\begin{eqnarray}
V_{v}&=&T_{v}\left(T_{q}Q\right)\simeq T_{q}Q,\qquad v\in T_{q}Q,\nonumber\\
V[p]&=&T_{p}\left(T^{\ast}_{q}Q\right)\simeq T^{\ast}_{q}Q,\qquad p\in T^{\ast}_{q}Q.\nonumber
\end{eqnarray}
Analytically, the basic vector fields of these distributions are given by
\begin{eqnarray}
V_{a}\left(q^{r},v^{r}\right)=\frac{\partial}{\partial v^{a}},\label{eq48}\\
V^{a}\left[q^{r},p_{r}\right]=\frac{\partial}{\partial p_{a}},\label{eq49}
\end{eqnarray}
$a=1,\ldots,n$}.

It is obvious that
\[
T_{v}TQ=H_{v}\oplus V_{v},\qquad T_{p}T^{\ast}Q=H[p]\oplus V[p].
\]
The induced metrics on $TQ$, $T^{\ast}Q$ are obtained as direct sums of $g$-metrics evaluated separately on vertical and horizontal components of tangent vectors. More precisely, by evaluation on horizontal components we mean simply the evaluation on $Q$-projected vectors; this term is evidently connection-independent. But the vertical components of vectors depend explicitly on $\Gamma$ and so do evaluations of the $g$-metric on them.

To write concisely the induced metric tensors on $TQ$ and $T^{\ast}Q$, it is convenient to use the dual co-base fields on the bundle manifolds.

On $TQ$ the co-frame field $H^{a}(q,v)$, $V^{a}(q,v)$, $a=1,\ldots,n$, dual to the above $H_{a}(q,v)$, $V_{a}(q,v)$, $a=1,\ldots,n$, is given by
\[
H^{a}\left(q^{r},v^{r}\right)=dq^{a},\qquad V^{a}\left(q^{r},v^{r}\right)=v^{k}\Gamma^{a}{}_{kr}(q)dq^{r}+dv^{a}.
\]
Similarly, in $T^{\ast}Q$ the co-frame $H^{a}[q,p]$, $V_{a}[q,p]$, $a=1,\ldots,n$, dual to $H_{a}[q,p]$, $V^{a}[q,p]$, $a=1,\ldots,n$, is given by
\[
H^{a}\left[q^{r},p_{r}\right]=dq^{a},\qquad V_{a}\left[q^{r},p_{r}\right]=-p_{k}\Gamma^{k}{}_{ar}(q)dq^{r}+dp_{a}.
\]
Let us concentrate on the phase space $T^{\ast}Q$. 

\bigskip

PROPOSITION 3. {\it It seems natural to consider the Riemannian metrics on $T^{\ast}Q$ given by
\begin{eqnarray}
&&G\left(p\in T^{\ast}_{q}Q\right)=G\left(q^{i},p_{i}\right)=\nonumber\\
&&=\alpha g_{ab}(q)H^{a}\left[q^{r},p_{r}\right]\otimes H^{b}\left[q^{r},p_{r}\right]+\beta g^{ab}(q)V_{a}\left[q^{r},p_{r}\right]\otimes V_{b}\left[q^{r},p_{r}\right]
\label{eq53}\\
&&=\alpha g_{ab}(q)dq^{a}\otimes dq^{b}+\beta g^{ab}(q)\left(dp_{a}-p_{k}\Gamma^{k}{}_{ar}(q)dq^{r}\right)\otimes \left(dp_{b}-p_{l}\Gamma^{l}{}_{bs}(q)dq^{s}\right),\nonumber
\end{eqnarray}
where $\alpha$, $\beta$ are constants. This is a natural construction in that, up to constant factors, the metric $G$ is evaluated on horizontal (more precisely, $Q$-projected) and vertical vectors in the sense of the corresponding $Q$-metric $g(q)\in T^{\ast}_{q}Q\otimes T^{\ast}_{q}Q$. The horizontal and vertical subspaces are automatically $G$-orthogonal}.

Obviously, the most natural choice is $\alpha=\beta=1$, nevertheless, some manipulation with constants $\alpha$, $\beta$ does not seem to contradict the general philosophy underlying the construction of $G$.

PROPOSITION 4. {\it Expressing $G$ explicitly in terms of independent $dq$-$dp$ tensor products we obtain
\begin{eqnarray}
G\left(q^{i},p_{i}\right)&=&
\left(\alpha g_{ab}(q)+\beta g^{rs}(q)p_{k}p_{l}\Gamma^{k}{}_{ra}(q)\Gamma^{l}{}_{sb}(q)\right)
dq^{a}\otimes dq^{b}\nonumber\\
&-&\beta g^{rb}(q)p_{k}\Gamma^{k}{}_{ra}(q)dq^{a}\otimes dp_{b}-
\beta g^{br}(q)p_{k}\Gamma^{k}{}_{ra}(q)dp_{b}\otimes dq^{a}\nonumber\\
&+&\beta g^{ab}(q)dp_{a}\otimes dp_{b}.
\label{eq54}
\end{eqnarray}
The structure of $n\times n$ coordinate blocks of this $2n$-metric is nicely readable}.

The corresponding formulae in $TQ$ read as follows:
\begin{eqnarray}
&&G\left(v\in T_{q}Q\right)=G\left(q^{i},v^{j}\right)=\label{eq55}\\
&&=\alpha g_{ab}(q)H^{a}\left(q^{r},v^{r}\right)\otimes H^{b}\left(q^{r},v^{r}\right)+\beta g_{ab}(q)V^{a}\left(q^{r},v^{r}\right)\otimes V^{b}\left(q^{r},v^{r}\right);
\nonumber
\end{eqnarray}
i.e., explicitly in terms of adapted coordinates:
\begin{eqnarray}
G\left(q^{i},v^{i}\right)&=&
\left(\alpha g_{ab}(q)+\beta g_{rs}(q)v^{k}v^{l}\Gamma^{r}{}_{ka}(q)\Gamma^{s}{}_{lb}(q)\right)
dq^{a}\otimes dq^{b}\nonumber\\
&+&\beta g_{rb}(q)v^{k}\Gamma^{r}{}_{ka}(q)dq^{a}\otimes dv^{b}+
\beta g_{br}(q)v^{k}\Gamma^{r}{}_{ka}(q)dv^{b}\otimes dq^{a}\nonumber\\
&+&\beta g_{ab}(q)dv^{a}\otimes dv^{b},
\label{eq56}
\end{eqnarray}
with the same comments as previously concerning the constants $\alpha$, $\beta$.

PROPOSITION 5. {\it Using these metrics one can measure distances in $T^{\ast}Q$ and $TQ$. According to the general principles of Riemann geometry one can also define the natural volume measures, $\mu_{TQ}$, $\mu_{T^{\ast}Q}$,
\begin{eqnarray}
d\mu_{T^{\ast}Q}\left(q^{i},p_{j}\right)&=&
\sqrt{\det\left[G_{zw}\left(q^{i},p_{j}\right)\right]}dq^{1}\cdots dq^{n}dp_{1}\cdots dp_{n},\label{eq57}\\
d\mu_{TQ}\left(q^{i},v^{j}\right)&=&
\sqrt{\det\left[G_{zw}\left(q^{i},v^{j}\right)\right]}dq^{1}\cdots dq^{n}dv^{1}\cdots dv^{n}.\label{eq58}
\end{eqnarray}
The same symbol $G$ was used here with two different meanings --- as the metric on $T^{\ast}Q$ and $TQ$ respectively; simply to avoid the superfluous multitude of characters. The indices $z,w=1,\ldots,2n$ label respectively the variables $\left(q^{i},p_{j}\right)$ and $\left(q^{i},v^{j}\right)$}.

Further on, having at our disposal the $G$-tensors we can restrict them to submanifolds of $T^{\ast}Q$ and $TQ$, and obtain the volume measures on these submanifolds, again according to the general Riemann prescription. 

From the purely geometric point of view the above metrics on $TQ$ and $T^{\ast}Q$ are interesting in themselves. Some natural questions appear concerning the compatibility of $g$-motivated Riemann $G$-metrics and the symplectic form $\gamma$. What is the relationship between the group of canonical transformations, i.e., symmetries of $\gamma$ and the group of isometries of $G$? Using the language of infinitesimals: which Hamiltonian vector fields are simultaneously Killing vectors of $G$-metrics and conversely? Having at our disposal two twice covariant tensors $G$ and $\gamma$, we can construct scalar quantities of the type
\[
{\rm Tr}\left({}^{\gamma}G^{p}\right),\qquad {\rm Tr}\left({}^{G}\gamma^{p}\right),
\]
where the exponents $p$ are integers and the mixed tensors ${}^{\gamma}G$, ${}^{G}\gamma$ are given by
\[
{}^{\gamma}G^{a}{}_{b}=\gamma^{ac}G_{cb},\qquad {}^{G}\gamma^{a}{}_{b}=G^{ac}\gamma_{cb};
\]
obviously ${}^{G}\gamma$ are inverses of ${}^{\gamma}G$. What is the meaning of these scalars? What is the meaning of their Hamiltonian vector fields? These are quite open questions.

Let us notice however that the metric tensor $G$ on $T^{\ast}Q$ is rather artificial from the point of view of statistical mechanics and, as a matter of fact, we do not need it at all. Besides, if the curvature tensor of $g$ is non-vanishing, the metric $G$ is terribly complicated. Only in the locally Euclidean (flat-space $(Q,g)$) case and in adapted coordinates, when $g_{ij}$ are constant, e.g.,
\[
g_{ij}=\delta_{ij},
\]
and therefore
\[
\Gamma^{i}{}_{jk}=0,
\]
is the formula for $G_{ab}$ simple and computationally useful. This has to do with certain formulations used, e.g., in ergodic theory. But, as a matter of fact, $G$ is rather poorly interpretable and essentially superfluous in the foundations of Gibbs theory. It is just a remainder of the configuration way of thinking. Any phase-space manifold $(P,\gamma)$, in particular, any $(T^{\ast}Q,d\theta)$ is endowed with a canonical volume measure independent of the assumed dynamical model and of the geometry of Q! It may be simply given by the formula analogous to the Riemannian volume, but based exclusively on $\gamma$:
\[
d\nu(z)=\sqrt{\det\left[\gamma_{ab}\right]}dz^{1}\cdots dz^{2n},
\]
where $z^{a}$, $a=1,\ldots,2n$, are coordinates in $P$. Using the adapted coordinates $q^{1},\ldots,q^{n}$, $p_{1},\ldots,p_{n}$ in $T^{\ast}Q$ (more generally, Darboux coordinates in $P$), we have
\[
d\nu(q,p)=dq^{1}\cdots dq^{n}dp_{1}\cdots dp_{n}.
\]
These are coordinate expressions, nevertheless, $\nu$ itself is a coordinate-indepen\-dent, intrinsic object. It is related to the differential $2n$-form
\[
\begin{array}{c}
\gamma^{\wedge n}\\
\
\end{array}
\begin{array}{c}
=\\
\
\end{array}
\begin{array}{c}
\underbrace{\gamma\wedge\gamma\wedge\cdots\wedge\gamma},\\
n {\rm \ factors}
\end{array}
\]
where $\gamma=dp_{i}\wedge dq^{i}$.

It is convenient to modify $\gamma^{\wedge n}$ by a constant multiplicative factor, namely,
\[
\Omega=\frac{1}{n!}(-1)^{n(n-1)/2}\gamma^{\wedge n},
\]
i.e., locally, in terms of adapted coordinates:
\[
\Omega=dp_{1}\wedge\cdots\wedge dp_{n}\wedge dq^{1}\wedge\cdots\wedge dq^{n}.
\]
This quantity has the physical dimension of the $n$-th power of the action. It is convenient to use the dimensionless form; this is possible because there exists an experimentally-determined physical constant of the dimension of the action, namely, the Planck constant $h=2\pi\hbar$. Therefore, one defines
\[
\mu=h^{-n}\Omega=\left(2\pi\hbar\right)^{-n}\Omega.
\]
The corresponding volume element is locally given by
\begin{equation}\label{eq69}
d\mu(q,p)=\left(2\pi\hbar\right)^{-n}dq^{1}\cdots dq^{n}dp_{1}\cdots dp_{n}.
\end{equation}
Unlike this, there is no canonical volume on the tangent bundle $TQ$. There, one must use either the above $g$-implied volume, or the $\mathcal{L}$-pull back of $\mu$; $\mathcal{L}$ denoting the Legendre transformation $\mathcal{L}:TQ\rightarrow T^{\ast}Q$. But the resulting form on $TQ$ always depends on something external like the $Q$-metric $g$ or the Hamiltonian $H$ (equivalently, on the Lagrangian $L$). Unlike this, the phase-space object $\mu$ is completely intrinsic. 

The above differential form $\gamma^{\wedge n}$ is but a special case of the sequence of integral invariants
\[
\begin{array}{c}
\gamma^{\wedge k}\\
\
\end{array}
\begin{array}{c}
=\\
\
\end{array}
\begin{array}{c}
\underbrace{\gamma\wedge\gamma\wedge\cdots\wedge\gamma},\\
k {\rm \ factors}
\end{array}
\qquad 
\begin{array}{c}
k=1,\ldots,n.\\
\
\end{array}
\]
They may be integrated over $2k$-dimensional submanifolds of $P$. In cotangent bundles there exist also $(2k+1)$-forms
\[
\theta\wedge \gamma^{\wedge k};
\]
with $\theta$ the Cartan one-form  given by (\ref{eq14}) and $\gamma$ given by (\ref{eq14}); they may be integrated over odd-dimensional submanifolds of $T^{\ast}Q$.

The existence of the canonical measure $\mu$ is crucial for the Gibbs statistical mechanics. Namely, by analogy with the entropy of probability distributions on discrete sets,
\[
S[p]=-\sum_{i}p_{i}\ln p_{i},
\]
one may introduce the Boltzmann-Gibbs-Shannon entropy \cite{B,G,Sh,56a} of probabilistic measures $\Pi$ on $P$ absolutely continuous with respect to $\mu$.

PROPOSITION 6. {\it If $\varrho$ is the Radon-Nikodym derivative,
\[
d\Pi(q,p)=\varrho(q,p)d\mu(q,p),
\]
then we put}
\[
S[\varrho]=-\int \varrho(q,p)\ln \varrho(q,p)d\mu(q,p).
\]

There are an infinity of functions that also have been considered \cite{56a}.
Both the measure $\mu$ and the density $\varrho$ are dimensionless quantities; thus, $S[\varrho]$ is well defined. Without the prescribed background measure $\mu$ there is no way to construct something like the entropy of probabilistic measures on continuous sets, in particular, on differential manifolds. The purely phase-space origin of $\mu$, its independence on any $Q$-objects like $g$ and on any particular dynamical model is an important argument in favour of $\mu$ as a fundamental tool of statistical mechanics. Let us observe that the Riemannian $g$-implied measure $\mu_{T^{\ast}Q}$ coincides with the Liouville-Poincar$\acute{e}$ measure $\mu$ (up to a constant multiplier) only if the Riemann tensor of $g$ vanishes.

There are some statistical and other problems where integration over submanifolds is essential. And here, apparently, we are faced with some difficulty when restricting ourselves to purely phase space concepts. Mainly, as we said above, the restriction of $\gamma$ to submanifolds $M\subset P$ may be degenerate and then the density $\det\left[\gamma\|M_{ab}\right]$ vanishes. In particular, this is the case when $M$ is a hypersurface, $\dim M=\dim P-1=2n-1$, because skew-symmetric tensors in odd-dimensional spaces are always singular. For many reasons, hypersurfaces, in particular energy shells, are of fundamental meaning for statistical mechanics. In general, when $\dim M=2n-m$, at points $p\in M$ the subspaces $T_{p}M^{\perp}\subset T_{p}P$ $\gamma$-dual to $T_{p}M$ are $m$-dimensional (non-singularity of $\gamma$), and their $P$-tangent subspaces $K_{p}M=T_{p}M^{\perp}\cap T_{p}M$ may have all possible dimensions $0\leq k \leq m$, with the proviso that $(m+k)$ is even. Then $T_{p}M/K_{p}M$ carries the natural symplectic structure given by the two-form $\gamma^{\prime}_{p}$ such that $\gamma_{p}|T_{p}M$ is  its pull-back under the projection of $T_{p}M$ onto its quotient space $T_{p}M/K_{p}M$. In the regular case, when $k$ does not depend on $p\in M$ (the generic situation in open subsets of $M$), the assignment
\[
M\ni p\mapsto K_{p}M\subset T_{p}M\subset T_{p}P
\]
is an integrable distribution with the integral foliation $K(M)$ built of $k$-dimen\-sional leaves, because $\gamma$ is closed. The quotient manifold $P^{\prime}(M)=M/K(M)$ has an even dimension $2\left(n-(m+k)/2\right)$ and is endowed with the natural symplectic two-form $\gamma^{\prime}$ such that $\gamma\|M$ is its pull-back under the projection
\[
\pi_{M}:M\rightarrow P^{\prime}(M)=M/K(M);
\]
one writes briefly
\[
\gamma\|M=\pi^{\ast}\cdot\gamma^{\prime}.
\]
Of course, these statements are based on the global topological assumption that the leaves of $K(M)$ are closed and so shaped that the quotient set $P^{\prime}(M)$ carries a natural structure of the $\left(2n-(m+k)\right)$-dimensional differential manifold. Then the family of all $\gamma^{\prime}_{p}$-s, $p\in M$, "glues" together the field $\gamma^{\prime}$ onto $P^{\prime}(M)$. The resulting symplectic structure $\left(P^{\prime}(M),\gamma^{\prime}\right)$ is called the reduced phase space. An extreme special case of the non-existence of a differential structure in $M/K(M)$ occurs, e.g., when the leaves of $K(M)$ are dense in $M$ ("the ergodic situation"). If $M$ is given by equations $F_{a}=0$, $a=1,\ldots,m$, i.e.,
\begin{equation}\label{eq78}
M=\left\{p\in P:F_{a}(p)=0,\ a=1,\ldots,m\right\},
\end{equation}
then
\[
m-k={\rm Rank}\left[\{F_{a},F_{b}\}|M\right].
\]
Let us remember the commonly accepted names of important special cases:
\begin{itemize}
\item[$(i)$] $k=m$ --- co-isotropic submanifold $M$; for any $p\in M$:
\[
T_{p}M^{\perp}\subset T_{p}M,\qquad K_{p}M=T_{p}M^{\perp}.
\]
Historical name: first-class submanifold.
\item[$(ii)$] $k=m=n$ --- Lagrangian submanifold $M$; for any $p\in M$:
\[
T_{p}M^{\perp}=T_{p}M=K_{p}M.
\]
These are co-isotropic submanifolds of the minimal possible dimension (the number of degrees of freedom).
\item[$(iii)$] $k=2n-m$ --- isotropic submanifold $M$; for any $p\in M$:
\[
T_{p}M\subset T_{p}M^{\perp},\qquad K_{p}M=T_{p}M,\qquad \gamma\|M=0.
\]
Lagrangian submanifolds are isotropic submanifolds of the maximal possible dimension (the number of degrees of freedom). They are simultaneously isotropic and co-isotropic.
\item[$(iv)$] There is the traditional term "the class of $M$",
\[
{\rm Cl}M=(k,m-k).
\]
Two extreme special cases are first-class constraints, ${\rm Cl}M=(m,0)$ ($M$-co-isotropic) and second-class constraints, ${\rm Cl}M=(0,m)$ ($M$-symplectic in the sense of $\gamma\|M$, thus, $m$-even) \cite{7}, \cite{8}, \cite{Dirac_wiezy}-\cite{15}, \cite{JJSa}.
\end{itemize}

It is clear that only for submanifolds $M$ of the second-class does the restricted symplectic two-form $\gamma\|M$ induce a volume form. In contrast, the configuration way of thinking and the use of $g$-induced volumes is misleading. The afore-mentioned analytical criteria for ${\rm Cl}M$ based on Poisson brackets provide a proper hint towards what to do to obtain an invariant measure on a submanifold. Namely, usually submanifolds $M$ are described by equations (\ref{eq78}), but $F_{a}$, $a=1,\ldots,m$, are not merely analytical tools, but rather the quantities of a profound physical interpretation. In Gibbs theory they are additive constants of the motion, the seven globally defined constants of motion (energy, linear momentum, angular momentum). Because of this, the phase-space functions $F_{a}$ are as physical as their value-surfaces. And this is a hint of how to construct natural measures on submanifolds independently of their class and without any use of the configurational $(Q,g)$-paradigm with its strange metric $G\left[g,T^{\ast}Q\right]$ and the corresponding measure $\mu_{T^{\ast}Q}[g]$. The idea is as follows:

Being a differential form of the maximal possible degree $2n$, the phase space volume form $\Omega$ is divisible by any simple differential form, in particular, by $dF_{1}\wedge\cdots\wedge dF_{m}$, thus,
\[
\Omega=dF_{1}\wedge\cdots\wedge dF_{m}\wedge\vartheta_{F},
\]
where $\vartheta_{F}$ is some $(2n-m)$-form. This form is not unique, however; its restriction
\[
\vartheta_{F,c}=\vartheta_{F}\|M_{F,c}
\]
to vectors tangent to the common value-surfaces of $F_{a}$
\[
M_{F,c}:=\left\{z\in P:F_{a}(z)=c_{a},\ a=1,\ldots,m\right\}
\]
is evidently unique (because the arbitrariness of $\vartheta_{F}$ is due to terms involving differentials $dF_{a}$). The $(2n-m)$-forms $\vartheta_{F,c}$ on $M_{F,c}$ define there, through the usual integration prescription, some measures $\mu(F, c)$.

\bigskip

PROPOSITION 7. {\it For probabilistic measures $\Pi$ on $M_{F,c}$ that are absolutely continuous with respect to $\mu(F, c)$,
\[
d\Pi(z)=\varrho(z)d\mu_{(F,c)}(z),
\]
entropy is defined by the usual prescription}
\[
S[\varrho]=-\int\varrho(z)\ln\varrho(z)d\mu_{(F,c)}(z).
\]

In statistical mechanics, one envisages having classical states given by probability distributions, $\varrho$, that are absolutely continuous with respect to measure $\mu$, and observables, $A$ that are continuous, or at least $\mu$ measurable, real-valued functions. Then we define the classical expectation value that "$A$ takes when measuring with respect to $\varrho$" as $\int A\varrho d\mu$. We thus obtain the $\mu$ measurable set that $A$ takes values between $\alpha_{1}$ and $\alpha_{2}$ and call it $[a_{1}, a_{2}]$. Then the probability that when measuring $A$ with respect to $\varrho$, $P[a_{1}, a_{2}]=\int \varrho d\mu$, $a_{1}\leq A\leq a_{2}$. We will have other options when we get to quantum mechanical measurement, as we will see we do here.

In various problems of statistical mechanics one deals with value-surfaces of the seven additive constants of motion of multi-body systems; i.e., the total energy, linear momentum and angular momentum are used as functions $F_{a}$ \cite{32,62}. By an appropriate choice of the reference frame one eliminates the total linear momentum and angular momentum; it is geometry of isoenergetic surfaces
\[
M_{H,E}:=\left\{p\in P:H(p)=E\right\}
\]
that is particularly relevant for statistical mechanics.

If manifolds $M_{F,c}$ are compact, then the corresponding measures may be normalized, e.g., to unity if interpreted as probability distributions.

In this way, all fundamental tools of Gibbs theory are essentially based on symplectic phase-space concepts. In spite of a rather complicated structure of the formula (\ref{eq54}) for the $g$-based metric tensor $G$ on $T^{\ast}Q$, an interesting fact that, the corresponding expression (\ref{eq57}) for the volume element $d\mu[G]_{T^{\ast}Q}$ on $T^{\ast}Q$ is, up to a non-essential constant factor, identical with the Liouville measure element $d\mu$ (\ref{eq69}). Indeed, the Levi-Civita connection is by definition symmetric, thus, in a neighbourhood of any fixed configuration $q\in Q$ there exist such local coordinates $q^{a}$ that at $q$ the components of $\Gamma^{i}{}_{jk}$ vanish, $\Gamma^{i}{}_{jk}(q)=0$. 

PROPOSITION 8. {\it From the block structure $G\left(T^{\ast}Q\right)$ and at any $p\in T_{q}^{}{\ast}Q$, then the volume element} (\ref{eq57}) {\it is equivalent to the $2n$-form 
\[
\alpha^{n}\beta^{n}\Omega_{p}=\alpha^{n}\beta^{n}dp_{1}\wedge\cdots\wedge dp_{n}\wedge dq^{1}\wedge\cdots\wedge dq^{n}|_{p}\in \wedge^{2n}T^{\ast}_{p}T^{\ast}Q
\]
at $p$. This fact, immediately visible in special coordinates, is evidently coordi\-nate-independent due to its tensorial character. Performing this procedure at all points of $Q$ we state that really the Riemannian $G$-volume on $T^{\ast}Q$ is independent of $g$ and essentially coincides with the Liouville measure}. 

For the very fundamentals of Gibbs theory the Riemann metric $G$ on $T^{\ast}Q$ is a superfluous concept, and is in fact a misleading one. The configuration metric $g$ is seemingly used, but, as a matter of fact, absent in fundamental formulae due to the mutual cancellation of $g$ and its contravariant inverse. It is interesting that such a cancellation does not occur in the formula (\ref{eq56}) for $G(TQ)$ --- the $g$-based metric on the tangent bundle. The corresponding Riemannian volume measure on the $\mu$-phase space depends explicitly on $g$. And it was to be expected, because the tangent bundle $TQ$, unlike $T^{\ast}Q$, does not carry any intrinsic volume.

The same problem of the superfluous or misleading use of the metric concepts appears when dealing with submanifolds $M\subset T^{\ast}Q$ of the cotangent bundle. This misunderstanding occurs often in considerations concerning microcanonical ensembles, ergodic hypothesis, etc. Namely, one then claims to use the measure element \cite{22}, \cite{Ko}
\[
\frac{d\sigma}{\|{\rm grad}H\|_{E}}
\]
on $M_{H,E}$, where $d\sigma$ is to be the hypersurface volume element on $M_{H,E}$ and $\|{\rm grad}H\|_{E}$ is the length (norm) of the gradient of $H$ taken at points of $M_{H,E}$. Obviously, when meant literally, all these objects are rather ill-defined and to construct them one must use something more then phase-space concepts; namely, one must use the $g$-implied metric tensor $G\left(T^{\ast}Q\right)$ on $T^{\ast}Q$. By comparison, for the value-surface $M_{A,a}$ of the function $A$, the corresponding $M_{A,a}$-supported distribution is given by $\theta_{A,a}$, where \cite{32}
\[
\theta_{A,a}\left(q^{i},p_{i}\right)=
\delta\left(A\left(q^{i},p_{i}\right)-a\right),
\]
and $\delta$ represents a limit of a $\delta$ sequence. It may be given by the usual symbolic formula:
\[
\delta(A-a)=\frac{1}{2\pi}\int^{\infty}_{-\infty}\exp\left(
ik(A-a)\right)dk.
\]
It gives rise to the non-normalised "probabilistic" measure assigning the non-normalised probabilities $p_{U}$ to subsets $U\subset P$ (e.g., $P=T^{\ast}Q$),
\[
p_{U}=\int_{U}\delta(A-a)\mu=\frac{1}{2\pi}\int^{\infty}_{-\infty}dk\int_{U}\exp\left(
ik(A-a)\right)d \mu;
\]
the order of integration (first over the measure element $d\mu$ corresponding to the $2n$-form $\mu$, then over the variable $k$) is essential here. If $M_{A,a}$ is compact, one can use the normalisation
\[
Z^{-1}\delta(A-a)=1
\]
such that
\[
\int\delta(A-a)d \mu=Z.
\]
If the Hamiltonian $H$ is substituted for $A$, and $a$ is a fixed energy value $E$, then, obviously, $\delta(H-E)$ is the microcanonical ensemble and $Z$ is its statistical sum.

The peculiarity of the $M_{A,a}$-supported Dirac "function" $\delta(A-a)$ is that, for any probabilistic measure absolutely continuous with respect to $\mu$ with the Radon-Nikodym derivative $\varrho$, the function $\alpha:\mathbb{R}\rightarrow\mathbb{R}$ given by
\[
\alpha(a)=\int\varrho\delta(A-a)d \mu
\]
equals the probability density for the result $a$ of the measurement of $A$ with respect to $\varrho$. Therefore, 
\[
p[a_{1},a_{2}]=\int^{a_{2}}_{a_{1}}\alpha(a)da
\]
is the probability of finding $a$ within the interval $[a_{1},a_{2}]$.

Let us take the subset
\[
M[A;a_{1},a_{2}]:=\left\{z\in P:a_{1}\leq A(z)\leq a_{2}\right\}
\]
and denote its characteristic function by $\chi_{M[A;a_{1},a_{2}]}$.

It is seen that
\begin{eqnarray}
p\left[a_{1},a_{2}\right]&=&\int\varrho\chi_{M[A;a_{1},a_{2}]}d \mu,\label{eq99}\\
\chi_{M[A;a_{1},a_{2}]}&=&\int^{a_{2}}_{a_{1}}\delta(A-a)da\nonumber\\
&=&
\frac{1}{2\pi}\int^{\infty}_{-\infty}dk\int^{a_{2}}_{a_{1}}\exp\left(
ik(A-a)\right)da\label{eq100}.
\end{eqnarray}
Here the integration order is essential.

It is interesting to see what happens when one performs the limit transition from the finite-thickness "shell" to the "membrane" in $P$. One can show that
\[
\delta(A-a)=\lim_{\varepsilon\rightarrow 0}\frac{1}{\varepsilon} \chi_{M[A;a-\varepsilon/2,a+\varepsilon/2]}.
\]

PROPOSITION 9. {\it If $A$ is a Hamiltonian $H$ and $a$ is an energy value $E$, then the above objects describe the microcanonical ensemble respectively with the finite range of energy $\left[E-\varepsilon/2,E+\varepsilon/2\right]$ and with the sharp energy $E$. It is independent on any metric concepts on} $Q$, $T^{*}Q$.

The characteristic function $\chi_{M[A;a_{1},a_{2}]}$ gives rise to the natural, geometrically distinguished measure on $P$ supported by $M[A;a_{1},a_{2}]$ and equivalent to the differential $2n$-form
\[
\chi_{M[A;a_{1},a_{2}]}\mu=\mu|M[A;a_{1},a_{2}].
\]
Similarly, the distribution $\delta(A-a)$ is equivalent to the measure $\nu_{(A,a)}$ on $M_{(A,a)}$ defined in the following way: If $V$ is an open subset of $M_{(A,a)}$ and $\overline{V}$ is an open subset of $P$ such that $V=\overline{V}\cap M_{(A,a)}$, then
\[
\nu_{(A,a)}(V)=\int_{\overline{V}}\delta(A-a)\mu=
\frac{1}{2\pi}\int^{\infty}_{-\infty}dk\int_{\overline{V}}\exp\left(
ik(A-a)\right)d\mu,
\]
where the integration order is essential. Obviously, the result does not depend on the super-set $\overline{V}$ of $V$. But we have also at our disposal the measure $\mu_{(A,a)}$ on $M_{(A,a)}$ given by
\[
\mu_{(A,a)}(V)=\int_{V}\vartheta_{(A,a)},\qquad V\subset P.
\]
The measures $\mu_{(A,a)}$, $\nu_{(A,a)}$ coincide (at least up to a constant multiplier). It is clear that $\mu_{(A,a)}=\nu_{(A,a)}$ and $\chi_{M[A;a_{1},a_{2}]}\mu$ are intrinsic and geometrically distinguished. If the functions $\varrho$ on $M_{(A,a)}$ or on $M_{[A;a_{1},a_{2}]}$ are Radon-Nikodym derivatives of some probabilistic measures supported by the corresponding subsets, then the maximum of the Boltzmann-Gibbs-Shannon entropy is attained when $\varrho=$const. Let us remind ourselves of the popular shorthand when $A$ is a Hamiltonian $H$ and $a_{1}$, $a_{2}$, $a$ are the corresponding boundary or sharply fixed energy values $E_{1}$, $E_{2}$, $E$: In a microcanonical ensemble all admitted phase-space points are "equally probable". Obviously, without the measures $\mu$ and $\mu_{(H,E)}$ based entirely on the phase space geometry and the energy concept, the statement "equally probable" would be meaningless.

It is clear that for a real-valued function $A:P\rightarrow\mathbb{R}$ the value-surface $M_{(A,a)}$ is really a "surface", i.e., $(2n-1)$-dimensional submanifold of $P$ only if $a$ is a regular value of $A$ and if $A$ is not constant in open subsets of $P$. (If $P$ is an analytic manifold and $A$ is an analytic function on $P$, this means simply that $A$ is non-constant). Then the induced measure $\mu_{(A,a)}$ on $M_{(A,a)}$ is equivalent to the $M_{(A,a)}$-concentrated Dirac distribution $\delta(A-a)$ on $P$. It is clear that $\delta(A-a)$ satisfies the following equations:
\begin{equation}\label{eq105}
A\delta(A-a)=a\delta(A-a),\qquad \left\{A,\delta(A-a)\right\}=0.
\end{equation}
The first equation expresses the fact that the physical quantity takes on without statistical spread a fixed value $a$ on the ensemble given by $\delta(A-a)$; the second equation tells us that this ensemble is invariant under the one-parameter group of canonical transformations generated by $A$ (more precisely, by the Hamiltonian vector field $X_{A}$). The structures underlying the very formulation of those equations are those induced by $C^{\infty}(P)$ respectively as the associative algebra under the pointwise product and the Lie algebra in the sense of Poisson bracket. The second of equations (\ref{eq105}) is equivalent to
\[
\pounds_{X_{A}}\vartheta_{(A,a)}=0,
\]
the invariance of the $M_{(A,a)}$-volume form under $X_{(A,a)}$ --- the restriction of $X_{A}$ to $M_{(A,a)}$. The latter is well defined because, at points of $M_{(A,a)}$, $X_{(A,a)}$ is tangent to $M_{(A,a)}$.

The microcanonical distribution $\delta(A-a)$ is not the only solution of equations
\begin{equation}\label{eq107}
A\varrho=a\varrho,\qquad \left\{A,\varrho\right\}=0.
\end{equation}
Indeed, the "eigenequation" is solved as well by any distribution of the form $F\delta(A-a)$, where $F$ is an arbitrary function on $P$ compatible with the probabilistic interpretation. The Poisson bracket condition implies in addition that
\[
\left\{F,A\right\}|_{M_{(A,a)}}=0,
\]
i.e., that $F|M_{(A,a)}$ is invariant under the Hamiltonian vector field $X_{A}$.

Nevertheless, among all statistical distributions satisfying (\ref{eq107}), $\delta(A-a)$ is peculiar in two respects:
\begin{itemize}
\item[$(i)$] As mentioned, it is a "membrane" limit of the "shell" microcanonical ensemble of finite "thickness", $a-\varepsilon/2\leq A\leq a+\varepsilon/2$. The latter ensemble maximizes the Boltzmann-Gibbs-Shannon entropy within the class of all probability distributions concentrated within the same range and absolutely continuous with respect to the Liouville measure, one of the fundamental objects of the phase space geometry.  
\item[$(ii)$] If integral curves of the Hamiltonian vector field $X_{A}$ are dense in value-surfaces $M_{(A,a)}$, i.e., if the corresponding dynamical systems on $M_{(A,a)}$ are "ergodic", then $F|M_{(A,a)}$ is constant, and $\delta(A-a)$ (up to a constant multiplier) is the only solution of (\ref{eq107}). The reason is that the only globally defined, one-valued and smooth "constants of motion" have the form $f(A)=f\circ A$, where $f:\mathbb{R}\rightarrow\mathbb{R}$ is smooth of the required class. As ergodicity is a rather generic property, the $\delta(A-a)$ are expected to play some particular role in statistical considerations \cite{LandL}, \cite{32}, \cite{62}.
\end{itemize}
This very special role of "microcanonical" distributions of $A$ and their obvious property (\ref{eq105}) fix our attention on the status of the joint condition (\ref{eq107}) in general. Apparently the two conditions in (\ref{eq107}) when taken separately are of a qualitatively different nature. The first of them, induced by the associative algebraic structure in $C^{\infty}(P)$, expresses the statistical-informational properties of $\varrho$ --- the lack of spread in the set of outcomes of $A$-measurements. The second sub-condition is based on the Poisson-Lie algebra structure in $C^{\infty}(P)$ and demands $\varrho$ to be invariant under the one-parameter group of canonical transformations (classical automorphisms) generated by $A$. Let us notice that this condition, although logically and qualitatively independent of the first informational one, is nevertheless somehow distinguished among the family of all solutions of the informational eigencondition $A\varrho=a\varrho$. Namely, it is quite natural to expect that some particular role will be played by such "eigendistributions" which are as closely as possible suited to the physical quantity $A$, e.g., inherit the symmetries of $A$. Of course, working in the phase space we mean symmetries belonging to the classical automorphism group, i.e., canonical transformations preserving $A$. They are infinitesimally generated by functions $B:P\rightarrow \mathbb{R}$ being in involution with $A$,
\begin{equation}\label{eq109}
\left\{B,A\right\}=0.
\end{equation}
But, if we admit the most general situations, including ones in which trajectories of the vector field $X_{A}$ are dense on the value-surfaces $M_{(A,a)}$ ("the generic" structure), then the only globally-defined, one-valued and smooth functions $B$ on $P$ satisfying (\ref{eq109}) are those of the form $B=f(A)=f\circ A$, where $f:\mathbb{R}\rightarrow\mathbb{R}$ is smooth of the required class \cite{2}, \cite{29}. The universally warranted symmetries of $A$ are those generated by $A$ itself and by expressions functionally built of $A$. Therefore, in general, "$\varrho$ inherits the symmetries of $A$" means:
\[
\left\{A,\varrho\right\}=0,
\]
i.e., the second subcondition of (\ref{eq107}). This is to be understood weakly (in the Dirac sense),
\[
\left\{A,\varrho\right\}|_{{\rm Supp}\varrho}=0,\qquad \left\{F,A\right\}|_{M_{(A,a)}}=0,
\]
when one deals with a single regular value $a\in A(P)$, or strongly, i.e., identically all over $P$, when the total set $A(P)\subset\mathbb{R}$ is concerned, i.e., when one deals with the foliation of $P$ by all value-surfaces $M_{(A,a)}$.

\section{Towards quantum mechanics and mechanical-optical analogy}

Quantum mechanics is, among other things, a theory in which states are positive, trace-class self-adjoint operators, $\hat{\varrho}$, of trace one on a Hilbert space $\cal{H}$. $\cal{H}$ is invariant under the action (on the left) of some symmetry group, $G$. Observables, $\hat{A}$, are a subset of the self-adjoint operators in $\cal{H}$. The quantum expectation of $\hat{A}$ in state $\hat{\varrho}$ is $\langle \hat{A}\rangle_{\hat{\varrho}}=\rm{Tr}(\hat{\varrho}\hat{A})$ and the quantum variance is $var_{\hat{\varrho}}\hat{A}=\rm{Tr}(\hat{\varrho}(\hat{A}-\langle \hat{A}\rangle_{\hat{\varrho}})^{2})$. 
(Physical) information on $\hat{\varrho}$ is confined to knowing $\rm{Tr}(\hat{\varrho}\hat{A})$ for various observables $\hat{A}$.

Now, when one remains on the purely classical level, the former reasoning unifying the information and invariance properties as above
\[
A\varrho=a\varrho,\qquad \left\{A,\varrho\right\}=0.
\]
might perhaps seem a little "scholastic". Nevertheless, it has an obvious counterpart in quantum theory and may be interpreted both as a forrunner of quantum structures and as their asymptotic expansion. Namely, if $\widehat{A}$ is a self-adjoint operator for some physical quantity which sharply takes on the value $a\in {\rm Spec}\widehat{A}$ on the state described by the density operator $\widehat{\varrho}$, i.e., $a$ is an eigenvalue of $\widehat{A}$, then
\begin{equation}\label{eq111}
\widehat{A}\widehat{\varrho}=a\widehat{\varrho}.
\end{equation}
This eigenequation implies the vanishing of the commutator quantum Poisson bracket,
\begin{equation}\label{eq112}
\frac{1}{i\hbar}\left[\widehat{A},\widehat{\varrho}\right]=0.
\end{equation}
Equation (\ref{eq111}) describes informational properties of $\widehat{\varrho}$, whereas the second equation gives an account of some invariance property of $\widehat{\varrho}$, namely,
\[
\widehat{U}(\tau)\widehat{\varrho}\widehat{U}(\tau)^{-1}=
\widehat{\varrho},\qquad \widehat{U}(\tau)=\exp\left(\frac{i}{\hbar}\widehat{A}\tau\right),\qquad \tau\in\mathbb{R}. 
\]
Hence $\widehat{\varrho}$ is symmetric under the one-parameter group of unitary transformations (quantum automorphisms) generated by $\widehat{A}$. And on the quantum level information of the type (\ref{eq111}) implies symmetry, because the Lie-algebraic operation (commutator, quantum Poisson bracket) is algebraically built from the associative-algebra operation (operator product). This fact is a kind of qualitative discontinuity of the limit transition from quanta to classics, where the Poisson bracket $\{A,B\}$ is not an algebraic function of the associative product $AB$. But nevertheless, as seen before, there exists a kind of "scholastics" which somehow joins $A\varrho=a\varrho$ with $\{A,\varrho\}=0$. $\widehat{A}$ need not have any eigenvalues, as is shown by the examples $\widehat{A}=p_{a}$ (the $a$-th coordinate of the momentum operator). If $\widehat{A}$ has no eigenvalues, then either we may go to the rigged Hilbert space formalism, or we may go to the phase space formulation of quantum mechanics which will be discussed later on.

To quantization, and quantum and quasi-classical problems we shall return later. Here let us consider the unification of information and symmetry suggested by "microcanonical" distributions $\delta(A-a)$ and expressed by (\ref{eq107}). For example, let us take $A=p_{1}$. The first component of the canonical momentum takes on a fixed value $b$ on the ensemble $\varrho\left(q^{i},p_{i}\right)=F\left(q^{i},p_{i}\right)
\delta\left(p_{1}-b\right)$. The invariance condition means that $\left\{p_{1},F\right\}$ vanishes on $M_{(p_{1},b)}$, at least weakly, and certainly strongly if one deals with the foliation given by manifolds $M_{(p_{1},b)}$, $b\in\mathbb{R}$. Therefore, we may put
\begin{eqnarray}
\varrho_{b}\left(q^{1},\ldots,q^{n};p_{1},\ldots,p_{n}\right)&=&
F\left(q^{2},\ldots,q^{n};p_{1},\ldots,p_{n}\right)\delta(p_{1}-b)\label{eq114}\\
&=&
F\left(q^{2},\ldots,q^{n};b,p_{2},\ldots,p_{n}\right)\delta(p_{1}-b).\nonumber
\end{eqnarray}
The invariance under canonical transformations generated by $p_{1}$ implies that $\varrho_{b}$ is invariant under translations along the $q^{1}$-coordinate line. Therefore, the $\varrho_{b}$-distribution is completely smeared out in the $q^{1}$-direction canonically conjugate to the spread-free $p_{1}$.

In statistical mechanics, when the physical Hamiltonian $H$ and its fixed (energy) value are used as $A$, $a$, the microcanonical ensemble $\delta(H-E)$ is a stationary statistical distribution in virtue of the vanishing Poisson brackets $\left\{H,\delta(H-E)\right\}=0$ and the classical Liouville-von Neumann equation
\[
\frac{\partial \varrho}{\partial t}=\left\{H,\varrho\right\}.
\]
Moreover, if one takes into account the existence of Hamiltonians $H$ with ergodic flows of $X_{H}$ on $M_{(H,E)}$ (and, in a sense, "generity" of such models), $\delta(H-E)$ is an essentially unique equilibrium ensemble under adiabatic external conditions.

Nevertheless, the above analysis based entirely on the phase space geometry is physically interpretable in a context wider than statistical mechanics. As we shall see, it is strongly related to the quasi-classical limit of quantum mechanics, the quantum-classical analogy and quantization (with all its ensuing problems).

In \cite{JJS} and papers quoted there we used the terms "proper ensembles" and "eigenensembles" of $A$ for statistical distributions $\varrho$ satisfying respectively the single condition
\begin{equation}\label{eq116}
A\varrho=a\varrho
\end{equation}
or the couple (\ref{eq107})
\[
A\varrho=a\varrho,\qquad \{A,\varrho\}=0.
\]
When $\varrho$ is fixed, then the set $\mathcal{E}(\varrho)\subset C^{\infty}(P)$ of smooth functions $\Phi$ satisfying
\begin{equation}\label{eq117}
\Phi\varrho=0
\end{equation}
is an ideal in $C^{\infty}(P)$ meant as an associative algebra under pointwise multiplication. This ideal consists of functions vanishing on ${\rm Supp}\varrho$ --- the support of $\varrho$. The set $\mathcal{E}(\varrho,a)\subset C^{\infty}(P)$ of functions satisfying (\ref{eq116}) is a coset of the ideal $\mathcal{E}(\varrho)$ in $C^{\infty}(P)$:
\[
\mathcal{E}(\varrho,a)=\mathcal{E}(\varrho)+a=
\left\{\Phi+a:\Phi\in\mathcal{E}(\varrho)\right\}.
\]
Obviously $\mathcal{E}(\varrho)$ is a maximal proper ideal of the pointwise product associative algebra $C^{\infty}(P)$ if and only if $\varrho$ is the Dirac distribution concentrated on a one-element subset $\{z\}\subset P$, $\varrho=\delta_{z}$. Then for any function on $P$ we have
\begin{equation}\label{eq119}
A\delta_{z}=A(z)\delta_{z},
\end{equation}
i.e., $\delta_{z}$ is a proper ensemble of any $A$ with the eigenvalue $A(z)$. In other words, tautologically speaking
\[
\left(A-A(z)\right)\in\mathcal{E}\left(\delta_{z}\right),\qquad A\in\mathcal{E}\left(\delta_{z},A(z)\right).
\]
(We do not distinguish graphically between the constant function and its value.)

The set $\mathcal{L}(\varrho)\subset C^{\infty}(P)$ of smooth functions $A$ being in involution with $\varrho$,
\[
\mathcal{L}(\varrho)=\left\{A\in C^{\infty}(P):\{A,\varrho\}=0\right\}
\]
is a Lie subalgebra of $C^{\infty}(P)$ in the Poisson-bracket sense. This is a general property of infinitesimal symmetries. Such functions are Hamiltonian generators of canonical transformations preserving $\varrho$. In particular, these transformations preserve ${\rm Supp}\varrho$, the support of $\varrho$. If $F,G\in\mathcal{L}(\varrho)$, then, obviously, the Hamiltonian vector fields $X_{F}$, $X_{G}$ are tangent to ${\rm Supp}\varrho$ and so is their Lie bracket $\left[X_{F},X_{G}\right]=-X_{\{F,G\}}$.

It was just mentioned that classical statistical ensembles are informationally optimal (all physical quantities are spread-free on them) when they are described by Dirac-delta probability distributions $\delta_{z}$ (point-concentrated measures). Then $\mathcal{E}(\delta_{z})$ is a maximal nontrivial ideal in the associative $C^{\infty}(P)$ and $\mathcal{E}(\delta_{z},a)$ are its maximal cosets.

The question arises as to the maximality of information in $L(\varrho)$ and the relationship between $\mathcal{L}(\varrho)$ and $\mathcal{E}(\varrho)$. By a symmetry of a state we mean a transformation which leaves the state fixed. Information is given in the set of statistical distributions, $\varrho$, by $\{\langle A\rangle_{\varrho}=\int A\varrho d\mu \ | \ A \in \mathcal{O}\}$ where $\mathcal{O}$ ia a subset of the set of all classical statistical functions (real-valued and $\mu$ - measurable). We will consider the case in which $\mathcal{O}$ is the singleton $\{A \}$. Then information in the sense of (\ref{eq116}) and symmetry of the state(s) contradict each other in view of the fact that the entropy is different for the two states. And this seems to be embarrassing and incompatible with everything said above about classical statistical mechanics, microcanonical ensembles, their motivation for classical "eigenconditions" (\ref{eq107}) (and more-so with the quantum rules mentioned in (\ref{eq111}), (\ref{eq112})). Let us formulate some heuristic qualitative remarks. The demand of maximal informational content, measurement without (or with a minimal) statistical spread of results, leads to maximally concentrated  probabilistic measures (statistical distributions). But those, being concentrated, have poor symmetries. If $P$ is compact (a rather academic situation) or if we admit non-normalisable distributions (only relative probabilities; comparison between different compact regions of $P$), then, obviously, the maximally symmetric ensembles are described by constant density functions $\varrho$. The Lie algebra $\mathcal{L}(\varrho)$ is then given by the total $C^{\infty}(P)$, the improper subalgebra. And such situations are maximally entropic (minimally informative). All canonical transformations are symmetries. Surface-supported distributions $\varrho$, concentrated on submanifolds $M\subset P$ have symmetry algebras $\mathcal{L}(\varrho)$ consisting of such functions $F$ that the corresponding Hamiltonian vector fields $X_{F}$ are tangent to $M$. This means that if $M$ is a common value-surface,
\[
M_{(A,a)}=\left\{p\in P:A_{i}(p)=a_{i}\in\mathbb{R},\ i=1,\ldots,m\right\},
\]
then
\begin{equation}\label{eq123}
\left\{F,A_{i}-a_{i}\right\}=\left\{F,A_{i}\right\}=
\lambda_{i}{}^{j}\left(A_{j}-a_{j}\right),
\end{equation}
where the functions $\lambda_{i}{}^{j}$ are smooth; more modestly said, they are smooth in some neighbourhood of $M_{(A,a)}$. In Dirac terms: Poisson brackets (\ref{eq123}) vanish weakly on $M_{(A,a)}$. It is clear that symmetries of any figure form a group and infinitesimal symmetries of anything in a manifold form a Lie algebra. Neverthe-less, one could also show directly that for any $F,G\in\mathcal{L}(\varrho)$ also $\{F,G\}\in\mathcal{L}(\varrho)$ holds. We shall not do this here.

Remark: A dangerous trap is hidden in the above reasoning. Namely, if canonical transformations generated by the above functions $F$ do preserve statistical density $\varrho$, then also its support ${\rm Supp}\varrho$ is preserved. But not conversely! And this has to do with classification of phase space submanifolds, the relationship between information and symmetry, and the difference between proper conditions and eigenconditions (\ref{eq116}), (\ref{eq117}). And finally some quasi-classical and prequantum structures are essential here.

Before going any further with these topics, let us give a warning concerning the simplest possible "statistical mechanics"; namely, one in the "phase space" given by a finite set $I$ consisting of "states" $i_{1},\ldots,i_{N}$. 

Probability distributions ("statistical states") of maximal information (minimal entropy) have the form $p_{K}$, where
\[
p_{K}\left(i_{L}\right)=\delta_{KL}.
\]
Obviously, for such states the Shannon entropy vanishes,
\[
S\left[p_{K}\right]=-\sum^{N}_{L=1}p_{K}\left(i_{L}\right)\ln p_{K}\left(i_{L}\right)=0.
\]
Let us also notice their idempotence property:
\[
p^{2}_{K}=p_{K}.
\]
Obviously, this property, together with the normalisation condition,
\[
\sum^{N}_{L=1}p_{K}\left(i_{L}\right)=1,\qquad p_{K}\left(i_{L}\right)=\delta_{KL},
\]
is just a kind of a joking allusion to properties of quantum density operators.

Physical quantities, i.e., random variables, are real-valued functions $A:I\rightarrow \mathbb{R}$; they form an associative commutative algebra under pointwise multiplication. Expectation values are given by the obvious formula:
\[
\langle A\rangle_{p}=\sum^{N}_{L=1}A\left(i_{L}\right)p\left(i_{L}\right).
\]

There is one very important difference between Gibbs statistical mechanics in symplectic manifolds and "statistical mechanics" in finite probabilistic spaces. Namely, in the linear space of random variables there is nothing like the Poisson bracket and, obviously, there is nothing like any differential structure in $I$. Nevertheless, the mutual relationship between information and symmetry of statistical ensembles still does exist. The difference is that realisations of symmetry are different. In differential-symplectic theory the same objects, namely phase-space functions, played a double role. They were random variables --- measurable quantities (informational aspect) and generators of physical automorphisms (symmetry aspect). In finite "phase spaces" only the informational aspect of physical quantities survives. There is no natural structure according to which they could generate automorphisms. Nevertheless, transformations and invariance aspects still exist and are somehow related to information, in spite of all the differences. Concerning information, the "pure states" $p_{K}$ of finite statistics also satisfy the "eigenequations" similar to (\ref{eq119}), 
\begin{equation}\label{eq129}
Ap_{K}=A\left(i_{K}\right)p_{K}.
\end{equation}
They satisfy these eigenequations for any physical quantity $A$. In this respect (\ref{eq129}) is different from its quantum counterpart (\ref{eq111}) which, also for pure states $\widehat{\varrho}$, holds only for exceptional quantities $\widehat{A}$.

Obviously, probability distributions $p_{K}$ concentrated at $i_{K}\in I$ are counterparts of Dirac distributions $\delta_{z}$ from the differential theory. If $p$ is a general probability distribution on $I$, then the linear subspace $\mathcal{E}(p)$ of functions $\Phi$ on $I$ satisfying
\[
\Phi p=0
\]
(pointwise multiplication) is an ideal. If $p$ has the form $p_{K}$ (is concentrated at $K$), this is a maximal non-trivial ideal. The set of functions
\[
\mathcal{E}(p,a)=\mathcal{E}(p)+a=\left\{\Phi+a:\Phi\in\mathcal{E}(p)\right\}
\]
is a coset of the ideal $\mathcal{E}(p)$. If $\mathcal{E}(p)$ is maximal, then cosets $\mathcal{E}(p,a)$ consist of such functions $A$ that the eigenequations hold:
\[
A\varrho=a\varrho
\]
for some fixed $a=A\left(i_{K}\right)$ if $\varrho=p_{K}$. 

Let us now consider a distinctly opposite case, namely, the probability distribution which is completely smeared out,
\begin{equation}\label{eq133}
p\left(i_{L}\right)=\frac{1}{N}\qquad {\rm for\ any}\ i_{L}\in I.
\end{equation}
Obviously, such statistical states carry no information; i.e., their Shannon entropy is maximal, and one easily obtains the Boltzmann-Gibbs formula:
\[
S[p]=-\sum^{N}_{L=1}p\left(i_{L}\right)\ln p\left(i_{L}\right)=\ln N.
\]
For such distributions the ideal $\mathcal{E}(p)$ is evidently improper and consists only of a single function, identically vanishing on $I$. Quite generally, it is clear that the all-nowhere vanishing functions do not belong to any proper ideal in the pointwise-product function algebra. Just like (normalised or not) constant density functions in symplectic manifolds, the smeared-out constant probability distributions (\ref{eq133}) have the maximal symmetry group, namely the total symmetric group $S^{(N)}$ permuting states in an arbitrary way. The distinctly opposite case is that $p$ as a function on $I$ is completely non-degenerate, i.e., perfectly distinguishes states:
\[
p\left(i_{K}\right)\neq p\left(i_{L}\right)\qquad {\rm if}\qquad K\neq L.
\]
Then the symmetry group is trivial; any nontrivial permutation of states changes the statistical ensemble. If there are some value-surfaces of more than one element, the symmetry group is larger, it is generated by subgroups $S^{(M)}\subset S^{(N)}$, $M<N$, preserving $M$-element subsets of $I$ on which $p$ takes some fixed values. It is seen that there is some interplay of information and symmetry; nevertheless, it is not very clear. Ensembles of maximal entropy, i.e., free of any diversity, have maximal symmetry groups, just the full groups of bijections of the "phase space." Ensembles of maximal diversity, i.e., distinguishing all states, have trivial, one-element symmetry groups.  But there is no well-defined mathematical measure of the order of diversity. Probability distribution $p$ may separate all states, but the differences $\left|p\left(i_{K}\right)-p\left(i_{L}\right)\right|$ may be "small". Nevertheless, its symmetry group will be still one-element, in spite of its almost-maximal entropy (almost $\ln N$). The pure statistical ensembles $p_{K}$ concentrated on $i_{K}$ have vanishing entropy and "relatively large" symmetry groups $S^{(N-1)}$. But the same symmetry group characterises  non-pure ensembles given by:
\[
p\left(i_{K}\right)=q,\qquad p\left(i_{L}\right)=x\neq q, \qquad L\neq K
\]
$K$-fixed. The normalisation condition implies
\[
x=\frac{1-q}{N-1}.
\]
One may show that the entropy of $p$ is given by
\begin{eqnarray}
S[p]&=&-q\ln q-(1-q)\ln (1-q)-(q-1)\ln (N-1)\nonumber\\
&=&(q-1)\ln \frac{1-q}{N-1}-q\ln q.\label{eq137}
\end{eqnarray}
If $q\neq 1$, all such ensembles have positive entropies smaller than $\ln N$ and depending on $q$, nevertheless they have the same, "relatively large" invariance group $S^{(N-1)}$ preserving $i_{K}\in I$. Passing to the special case $q=1/N$, we catastrophically jump to $S^{(N)}$, the improper subgroup of itself, and $S[p]=\ln N$, just the Gibbs entropy for equally probable microstates.

Let us now go back to the symplectic phase space setting. As mentioned, a weak link between symmetry and information existed there as well. Let us notice also, there was some contradiction, some discrepancy between them. Maximally informative Dirac distributions $\delta_{z}$ were invariant under canonical transformations which did not affect their concentration points $z\in P$. Infinitesimally this leads to the Lie algebra of symmetries with Hamiltonian generators $A$ the differentials of which vanish at $z$,
\[
\mathcal{L}\left[\delta_{z}\right]=\left\{A\in C^{\infty}(P):dA_{z}=0\right\}.
\]
There are symmetry properties implied by informative ones. There is an obvious analogy with the fact from probability on finite sets: the informationally optimal $p_{K}$ is invariant under the "large" groups $S^{(N-1)}_{K}$ preserving $i_{K}\in I$; as a matter of fact this is the maximal nontrivial subgroup of $S^{(N)}$.

Nevertheless in Gibbs' statistical mechanics there is no direct rule according to which information properties would simply logically imply some invariance properties. And moreover as mentioned above both in symplectic manifolds and in finite sets there is a kind of competition between two properties: concentrating statistical distributions to make them informationally better, we make them less symmetric. Some compromise however does exist in both: let us mention just the above examples and those of microcanonical ensembles and in general, distributions satisfying the couple (\ref{eq107}). Obviously, if one is aware of quantum mechanics and the implication between (\ref{eq111}), (\ref{eq112}), everything is clear. But we just saw that in Gibbs' theory and phase-space geometry there was something suggesting the couple (\ref{eq107}) as something more natural than (\ref{eq116}) alone \cite{G}.

Let us now formalise the above philosophical "prophecy". We start with recalling the concepts of some ideals of functions on a symplectic manifold \cite{35}. It is clear that in any differential manifold $M$ the following linear subspaces of $C^{\infty}(M)$ are ideals (in the pointwise-product-algebra sense) \cite{33}:
\begin{eqnarray}
V(N)&:=&\left\{f\in C^{\infty}(M):f|_{N}=0\right\},\nonumber\\
V^{1}(N)&:=&\left\{f\in C^{\infty}(M):f(x)=0,\ df_{x}=0,\ {\rm for\ any}\ x\in N\right\},\nonumber\\
\vdots&&\vdots\label{eq139}\\
V^{k}(N)&:=&\left\{f\in C^{\infty}(M):\partial^{m}f_{x}=0,\ m\leq k,\ {\rm for\ any}\ x\in N\right\}.\nonumber
\end{eqnarray} 
In these formulae, $N\subset M$ is some fixed subset and $\partial^{m}f_{x}$ is the system of $m$-th order partial derivatives of $f$ at $x$ (the system of $\partial^{m}f_{x}$, $m\leq k$, is what is usually called the $k$-th order jet of $f$ at $x$). Then
\[
V^{k}(N)\subset V^{l}(N)\qquad {\rm if}\qquad k>l.
\]
Maximal ideals have the form
\[
V_{x}:=V\left(\{x\}\right);
\]
any of them consist of all functions vanishing at some fixed point $x\in M$ (and maybe, but not necessarily, also somewhere else).

Now let $J\subset C^{\infty}(M)$ be an arbitrary associative ideal. The set of its zeros will be denoted by $N(J)$:
\[
N(J):=\left\{x\in M:f(x)=0\ {\rm for\ any}\ f\in J\right\};
\]
i.e.,
\[
N(J)=\bigcap_{f\in J}f^{-1}(0).
\]

From the point of view of informational analysis of random variables on $M$, particularly interesting are ideals $J$ satisfying:
\begin{equation}\label{eq143}
J=V\left(N(J)\right).
\end{equation}
Now, for any pointwise-product-ideal $J$ in $C^{\infty}(M)$ the following inclusion holds:
\[
J\subset V\left(N(J)\right).
\]
Therefore, (\ref{eq143}) is an extreme situation. This peculiar special case will be referred to as a probabilistic ideal, or sometimes, a radical ideal. Mathematically they are exceptional in that it is uniquely and without any additional restrictions, determined by its radical, i.e., set of zeros. Physically, when functions on $M$ are interpreted as random variables, $V(N)$ is informationally peculiar as the set of all random variables which spread-freely give the result zero when measured on statistical ensembles supported by $N$. Shifting $V(N)$ by constant functions we obtain affine cosets of random variables taking dispersion-free some values when measured on all ensembles supported by $N$.

Analytically $N$ is usually given by the system of equations:
\[
F_{r}(x)=0,\qquad r=1,\ldots,p,
\]
i.e.,
\begin{equation}\label{eq146}
N=\left\{x\in M:F_{r}(x)=0,\ r=1,\ldots,p\right\}.
\end{equation}
It is assumed that the $F_{r}$ are functionally independent in some neighbourhood of $N$; i.e., there exists $\varepsilon(x)>0$ such that
\[
{\rm Rank}\left[\frac{\partial F_{r}}{\partial x^{i}}(x)\right]=p\qquad {\rm if}\qquad \left|F_{r}(x)\right|<\varepsilon(x). 
\]
Then $\dim N=m-p=\dim M-p$. The ideal $V(N)$ is then generated by functions $F_{r}$; i.e., for any $F\in V(N)$
\[
F=\lambda^{r}F_{r},
\]
where $\lambda^{r}$ are arbitrary smooth functions and the summation convention is meant for the index $r$.

General value-surfaces of random variables are obtained by putting $F_{r}=A_{r}-a_{r}$, $a_{r}\in A_{r}(M)$, denoting constants. Let us denote as usual:
\begin{equation}\label{eq149}
N_{(A,a)}:=\left\{x\in M:A_{r}(x)=a_{r},\ r=1,\ldots,p\right\}.
\end{equation}
All random variables $A_{r}$ take fixed values $a_{r}$ when measured on statistical ensembles supported by $N_{(A,a)}$, without any statistical spread. In practical applications usually the functions $A_{r}$ themselves are treated as something primary and one considers foliations of $M$ by the family of all value-surfaces (\ref{eq149}).

The manifold $M$ was general, but now we return to the main subject of our interest, i.e., to the symplectic phase space $P$. Everything said above may be repeated, but in addition, the Poisson bracket Lie algebraic structure in $C^{\infty}(P)$ and the resulting symmetry properties introduce new qualities. The same might be said of course about the existence of the Liouville measure and other volume structures which have been mentioned previously. In analytic manifolds $P$, we are interested mainly in the subspace of analytic functions $C^{\omega}(P)$. Smooth functions on $P$ have two properties: they are measured as physical quantities (random variables) and they generate one-parameter groups of canonical transformations. The compromise between informational content and symmetry is attained when $\mathcal{E}(\varrho)$ and $\mathcal{L}(\varrho)$ are closely related to each other. And this is possible because the associative and Poisson-Lie structures in $(P,\gamma)$ are compatible in the sense that the Poisson bracket with a fixed function, ${\rm ad}_{F}=\{F,\cdot\}$ is a differentiation of the pointwise product associative algebra. In classical statistical mechanics it is mainly $C^{\omega}(P)$ that is used as the set of physical quantities (random variables). As mentioned, in analytical manifolds $P$, the space $C^{\infty}(P)$ is perhaps more natural. If $P$ is endowed with an affine structure, one uses also $W(P)$ --- the linear manifold of all polynomials on $P$. Obviously, some non-smooth random variables are also convenient in certain problems; however they usually may be approximated by smooth functions or obtained as limits of their sequences. In any case, it is intuitively obvious and compatible with realistic models that "true" physical quantities are polynomially or analytically built of Darboux canonical coordinates $q^{i}$, $p_{i}$. All the function spaces $C^{\infty}(P)$, $C^{\omega}(P)$, $W(P)$ are Poisson bracket Lie algebras.

DEFINITION 1. {\it An associative ideal $J$ in $C^{\infty}(P)$ or $C^{\omega}(P)$ or $W(P)$ is said to be self-consistent if it is also a Lie algebra in the Poisson-bracket sense}. 

One can easily show the following proposition relating these concepts to the classification of submanifolds in $(P,\gamma)$:

PROPOSITION 10. $V(N)$ {\it is a self-consistent ideal (of $C^{\infty}(P)$, $C^{\omega}(P)$, $W(P)$) if and only if $N$ is a co-isotropic submanifold of} $(P,\gamma)$.

Analytically, if $N$ is given by equations (\ref{eq146}), this means that Poisson brackets $\left\{F_{a},F_{b}\right\}$ vanish weakly; i.e.,
\[
\left\{F_{a},F_{b}\right\}=C^{r}{}_{ab}F_{r};\qquad {\rm i.e.,}\qquad \left\{F_{a},F_{b}\right\}|N=0,
\]
for some smooth functions $C^{r}{}_{ab}$. If one uses the value-surface (\ref{eq149}), then of course
\begin{equation}\label{eq151}
\left\{A_{r},A_{s}\right\}=\left\{A_{r}-a_{r},A_{s}-a_{s}\right\}=
C^{z}{}_{rs}\left(A_{z}-a_{z}\right).
\end{equation}
If, as typical in applications, functions $A_{r}$ are primary objects and $P$ is foliated by the family of co-isotropic submanifolds $N_{(A,a)}$, then the Poisson brackets (\ref{eq151}) vanish in the strong sense,
\[
\left\{A_{r},A_{s}\right\}=0.
\]

This is obviously the classical counterpart of the correlation-free commensurability of commuting observables. But we decided here to avoid explicitly quantum argumentation as far as possible.

Self-consistent ideals $V(N)$ give rise to statistical ensembles concentrated on co-isotropic manifolds $N$.  Those ensembles are informationally as valuable as possible when $N$ are co-isotropic submanifolds of minimal possible dimension, i.e., Lagrangian submanifolds. Let us remind ourselves that their dimension equals the number of degrees of freedom $n=(1/2)\dim P$.

If we start from probabilistic distributions $\varrho$ and take $N={\rm Supp}\varrho$, then the co-isotropic class of ${\rm Supp}\varrho$ implies that there exists such a system $\{F_{i}\}$ of independent generators of $\mathcal{E}(\varrho)$ that not only have (by definition)
\[
F_{i}\varrho=0,
\]
but also
\[
\left\{F_{i},\varrho\right\}=0;
\]
i.e., $\varrho$ is not only a proper ensemble of all $\Phi_{i}$ (with vanishing "proper values") but also their eigenensemble; i.e., in a sense, information implies symmetry. This is again the mentioned "compromise" between apparently incompatible demands of information ("sharp", concentrated $\varrho$) and symmetry ("homogeneous", smeared out $\varrho$).

Obviously, by $\{F_{i}\}$ being generators of $\mathcal{E}(\varrho)$ we mean that any other $F\in\mathcal{E}(\varrho)$ has the form
\[
F=\sum_{i}K^{i}F_{i},
\]
where $K^{i}$ are some smooth functions on $P$.

In spite of our provisos about avoiding too direct a motivation based on the quantum analogy, let us consider some. We mean one based on the properties of integrable Hamiltonian systems and polarizations, i.e., foliation of a symplectic manifold of dimension $2n$ by a smooth $n$-parameter family of ($n$-dimensional) Lagrange manifolds. Let those manifolds be given by the following system of equations:
\begin{equation}\label{eq178}
A_{i}(q, p)-a_{i}=0,
\end{equation}
where $A_{i}$, $i=1, \cdots, n$ is a system of functionally independent physical quantities in involution, 
\[
\{A_{i}, A_{j}\}=0.
\]

Then, (\ref{eq178}) describes a polarization, i.e., some foliation of $P$ by the family of Lagrangian manifolds $\mathfrak{m}_{a}$. The system of $n$ eigenequations for the probabilistic distribution $\varrho$, 
\begin{equation}\label{eq180}
A_{i}\varrho=a_{i}\varrho
\end{equation}
is, up to over-all normalization, uniquely solved by
\begin{equation}\label{eq181}
\varrho(q, p)=\delta(A(q, p)-a)=\delta(A_{1}(q, p)-a_{1})\cdots \delta(A_{n}(q, p)-a_{n}).
\end{equation}

They also automatically satisfy the following invariance conditions under $X[A_{i}]$, i.e., the Hamiltonian vector field generated by $A_{i}$ \cite{46,47}, 
\begin{equation}\label{eq182}
\{A_{i}, \varrho\}=0.
\end{equation}

Let us observe that when dealing with polarization, the system (\ref{eq180}) implies (\ref{eq182}), rather than for a single function $A$ and the proper equation
\[
A\varrho=a\varrho.
\]
The solution (\ref{eq181}) is unique up to multiplication by a factor depending only on constants $a_{i}$, $i=1, \cdots, n$. This is the $n$-th order microcanonical ensemble for the quantities $A_{i}$, $i=1, \cdots, n$. Obviously, when the manifolds 
$\mathfrak{m}_{a}$ are transversal to the fibres $T^{*}Q$ of the cotangent bundle, they may be analytically represented by the following system of equations:
\[
p_{i}-\frac{\partial S}{\partial q^{i}}(q, a)=0, \quad i=1, \cdots, n,
\]
and the function $S:Q\times A \rightarrow \mathbb{R}$ is a common solution, just the complete integral, of the system:
\[
A_{i}\left(\cdots, q^{j}, \cdots; \cdots, \frac{\partial S}{\partial q^{j}}(q, a), \cdots\right)=a_{i}, \quad i=1, \cdots, n.
\]

\bigskip

PROPOSITION 11. {\it After some elementary manipulations with the Dirac distribution one obtains finally that}:
\[
\delta(A(q, p)-a)=\delta(A_{1}(q,p)-a_{1})\cdots \delta(A_{n}(q,p)-a_{n})=
\]
\begin{equation}\label{eq186}
=\left|{\rm det}\left[\frac{\partial^{2} S}{\partial q^{i}\partial a^{j}}\right]\right|\delta \left(p_{1}-\frac{\partial S}{\partial q^{1}}(q, a)\right) \cdots \delta \left(p_{n}-\frac{\partial S}{\partial q^{n}}(q, a)\right)=
\end{equation}
\[
\left|{\rm det}\left[\frac{\partial^{2} S}{\partial q^{i}\partial a^{j}}\right]\right|\delta \left(p-\nabla_{q}S(q, a)\right).
\]

\bigskip
Here the quantity:
\begin{equation}\label{eq187}
{\rm det}\left[\frac{\partial^{2} S}{\partial q^{i}\partial a^{j}} \right]
\end{equation}
is known as the Van Vleck determinant \cite{56}. It is uniquely, up to a constant multiplier, assigned to any pair of mutually transversal polarizations; in this case they are built respectively of the leaves $T_{q}{}^{*}Q$, $\mathfrak{m}_{a}$. This quantity appeared in the co-called quasi-classical WKB-analysis of the Schr\"{o}dinger equation \cite{23}, \cite{m3}, \cite{m5}, \cite{AM}, \cite{m6}. Namely, let us assume that the function $S:Q \times A \rightarrow \mathbb{R}$ is a complete solution of the Hamilton-Jacobi equation \cite{AM2}, \cite{m4}, \cite{m2}
\[
H\left( \cdots, q^{i}, \cdots; \cdots, \frac{\partial S}{\partial q^{i}}(q,a), \cdots\right)=E(a_{1}, \cdots, a_{n}).
\]

If there are no turning points, then the quasi-classical wave functions:
\begin{equation}\label{eq189}
\Psi_{a}(q):=\sqrt{\left|{\rm det}\left[\frac{\partial^{2} S}{\partial q^{i}\partial a^{j}} \right]\right|}{\rm exp}\left(\frac{i}{\hbar}S(q,a)\right)
\end{equation}
are the famous WKB-solutions of the Schr\"{o}dinger equation:
\begin{equation}\label{eq190}
\hat{H}\Psi=E\Psi.
\end{equation}
The eigenvalue of energy $E$ is a function of the constants of motion $a_{1}, \cdots, a_{n}$, $E(a_{1}, \cdots,$ $a_{n})$ \cite{4}. 

\bigskip

PROPOSITION 12. {\it Using explicitly the time variable $t$, we have the solution 
\begin{equation}\label{eq191}
\Psi_{a}(t,q):=\sqrt{\left|{\rm det}\left[\frac{\partial^{2} S}{\partial q^{i}\partial a^{j}} \right]\right|}{\rm exp}\left(-\frac{i}{\hbar}\left( E(a)t-S(q,a)\right)\right)
\end{equation}
for the Schr\"{o}dinger equation with time}:
\begin{equation}\label{eq192}
\hbar i \frac{\partial \Psi}{\partial t}=\hat{H}\Psi.
\end{equation}

Remark: we do not delve here into the problem of defining $\hat{H}$ by $H$, a rather complicated problem. In all practical problems it is somehow solvable. 

If there are no classical turning points, or far from them, if they do exist, (\ref{eq189}) is a WKB-solution of (\ref{eq190}), and (\ref{eq191}) is a WKB-approximation to the solution of (\ref{eq192}). Also, when the classically accessible region is topologically $\mathbb{R}^{n}$, one deals with continuous spectra of the $a_{i}$-s. However, if this region is compact; i.e., if one considers the motion on the torus $T^{n}$, then the $a_{i}$-s become quantized, so that the integrals \cite{9}, \cite{10}
\[
\oint_{\tau}p_{i}dq^{i}=nh
\]
over closed loops on the manifolds $\mathfrak{m}_{a}$ are integer multiplies of $h$.

In spite of our using a quantum language here, the above concepts are obviously classical and except for the existence of the Planck constants $\hbar$, they might be considered as a mechanical-optical analogy, and to large extent, they were indeed studied in $XIX$-th century physics. Quantum mechanics, with its introduction of the $\hbar$-constant, gave them a new interpretation, especially due to achievements by Planck, Heisenberg, Bohr, Schr\"{o}dinger and de Broglie. In our treatment this was the statistical unification of information and symmetry of statistical ensembles. Let us mention that the quantity (\ref{eq187}), although written in coordinates, has an invariant global meaning if one deals with two complementary polarizations of $P$, i.e., its foliations by the family of Lagrange manifolds. In the example above, those were Lagrange manifolds $T_{q}{}^{*}Q$, $\mathfrak{m}_{a}$, but it only matters in their complementary intersection. Also in spite of the analytical character of the expression (\ref{eq187}), this object has an important geometrical meaning. Namely, for any $q\in Q$, $a\in A$, it is a doubled-type geometric quantity: the scalar density of weight two in $T_{q}Q$ and $T_{a}A$. Therefore, the square root is a scalar density of weight one in $T_{q}Q$ and $T_{a}A$, i.e., something that admits an invariantly defined integration over the $(q, a)$-variables. It is assumed of course that there are no turning points; the quantity (\ref{eq187}) does not vanish; and because of this, the square root in (\ref{eq189}), (\ref{eq191}) is well-defined. 

PROPOSITION 13. {\it For any pair of regions $X\subset Q$, $Y\subset A$, the quantity
\[
P(X,Y)=\int_{Q\times A}\sqrt{\left|{\rm det} \left[\frac{\partial^{2}S}{\partial q^{i}\partial a^{j}} \right]\right|}dq^{1}\cdots dq^{n}da^{1}\cdots da^{n}
\]
is the non-normalized, ralative probability that the particle created in the region $X\subset Q$ will be detected in the region $Y\in A$, and conversely. This is in principle the non-normalized, relative probability, and because of this, for any two compact regions $X_{1}$, $X_{2}$ in $Q$ and for any region $Y$ in $A$, the number
\[
P(X_{1}, Y) / P(X_{2}, Y) 
\]
tells us what the ratio of detections is in $X_{1}$, $X_{2}$ when the particle / object was created at $Y$ in $A$}. 

\section{The message of the Weyl-Wigner-Moyal-Ville formalism}

It is seen from the above reasoning and from the formulas (\ref{eq181}), (\ref{eq186})-(\ref{eq189}) and (\ref{eq191}) that there exists a relationship between probability distributions concentrated on Lagrangian submanifolds and WKB wave functions. Our arguments above were based on a general symplectic manifold and integrable Hamiltonian systems in it. But in affine phase spaces there are also other arguments based on the Weyl-Wigner-Moyal-Ville distributions based on the Weyl prescription \cite{5}, \cite{41}, \cite{Sch}, \cite{61}.

So, let $(P, \Gamma)$ be a $2n$-dimensional affine phase space. $P$ is its underlying set and $\Gamma$ is the symplectic two-form on the linear space $\Pi$ of translations in $P$. With any translation $\overline{z}\in \Pi$ we associate some linear transformation $\mathbb{W}(\overline{z})$ acting on functions on $P$. Incidentally, it is given by
\[
\mathbb{W}[\overline{\alpha},\underline{\pi}]:={\rm exp}\left(\frac{i}{2\hbar}\underline{\pi}\cdot \overline{\alpha} \right)
\mathcal{W}(\overline{\alpha},\underline{\pi}),
\]
where the meanings of the symbols are as follows:
\begin{itemize}
\item[$(i)$] We put
\[
\Pi=V\times V^{*}
\]
where $V$ is a linear $n$-dimensional space of translations in the configuration space $Q$, $V^{*}$  is the dual of $V$ and 
\[
P=Q\times V^{*}
\]
\item[$(ii)$] $\mathcal{W}$ is defined as 
\[
\mathcal{W}(\overline{\alpha}, \underline{\pi}):=U(\overline{\alpha})V(\underline{\pi}), \ (U(\overline{\alpha})\Psi)(\overline{x})=\Psi(\overline{x}-\overline{\alpha}), \ (V(\underline{\pi})\hat{\Psi})[\underline{p}]:=\hat{\Psi}[\underline{p}-\underline{\pi}]
\]
where $\hat{\Psi}$ is the Fourier transform of $\Psi$. Therefore,
\[
(V(\underline{\pi})\Psi)(\overline{x})={\rm exp}\left(\frac{i}{\hbar}\underline{\pi}\cdot \overline{x} \right)\Psi(\overline{x}).
\]
\end{itemize}

For the Fourier transforms the following convention is used:
\[
\Psi(\overline{x})=\frac{1}{(2\pi \hbar)^{n}}\int \hat{\Psi}[\underline{p}]{\rm exp}\left(\frac{i}{\hbar}\underline{p}\cdot \overline{x} \right)d_{n}\underline{p}, \ \hat{\Psi}[\underline{p}]=\int\Psi(\overline{x}){\rm exp}\left(-\frac{i}{\hbar}\underline{p}\cdot \overline{x} \right)d_{n}\overline{x}.
\]
$Q$ is identified here with $V$, $P$ is identified with $\Pi=V\times V^{*}$, and $\Gamma$ is then expressed as 
\[
\Gamma((\overline{v}_{1}, \underline{\pi}_{1}), (\overline{v}_{2}, \underline{\pi}_{2}))=\underline{\pi}_{1}\cdot \overline{v}_{2}-\underline{\pi}_{2}\cdot \overline{v}_{1}=\pi_{1i}v_{2}{}^{i}-\pi_{2i}v_{1}{}^{i}.
\]
Then the assignment
\[
\overline{z}\mapsto \mathbb{W}[\overline{z}]
\]
does satisfy:
\begin{equation}\label{eq204}
\mathbb{W}[\overline{z}_{1}]\mathbb{W}[\overline{z}_{2}]={\rm exp}\left( \frac{i}{2\hbar}\Gamma(\overline{z}_{1}, \overline{z}_{2})\right)\mathbb{W}[\overline{z}_{1}+\overline{z}_{2}]
\end{equation}
and 
\[
\mathbb{W}[\overline{z}]^{-1}=\mathbb{W}[-\overline{z}].
\]
Condition (\ref{eq204}) means that the assignment is a projective representation of $\mathbb{W}$ \ on  $P$ \cite{6}, \cite{42}, \cite{43}, \cite{50}, \cite{57}-\cite{59}. Incidentally, there is no other representation of the $2n$-dimensional Abelian group by canonical transformations in the $n$-dimensional symplectic space. The maximal admissible dimension of the momentum mapping for the Abelian group equals $n$ in the $2n$-dimensional phase space.

The following holds for the group commutators:
\[
W[\overline{z}_{1}]W[\overline{z}_{2}]W[-\overline{z}_{1}]W[-\overline{z}_{1}]={\rm exp}\left( \frac{i}{\hbar}\Gamma(\overline{z}_{1}, \overline{z}_{2})\right){\rm Id},
\]
where, obviously, ${\rm Id}$ denotes the identity operator. The Weyl prescription, i.e., the correspondence between phase-space functions $A(\overline{\alpha}, \underline{\pi})$ and the quantum-like operators $A$, has the form
\[
{\bf A}=\int \hat{A}(\overline{\alpha}, \underline{\pi})\mathbb{W}[\overline{\alpha}, \underline{\pi}]d_{n}\overline{\alpha}\frac{d_{n}\underline{\pi}}{(2\pi\hbar)^{n}}=
\]
\[
\ \ \ \ \ \ \quad \quad \  \int \hat{A}(\overline{\alpha}, \underline{\pi}){exp}\left(\frac{i}{\hbar}(\pi_{a} {\bf Q}^{a}+\alpha^{a} {\bf P}_{a}) \right)d_{n}\overline{\alpha}\frac{d_{n}\underline{\pi}}{(2\pi\hbar)^{n}},
\]
where $\hat{A}$ is the Fourier transform of $A$,
\[
A(\overline{Q}, \underline{P})=\int \hat{A}(\overline{\alpha}, \underline{\pi}){exp}\left(\frac{i}{\hbar}(\pi_{a} Q^{a}+\alpha^{a} P_{a}) \right)d_{n}\overline{\alpha}\frac{d_{n}\underline{\pi}}{(2\pi\hbar)^{n}}.
\]

The product of operators ${\bf A B}$ and the quantum Poisson bracket $\{{\bf A}, {\bf B} \}_{quant}=({\bf A B}-{\bf BA})/ i\hbar$ are represented by the non-local operations,
\[
(A*B)(\overline{z})=2^{2n}\int {\rm exp}\left( \frac{2i}{\hbar}\Gamma(z-z_{1}, z-z_{2})\right)A(z_{1})B(z_{2})d\mu (z_{1})d\mu (z_{2}),
\]
\[
\{ A,  B \}_{quant}=\frac{1}{\hbar i}(A*B-B*A)
\]
where, obviously, 
\[
d\mu(z)=d\mu(q, p)=\frac{1}{(2\pi \hbar)^{n}}dq^{1} \cdots dq^{n}dp_{1} \cdots dp_{n}.
\]

PROPOSITION 14. {\it One can show that if $A$ is represented by the kernel $A[q, q']$ of the integral operator,
\[
({\bf A}\Psi)(q)=\int A[q, q']\Psi(q')d_{n}q',
\]
then the following holds}:
\begin{equation}\label{eq212}
A[q, q']=\int{\rm exp}\left( \frac{i}{\hbar}\underline{p}(\overline{q}-\overline{q}')\right)A\left(\frac{1}{2}(\overline{q}+\overline{q}'), \underline{p} \right)\frac{d_{n}\underline{P}}{(2\pi\hbar)^{n}}
\end{equation}
\[
A(q, p)=\int{\rm exp}\left( -\frac{i}{\hbar}\underline{p}\cdot \overline{\alpha}\right)A\left[ \overline{q}+\frac{\overline{\alpha}}{2}, \overline{q}-\frac{\overline{\alpha}}{2}\right]d_{n}\overline{\alpha}.
\]

This non-local "star multiplication", or the "Weyl-Wigner-Moyal-Ville product" has obviously, all the properties which the multiplication of operators has. It is associative, bilinear, and the complex conjugation satisfies:
\[
\overline{A*B}=\overline{B}*\overline{A}
\]
representing the Hermitian conjugation of operators. It is also invariant under the symplectic-affine group; in particular it is translationally-invariant. The trace and scalar products, when they exist, satisfy the obvious rules for operators. Thus, e.g., 
\[
{\rm Tr} {\bf A}=\int A(\overline{z})d\mu(\overline{z}),
\]
\[
({\bf A}, {\bf B})={\rm Tr}({\bf A}^{+}{\bf B })=\int \overline{A}(z)B(z)d\mu(\overline{z})=(A, B),
\]
\begin{equation}\label{eq213}
(C*A, B)=(A, \overline{C}*B)=(C, B*\overline{A}),
\end{equation}
but in general
\[
\int A*B*Cd\mu\neq \int ABCd\mu.
\]
Besides,
\[
1*A=A*1=A, \quad \overline{A}*A\neq 0
\]
unless $A$ vanishes almost everywhere. 

The "density operators" are represented by quasi-probability distributions which need not be positive, although after  coarse-graining over regions, the $\mu$-volume which is much larger than $1$, they make the impression of being positive-the negative contributions cancel in the integration procedure. The formulas for expectation values and detection probabilities are like in classical statistical mechanics; i.e., they are just based on (\ref{eq213}):
\begin{eqnarray}\label{eq215}
\langle A\rangle_{\varrho}=\int A\varrho d\mu, \\
P(\varrho ', \varrho)=\int \varrho ' \varrho d\mu \nonumber
\end{eqnarray}
if $\varrho$ is a pure state, etc.

For pure states $\varrho=|\Psi\rangle\langle\Psi| $  it may be shown that
\[
\varrho(\overline{q}, \underline{p})=\frac{1}{(2\pi)^{n}}\int \overline{\Psi}\left( \overline{q}-\frac{1}{2}\hbar \overline{\tau}\right){\rm exp} (-i \overline{\tau}\cdot \underline{p})\Psi\left( \overline{q}+\frac{1}{2}\hbar \overline{\tau}\right)d_{n}\overline{\tau}.
\]
Pointwise, this expression, although real (because, $ \boldsymbol{\varrho }^{+}=\boldsymbol{\varrho}$, therefore $\overline{\varrho}=\varrho$ ) is not positive, except for some very special situation, explicitly the ground state of the harmonic oscillator \cite{23}. It is only positive in the non-local sense of positive values of positive physical quantities, $\overline{A}*A$:
\[
(\varrho, \overline{A}*A)=\int \varrho(\overline{z})(\overline{A}*A)(\overline{z})d\mu(\overline{z})>0.
\]

Let us notice, however, that the marginals of $\varrho$ are always positive:
\[
\int \varrho(\overline{q}, \underline{p})d_{n}\overline{q}=|\hat{\Psi}(\underline{p})|^{2}
\]
\[
\int \varrho(\overline{q}, \underline{p})\frac{d_{n}\underline{p}}{(2\pi\hbar)^{n}}=|\Psi(\overline{q})|^{2},
\]
and similarly for the mixed states. 

Now we go to a rigged Hilbert space formalism to proceed. Rigged Hilbert spaces are discussed at length in \cite {4}. In there, for $H$ a Hilbert space, you construct $H^{-}$ as a subHilbert space and $H^{+}$ which is not a Hilbert space but includes $H$.

Let us take a complete system $|i\rangle$ and the corresponding $"H^{+}$-algebraic complete systems" \cite{33}, \cite{38}, \cite{39}
\[
\boldsymbol{\varrho }_{ij}=|i\rangle \langle j|=\boldsymbol{\varrho }_{ji}{}^{+}, \quad \varrho_{ij}=\overline{\varrho_{ji}},
\]
\[
\boldsymbol{\varrho }_{ij}\boldsymbol{\varrho }_{kl}=\delta_{jk}\boldsymbol{\varrho }_{il}, \quad \varrho_{ij}\boldsymbol{*}\varrho_{kl}=\delta_{jk}\varrho_{il}.
\]

PROPOSITION 15. {\it It is interesting to take the continuous value case for states of definite positions and momenta, when $|i\rangle=|\overline{q}\rangle$ or $|i\rangle=|\underline{p}\rangle$},
\[
\varrho_{\overline{q}_{1} \overline{q}_{2}}(\overline{q}, \underline{p})=\delta\left( \overline{q}-\frac{1}{2}(\overline{q}_{1}+\overline{q}_{2})\right){\rm exp}\left(\frac{i}{\hbar}\underline{p}\cdot (\overline{q}_{2}-\overline{q}_{1}) \right),
\]
\[
\varrho_{\underline{p}_{1}\underline{p}_{2}}(\overline{q}, \underline{p})=\delta\left( \underline{p}-\frac{1}{2}(\underline{p}_{1}+\underline{p}_{2})\right){\rm exp}\left(\frac{i}{\hbar} (\underline{p}_{1}-\underline{p}_{2})\cdot \overline{q} \right).
\]

The corresponding kernels of the operator $\bf{A}$ satisfy
\[
{\bf A}=\int A[\overline{q}_{1}, \overline{q}_{2}]\varrho_{\overline{q}_{1} \overline{q}_{2}}d_{n}{\overline{q}_{1}}
d_{n}{\overline{q}_{2}},
\]
\[
{\bf A}=\int A[\underline{p}_{1},\underline{p}_{2}]\varrho_{\underline{p}_{1}\underline{p}_{2}}\frac{d_{n}\underline{p}_{1}}{(2\pi \hbar)^{n}}\frac{d_{n}\underline{p}_{2}}{(2\pi \hbar)^{n}}.
\]
Obviously, in spite of our quantum-like arguments, all this has an important classical meaning. And it is again here where quasiprobability distributions concentrated on Lagrange manifolds appear. 

PROPOSITION 16. {\it The diagonal matrices of "continuous $H$-bases" have the forms 
\[
\varrho_{\overline{\alpha}\overline{\alpha}}(\overline{q}, \underline{p})=\delta(\overline{q}-\overline{\alpha}), \quad \varrho_{\underline{\pi}\underline{\pi}}(\overline{q}, \underline{p})=\delta(\underline{p}-\underline{\pi}).
\]
They are evidently concentrated on Lagrange manifolds of definite positions and definite momenta} \cite{LandL}. 

For a general WKB phase space function $\varrho[D, S]$ of the "quantum" pure state with the function 
\begin{equation}\label{eq223}
\Psi(\overline{q})=\sqrt{|D(\overline{q})|}{\rm exp}\left( \frac{i}{\hbar}S(\overline{q})\right),
\end{equation}
this is not the case. 

PROPOSITION 17. {\it In the WKB-limit, when the function $S$ is quickly-varying, what is formally (although non-precisely) modelled by the asymptotic $\hbar \rightarrow 0$ procedure, when $D, S$ themselves are assumed to be $\hbar$-independent, it turns out that}
\[
\varrho_{cl}[D, S]={\rm lim}_{\hbar \rightarrow 0}\varrho[D, S]=D(\overline{q})\delta\left(p_{1}-\frac{\partial S}{\partial q^{1}} \right)\cdots \delta\left(p_{n}-\frac{\partial S}{\partial q^{n}} \right). 
\]
In other words, this is practically (\ref{eq186}). Notice that, in consequence of (\ref{eq212}), (\ref{eq223}), one obtains 
\[
(A\Psi)(q)=A\left(q^{i},\frac{\partial S}{\partial q^{i}} \right)\Psi+\frac{\hbar}{i}\left( \pounds_{v} f\right){\rm exp}\left( \frac{i}{\hbar}S(q)\right)
\]
up to terms of higher order in $\hbar$. $\pounds_{v}$ denotes the Lie derivative with respect to the vector field $v[A, S]$ given by 
\[
v^{i}=\frac{\partial A}{\partial p_{i}}\left(q^{j}, \frac{\partial S}{\partial q^{j}} \right),
\]
and $|D(\overline{q})|=f(\overline{q})^{2}$. This velocity field is obtained by projecting onto $Q$ the Hamiltonian vector field
\[
X[A]=\frac{\partial A }{\partial p_{i}}\frac{\partial }{\partial q^{i}}-\frac{\partial A }{\partial q^{i}}\frac{\partial }{\partial p_{i}}
\]
restricted to  $\mathfrak{m}_{s}$ given by equations
\[
p_{i}=\frac{\partial S}{\partial q^{i}}.
\]
(This restriction is well-defined became $X[A]$ is tangent to $\mathfrak{m}_{s}$). One can show that 
\[
\pounds_{v}f=v^{i}\frac{\partial f}{\partial q^{i}}+\frac{1}{2}\frac{\partial v^{i}}{\partial q^{i}}f.
\]
The quadratic structure of $D$ in $f$ implies that 
\[
\pounds_{v}D=v^{i}\frac{\partial D}{\partial q^{i}}+\frac{\partial v^{i}}{\partial q^{i}}D=\frac{\partial}{\partial q^{i}}(Dv^{i}).
\]

PROPOSITION 18. {\it Therefore, the Schr\"{o}dinger equation
\[
\hbar i\frac{\partial \Psi}{\partial t}=\hat{{\bf H}}\Psi
\]
implies that
\[
\frac{\partial S}{\partial t}+H\left( q^{i}, \frac{\partial S}{\partial q^{i}}, t\right)=0,
\]
\begin{equation}\label{eq231}
\end{equation}
\[
\frac{\partial D}{\partial t}+\frac{\partial j^{i}}{\partial q^{i}}=0,
\]
where the current $j^{i}$ is given by}:
\[
j^{i}=Dv[H, S]^{i}=D\frac{\partial H}{\partial p_{i}}\left( q, \frac{\partial S}{\partial q}\right).
\]

Therefore, the second of equations (\ref{eq231}) may be written as
\[
\frac{\partial D}{\partial t}+\pounds_{v[H, S]}D=0.
\]

We see then, that without rigged Hilbert spaces, some of this would make a limited sense. When there are classical turning points, the above solutions fail there and must be somehow combined with Airy functions. In a noncompact space and without turning points, these expressions correspond to quantum states of continuous spectrum. Usually one deals with a complete integral for $S$ and $D$ is then given by the Van Vleck determinant (\ref{eq187}) above. It is non-vanishing in generic and may be taken positive, at least in open domains. If there are turning points, they are critical and the mentioned combination with Airy functions must be made. So, in principle $D$ is positive and in non-relativistic theory $j^{i}$ is the current density vector \cite{32}, \cite{Mack}, \cite{34}, \cite{40}, \cite{41}, \cite{SiWo}.

\section{Lagrange and Legendre submanifolds}

a) Lagrange spaces and their projections \\
Above we have found a relationship between information and symmetry of probabilistic distributions in a symplectic manifold through the concept of polarizations. We begin with purely algebraic symplectic concepts, and later on we pass to the manifolds framework. Let us assume for a moment that $(P, \gamma)$ is a linear symplectic space, and let $\Delta(P)$ denote the set of all Lagrangian, thus $n={\rm dim}P/2$-dimensional, manifolds $\varepsilon$ such that $\gamma||\varepsilon=0$, i.e., $\gamma(u, v)=0$ for any pair of vectors $u, v\in \varepsilon$. Now, let $M\subset P$ be a linear co-isotropic subspace of $P$, $M^{\perp}\subset M$. Therefore, $M$ contains some Lagrange subspaces. Let $K(M)=M^{\perp}\subset M$. Let us quote without proofs some important relationships between Lagrange and first-class subspaces:
\[
A\delta(A-a)=a\delta(A-a),\qquad \left\{A,\delta(A-a)\right\}=0.
\]
The set of all Lagrange subspaces of $P$ contained in $M$ will be denoted by $\Delta(M)$.

For any co-isotropic subspace $M\subset P$ and any Lagrange subspace $\varepsilon\subset P$, the subspace $\varepsilon\cap M$; that is the set-theoretical intersection of $\varepsilon$  and $M$ is a non-empty isotropic subspace of $P$ and there exists exactly one Lagrange subspace $\widetilde{\varepsilon}\subset M$ passing through $\varepsilon \cap M$; it is:
\[
\widetilde{\varepsilon}:=\varepsilon\cap M +M^{\perp}.
\]
This gives rise to the mapping
\[
E_{M}:\Delta(P)\rightarrow \Delta(M).
\]
The mapping $E_{M}$ is a retraction of $\Delta(P)$ onto $\Delta(M)$:
\begin{equation}\label{eq236}
E_{M}|_{\Delta(M)}= {\rm id}_{\Delta(M)}.
\end{equation}

PROPOSITION 19. {\it Therefore, it is also a projection,
\[
E_{M}\circ E_{M}=E_{M}.
\]

An important point is the relationship between various $E_{M}$-operations. If both $M, N$ are not only co-isotropic but also mutually compatible in the sense that $M \cap N$ is also co-isotropic, then 
\begin{equation}\label{eq238}
E_{M}\circ E_{N}=E_{N}\circ E_{M}=E_{M\cap N}.
\end{equation}

This statement has also an inverse. Namely, if $M, N$ are co-isotropic and the corresponding retractions do commute, i.e., if the following holds, 
\[
E_{M}\circ E_{N}=E_{N}\circ E_{M},
\]
then $M, N$ are mutually compatible, i.e., $M\cap N$ is also co-isotropic and} (\ref{eq238}) {\it holds.

And, finally, we conclude with the statement that for any symplectic mapping $f\in Sp(P, \gamma)$ the following property is satisfied:
\begin{equation}\label{eq239}
E_{f(M)}=F\circ E_{M}\circ F^{-1},
\end{equation}
where $F:\Delta(P) \rightarrow \Delta(P)$ is the transformation of $\Delta(P)$ induced by $f$}.

Let us illustrate this review of properties of $E_{M}$ by an intuitive example. Namely, let $(q_{1}, \cdots, q_{n}; p^{1}, \cdots p^{n})$ be mutually dual symplectic bases in $P, P^{*}$, and let us take the linear subspace $M:={\rm Ker} \ q_{1}$ in $P$.

Obviously, $M$ is co-isotropic, as any hypersurface is. And furthermore, $M^{\perp}=\mathbb{R}p_{1}$. Now, let us take the Lagrange subspace $\varepsilon:={\rm Ker} \ p_{1} \cap \cdots \cap {\rm Ker} p_{n}$ of sharply defined momentum variables. Then $\varepsilon \cap M$ is an isotropic manifold ${\rm Ker} \ q^{1} \cap {\rm Ker} \ p_{1}\cap \cdots {\rm Ker} p_{n}$. Then, one has 
\[
E_{M}(\varepsilon)=\varepsilon\cap M+M^{\perp}=\mathbb{R}p^{1}\oplus \mathbb{R}q_{2}\oplus \cdots \oplus \mathbb{R}q_{n}.
\]
$E_{M}(\varepsilon)$ is the linear span of $p^{1}, q_{2}, \cdots, q_{n}$; i.e., the following holds:
\[
E_{M}(\varepsilon)={\rm Ker} \ q^{1} \cap {\rm Ker} \ p_{2}\cap \cdots {\rm Ker} p_{n}.
\]

This means that projecting the Lagrange subspace of sharply fixed linear momentum onto the subspace with a sharply defined value of $q^{1}$, we obtain the Lagrange subspace with the non-restricted and quite arbitrary value of $p_{1}$ and non-disturbed fixed values of the other components of linear momentum. Fixing $q^{1}$ results in diffusion and indeterminancy of $p_{1}$. \\

 b) Affine symplectic manifolds and projections of affine Legendre subspaces\\
We have discussed rigorously in $a)$ this very special case. This rigour is no longer the case in a general situation. Of course, one can relatively easy extend those definitions onto the slightly more general situation of affine spaces. So, let $(P, \Pi, \mu, \Gamma)$ be an affine-symplectic space. $(P, \Pi, \mu)$ is a $2n$-dimensional affine space; $P$ is its underlying set, $\Pi$ is the linear space of translations in it, and $\mu$ is the operation $\mu:P\times P \rightarrow \Pi$ which to any pair of points $z, w\in P$ assigns the radius vector $\mu(z, w)$ of $w$ with respect to $z$. Obviously, $\Gamma$ is the symplectic two-form on $\Pi$. It gives rise in a natural way to the differential two-form $\gamma$ on $P$. Let $M$ be an affine co-isotropic (first-class) submanifold in $P$, and $\mathfrak{m}$ be an affine Lagrange submanifold of $P$. The linear subspace of translations in $M$ induced by $\Pi$ will be denoted by $L[M]$ and the corresponding radius-vector operation by $\mu '_{M}$. The singular foliation $\mathcal{K}(M)$ of $M$ consists of affine submanifolds parallel to their common translation space $L[M]^{\perp}\cap L[M]$. In the special case of co-isotropic submanifolds, we have $L[M]^{\perp}\subset L[M]$; and thus $L[M]^{\perp}\cap L[M]=L[M]^{\perp}$. The natural projection of $M$ onto $M/\mathcal{K}(M)$, i.e., onto the reduced affine space $(P'(M), L'[M], \mu '_{M}, \Gamma ')$, will be denoted by $\pi_{M}:M \rightarrow P'(M)$. If $\mathfrak{m}$ intersects $M$, then it gives rise to the new affine-Legendre submanifold $\widetilde{\mathfrak{m}} \subset M$ given by
\begin{equation}\label{eq242}
\widetilde{\mathfrak{m}}=\Lambda_{M}(\mathfrak{m})=\pi^{-1}_{M}(\pi_{M}(\mathfrak{m}\cap M)).
\end{equation}
In this way one operates mainly between affine Legendre submanifolds. Nevertheless, to make the prescription globally valued within this range, we have to accept some additional conventions. Namely, we add the empty set $\phi$ to the family of all Lagrange submanifolds of $P$. We also put:
\[
\Lambda_{M}(\phi)=\phi, \quad \Lambda_{M}(\mathfrak{m})=\phi,
\]
if $M$ and $\mathfrak{m}$ are disjoint. One can also complete this by putting $ \Lambda_{M}(\mathfrak{m})=\phi$ if $M$ is not co-isotropic. When defined in this way, the operation $\Lambda_{M}$ satisfies again the above properties (\ref{eq236})-(\ref{eq239}) of the purely algebraic operation $E_{M}$ in symplectic spaces. Admitting the empty set for the parallel affine subspaces we make clear what does it mean "mainly" in the above definition. Linear subspaces always intersect; the affine ones need not, they may be parallel. And on the quasi-classical level the latter situation is a model of orthogonality.\\

c) General symplectic manifolds and projections of Lagrangian manifolds\\
Now let us introduce a final generalization. We assume that $M$ is a simple submanifold of the phase space $(P, \gamma)$; i.e., that it is co-isotropic and globally regular. By this we mean that the singular fibres of $\mathcal{K}(M)$ have constant dimension equal to $m={\rm codim}M$, and that the foliation is such that the quotient set $M/\mathcal{K}(M)$ carries a natural differential structure of dimension $2(n-m)$. Let us consider the set $\mathcal{R}(M)\subset D(P, \gamma)$ of Lagrangian submanifolds of $(P, \gamma)$ which intersect $M$ in a regular way, i.e., like linear spaces. Therefore, for any $p\in \mathfrak{m}\cap M$, $\mathfrak{m}\in \mathcal{R}(M)$, we have
\[
T_{p}(\mathfrak{m}\cap M)=T_{p}\mathfrak{m}\cap T_{p}M.
\]

The operation $\Lambda_{M}$ is defined in such a way that 
\begin{equation}\label{eq245}
\Lambda_{M}(\mathfrak{m})=\pi_{M}^{-1}(\pi_{M}(\mathfrak{m}\cap M)),
\end{equation}
(just like in (\ref{eq242})). And like in (\ref{eq242}) we assume that 
\[
\Lambda_{M}(\phi)=\phi, \quad \Lambda_{M}(\mathfrak{m})=\phi
\]
if $\mathfrak{m}\cap M=\phi$. Nevertheless, there are obvious differences with the definition (\ref{eq242}). Namely, $\Lambda_{M}$ defined like in (\ref{eq245}) does not map $D(P, \gamma)$ onto the whole $D(M)$ but rather onto the subfamily $D_{st}(M)$ which consists of standard Lagrangian subsets of $M$, i.e., ones maximal along $\mathcal{K}(M)$. To finish with this not very rigorous definition, let us repeat the general properties of the operation $\Lambda_{M}:D(P, \gamma) \rightarrow D(M)$:

PROPOSITION 20. {\it $\Lambda_{M}$ is a retraction of $D(P, \gamma)$, or rather, of $\mathcal{R}(M)$, onto $D_{st}(M)$,
\[
\Lambda_{M}|_{D_{st}(P, \gamma)}={\rm Id}_{D_{st}(P, \gamma)}
\]
Therefore, it is also idempotent:
\[
\Lambda_{M}\circ \Lambda_{M}=\Lambda_{M}.
\]
If submanifolds $M, N$ are closed and compatible, then:
\begin{equation}\label{eq249}
\Lambda_{M}\circ \Lambda_{N}=\Lambda_{N}\circ \Lambda_{M}=\Lambda_{N\cap M}.
\end{equation}
And conversely, if $\Lambda_{M}$ commutes with $\Lambda_{N}$, then $M$ is compatible with $N$ and} (\ref{eq249}) {\it is satisfied.

Finally, denoting by $F$ the mapping of $D(P, \gamma)$ onto itself, generated by the canonical transformation $f$ of $P, \gamma$ onto itself, we have:}
\[
\Lambda_{f(M)}=F\circ \Lambda_{M}\circ F^{-1}.
\]

In spite of the formal similarity to (\ref{eq236})-(\ref{eq239}), there are also certain differences, but we choose not to discuss them here. 

\section{Geometrization of the Huygens prescription}

Roughly speaking, for any co-isotropic manifold $M$ the operation $\Lambda_{M}$ produces from any Lagrange manifold $\mathfrak{m}$ its "projection" onto $M$. This is particularly striking when we deal with some co-isotropic foliation $\{ M_{a}: a\in A, {\rm dim }A=m={\rm codim}\ M_{a}\}$. But on the purely symplectic level of $P$ it is impossible to answer, and even to formulate the following questions: "what is the contribution of $\Lambda_{M_{a}}(\mathfrak{m})$, for various $a\in A$, to the total $\mathfrak{m}$?," and "when some co-isotropic submanifold $M \subset P$ is a disjoint sum of co-isotropic submanifolds $M_{a}$, $M= \cup_{a\in}M_{a}$, $M_{a}\cap M_{b}=\phi$, then what are contributions of $\Lambda_{M_{a}}(\mathfrak{m})$ to $\Lambda_{M}(\mathfrak{m})$?" The point is that because of the non-linear structure of the Hamilton-Jacobi equation, $\Lambda_{M_{a}}(\mathfrak{m})$ do not superpose to obtain $\mathfrak{m}$ or  $\Lambda_{M}(\mathfrak{m})$ when $M=\cup_{a\in A}M_{a}$. But it is known that they superpose nonlinearly, in the envelope-wise sense. The envelope of diagrams of a continuous family of solutions of the Hamilton-Jacobi equation is again a solution of the same equation. And it is just here where the concept of superposition coefficients and the weight value of various contributions appears in the "$U(1)$-or its universal covering $\mathbb{R}$-sense."

Let us repeat briefly this envelope superposition principle. Let $S_{a}:Q \rightarrow \mathbb{R}$ be a continuous family of solutions of the Hamilton-Jacobi equation
\begin{equation}\label{eq251}
F\left(q^{i},\frac{\partial S}{\partial q^{i}} \right)=0.
\end{equation}
At this moment it does not matter if $q^{i}$ are the usual coordinates in the configuration space, or space-time variables or something more general. We assume that there are $n$ coordinates $q^{i}$, $i=1, \cdots, n$ and therefore, ${\rm dim}T^{*}Q=2n$. Equation (\ref{eq251}) fixes some $(2n-1)$-dimensional submanifold $M\subset T^{*}Q$ and one asks for Lagrangian submanifolds placed on it given by equations:
\[
\mathfrak{m}: p_{i}-\frac{\partial S}{\partial q^{i}}=0.
\]
Its complete integral is given by a function $S:Q\times A \rightarrow \mathbb{R}$, where ${\rm dim}A=n-1$. The point is that there is no algebraic $S$ itself in (\ref{eq251}); because of this one of the constants may be always declared additive. We assume that the diagrams of functions $S_{a}=S(\cdot, a):Q \rightarrow \mathbb{R}$ are tangentially disjoint, thus that their phase-space lifts; i.e., the manifolds
\[
\mathfrak{M}_{a}:=\{(q, S_{a}(q), dS_{aq}):q\in \mathbb{R}\},
\]
are mutually disjoint in $M \times \mathbb{R}$ (or $M \times U(1)$). Their union 
$\cup_{a\in A}\mathfrak{M}_{a}$ is an image of some cross-section of $M \times \mathbb{R}$ (or $M \times U(1)$) over $M$. 
Sometimes one considers simultaneously an isotropic foliation of $T^{*}Q$ given by:
\[
F(q^{i}, p_{i})=a_{i}
\]
and the corresponding congruence of Hamilton-Jacobi equations. Then the corresponding union will be an image of some cross-section of the total $T^{*}Q \times \mathbb{R}$ over $T^{*}Q$. 

The general solution of Hamilton-Jacobi equations or their systems depends on arbitrary functions. But the particular role of complete integrals consists in that those arbitrary functions may be just functions of the complete-integral parameters; explicitly arbitrary ones. This is a quasiclassical remainder of the usual superposition principle for partial linear equations.

Unlike what people often claim about the role of the optical-mechanical analogy in formulating quantum mechanics, it turns out that it is rather quantum theory that suggests to us some serious reviewing of the classical phase-space concepts.

The superposition principle is a characteristic feature and powerful tool of the theory of linear differential equations. It is very important in field theory, and in particular in the theory of linear waves. 

If $\Psi_{n}$, $n=1, \cdots, N$ are solutions of a system of linear equations, then any linear combination
\[
\Psi=\sum_{n=1}^{N}c_{n}\Psi_{n}
\]
is a solution too; $c_{n}$ denote here arbitrary constants. If one deals with an infinite countable family of solutions $\Psi_{n}$, then the function series
\[
\Psi=\sum_{n=1}^{\infty}c_{n}\Psi_{n}
\]
is a solution as well, provided of course that the series is convergent in an appropriate sense and that it is differentiable term by term. If some family of solutions $\{\Psi_{a}\}$ is labelled by a continuous "index" a running over some arithmetic space $\mathbb{R}^{n}$ or its $n$-dimensional domain, then the "continuous superpositions"
\[
\Psi=\int c(a)\Psi_{a}da^{1} \cdots da^{n}
\]
are solutions, again if the integral does exist and the underlying differential operators do commute with the integration procedure.

If solutions are labelled by points of some differentiable manifold $A$, then one can use generalized superpositions of the form
\[
\Psi=\int \Psi_{a}d\mu(a),
\]
where $d\mu(a)$ is a measure on $A$ including a measure on a finite set, a countably infinite set or a continuous set, and where we presume that the integral exists and the differential operators commute with the integral.

In linear field theories, in particular in theory of linear waves, a very important role is played by the concepts of Green's functions and propagators. They are used to produce special solutions from the source terms (non-homogeneous terms in field equations) and from some initial or boundary data. They are also useful in nonlinear theories with a perturbative structure of nonlinearity; i.e., such that the nonlinear term is a "small" correction to the linear background. This correction is controlled by some "coupling parameter", and certain information about solutions may be obtained from expressing the field as a power series of this "small" parameter. One substitutes this expansion in field equations and solves step by step the resulting hierarchy of equations. This hierarchy is obtained by collecting the terms of the same powers of the parameter. At every stage one deals with non-homogeneous linear differential equations. This leads in general to some asymptotic series representing the solution. The use of propagators in linear wave equations is a mathematical expression of the qualitative idea due to Huyghens about the mechanism of propagation as a superposition of elementary waves radiated from points approached by the wave fronts. As a rule, propagators have distribution-like singularities at the radiating points. We would say that the traditional Huyghens idea, although qualitatively based on the wave picture, was still close to the Newton corpuscular language, because of its stress on envelopes and propagation of wave fronts along geometric rays. Nevertheless, 
the envelope picture describes the wave propagation in the asymptotic range of short waves. The eikonal equation in optics and mechanical Hamilton-Jacobi equation describe the propagation of the phase (eikonal) of short waves. The exciting and mysterious optical-mechanical analogy was the prophecy of quantum mechanics with its convolution of wave and particle concepts. Basing on the geometric concept of envelope one can construct "Huyghens-Fresnel propagators". 

Those propagators also show some kind of singularity at the point of radiation (source of propagation). One deals there with the crises of tangency and envelope concepts. Those critical points resemble distribution-like singularities of propagators and in fact they give an account of the short-wave asymptotics of propagator properties. It turns out however that all singularities disappear, become illusory, if one reformulates the problem from the configuration space/space-time to the appropriate phase space or contact space. 

Let us take a set of Legendre submanifolds corresponding to the complete integral $\{S_{a}:a\in A\}$ of the Hamilton-Jacobi equation (\ref{eq251}). Any solution may be represented by its diagram in $Q\times \mathbb{R}$,
\[
{\rm Graph}{} \ S_{a}:=\{(q, S_{a}(q)):q\in Q\}.
\]
Obviously, the independence of (\ref{eq251}) on the variable $S$ implies that any $S_{a}+t(a)$, where $t(a)\in \mathbb{R}$, is also a solution. But equation (\ref{eq251}) imposes only conditions on the tangent elements of functions; therefore, it is well known that the envelope of diagrams $\{(q, S_{a}(q)+t(a)):q\in Q\}$, i.e.,
\begin{equation}\label{eq260}
{\rm Env}_{a\in A}\{(q, S_{a}(q)+t(a)):q\in Q\}
\end{equation}
also represents some solution. And the arbitrariness of those solutions is that of the choice of functions $t:A \rightarrow \mathbb{R}$. (One can also consider $U(1)$-valued functions). 

PROPOSITION 21. {\it Let us repeat tha}t (\ref{eq260}) {\it is a diagram of
\[
\{(q, S(q)):q\in Q\},
\]
where $S$ is obtained from the family of functions $S_{a}$ and $t$ in the following way}:

1) {\it We write the equations
\[
\frac{\partial}{\partial a_{i}}(S_{a}(q)+t(a))=0
\]
and solve them with respect to $a$. One obtains then some $q$-dependent solution, $a(q)$}.

2) {\it We substitute this solution in $S_{a}(q)+t(a)$, obtaining the expression denoted by the Stat-symbol,}
\begin{equation}\label{eq263}
S(q)=S_{a(q)}(q)+t(a(q))={\rm Stat}_{a\in A}(S_{a}(q)+t(a)).
\end{equation}
More precisely: for a differentiable function $f$, ${\rm Stat}f$ is the set of values of $f$ at its all stationary points,
\[
{\rm Stat} f:=\{ f(x):df_{x}=0\}.
\]
Elements $y$ of ${\rm Stat}f$ are stationary values of $f$. When $f$ is a function on the manifold $M$, it is  sometimes convenient to write ${\rm Stat}f$ alternatively as 
\[
{\rm Stat}_{M}f={\rm Stat}_{x \in M}f(x).
\]

This $S$ is a new solution, labelled by an arbitrary function $t: A\rightarrow \mathbb{R}$ (or $t: A \rightarrow U(1)$). Obviously, everything is simple when the solution for $a(q)$ is unique. If there is a connected family of solutions, it is also good, because (\ref{eq263}) does not depend on the choice of $a(q)$. However if there are a few discrete solutions, then $S(q)$ is a multivalued function of $q$. 

This fact belongs to the theory of Hamilton-Jacobi equations. However, it may be also "derived" from the continuous superpositions of wave functions satisfying the Schr\"{o}dinger or other linear wave equations, by performing the WKB-limit transition $\hbar \rightarrow 0$ on the following continuous superposition
\begin{equation}\label{eq264}
\int w(a){\rm exp}\left( \frac{i}{\hbar}t(a)\right)D(q, a){\rm exp}\left( \frac{i}{\hbar}S(q, a)\right)d_{n}a.
\end{equation}
One must stress again that there are strong arguments against manipulating constants like $\hbar$. Therefore, one must stress carefully that the above limiting procedure is a shorthand for considering rapidly-oscillating functions.

The same concerns the scalar product of wave functions. The phase of its quasi-classical expression may be obtained as follows in the WKB-limit. Let $\Psi_{1}$, $\Psi_{2}$ be two quickly oscillating wave functions. 
\[
\Psi_{1}=\sqrt{D_{1}}{\rm exp}\left( \frac{i}{\hbar}S_{1}\right), \quad \Psi_{2}=\sqrt{D_{2}}{\rm exp}\left( \frac{i}{\hbar}S_{2}\right).
\]
Let us stress, they need not be just quantum wave functions, they may be as well any quickly oscillating amplitudes of some wave processes, and $\hbar$ may be as well any parameter determining that they are "almost geometrical". 

PROPOSITION 22. {\it Take the scalar product
\[
\langle\Psi_{1}|\Psi_{2}\rangle=\sqrt{D}{\rm exp}\left( \frac{i}{\hbar}S\right)=\int \overline{\Psi_{1}}(q)\Psi_{2}(q)d_{n}q.
\]
After calculating the WKB-limit, one obtains
\begin{equation}\label{eq267}
S={\rm Stat}(S_{2}-S_{1})=(S_{2}-S_{1})(q_{0})
\end{equation}
where}
\[
d(S_{2}-S_{1})_{q_{0}}=0.
\]

Again we get the same thing we were faced with in the quasi-classical superposition: Everything is very well when there is a single such $q_{0}$, or when they form a connected subset; then (\ref{eq267}) is unique. In other cases, when there is a finite or countable number of such $q_{0}$'s, one obtains a whole family of the values $(S_{2}-S_{1})(q_{0})$.

\section{The message of contact geometry}

Let us now describe the quasiclassical superposition of scalar products and $U(1)$-gauge transformations in geometric terms. To do that, we shall use the contact geometry, i.e., the odd-dimensional companion of a symplectic manifold \cite{JJS}. We were already dealing with such structures in non-conservative mechanics with time-dependent Hamiltonians $H(t; q, p)$. Namely, the contact manifold then was defined as the direct product $C=\mathbb{R}\times P$ with the one-form given locally by 
\[
\Omega_{H}=p_{i}dq^{i}-H(t,q,p)dt.
\]
The corresponding presymplectic structure was given by
\[
\Gamma_{H}=d\Omega=dp_{i}\wedge dq^{i}-dH \wedge dt,
\]
and trajectories were given by its singular vector field which in these coordinates has the form:
\[
X_{H}=\frac{\partial }{\partial t}+\frac{\partial H}{\partial p_{i}}\frac{\partial }{\partial q^{i}}-\frac{\partial H}{\partial q^{i}}\frac{\partial }{\partial p_{i}}.
\]
The proof of the connection between the Hamilton-Jacobi equation and canonical Hamilton equations is based on introducing new variables in which $\Omega_{H}$ takes the form
\begin{equation}\label{eq272}
\Omega=p_{i}dq^{i}-dz.
\end{equation}
The form (\ref{eq272}) is the typical local expression of the idea of contact geometry, but it is not a formal definition. Let us define formally the contact space: We say that in a manifold $C$ the contact structure is given when $C$ is a principal fibre bundle over the symplectic manifold (phase space)($P, \gamma$) with the one-dimensional structure group $\mathbb{R}$ (additive) or perhaps $U(1)$ (multiplicative) and with the connection one-form $\Omega$ for which the curvature two-form $\Gamma$ is a pull-back of $\gamma$ under the projection $\pi:C \rightarrow P$, $\Gamma=d\Omega=\pi^{*}\gamma$.

Obviously, this means that, locally, $\Omega$ is given by (\ref{eq272}) and the principal vector field is given by
\[
k=-\frac{\partial }{\partial z}.
\]

This is common to geometric quantization, getting polarization conditions, etc. We won't go into those topics, however. Instead, we send the reader to the known books by Simms and Woodhouse, \'{S}niatycki, Guillemin and Sternberg, Kostant, Tulczyjew and others \cite{GuiSte,20,Ko,SiWo,Sn}.

Nevertheless, with a given $(P, \gamma)$ there are usually various topologically inequivalent contact raisings of $(P, \gamma)$ to $(C, \Omega)$. We won't deal with such problems. To avoid them, we assume that $C$ is primary and $\gamma$ a secondary object, or just put locally $C=\mathbb{R}\times P$. The contact manifold $(C, \Omega)$ provides a linearization of the second-order Pfaff problem for $\gamma$ and introduces the one-dimensional action of the vertical structure group $\mathbb{R}$ or $U(1)$. Integral surfaces of $\Omega$ are called horizontal. Any isotropic submanifold $\mathfrak{m}$ of $P$ admits a family of one-dimensional horizontal lifts foliating $\pi^{-1}(\mathfrak{m})$. The structure group transforms the various lifts $\mathfrak{M}$ into each other. In particular, this concerns maximally-dimensional, i.e., $n$-dimensional horizontal submanifolds of $C$, i.e., Legendre submanifolds. It was said above and also in \cite{JJS} that Lagrange submanifolds in a phase space do correspond to supports of quasiclassical probability distributions. But they do not feel the action of the structure group of $\mathbb{C}$. In particular, they cannot be "multiplied" by numbers and superposed. Legendre submanifolds admit this, as it was seen in (\ref{eq263}), (\ref{eq264}). And there exists their scalar product in the sense of (\ref{eq267}). Let us describe those structures in terms of the geometry of $(C, \Omega)$. The set of all Legendre submanifolds will be denoted by $\mathcal{H}(C)$ or briefly by $\mathcal{H}$ when it is clear what $C$ is.

All Lagrange concepts may be lifted horizontally from $P$ to $C$. Constraints $M\subset P$ are lifted to $C$ as $\pi^{-1}(M)$. Their singular foliation $\mathcal{K}(M)$ may be lifted horizontally from $M$ to $\pi^{-1}(M)$. There they become 
\[
\mathcal{K}^{\Omega}(M)=h \ lift \mathcal{K}(M).
\]
(We mean co-isotropic constraints in $P$). But as it concerns Lagrangian submanifolds in $(P, \gamma)$, one considers them to be byproducts of Legendre submanifolds in $C$, $n$-dimensional horizontal submanifolds. Those Lagrangian submanifolds, (quasiclassical wave functions) may to be translated by the group $\mathbb{R}$ (additive) or $U(1)$, (multiplicative) when $S$ is taken modulo $h=2\pi \hbar$.

Let $\mathfrak{M}_{1}$, $\mathfrak{M}_{2}$, be two Legendre submanifolds such that their projections $\mathfrak{m}_{1}$, $\mathfrak{m}_{2}$ from $C$ onto $P$ intersect at a single point or along some connected and simply-connected region in $P$. Then there is exactly one element $t$ of the structural group such that $g_{t}\mathfrak{M}_{1}\cap \mathfrak{M}_{2}\neq \phi$. If the above natural assumption about the intersection of $\mathfrak{m}_{1}$, $\mathfrak{m}_{2}$ is not satisfied, e.g., if this intersection consists of a finite or discrete number of connected components, then in general there will be a finite or discrete number of the group elements $t$.

DEFINITION 2. {\it The Huygens scalar product or vertical distance $[\mathfrak{M}_{1}| \mathfrak{M}_{2}]$ of $\mathfrak{M}_{1},\mathfrak{M}_{2}\in \mathcal{H}(C)$ (the set of all Lagrange submanifolds in $C$) is defined as a subset of the structural group consisting of such $t\in [\mathfrak{M}_{1}| \mathfrak{M}_{2}]$ that $\mathfrak{M}_{2}\cap g_{t} \mathfrak{M}_{1}\neq \phi$. We say that $\mathfrak{M}_{1}$, $\mathfrak{M}_{2}$ are orthogonal if the set $[\mathfrak{M}_{1}| \mathfrak{M}_{2}]$ is empty}.

Any symplectic (canonical) transformation of $P$ onto itself, $\varphi:P \rightarrow P$ may be uniquely lifted to the mapping $\overline{\varphi}:C \rightarrow C$ which projects to $P$ onto $\varphi$ and preserves $\Omega$,
\[
\pi\cdot\overline{\varphi}=\varphi\cdot \pi, \quad \overline{\varphi}^{*}\Omega=\Omega.
\]

Such mappings $\overline{\varphi}$ are called special contact transformations. To be more precise, it is sufficient to assume only the second condition; the first is then an automatic consequence.

It is easy to see that all such transformations are unitary mappings of the set of Legendre manifolds; i.e., they preserve their scalar products,
\[
[\varphi\mathfrak{M}_{1}|\varphi\mathfrak{M}_{2}]= [\mathfrak{M}_{1}| \mathfrak{M}_{2}]
\]
for any pair $\mathfrak{M}_{1}$, $\mathfrak{M}_{2}\in \mathcal{H}$. There are also linear superpositions. First, one must define the superposition and projector, in the action on Legendre manifolds.

Let us take a differential submanifold $N\subset C$. We say that its determinant set, or characteristic set $\Sigma (N)$ is the set which consists of such points $z\in N$ that
\begin{equation}\label{eq276}
\Omega_{z}|_{T_{z}N}=0.
\end{equation}  

DEFINITION 3. {\it Let us now assume some family $\{\mathfrak{M}_{a}:a\in A\}$ of Legendre submanifolds of $C$, $\{\mathfrak{M}_{a}\in  \mathcal{H}(C)\}$. The superposition of $\mathfrak{M}_{a}$-s, denoted by 
\[
\mathfrak{M}=E_{a\in A}\mathfrak{M}_{a},
\]
is defined as the maximal element of $\mathcal{H}(C)$ contained in the characteristic set $\Sigma(\cup_{a\in A}\mathfrak{M}_{a})$ defined in} (\ref{eq276}).

Let us give a few examples:
\begin{itemize}
\item [$1)$] In the contact space $T^{*}Q\times \mathbb{R}$ we take the family of Legendre manifolds with definite positions, $\mathfrak{M}_{q}:=(T_{q}{}^{*}Q, 0)$, and the manifold $\mathfrak{M}_{S}$ given by equations $p_{i}=\partial S / \partial q^{i}$, $i=1, \cdots, n$; i.e.,
\[
\mathfrak{M}_{S}=\{(dS_{q}, S(q)):q\in Q\}.
\]
Then, the following holds:
\[
\mathfrak{M}_{S}=E_{q\in Q}[S(q)]\mathfrak{M}_{q},
\] 
where for any number $t$, $[t]\mathfrak{M}$ denotes $\mathfrak{M}$ raised by $t$ in the $z$-direction.

\item [$2)$] Consider again the contact space $(C, \Omega)=T^{*}Q \times \mathbb{R}$ with the usual contact form $\Omega_{Q}$. Take some function $S:Q\times A \rightarrow \mathbb{R}$, $f:A \rightarrow \mathbb{R}$ and the family of Legendre manifolds,
\[
\mathfrak{M}_{a}:=\mathfrak{M}_{S(\cdot, a)}=\{ (dS(\cdot, a)_{q}, S(q, a)):q\in Q\}.
\]  
Assume that 
\[
\mathfrak{M}_{S}=E_{a\in A}[f(a)]\mathfrak{M}_{a}=\{ (dS_{q}, S(q)):q\in Q\}.
\]
Then
\[
S(q)={\rm Stat}_{a \in A}(S(q, a)+f(a)).
\]

The projection of $\mathfrak{M}_{S}$ onto $Q\times \mathbb{R}$ (or $Q \times U(1)$), i.e.,
\[
\varepsilon_{S}:=\{(q, S(q)):q \in Q\}\subset Q \times \mathbb{R},
\]
is the usual envelope of the set of surfaces
\[
\varepsilon_{a}:=\varepsilon_{S(\cdot, a)}=\{(q, S(q, a)):q \in Q\}\subset Q \times \mathbb{R}.
\]
This is a regular situation. Unlike this, in the former example we were dealing with  a singular situation. There $\mathfrak{M}_{S}$ was the "envelope" of the $n$-parameter family of $0$-dimensional manifolds $\{(dS_{q}, S(q))\}$. Nevertheless, in the phase-space language that situation was just as regular as the present one.

\item [$3)$] Next we assume that $Q$ is an $n$-dimensional linear space $V$, therefore $T^{*}Q$ becomes $V \times V^{*}$ and $C$ becomes $V \times V^{*} \times \mathbb{R}$ (or $V \times V^{*} \times U(1)$). Let us consider the following families of Legendre submanifolds of the fixed positions or momenta 
\[
\mathfrak{M}[x]=\{(x,p,0):p \in V^{*}\}, \quad \mathfrak{M}[p]=\{(x,p,\langle p, x\rangle):x \in V\}.
\]
Then the following Fourier rules are satisfied:
\[
\mathfrak{M}[p]=E_{x \in V}[\langle p, x\rangle] \mathfrak{M}[x], \quad \mathfrak{M}[x]=E_{p \in V^{*}}[ - \langle p, x\rangle] \mathfrak{M}[p]
\]
and for any (sufficiently smooth) function $S:V \rightarrow \mathbb{R}$
\[
\mathfrak{M}_{s}=E_{x \in V}[S(x)] \mathfrak{M}[x]=E_{p \in V^{*}}[\hat{S}(p)]\mathfrak{M}[p],
\]
where the following relationships are satisfied:
\[
\hat{S}[p]={\rm Stat}_{x \in V}(S(x)-\langle p, x\rangle), \quad S[x]={\rm Stat}_{p \in V^{*}}(\hat{S}(p)+\langle p, x\rangle).
\] 
\end{itemize}
It turns out that superposition of Legendre submanifolds behave in a "linear" way under special contact transformations, i.e., under the $C$-lifts of canonical mappings. More precisely:

Let $F: \mathcal{H}(C) \rightarrow \mathcal{H}(C)$ be a mapping of the set of Legendre transformations onto itself, induced by some special contact transformation $f: C \rightarrow C$; $\Omega=f^{*} \Omega$. Then for any system of coefficients $t_{a}$, the following holds:
\[
FE_{a \in A}[t_{a}]\mathfrak{M}_{a}=E_{a \in A}[t_{a}]F\mathfrak{M}_{a}.
\]
Similarly, the vertical distance of Legendre manifolds is also preserved by contact transformations:
\[
[U\mathfrak{M}_{1}|U\mathfrak{M}_{2}]=[\mathfrak{M}_{1}|\mathfrak{M}_{2}].
\]

Finally, let us define the projector $\Pi_{M}: \mathcal{H}(C) \rightarrow \mathcal{H}_{M}(C)$, where $\mathcal{H}_{M}(C)$ denotes the set of Legendre submanifolds  of $C$ contained in $\pi^{-1}(M)$, where $M \subset P$ are co-isotropic constraints. This projector is defined by the following pair of conditions:
\[
\Pi \circ \Pi_{M}=\Lambda_{M} \circ \Pi
\]
\[
(\Pi_{M}\mathfrak{M})\cap \mathfrak{M}=(\pi^{-1}(M))\cap \mathfrak{M}.
\]
The operation $\Pi_{M}$ has natural properties strongly related to those of $\Lambda_{M}$ above, in particular:

 PROPOSITION 23. 1) {\it $\Pi_{M}$ is a retraction of $\mathcal{H}(C)$ onto $\mathcal{H}_{stM}(C)$, i.e., onto the set of "saturating" Legendre submanifolds of $\pi^{-1}(M)$, i.e., ones containing the whole fibres of the horizontal lift of $\mathcal{K}(M)$. In particular $\Pi_{M}$ is idempotent}
\[
\Pi_{M} \circ\Pi_{M}=\Pi_{M}. 
\]

2) {\it If $M, N$ are compatible co-isotropic constraints in $P$, then} 
\begin{equation}\label{eq292}
\Pi_{M} \circ\Pi_{N}=\Pi_{N} \circ \Pi_{M}=\Pi_{M\cap N}. 
\end{equation}

3) {\it If the commutativity
\[
\Pi_{M} \circ\Pi_{N}=\Pi_{N} \circ \Pi_{M},
\]
holds, then $M, N$ are compatible and} (\ref{eq292}) {\it is satisfied.}

4) {\it If $F$ is a mapping of $\mathcal{H}(C)$ onto itself induced by the special contact transformation $f$, then} 
\[
\Pi_{f(M)}= F\circ \Pi_{M} \circ F^{-1}.
\]

5) {\it Projection operators $\Pi_{M}$ are "linear" in the Huygens form; therefore,}
\[
\Pi_{M}E_{a \in A}[T_{a}]\mathfrak{M}_{a}=E_{a \in A}[T_{a}]\Pi_{M}\mathfrak{M}_{a}.
\]

{\it The last property is a new one in comparison with the previous ones; its essential novelty is just the occurence of the superposition operation.

It may happen that a family of subsets $\{\mathfrak{M}_{a}:a \in A\}\subset \mathcal{H}(C)$ has the following property. The subsets $\mathfrak{m}_{a}=\pi(\mathfrak{M}_{a})$ are leaves of a polarization of $(P, \gamma)$, i.e., they foliate regularly $P$, resulting in an $n$-dimensional quotient manifold $A$. Then, under certain additional conditions, for simplicity omitted here, we have
\[
\mathfrak{M}=E_{a \in A}[\mathfrak{M}_{a}|\mathfrak{M}]\mathfrak{M}_{a},
\]
i.e., the $\mathfrak{M}_{a}$'s form a "basis" with respect to which $\mathfrak{M}$ may be "orthogonally expanded"}.

All those statements concerning superpositions in $C$, i.e., envelopes in $Q \times \mathbb{R}$, are a bit incorrect. This is typical for any statements concerning envelopes. To make them rigorous, one has either to restrict the class of considered objects, e.g., to affine ones, or to be more precise, to quadratic function $S$, or, to commit a rather heavy research. We shall not do it here; instead we finish with a few "non-rigorous formulas." 

PROPOSITION 24. {\it For example, let $\{\mathfrak{M}_{q}:a \in Q\}$ be a basis of $\mathcal(C)$ and take a contact transformation $u$ in $C$, i.e., a lift of some symplectic transformation in $P$. Let us assume that $u$ is transversal with respect to this basis, i.e., that it projects onto such a symplectic transformation $\varphi$ of $P$ onto itself that for any projection $\mathfrak{m}_{q}=\pi(\mathfrak{M}_{q})$ onto $P$ the following holds:
\[
\varphi(\mathfrak{m}_{q})\cap \mathfrak{m}_{q'}
\]
is a one-element set for any $q, q'\in Q$. Then the following holds:
\[
U\mathfrak{M}_{q}=E_{q\in Q}U(q', q)\mathfrak{M}_{q'}, \quad U(q', q)=[\mathfrak{M}_{q'}|U\mathfrak{M}_{q}]
\]
and for any
\[
\mathfrak{M}=E_{q \in Q}[S(q)] \mathfrak{M}_{q},
\]
we have 
\[
U\mathfrak{M}=E_{q \in Q}[S'(q)] \mathfrak{M}_{q}=E_{q \in Q}[S(q)]U\mathfrak{M}_{q},
\]
where
\[
S'[q]={\rm Stat}_{q' \in Q}(U(q, q')+S(q')).
\]
$W(q, q')$ is the $W$-type generating function for $U$. If the above conditions are not literally satisfied, it may be still interpretable as a generalized generating function} \cite{JJS}.

\section{Relation \  \ between \ contact \ geometry and \\ Hamilton-Jacobi equations}

Only very roughly, without a sufficient mathematical rigour, we mention now the relationships between the concepts of contact geometry and the Cauchy (or boundary) problems for the systems of Hamilton-Jacobi equations. The analogy with the corresponding problems for linear partial differential equations, first of all the quantum-mechanical ones becomes visible \cite{1}, \cite{9}, \cite{10}, \cite{11}, \cite{16}, \cite{17}, \cite{GuiSte}, \cite{20}, \cite{36}, \cite{37}.

Let $X$ be a manifold, e.g., space-time; it does not matter if Galilean, Minko-\newline wskian, generally-relativistic, or even some more general one. In the cotangent bundle $T^{*}X$ some first-class submanifold $M$ is fixed, given by the system of equations $F_{a}=0$, $a=1, \cdots, m$. Here $F_{a}$ are phase-space functions at least weakly in involution. The corresponding system of Hamilton-Jacobi equation is locally given by equations
\[
F_{a}\left(\cdots, x^{\mu}, \cdots; \cdots, \frac{\partial S}{\partial x^{\nu}}, \cdots \right)=0.
\]

Any fiber $T_{x}{}^{*}X$ of the cotangent bundle may be $\Lambda_{M}$-projected onto the set $D(M)$ of Lagrange submanifolds of $M$, resulting in some manifold
\[
\mathfrak{m}_{x}:=\Lambda_{M}(T_{x}{}^{*}X).
\]
The $(N+1)$-dimensional submanifold $\pi^{-1}(\mathfrak{m}_{x})$ of the contact space $C$ $({\rm dim}X=N)$ is foliated by the family of Legendre manifolds-horizontal lifts of $\mathfrak{m}_{x}$. If $C=T^{*}X\times \mathbb{R}$ or $C=T^{*}X\times U(1)$, this is the family of manifolds
\[
\mathfrak{M}_{(x,c)}=\Pi_{M}(T_{x}{}^{*}X, c).
\]
Then locally we have
\[
\mathfrak{m}_{x}\cap T_{y}{}^{*}X=\{d\sigma (x, \cdot)_{y}\},
\]
\[
\mathfrak{M}_{x}\cap (T_{y}{}^{*}X \times \mathbb{R})=\{(d\sigma (x, \cdot)_{y}, \sigma(x, y))\},
\]
where the two-argument "space-time" function $\sigma$ is uniquely defined. It is called a two-point characteristic function or a propagator of $M$, or rather - of the corresponding Hamilton-Jacobi system of equations. And the function $\sigma(x, \cdot)$ will be referred to as a propagator of our system at $x \in X$. One can also use the term "fundamental solution" of the system at $x \in X$.

Let $x, y$ be two points of $X$ and let us consider a family of curves in $M$ joining the fibres $T_{x}{}^{*}X$, $T_{y}{}^{*}X$ and placed entirely in a characteristic band through $x, y$. Therefore, their tangent vectors $u$ satisfy
\[
u\rfloor \gamma || M=0, \quad i.e.,\quad  \gamma(u, v)=0
\]
for any vector $v$ tangent to $M$. In other words, they are solutions of the homogeneous dynamics given by constraints $M$ in $T^{*}X$. Of course, we have in mind mainly a hypersurface situation, when ${\rm dim}M=2N-1=2{\rm dim}X-1$, but the concepts are applicable in a more general situation as well. One can show that for any pair of points $x, y \in X$ we have:
\[
\sigma_{M}(x, y)=\int _{l(x, y)}\omega=\int _{l(x, y)}p_{\mu}dx^{\mu},
\]
where $l(x, y)$ denotes any of the mentioned curves from $x$ to $y$. In a hypersurface case, for sufficiently close points $x, y$ there is, obviously, exactly one curve $l$ of this type. With the obvious exception of causality-type restrictions which, in the non-relativistic case forbid $x, y$ to be simultaneous events, and in the Minkowski case they forbid the space-like mutual relationship between them. If $x, y$ are "finitely-separated," then it may happen that $l(x, y)$ is not unique and there are several branches of values for $\sigma$. One can show that if
\[
\Pi_{M}\mathfrak{M}_{S}=\mathfrak{M}_{S'}, \quad \Lambda_{M}\mathfrak{m}_{S}=\mathfrak{m}_{S'},
\]
then
\[
S'(x)={\rm Stat}_{y}(S(y)+\sigma_{M}(y, x)).
\]

PROPOSITION 25. {\it The idempotence property of $\Pi_{M}$, $\Lambda_{M}$ implies that
\[
\sigma_{M}(x, y)={\rm Stat}_{z}(\sigma_{M}(x, z)+\sigma_{M}(z, y)).
\]
And more generally, we have the "Feynmann rule":}
\[
\sigma_{M}(x, y)={\rm Stat}_{(z_{1}\cdots z_{k})}(\sigma_{M}(x, z_{1})+\sigma_{M}(z_{1}, z_{2})+ \cdots +\sigma_{M}(z_{k}, y)).
\]

Let us now assume that $\Sigma \subset X$ is a Cauchy surface, whatever should it mean, for the homogeneous dynamics. Then the unique solution of the Cauchy problem has the form
\[
S(x)={\rm Stat}_{q \in \Sigma}(f(q)+\sigma_{M}(q, x)),
\]
where $f:\Sigma \rightarrow \mathbb{R}$ are "initial data". So the two-point characteristic function is a Hamilton-Jacobi propagator.

Let us quote an example, namely the dynamics of the free material point in Galilean space-time. Then one can show that the corresponding phase of the Schr\"{o}dinger propagator equals 
\[
\frac{1}{\hbar}\sigma_{M}(x, y)=\frac{1}{\hbar}S(a, z; q, t)=\frac{m}{2\hbar(t-z)}g_{ij}(q^{i}-a^{i})(q^{j}-a^{j}).
\]
The corresponding Van Vleck determinant,
\begin{equation}\label{eq296+14}
{\rm det}\left[ \frac{\partial ^{2}S}{\partial q^{i}\partial a^{j}}\right]
\end{equation}
also essentially corresponds to the quasiclassical expression for the modulus of the Schr\"{o}dinger propagator.

When multiplying the square root of (\ref{eq296+14}) by an appropriate normalization constant, one obtains from the complex scalar density of weight one,
\[
\sqrt{\left[ \frac{\partial^{2}S}{\partial q^{i}\partial a^{j}}\right]}{\rm exp}\left(\frac{im}{2\hbar(t-z)} g_{kl}(q^{k}-a^{k})(q^{l}-a^{l})\right)
\]
the following expression:
\[
\mathcal{K}(\overline{\xi}, \tau)=\left(\frac{m}{2\pi i \hbar \tau} \right)^{n/2}{\rm exp}\left(\frac{im}{2\hbar\tau} \overline{\xi}^{2}\right)
\]
where
\[
\tau=t-z, \quad \xi^{k}=q^{k}-a^{k}, \quad \overline{\xi}^{2}=g_{kl}\xi^{k}\xi^{l}.
\]
The mentioned normalization means that
\[
\lim_{\tau\rightarrow 0}\mathcal{K}(\overline{\xi}, \tau)=\delta(\overline{\xi}).
\]

Let us observe that the $\mathcal{K}$ - function is the usual propagator of initial conditions for the Schr\"{o}dinger equation for a free particle in $n$ dimensions of the Euclidean space:
\[
\hbar i \frac{\partial \Psi}{\partial t}=-\frac{\hbar^{2}}{2m}\Delta\Psi=-\frac{\hbar^{2}}{2m}g^{ij}\partial_{i}\partial_{j}\Psi.
\]
But the above construction is purely classical. It is simply the classical Hamilton-Jacobi equation for the free particle, its complete integral given by the two-point characteristic function and the canonical Van Vleck probabilistic density built of the mentioned complete integral. Everything might be found and even physically interpreted practically without any knowledge of quantum mechanics. It is just the Huygens-Fresnel optico-mechanical analogy based entirely on the Hamilton-Jacobi eikonal equation.

\section{Hamiltonian systems on Lie groups and their coadjoint orbit representations}

As mentioned above, every Hamiltonian system with some symmetry Lie group $G$ acting symplectically on its phase space (and preserving a Hamiltonian) might be realized in some coadjoint orbit of this group or otherwise in the disjoint union of orbits of $G/H$, $H$ denoting a closed subgroup of $G$. Therefore, an interesting class of examples is obtained when one assumes that the configuration space $Q$ may be identified with some Lie group $G$ or its homogeneous space $G/H$. The original phase space is then traditionally given by the cotangent bundle $T^{*}Q$; however, there is an important class of models where at a certain stage one can "forget" about this and concentrate on the coadjoint orbit description. This covers, as particular special cases and interesting examples, certain problems in rigid body mechanics, a classical spinning particle, or the mechanics of deformable bodies. Of course, there are important physical models like the Galilei group, when the model of the co-adjoint orbit rather fails. It is interesting that the language of the co-adjoint orbits in the central extension be an alternative with respect to the system of disjoint co-adjoint orbits. 

Let $G$ be a real Lie group. We are dealing only with linear groups; so it is assumed that $G$ is a subgroup of ${\rm GL}(N, \mathbb{R})$ or ${\rm GL}(N, \mathbb{C})$. Notice that, e.g., the unitary group $U(N)\subset {\rm GL}(N, \mathbb{C})$) is a real Lie group with a real Lie algebra, in spite of having complex matrix entries. Sometimes it is convenient to write in a more "sophisticated" way that $G\subset {\rm GL}(W)$, $W$ denoting some real or complex linear space of finite dimension.

The group $G$ acts on itself through two transformation groups consisting of left and right regular translations.

The corresponding actions of $k\in G$ are denoted respectively as 
\[
G\ni g \mapsto L_{k}(g)=kg
\]
\[
G\ni g \mapsto R_{k}(g)=gk.
\]
If $G$ is non-Abelian, those are different transformation groups, although not always disjoint; if $k$ belongs to the centre $Z\subset G$, then obviously $L_{k}=R_{k}$.

In the case of a homogeneous space realized as $G/H$, the manifolds of left cosets, $L_{k}$ acts as 
\[
L_{k}(gH)=kgH.
\]

Sometimes, it is not recognized that some right-acting transformations also may exist. Namely, let $N\subset G$ be the normalizer of $H$, i.e., the maximal subgroup such that $H\subset N$ and $H$ is a normal subgroup of $N$. Then any $l\in N$ acts on the right as follows:
\begin{equation}\label{eq304}
G/H \ni gH \rightarrow glH=gHl.
\end{equation}

Realization of the homogeneous space in terms of the left or right coset manifolds, $G/H$ or $H \backslash G$ formally is a matter of convention, although it happens that there are some physical reasons for choosing a particular one of two conventions. 

If we use the language of linear groups, then the Lie algebra $G'$ is defined as a linear subspace of $L(N, \mathbb{R})$ or $L(N, \mathbb{C})$, sometimes written as $L(W)$, tangent to the "surface" $G \subset {\rm GL}(W)$ at the identity element $Id_{W}$:
\[
G'=T_{Id_{W}}G\subset L(W).
\]
Here $W$ is a complex or real linear space of dimension $N$.

Let $q^{1}, \cdots, q^{n}$ be coordinates on $G$; usually they are chosen in such a way that their vanishing values correspond to the identity element, $g(0, \cdots, 0)=Id_{W}$ (analytically the identity $N \times N$ matrix). The natural basis of $G'$ is given by linear mappings/matrices
\[
E_{a}=\frac{\partial g}{\partial q^{a}}(0, \cdots, 0).
\]
The Lie-algebraic property of $G'$ is that it is closed under the matrix commutator, so that 
\[
[G', G']\subset G', \quad [E_{a}, E_{b}]=C^{d}{}_{ab}E_{d},
\]
$C^{i}_{jk}$ denoting structure constants with respect to a given basis dual to the coordinates $q^{i}$: $E_{a}(q^{i})=\delta^{i}{}_{a}$.

Very often, although not always, one uses canonical coordinates of the first kind,
\[
g(q^{1}, \cdots, q^{n})=\exp (q^{a}E_{a}).
\]

For example, the rotation vector used in the analytical description of ${\rm SO}(3, \mathbb{R})$ belongs to this class of coordinates, but commonly used Euler angles do not.

Motion is described by smooth curves in $G$,
\[
\mathbb{R}\ni t \mapsto g(t) \in G,
\]
and $dq^{a}/ dt$ are generalized velocities.

In many problems the analysis of motion in homogeneous spaces $G/H$ may be considered a byproduct of analysis in the group itself. For example, motion of a material point on the sphere ${\rm S}^{2}(0, 1)={\rm SO}(3, \mathbb{R})/{\rm SO}(2, \mathbb{R})$ maybe considered as a byproduct of motion in ${\rm SO}(3, \mathbb{R})$, i.e., of the rigid body dynamics in $\mathbb{R}^{3}$ (without translational motion). So let us remain on the level of group $G$. It is both computationally convenient and theoretically deeply justified to replace $dq/dt$ or $dq^{i}/dt$ by Lie-algebraic objects of two alternative types
\[
\Omega=\frac{dg(t)}{dt}g(t)^{-1}, \quad \hat{\Omega}=g(t)^{-1}\frac{dg(t)}{dt};
\]
obviously,
\[
\Omega=g\hat{\Omega}g^{-1},
\]
where all quantities are taken at the same instant of time. $\Omega$, $\hat{\Omega}$ are elements of the Lie algebra $G'$ \cite{24}, \cite{KarMas}, \cite{26}, \cite{27}, \cite{JJS}-\cite{JJSc}. 

In rigid body mechanics, where $G={\rm SO}(n, \mathbb{R})$, and $G'={\rm SO}(n, \mathbb{R})'$ consists of skew-symmetric matrices, $\Omega$ and $\hat{\Omega}$ represent the angular velocity respectively with respect to the space-fixed and body-fixed system of axes. We will see that in the case of classical versus quantum mechanics, this difference and the corresponding difference between $L_{k}$ and $R_{k}$ is also crucial. In the physical case $n=3$, $\Omega$ and $\hat{\Omega}$ each have three independent components from which one builds axial vectors of angular velocity. This property is a peculiarity of dimension three. And in dimension $n=2$ (planar rotator), $\Omega$ and $\hat{\Omega}$ have only one independent pseudo-scalar component.

The group translations are 
\[
G \ni g \mapsto (L_{k}R_{h})(g)=kgh.
\]
They affect the quantities $\Omega$, $\hat{\Omega}$ as follows:
\[
\Omega \mapsto k\Omega k^{-1}=Ad_{k}\Omega, \quad \hat{\Omega} \mapsto h^{-1}\hat{\Omega} h=Ad_{h^{-1}}\hat{\Omega};
\]
this is a mixture of the adjoint rules and invariance.

Sometimes one uses the analytical representation:
\[
\Omega=\Omega^{a}E_{a}, \quad \hat{\Omega}=\hat{\Omega}^{a}E_{a}; \quad \Omega^{a}=\Omega^{a}{}_{i}(q)\frac{dq^{i}}{dt}, \quad \hat{\Omega}^{a}=\hat{\Omega}^{a}{}_{i}(q)\frac{dq^{i}}{dt}.
\]

We indicate here explicitly the dependence of $\Omega^{a}{}_{i}$, $\hat{\Omega}^{a}{}_{i}$ on coordinates $q^{i}$. If $G$ is non-Abelian, e.g. Galilei, Poincare, or rotations, this dependence is unavoidable and there are no generalized coordinates $Q^{a}$ for which $\Omega^{a}$ would be time derivatives; similarly for  $\hat{\Omega}^{a}$. In this sense $\Omega$, $\hat{\Omega}$ are non-holonomic velocities, or quasivelocities based on the group $G$. Unlike the general tangent bundle $TQ$, the tangent bundle $TG$ may be trivialized in the sense of the above construction in two canonical ways:
\[
TG\simeq G\times G',
\]
in the sense that the elements of $TG$ are represented by pairs
\[
(g, \Omega) \quad or \quad (g, \hat{\Omega}),
\]
i.e., configuration and generalized angular velocity, respectively in "spatial" or "co-moving" representations. The other way, the cotangent bundle splits:
\[
T^{*}G\cong G \times G'^{*}
\]
and is alternatively represented by pairs
\[
(g, \Sigma), \quad (g, \hat{\Sigma})
\]
consisting of configuration $q$ and "generalized angular momentum" $\Sigma$ or $\hat{\Sigma}$, respectively in the spatial and co-moving representations \cite{24}, \cite{KarMas}, \cite{26}, \cite{27}, \cite{JJS}-\cite{JJSc}. The quantities $\Sigma$, $\hat{\Sigma}$ are "momentum mappings" \cite{1,2,36,37} corresponding to the action of groups $L_{G}$, $R_{G}$ lifted to the phase space. They are Hamiltonian generators of these groups. In any case, with the non-trivial cohomology $H$-groups there is no isomorphism of the action of Lie algebras into the Poisson-bracket Lie algebra of functions. There are only ones modified by constants. But it is not the case when dealing with the semi-simple algebras. We use the expansion
\[
\Sigma=\Sigma_{a}E^{a}, \quad \hat{\Sigma}=\hat{\Sigma}_{a}E^{a},
\]
where $(\cdots, E^{a}, \cdots)$ form the dual basis of  $(\cdots, E_{a}, \cdots)$. Furthermore we expand:
\[
\Sigma_{a}=p_{i}\Sigma^{i}{}_{a}(q), \quad \hat{\Sigma}_{a}=p_{i}\hat{\Sigma}^{i}{}_{a}(q),
\]
where $p_{i}$ are canonical momenta conjugate to generalized coordinates $q^{i}$, or rather $p_{i}$ are dual to virtual generalized velocities $\dot{q^{i}}$. $\Sigma_{a}$, $\hat{\Sigma}_{a}$ are conjugate to the non-holonomic velocities $\Omega^{a}$, $\hat{\Omega}^{a}$; thus
\[
\Sigma_{a}\Omega^{a}=\hat{\Sigma}_{a}\hat{\Omega}^{a}=p_{i}\dot{q}^{i}.
\]
Consequently, 
\[
\Sigma^{i}{}_{a}\Omega^{a}{}_{j}=\delta^{i}{}_{j}, \quad \hat{\Sigma}^{i}{}_{a}\hat{\Omega}^{a}{}_{j}=\delta^{i}{}_{j}.
\]
Then, $\Sigma_{a}$, $\hat{\Sigma}_{a}$ are non-holonomic momenta in the sense that they do not Poisson-commute if $G$ is non-Abelian and no change of coordinates may make $\Sigma^{i}{}_{a}$ constant. As expected, their Poisson brackets are directly built of structure constants of $G$,
\[
\{\Sigma_{a}, \Sigma_{b}\}=C^{d}{}_{ab}\Sigma_{d}, \quad \{\hat{\Sigma_{a}}, \hat{\Sigma_{b}}\}=-C^{d}{}_{ab}\hat{\Sigma_{d}}, \quad \{\Sigma_{a}, \hat{\Sigma_{b}}\}=0.
\]

The difference in signs in Poisson brackets for $\Sigma_{a}, \hat{\Sigma_{a}}$ follows from the fact that $L_{G}$ is a realization of $G$, whereas $R_{G}$ is its anti-realization (respectively the left and right actions). The vanishing of the mixed Poisson bracket $\{\Sigma_{a}, \hat{\Sigma_{b}}\}$ is due to the fact that the left and right translations mutually commute. 

Remark: concerning the difference in sign: One is faced with this problem in classical and quantum mechanics of rigid bodies (e.g., molecules). Namely, the co-moving angular momenta and spatial angular momenta have opposite signs on the right hand sides of their Poisson/commutator brackets.

For many linear Lie groups, their Lie co-algebras may be simply identified with the Lie algebras themselves, namely in the sense of the trace formula
\[
\langle \Sigma, \Omega \rangle={\rm Tr}(\Sigma\Omega)={\rm Tr}(\hat{\Sigma}\hat{\Omega})=\langle\hat{\Sigma},\hat{\Omega}\rangle.
\]
It is certainly a fact for ${\rm GL}(n, \mathbb{R})$, ${\rm SL}(n, \mathbb{R})$, ${\rm SO}(n, \mathbb{R})$, ${\rm U}(n)$, ${\rm SU}(n)$ and for the connected components of ${\rm SO}(k, n-k)$. Then the coadjoint action of $L_{G}R_{G}$ on non-holonomic momenta has the form analogous to the action on non-holonomic velocities,
\[
L_{k}\circ R_{h}: \Sigma \mapsto k \Sigma k^{-1}, \quad \hat{\Sigma} \mapsto h^{-1}\Sigma h.
\]
If $f$ is a function depending only on the configuration $g\in G$, i.e., only on generalized coordinates $q^{1}, \cdots, q^{n}$, then  
\[
\{\Sigma_{a}, f\}=-\mathcal{L}_{a}f, \quad \{\hat{\Sigma}_{a}, f\}=-\mathcal{R}_{a}f,
\]
where $\mathcal{L}_{a}$, $\mathcal{R}_{a}$ are first-order differential operators generating respectively left and right regular translations,
\[
\frac{\partial}{\partial x^{a}}f\left( \exp \left(x^{a}E_{a}\right)g\right)|_{x=0}=(\mathcal{L}_{a}f)(g)
\]
\[
\frac{\partial}{\partial x^{a}}f\left(g \exp \left(x^{a}E_{a}\right)g\right)|_{x=0}=(\mathcal{R}_{a}f)(g).
\]
One can show that \cite{JJS1,JJS2,JJS3}
\[
\mathcal{L}_{a}=\Sigma^{i}{}_{a}(q)\frac{\partial }{\partial q^{i}}, \quad \mathcal{R}_{a}=\hat{\Sigma^{i}{}_{a}}(q)\frac{\partial }{\partial q^{i}}.
\]
These are just formulas for the Hamiltonian generators $\Sigma_{a}$, $\hat{\Sigma}_{a}$ with the conjugate momenta $p_{i}$ replaced by the operators $\partial / \partial q^{i}$ (but put on the right of $\Sigma^{i}{}_{a}$, $\hat{\Sigma}^{i}{}_{a}$!).

Obviously, for any pair of functions $f$, $h$ depending only on the configuration variable $g$, we have
\[
\{f, h\}=0.
\]
The above-quoted system of Poisson brackets is basic and is sufficient to calculate easily any other Poisson bracket, with the use of standard properties of this operation.

The commutation rules for the operators $\mathcal{L}_{a}$, $\mathcal{R}_{a}$ have the following usual form:
\[
[\mathcal{L}_{a}, \mathcal{L}_{b}]=C^{d}{}_{ab}\mathcal{L}_{d}, \quad [\mathcal{R}_{a}, \mathcal{R}_{b}]=-C^{d}{}_{ab}\mathcal{R}_{d}, \quad [\mathcal{L}_{a}, \mathcal{R}_{b}]=0.
\]

PROPOSITION 26. {\it Expressing some phase space functions in terms of $(q^{a}, \Sigma_{a})$ or $(q^{a}, \hat{\Sigma}_{a})$ as independent functions, we may summarize the above Poisson brackets in two alternative uniform ways}:
\[
\{A, B\}=\Sigma_{d}C^{d}{}_{ab}\frac{\partial A}{\partial \Sigma_{a}}\frac{\partial B}{\partial \Sigma_{b}}-\frac{\partial A}{\partial \Sigma_{a}}\mathcal{L}_{a}B+(\mathcal{L}_{a}A)\frac{\partial B}{\partial \Sigma_{a}},
\]
\[
\{A, B\}=-\hat{\Sigma}_{d}C^{d}{}_{ab}\frac{\partial A}{\partial \hat{\Sigma}_{a}}\frac{\partial B}{\partial \hat{\Sigma}_{b}}-\frac{\partial A}{\partial \hat{\Sigma}_{a}}\mathcal{R}_{a}B+(\mathcal{R}_{a}A)\frac{\partial B}{\partial \hat{\Sigma}_{a}}.
\]

These uniform expressions are very convenient in all calculations. Besides, they have a lucid geometric structure, are formulated as invariantly as possible and involve only globally defined quantities. If functions $A,B$ are configuration-independent, they reduce to the well-known formulas for coadjoint algebras. Only the first terms survive then. If the Hamiltonian also is configuration-independent (geodetic models), then the equations of motion reduce to those on coadjoint orbits with the Hamiltonians obtained by the restriction of the original geodetic Hamiltonian.

Lagrangians of non-dissipative systems are functions on the tangent bundle $L: TG \rightarrow \mathbb{R}$, analytically represented as $L(q^{i}, \dot{q}^{i})$. When making use of the identifications of $TG$ with $G \times G'$ we identify them analytically as $L(q^{a}, \Omega^{a})$ or $L(q^{a}, \hat{\Omega}^{a})$. This is of course a simplified way of writing; strictly speaking, one should have used some symbols different from the original $L$. 

Similarly, the Legendre transformation $\mathcal{L}:TG \rightarrow T^{*}$, analytically described by 
\[
(q^{i}, \dot{q}^{i}) \mapsto (q^{i}, p_{i})=\left( q^{i}, \frac{\partial L}{\partial \dot{q}^{i}}\right),
\]
may be represented as mappings from $G \times G'$ to $G \times G'^{*}$ analytically given by
\[
(q^{a}, \Omega^{a}) \mapsto (q^{a}, \Sigma_{a})=\left( q^{a}, \frac{\partial L}{\partial \Omega^{a}}\right),
\]
\[
(q^{a},\hat{ \Omega}^{a}) \mapsto (q^{a}, \hat{\Sigma}_{a})=\left( q^{a}, \frac{\partial L}{\partial \hat{ \Omega}^{a}}\right).
\]
The energy function, analytically given by the traditional formula
\[
E=\dot{q}^{i} \frac{\partial L}{\partial \dot{q}^{i}}-L,
\]
may be expressed in two ways as a function on $G \times G'$; analytically
\[
E=\Omega^{a} \frac{\partial L}{\partial \Omega^{a}}-L=\hat{\Omega}^{a}\frac{\partial L}{\partial \hat{\Omega}^{a}}-L.
\]
Here again we use the same shorthand, namely the same symbol $E$ for logically different things, but it is clear from the context what is meant.

Assuming that the Legendre transformation is invertible (as in the "usual" mechanical systems), we can express $\Omega^{a}$, $\hat{\Omega}^{a}$ as functions of $\Sigma_{a}$, $\hat{\Sigma}_{a}$, and perhaps of $g$, but not in problems of present interest for us. Then, substituting the resulting expression for $E$, we obtain the Hamiltonian $H:T^{*}G \rightarrow \mathbb{R}$, analytically $H(q^{i}, p_{i})$, also represented as a function of $(q^{a}, \Sigma_{a})$ or $(q^{a}, \hat{\Sigma}_{a})$.

The convolution of phase-space-geometry and Lie-group-based degrees of freedom is particularly convenient and efficient when deriving equations of motion. The direct use of the variational principle and Euler-Lagrange equations
\[
\delta\int Ldt=0, \quad \frac{\partial L}{\partial q^{i}}-\frac{d}{dt}\frac{\partial L}{\partial \dot{q}^{i}}=0
\]
leads usually to very complicated and non-readible equations. The phase-space-description together with group-theoretic symmetry principles is much more effective. Hamiltonian equations written in the form
\[
\frac{dF}{dt}=\{F,H\}
\]
are much clearer, when $F$ runs over some system of appropriately chosen Jacobi-independent functions. Usually one chooses just the generators $\Sigma_{a}$, $\hat{\Sigma}_{a}$ among the functions $F$. One obtains the balance laws of generalized angular momenta:
\[
\frac{d\Sigma_{a}}{dt}=\{\Sigma_{a},H\}, \quad \frac{d\hat{\Sigma}_{a}}{dt}=\{\hat{\Sigma}_{a},H\},
\]
i.e., generalized Euler equations \cite{JJS}-\cite{JJSc}. Even if not directly solvable, they give a deep insight into the dynamics and enable one at least to understand the problem qualitatively. Substituting here $\Sigma$, $\hat{\Sigma}$ as functions of $\Omega$, $\hat{\Omega}$ (Legendre transformation) and expressing $\Omega$, $\hat{\Omega}$ through $(q^{i}, \dot{q}^{i})$, one obtains the second-order differential equations for $q^{i}$ as functions of time. They are, however, rarely explicitly solvable and almost never qualitatively readible.

Let us begin with traditional potential models
\[
L=T-\mathcal{V},
\]
where $\mathcal{V}$ is the potential energy depending only on the configuration $g$, i.e., on $(\cdots, q^{a}, \cdots)$, and $T$ is the kinetic energy form \cite{AM2}, \cite{m3}, \cite{m6}; in nonrelativistic models it is a quadratic form of velocities with coefficients depending on the configuration. These coefficients may be interpreted as components of some metric tensor on $G$,
\[
T=\frac{1}{2}g_{ij}(q)\frac{dq^{i}}{dt}\frac{dq^{j}}{dt}.
\]
If $\mathcal{V}=0$, one is dealing with geodetic models; there are no external forces and motion is purely inertial. When one tries to be too general, then usually nothing really interesting may be obtained. The very taste of systems with degrees of freedom ruled by groups is when their dynamics, or at least kinetic energy are somehow suited to the geometry of degrees of freedom, i.e., to the group of kinematical symmetries $G$. Such models are also geometrically interesting and practically useful in the mechanics of rigid bodies and incompressible ideal fluids, for example.

PROPOSITION 27. {\it Left invariant kinetic energies have the form:
\[
T=\frac{1}{2}\gamma_{ab}\hat{\Omega}^{a}\hat{\Omega}^{b},
\]
where $\gamma_{ab}$ are constants. They are components of some metric tensor on the Lie algebra $G'$, $\gamma\in G'^{*} \otimes  G'^{*}$. If $G$ is non-Abelian, the corresponding metric $g$ is curved and essentially Riemannian (its Riemann tensor is nonvanishing). One example is a rigid body (in the usual, metrical sense).

Right-invariant kinetic energies have the form
\[
T=\frac{1}{2}\gamma_{ab}\Omega^{a}\Omega^{b},
\]
again with constant $\gamma_{ab}$; geometrically} $\gamma\in G'^{*} \otimes G'^{*}$.

Of course, from the very formal point of view, the two models are mirror-identical and the difference between them seems to be only the convention of how to define the superposition of mappings. But physically this is not the case. It happens often, that with a fixed convention concerning superposition, both types of invariance have their own physical meaning. For example, if superposition is defined according to the more popular convention, $(f \circ g)(x)=f(g(x))$ (not $(f \circ g)(x)=g(f(x))$), then in rigid body mechanics and in elasticity, including mechanics of affinely-rigid bodies, the left invariance has to do with the isotropy of the physical space. Unlike this, the right invariance describes symmetries of the material or of how the body is shaped and its mass distributed (inertial tensor). In mechanics of ideal incompressible fluids the physical difference between left and right transformations and symmetries is even more important and drastic. The difference will be essential in the classical mechanics on a Hilbert space versus quantum mechanics.

A very important point is that quite often one considers left/right invariant geodetic models which are simultaneously invariant under the right/left action of some subgroup $H\subset G$. This interplay of both types of symmetries has to do with many physical problems and leads to various kinds of balance equations for momentum mappings, i.e., generators in the phase space.

Performing the Legendre transformations, we have the following expressions for the kinetic energies above in terms of $\Sigma_{a}, \hat{\Sigma}_{a}$. 

PROPOSITION 28. {\it There the kinetic geodetic Hamiltonians respectively for the right-invariant and left-invariant models are given by
\[
\mathcal{T}_{{\rm right}}=\frac{1}{2}\gamma^{ab}\Sigma_{a}\Sigma_{b}, \quad \mathcal{T}_{{\rm left}}=\frac{1}{2}\gamma^{ab}\hat{\Sigma}_{a}\hat{\Sigma}_{b},
\]
where $\gamma^{ab}$ represents the contravariant inverse of} $\gamma_{ab}$, $\gamma^{ac}\gamma_{cb}=\delta^{a}{}_{b}$. 

If potentials are admitted, the corresponding Hamiltonians are given respectively by
\[
H_{r}=\frac{1}{2}\gamma^{ab}\Sigma_{a}\Sigma_{b}+\mathcal{V}(q), \quad H_{l}=\frac{1}{2}\gamma^{ab}\hat{\Sigma}_{a}\hat{\Sigma}_{b}+\mathcal{V}(q).
\]

PROPOSITION 29. {\it Using the equations of motion in the Poisson bracket form we obtain for the system with the $\mathcal{T}_{{\rm right}}$ kinetic energy:
\[
\frac{d\Sigma_{a}}{dt}=\gamma^{bc}\Sigma_{c}\Sigma_{d}C^{d}{}_{ab}-\mathcal{L}_{a}\mathcal{V},
\]
or equivalently},
\[
\frac{d\hat{\Sigma}_{a}}{dt}=-\mathcal{R}_{a}\mathcal{V}.
\]
These equations become closed when one substitutes
\[
\Sigma_{a}=\gamma_{ab}\Omega^{b}, \quad \frac{dg}{dt}=\Omega g.
\]
The quantities $-\mathcal{L}_{a}\mathcal{V}$, $-\mathcal{R}_{a}\mathcal{V}$ are generalized torques (generalized moments of forces), respectively in the "spatial" and co-moving representations.

Let us denote those generalized forces by $\mathcal{N}$
\[
\mathcal{N}_{a}=-\mathcal{L}_{a}\mathcal{V}, \quad \mathcal{\hat{N}}_{a}=-\mathcal{R}_{a}\mathcal{V}.
\]
They are here derived from the potential energy $\mathcal{V}$, but once derived, the above equations of motion may be generalized so as to admit dissipative forces; namely the above expressions for $\mathcal{N}_{a}$, $\mathcal{\hat{N}}_{a}$ should be made general by adding to $-\mathcal{L}_{a}\mathcal{V}$, $-\mathcal{R}_{a}\mathcal{V}$ some phenomenological viscous terms, usually linear in velocities $\Omega$, $\hat{\Omega}$. Let us notice that at any point $g(q^{1}, \cdots, q^{n})\in G$, we have 
\[
\Omega^{a}=({\rm Ad}_{g})^{a}{}_{b}\hat{\Omega}^{b}, \quad \Sigma_{a}=\hat{\Sigma}_{b}({\rm Ad}_{g}{}^{*}), \quad \mathcal{\hat{N}}_{a}=\mathcal{N}_{b}({\rm Ad}_{g}{}^{-1})^{b}{}_{a}.
\]

PROPOSITION 30. {\it For systems with the left-invariant kinetic energy, we have
\[
\frac{d\Sigma_{a}}{dt}=\mathcal{N}_{a},
\]
or equivalently
\[
\frac{d\hat{\Sigma}_{a}}{dt}=-\gamma^{bc}\hat{\Sigma}_{c}\hat{\Sigma}_{d}C^{d}{}_{ab}+\mathcal{\hat{N}}_{a},
\]
where for the potential forces, $\mathcal{N}_{a}$ and $\mathcal{\hat{N}}_{a}$ are given again by the same formulas as previously}.

These equations also become closed rules of motion when considered jointly with the Legendre transformation
\[
\hat{\Sigma}_{a}=\gamma_{ab}\hat{\Omega}^{b}
\]
and the definition of $\hat{\Omega}$:
\[
\frac{dg}{dt}=g\hat{\Omega}.
\]
Particularly interesting is the special case of the double isotropy of kinetic energy; i.e., one invariant under $L_{G}R_{G}$. This occurs when, e.g., $G$ is semisimple and $\gamma$ coincides with or is proportional to the Killing tensor $\Gamma\in G'^{*}\otimes G'^{*}$, 
\[
\Gamma_{ab}=C^{d}{}_{ea}C^{e}{}_{db}.
\]
This is an analogue of the spherical rigid body. It is clear that all non-dynamical terms on the right hand sides of the balance laws vanish then and we obtain the balance laws for Hamiltonian generators $\Sigma_{a}, \hat{\Sigma}_{a}$ (momentum mappings, generalized angular momenta),
\[
\frac{d\Sigma_{a}}{dt}=\mathcal{N}_{a}, \quad \frac{d\hat{\Sigma}_{a}}{dt}=\mathcal{\hat{N}}_{a},
\]
or, in the geodetic case, simply the conservation laws,
\[
\frac{d\Sigma_{a}}{dt}=0, \quad \frac{d\hat{\Sigma}_{a}}{dt}=0.
\]

Let us observe that one has the same equations for simple groups. Semisimple groups may be decomposed into direct products of simple ones $G^{(\kappa)}$; any of them has its own Killing tensor $\Gamma^{(\kappa)}$ on $G^{(\kappa)'}$; and instead of $\Gamma$, we may take any linear combination of tensors $\Gamma^{(\alpha)}$ which are Killing on $G^{(\kappa)'}$ and vanish on $G^{(\rho)'}$, $\rho\neq \kappa$. Similarly, when $G'$ is not semisimple, but, e.g., is the direct product of some semisimple algebra and a one-dimensional centre $\mathbb{R}$, with its own natural metric. And finally, if $G$ is not semisimple, nothing is essentially changed when $\Gamma_{\alpha\beta}$  (or the mentioned combination of $\Gamma^{(\kappa)}_{\alpha\beta}$' s) is replaced by 
\[
\tilde{\Gamma}_{ab}=\lambda C^{d}{}_{ea}C^{e}{}_{db}+\mu C^{d}{}_{da}C^{e}{}_{eb}.
\]
It is clear that for semisimple groups $G$ the second term always vanishes.

These equations are generalized Euler equations; i.e., they become historical Euler equations when $G={\rm SO}(3, \mathbb{R})$; i.e., in rigid body mechanics (left invariant kinetic energy on $G$).

For the purely geodetic case, when there are no external torques, one obtains geodetic Euler equations. Thus
\[
\frac{d\hat{\Sigma}_{a}}{dt}=-\gamma^{bc}\hat{\Sigma}_{c}\hat{\Sigma}_{d}C^{d}{}_{ab}, \quad \frac{d\Sigma_{a}}{dt}=0
\]
for the left-invariant models and
\[
\frac{d\Sigma_{a}}{dt}=\gamma^{bc}\Sigma_{c}\Sigma_{d}C^{d}{}_{ab}, \quad \frac{d\hat{\Sigma}_{a}}{dt}=0
\]
for the right-invariant models.
When dealing with generalized Euler equations, i.e., those on the left-hand sides, we immediately see that they are equations defined purely on the Lie co-algebras $G'^{*}$. So, we can simply forget about $G$ and consider them as dynamical systems on $G'^{*}$. They may be reduced to Hamiltonian systems on the coadjoint orbits $N\subset G'^{*}$, i.e., orbits of the co-adjoint representations. One uses the symplectic structures on $N$ induced by the natural Poisson structure of $G'^{*}$, and the Hamiltonians $H_{(N)}$ obtained as the restriction
\[
H(N)=H|_{N}
\]
of the original Hamiltonian on $G'^{*}$ (the last is well-defined, because it is independent of the configuration $g(q)\in G$).

The natural question arises as to "solutions without solutions", i.e., to the general form of possible solutions to be obtained without detailed calculations. 

If $\gamma$ is the Killing tensor $\Gamma$ on $G'$, or the mentioned "deformed" Killing tensor $\tilde{\Gamma}$ (when $G'$ is semisimple, but not simple), then on $T^{*}G$ the general solution is given by the system of one-parameter subgroups and their cosets in $G$. On the level of $G'$, $G'^{*}$ and co-adjoint orbits, this means that the general solutions consist of equilibria, $\hat{\Omega}={\rm const}$, $\hat{\Sigma}={\rm const}$. On the level of the manifold $G$, solutions are given by
\[
g(t)=g_{0}\exp (\hat{F}t)=\exp (Ft)g_{0},
\]
where $F$, $\hat{F}$ are constant, $g_{0}\in G$ is an arbitrary initial condition, and
\[
F=g_{0}\hat{F}g_{0}{}^{-1}, \quad \Omega(t)=F, \quad \hat{\Omega}(t)=\hat{F},
\]
for any $t\in \mathbb{R}$. This solution is the general solution with $2n$ independent parameters $g_{0}$ and $\hat{F}$. On the level of the Lie co-algebra $G'^{*}$, all its points represent constant solutions. The situation becomes more complicated when the system is invariant on the left, but is not invariant on the right. One can ask if there are solutions given by one-parameter groups and their cosets. If $\mathbb{R}\ni t \rightarrow g(t)\in G$ is a solution, then, due to the left invariance, $\mathbb{R}\ni t \mapsto g_{0}g(t)\in G$ is also a solution for any $g_{0}$. The question is if there exist exponential solutions at all. The answer is that such solutions are exceptional in the sense that they exist in general only for some special values of the exponent. If we assume that $g(t)=g_{0}\exp (\hat{F}t)$ and substitute this to the equations of motion of a right-invariant geodetic system of generalized Euler equations, then we obtain the following algebraic conditions for $F$: 
\[
\hat{F}^{c}\gamma_{cd}C^{d}{}_{ab}\hat{F}^{b}=0,
\]
where $\hat{F}^{a}$ are components of $F$ with respect to some fixed basis $(\cdots, E_{a}, \cdots)$
\[
\hat{F}=\hat{F}^{a}E_{a}.
\]
A symmetric result may be obtained for right-invariant systems. Then the solutions of the type $g(t)=\exp (Ft)g_{0}$ do exist if $F$ satisfies an analogous condition,
\[
F^{c}\gamma_{cd}C^{d}{}_{ab}F^{b}=0,
\]
where
\[
F=F^{a}E_{a}.
\]
In all solutions of this type, $g_{0}$ is completely arbitrary due to the left and right invariance respectively. But the system of $n$ quantities $F^{a}$ satisfies a system of algebraic quadratic equations. Its general solution, and the number of independent solutions depends on the details of $\gamma_{ab}$, $C^{d}{}_{ab}$ and their mutual relationships. For example, if $\gamma_{ab}$ is the Killing metric $\Gamma_{ab}$, and the algebra $G'$ is semisimple, then the equations for $F$ become identities and do not restrict $F$ at all, though one is dealing then with the doubly-invariant geodetic model. If we forget about $G$ and are thinking only on the level of the coadjoint algebra, then the stationary solutions, i.e., relative equilibria described above, may be obtained in the following way: On any fixed co-adjoint orbit $\mathcal{N}\subset G'^{*}$ we take the Hamiltonian $H(N):=H|_{N}$ and the corresponding symplectic two-form $\gamma(N)$ on $N$. We obtain some reduced Hamiltonian system. One can show that the above stationary solutions for
\[
\hat{\Sigma}_{a}=\gamma_{ab}\hat{\Omega}^{b},
\]
(left-invariant), or
\[
\Sigma_{a}=\gamma_{ab}\Omega^{b}
\]
(right-invariant),
may be found as critical points of $H(N)$.

Consider as an example a rigid body without translational motion. Its configuration space $G$ is the special orthogonal group ${\rm SO}(n, \mathbb{R})$ in $n$ real dimensions. It consists of real matrices satisfying $\varphi^{T}\varphi=\rm{I}$, i.e., $\varphi^{T}=\varphi^{-1}$, and having the unit determinants, $\rm {det}\varphi=1$ (the first condition implies only that $|\rm {det}\varphi|=1$, i.e., $\rm {det}\varphi=\pm 1$.) ${\rm SO}(n, \mathbb{R})$ is a connected subgroup of ${\rm GL}^{+}(n, \mathbb{R})\subset {\rm L}(n, \mathbb{R})$, i.e., of the group of real $n \times n$ matrices with positive determinants. The latter group is an open connected submanifold of ${\rm L}(n, \mathbb{R})$, the linear space of all real $n\times n$ matrices; and ${\rm L}(n, \mathbb{R})$ is canonically identical with ${\rm GL}(n, \mathbb{R})'$, the Lie algebra of ${\rm GL}^{+}(n, \mathbb{R})$. ${\rm SO}(n, \mathbb{R})$ is an $n(n-1)/2$-dimensional surface in ${\rm L}(n, \mathbb{R})$ and its Lie algebra ${\rm SO}(n, \mathbb{R})'$; i.e., the tangent space at the identity element $\rm {I}_{n}$, consists of all real skew-symmetric matrices, $\Omega=-\Omega^{T}$. The elements of ${\rm SO}(n, \mathbb{R})$ may be written in the form $g=\exp (\Omega)$; the matrices $\Omega$ are canonical coordinates of the first kind on ${\rm SO}(n, \mathbb{R})$. The total orthogonal group ${\rm O}(n, \mathbb{R})$ consists of all matrices $\varphi$ satisfying $\varphi^{T}\varphi=\rm {I}$, including ones with determinant $(-1)$ \cite{46}, \cite{60}. This total group is not-connected and is a disjoint union of two connected components, namely the subgroup ${\rm SO}(n, \mathbb{R})$ and its coset consisting of orthogonal matrices with determinants $(-1)$. Elements of ${\rm O}(n, \mathbb{R})$ are linear isometries of $\mathbb{R}^{n}$ onto itself preserving the scalar product 
\[
(u, v):=\sum_{a=1}^{n}u^{a}v^{a}=\delta_{ab}u^{a}v^{b}.
\]
Elements of ${\rm SO}(n, \mathbb{R})$ are rotations, and elements of ${\rm O}(n, \mathbb{R})$ with the $(-1)$ determinants are improper rotations, i.e., rotations combined with reflections with respect to some arbitrarilly chosen $(n-1)$-dimensional plane in $\mathbb{R}^{n}$, e.g. 
\[
(x^{1}, \cdots, x^{n-1}, x^{n}) \mapsto (x^{1}, \cdots, x^{n-1}, -x^{n}).
\]
The spatial and co-moving representations of the angular velocity are respectively given by
\[
\Omega=\frac{d\varphi}{dt}\varphi^{-1}, \quad \hat{\Omega}=\varphi^{-1}\frac{d\varphi}{dt}, \quad \Omega=\varphi\hat{\Omega}\varphi^{-1}.
\]

Let us consider the special, physical, case of $n=3$. Skew-symmetric tensors of angular velocity may be identified with axial vectors; this is a peculiarity of dimension three. We write laboratory and co-moving angular velocities as follows:
\begin{eqnarray}\label{eq370}
\Omega=\left[\begin{array}{ccc}
0 & -\Omega_{3} & \Omega_{2}  \\
 \Omega_{3} & 0 &-\Omega_{1}  \\
  -\Omega_{2} & \Omega_{1} & 0 
\end{array}\right], \quad 
\hat{\Omega}=\left[\begin{array}{ccc}
0 & -\hat{\Omega}_{3} & \hat{\Omega}_{2}  \\
 \hat{\Omega}_{3} & 0 &-\hat{\Omega}_{1}  \\
 - \hat{\Omega}_{2} & \hat{\Omega}_{1} & 0 
\end{array}\right].
\end{eqnarray}
Similarly, for laboratory and comoving angular momentum (spin) we have
\begin{eqnarray}\label{eq371}
\Sigma=\left[\begin{array}{ccc}
0 & \Sigma_{3} & -\Sigma_{2}  \\
 -\Sigma_{3} & 0 &\Sigma_{1}  \\
  \Sigma_{2} & -\Sigma_{1} & 0 
\end{array}\right], \quad 
\hat{\Sigma}=\left[\begin{array}{ccc}
0 & \hat{\Sigma}_{3} & -\hat{\Sigma}_{2}  \\
 -\hat{\Sigma}_{3} & 0 &\hat{\Sigma}_{1}  \\
  \hat{\Sigma}_{2} & -\hat{\Sigma}_{1} & 0 
\end{array}\right].
\end{eqnarray}
Kinetic energy is given by
\[
T=\sum_{a=1}^{3}\frac{1}{2I_{a}}(\hat{\Sigma}_{a})^{2},
\]
where $I_{a}$ are the main moments of inertia (comoving ones) and thus constant. Poisson brackets have the form
\[
\{\hat{\Sigma}_{a}, \hat{\Sigma}_{b}\}=-\epsilon_{abc}\hat{\Sigma}_{c},
\]
or, in a spatial (laboratory) representation
\[
\{\Sigma_{a}, \Sigma_{b}\}=\epsilon_{abc}\Sigma_{c},
\]
and
\[
\{\Sigma_{a}, \hat{\Sigma}_{b}\}=0.
\]
If there exists some potential $V(R)$ on the rotation group, the total Hamiltonian on $T^{*}G$ has the form 
\[
H=\mathcal{T}+V(R), \quad R\in {\rm SO}(3, \mathbb{R}).
\]

Making use of Poisson brackets, we have the following balance equations for the co-moving spin, i.e., Euler equations:
\[
\frac{d\hat{\Sigma}_{1}}{dt}=\left( \frac{1}{I_{3}}-\frac{1}{I_{2}}\right)\hat{\Sigma}_{2}\hat{\Sigma}_{3}+\hat{\mathcal{N}}_{1},
\]
\[
\frac{d\hat{\Sigma}_{2}}{dt}=\left( \frac{1}{I_{1}}-\frac{1}{I_{3}}\right)\hat{\Sigma}_{3}\hat{\Sigma}_{1}+\hat{\mathcal{N}}_{2},
\]
\[
\frac{d\hat{\Sigma}_{3}}{dt}=\left( \frac{1}{I_{2}}-\frac{1}{I_{1}}\right)\hat{\Sigma}_{1}\hat{\Sigma}_{2}+\hat{\mathcal{N}}_{3}.
\]
The Legendre transformations
\[
\hat{\Sigma}_{a}=I_{\underline{a}}\hat{\Omega}^{\underline{a}}
\]
enable one to write down Euler equations in the form:
\[
I_{1}\frac{d\hat{\Omega}_{1}}{dt}=(I_{2}-I_{3})\hat{\Omega}_{2}\hat{\Omega}_{3}+\hat{N}_{1},
\]
\[
I_{2}\frac{d\hat{\Omega}_{2}}{dt}=(I_{3}-I_{1})\hat{\Omega}_{3}\hat{\Omega}_{1}+\hat{N}_{2},
\]
\[
I_{3}\frac{d\hat{\Omega}_{3}}{dt}=(I_{1}-I_{2})\hat{\Omega}_{1}\hat{\Omega}_{2}+\hat{N}_{3}.
\]
One always has
\[
\frac{d\Sigma_{a}}{dt}=N_{a},
\]
and for the spherical rigid body
\[
\frac{d\hat{\Sigma}_{a}}{dt}=\hat{N}_{a}.
\]
In the above equations $N_{a}$ and $\hat{N}_{a}$ denote respectively the spatial and
co-moving moments of forces, torques.

In the case of a Killing metric, when the system has a doubly-invariant kinetic energy, the non-dynamical terms on the right-hand side of the Euler equations disappear because then $I_{1}=I_{2}=I_{3}$. This is the spherical rigid body. If the rigid body is symmetric, e.g., $I_{1}=I_{2}=B$, but not necessarily $I_{3}=A=B$, then 
\[
\frac{d\hat{\Sigma}_{1}}{dt}=\left( \frac{1}{A}-\frac{1}{B}\right)\hat{\Sigma}_{2}\hat{\Sigma}_{3}+\hat{\mathcal{N}}_{1},
\]
\[
\frac{d\hat{\Sigma}_{2}}{dt}=\left( \frac{1}{B}-\frac{1}{A}\right)\hat{\Sigma}_{1}\hat{\Sigma}_{3}+\hat{\mathcal{N}}_{2},
\]
\[
\frac{d\hat{\Sigma}_{3}}{dt}=\hat{\mathcal{N}}_{3}.
\]
In the free geodetic case one obtains simply: $\hat{\Sigma}_{3}=\rm {const}$,
\[
\frac{d\hat{\Sigma}_{1}}{d\Sigma_{2}}=-\frac{\hat{\Sigma}_{2}}{\hat{\Sigma}_{1}}.
\]

In the completely anisotropic case without external torques (the geodetic left-invariant case) the problem reduces to the Lie co-algebra and the only solutions are stationary rotations about principal axes of inertia. So it is from the point of view of $G$ and $T^{*}G$. In coadjoint orbits we have three branches of solutions
\[
\hat{\Sigma}_{1}=0, \ \ \ \ \  \ \  \quad \hat{\Sigma}_{2}=0, \ \ \ \quad \hat{\Sigma}_{3}-arbitrary
\]
\[
\hat{\Sigma}_{1}=0, \ \ \ \ \ \ \ \quad \hat{\Sigma}_{2}-arbitrary, \ \ \ \quad \hat{\Sigma}_{3}=0
\]
\[
\hat{\Sigma}_{1}-arbitrary, \ \ \ \ \ \ \quad \hat{\Sigma}_{2}=0, \ \ \ \quad \hat{\Sigma}_{3}=0,
\]
and (via Legendre transformations between $G'$ and $G'^{*}$, the same for $\hat{\Omega}_{1}$, $\hat{\Omega}_{2}$, $\hat{\Omega}_{3}$.

In the language of $G$, $T^{*}G\simeq G \times G'^{*}$, we obtain uniform rotations about principal axes of inertia placed in an arbitrary way in the physical space. From the point of view of $G'^{*}$, when there is no configuration-dependent potential, those stationary rotations are obtained as stationary points of $H(N)=H|_{N}$,
\[
H=\sum_{a=1}^{3}\frac{1}{2I_{a}}\hat{\Sigma}_{a}^{2}
\]
on the co-adjoint orbits $N$ given by equations
\[
F=\sum_{a=1}^{3}(\hat{\Sigma}_{a})^{2}-s^{2}=0.
\]
Here $s\geq 0$ is the fixed magnitude of "spin" (internal/relative angular momentum). This is almost a "school" exercise in constrained (conditional) extrema. One solves it using Lagrange multipliers, i.e., Lusternik's theorem. One obtains a system of four equations for the quadrupole of variables $\hat{\Sigma}_{a}$, $\lambda$, $a=1, 2, 3$,
\[
\frac{\partial H_{\lambda}}{\partial \hat{\Sigma}_{a}}=\frac{\partial }{\partial \hat{\Sigma}_{a}}\left(\sum_{a=1}^{3}(\hat{\Sigma}_{a})^{2}-\lambda F \right)=0,
\]
\[
\sum_{a=1}^{3}(\hat{\Sigma}_{a})^{2}-s^{2}=0.
\]

Solving this system one obtains just the formerly quoted statement; e.g.,
\[
\hat{\Sigma}_{1}=0, \quad \hat{\Sigma}_{2}=0, \quad \hat{\Sigma}_{3}=\pm s,
\]
or, in terms of angular velocities,
\[
\hat{\Omega}_{1}=0, \quad \hat{\Omega}_{2}=0, \quad \hat{\Omega}_{3}=\pm \frac{s}{I_{3}},
\]
and so on by cyclic permutations of $(1, 2, 3)$. 

If inertia is once degenerate, e.g., $I_{1}=I_{2}=B$, $I_{3}=A\neq B$, then the method of conditional stationary points on the orbits gives two possible kinds of solutions:
\[
(i) \quad  \hat{\Sigma}_{1}=0, \quad \hat{\Sigma}_{2}=0, \quad \hat{\Sigma}_{3}=\pm s,
\]
\[
(ii) \quad \ \ \ \ \hat{\Sigma}_{1}{}^{2}+\hat{\Sigma}_{2}{}^{2}=s^{2}, \quad \hat{\Sigma}_{3}=0.
\]

So, there is one-parameter family of stationary solutions on the co-adjoint orbit, namely the circle of radius $s$ in the $(1, 2)$-plane in ${\rm SO}(3, \mathbb{R})'^{*}\simeq {\rm SO}(3, \mathbb{R})'\simeq \mathbb{R}^{3}$, and in addition a separate pair of solutions on the orthogonal third axis. Everything is the same for the cyclic permutation of axes. In $T^{*}G$-language, the rotation axes keep arbitrary but fixed positions in the physical space (a consequence of the left-invariance).

If the body is completely degenerate, then the system of stationary solutions coincides with the total sphere ${\rm S}^{2}(0, s)\subset  {\rm SO}(3, \mathbb{R})'\simeq \mathbb{R}^{3}$, given by equations
\[
\hat{\Sigma}_{1}{}^{2}+\hat{\Sigma}_{2}{}^{2}+ \hat{\Sigma}_{3}{}^{2}=s^{2}.
\]
This is the general solution in ${\rm SO}(3, \mathbb{R})'^{*}$, or equivalntly in ${\rm SO}(3, \mathbb{R})'$:
\[
(\hat{\Omega}_{1})^{2}+(\hat{\Omega}_{2})^{2}+ (\hat{\Omega}_{3})^{2}=\frac{1}{I}s^{2}, \quad I_{1}=I_{2}=I_{3}=I.
\]

This is a $2$-parameter family; the third parameter is $s$, and when working in $G$, $T^{*}G$, three additional parameters are there, namely the coordinates in ${\rm SO}(3, \mathbb{R})$ (because the problem is left-invariant on this simple group). So together we have six parameters which label the general solution, just as it should be for mechanical systems with three degrees of freedom.

Obviously, the origin in $\mathbb{R}^{3}$ is an orbit in itself: $s=0$. This is the simplest example that the symplectic leaves in general do not form a regular foliation with a constant dimension fixed all over. The jumps of rank are possible and even geometrically interesting. 

To finish this example let us introduce the convenient Darboux coordinates on two-dimensional leaves of $G'^{*}$. We are given coordinates $\hat{\Sigma}_{1}, \hat{\Sigma}_{2}, \hat{\Sigma}_{3}$ on $G'^{*}$. Let us introduce the variables $q, p$ in the region where $\hat{\Sigma}_{1}{}^{2}+\hat{\Sigma}_{2}{}^{2}+ \hat{\Sigma}_{3}{}^{2}>0$, namely:
\[
q:=\arctan \hat{\Sigma}_{2}/\hat{\Sigma}_{1}, \quad p:=\hat{\Sigma}_{3}.
\]

It is clear that together with the Casimir invariant or rather its square root
\[
z:=\sqrt{(\hat{\Sigma}_{1})^{2}+(\hat{\Sigma}_{2})^{2}+ (\hat{\Sigma}_{3})^{2}},
\]
they form the system of Darboux coordinates on $G'^{*}$:
\[
\{q, p\}=1, \quad \{q, z\}=0, \quad \{p, z\}=0.
\]

Coadjoint orbits are given by spheres of the fixed radius value $z=s\neq 0$. The exceptional value $z=0$ corresponds to the origin of coordinates, i.e., to the singular one-element co-adjoint orbit. 

\section{The rigid body and the affine body from the symplectic and Poisson point of view}

As another class of examples on coadjoint orbits, we discuss some facts about rigid bodies and related objects like affinely-rigid bodies in the general $n$-dimen-\newline sional space $\mathbb{R}^{n}$. To avoid certain non-desirable artefacts of $\mathbb{R}^{3}$ it may be more convenient to consider those problems in some $n$-dimensional linear space $V$ endowed (or not) with a metric tensor $g\in V^{*} \otimes V^{*}$. If translational degrees of freedom are admitted, we work in an affine space $M$ with the linear space of translations $V$. Analytically, everything reduces to the $\mathbb{R}^{n}$-language when some $g$-orthonormal basis is fixed in $V$ and some origin (reference point) is fixed in $M$.

Let us begin with introducing some notation. If $e_{a}$, $a=1, \cdots, n$ are vectors of some fixed basis in $V$ and $e^{a}$, $a=1, \cdots, n$ are elements of the dual basis in $V^{*}$,
\[
e^{a}(e_{b})=\langle e^{a}, e_{b}\rangle=\delta^{a}{}_{b},
\]
then it is natural to use in $ V \otimes V$,  $V^{*} \otimes V^{*}$,  $V \otimes V^{*}$,  $V^{*} \otimes V$, etc., respectively the following basic elements:
\[
e_{a} \otimes e_{b}, \quad e^{a} \otimes e^{b}, \quad  e_{a} \otimes e^{b}, \quad e^{a} \otimes e_{b}.
\]

Linear spaces $V \otimes V^{*}$,  $V^{*} \otimes V$ are canonically isomorphic respectively with ${\rm L}(V)$, ${\rm L}(V^{*})$, the algebras of linear mappings of $V$ into $V$ and $V^{*}$ into $V^{*}$. And ${\rm L}(V)$, ${\rm L}(V^{*})$ are canonically isomorphic with ${\rm GL}(V)'$, ${\rm GL}(V^{*})'$, Lie algebras of ${\rm GL}(V)$, ${\rm GL}(V^{*})$ with the usual commutator as a Lie bracket.

We shall use the symbols
\begin{equation}\label{eq398}
E_{a}{}^{b}:=e_{a}\otimes e^{b}, \quad E^{a}{}_{b}:=e^{a} \otimes e_{b}.
\end{equation}

The basic linear mappings of $V$ into $V$ act on the basis of $V$ as follows:
\begin{equation}\label{eq399}
E_{a}{}^{b}e_{c}=\delta^{b}{}_{c}e_{a}
\end{equation}
and satisfy the obvious commutation rules:
\[
\left[E_{a}{}^{b}, E_{c}{}^{d}\right]=\delta^{b}{}_{c}E_{a}{}^{d}-\delta_{a}{}^{d}E_{c}{}^{b}.
\]
Explicitly in terms of structure constants:
\[
\left[E_{a}{}^{b}, E_{c}{}^{d}\right]=\left(\delta^{b}{}_{c}\delta_{a}{}^{k}\delta^{d}{}_{l}-\delta_{a}{}^{d}
\delta_{c}{}^{k}\delta^{b}{}_{l}\right)E_{k}{}^{l}.
\]
Matrix elements of $L\in {\rm L}(V)\simeq V \otimes V^{*}$ coincide with expansion coefficients of $L$ with respect to the $E$-basis,
\begin{equation}\label{eq402}
L=L^{a}{}_{b}E_{a}{}^{b}, \quad Le_{a}=e_{b}L^{b}{}_{a}.
\end{equation}

With this convention the matrix of the commutator is identical with the commutator of matrices without reversal of sign; i.e., if
\begin{equation}\label{eq403}
\underset{1}{L}=\underset{1}{L}^{a}{}_{b}E_{a}{}^{b}, \quad \underset{2}{L}=\underset{2}{L}^{a}{}_{b}E_{a}{}^{b},
\end{equation}
then
\[
[\underset{1}{L}, \underset{2}{L}]=\left(\underset{1}{L}^{a}{}_{c}\underset{2}{L}^{c}{}_{b}-\underset{2}{L}^{a}{}_{c}\underset{1}{L}^{c}{}_{b}\right)E_{a}{}^{b}.
\]

Let ${\rm O}(V, g)\subset {\rm GL}(V)$ denote the subgroup of $g$-isometries ($g$-orthogonal transformations), i.e., linear transformations $L$ preserving $g$,
\[
L^{*}g=g, \quad g_{cd}L^{c}{}_{a}L^{d}{}_{b}=g_{ab}.
\]

As usual, its subgroup of orientation-preserving mappings (ones with determinant $\rm {det}L=1$) will be denoted by ${\rm SO}(V, g)$. If $g$ is positively/negatively definite, ${\rm SO}(V, g)$ coincides with the connected component of unity in ${\rm O}(V, g)$. (It is not the case if $g$ is non-definite). The corresponding Lie algebra consists of $g$-skew-symmetric mappings $\Omega$, i.e., ones satisfying
\[
\Omega^{a}{}_{b}=-\Omega_{b}{}^{a}=-g_{bc}g^{ad}\Omega^{c}{}_{d}, \quad g(\Omega x, y)=-g(x, \Omega y).
\]

It is convenient to use in ${\rm SO}(V, g)'$ the redundant "basis" consisting of linear mappings:
\[
\epsilon^{ab}=E^{ab}-E^{ba}=g^{ac}E_{c}{}^{b}-g^{bc}E_{c}{}^{a}=-\epsilon^{ba}.
\]
One can also use the following conventions:
\[
\epsilon_{ab}=g_{ac}g_{bd}\epsilon^{cd}, \quad \epsilon^{a}{}_{b}=\epsilon^{ac}g_{cb}, \quad \epsilon_{a}{}^{b}=g_{ac}\epsilon^{cb}.
\]
We have used the term "redundant basis" because the system of $\epsilon^{ab}$'s is not linearly
independent as a consequence of the skew-symmetry in its labels. The basis in a literal sense would be given e.g. by $\epsilon^{ab}$, $a<b$. However, it is more convenient and certainly more "elegant" to use the total system $\epsilon^{ab}$,
with the convention that the expansion coefficients are also skew-symmetric. Therefore, canonical parametrization of the first kind is meant in the sense:
\[
L(\omega)=\exp \left(\frac{1}{2}\omega_{ab}\epsilon^{ab}\right)
\]
where
\[
\omega_{ab}=-\omega_{ba}.
\]

The alternative conventions $\epsilon_{ab}$, $\epsilon^{a}{}_{b}$, $\epsilon_{a}{}^{b}$ for the redundant basis of ${\rm SO}(V, g)'$ are associated with the alternative conventions for redundant coordinates, $\omega^{ab}$, $\omega_{a}{}^{b}$, $\omega^{a}{}_{b}$. Obviously, in all expressions the shift of indices is meant in the sense of $g$. If $g$ is positively definite and an orthonormal basis is used so that $g_{ab}=\delta_{ab}$, then analytically all the above expressions are identical.

The basic commutation relations in ${\rm SO}(V, g)'$ have the form:
\[
\left[\epsilon^{ab}, \epsilon^{cd}\right]=g^{ad}\epsilon^{bc}+g^{bc}\epsilon^{ad}- g^{ac}\epsilon^{bd}-g^{bd}\epsilon^{ac},
\]
or, factorizing explicitly the structure constants,
\[
\left[\epsilon^{ab}, \epsilon^{cd}\right]=\left(g^{ad}\delta^{b}{}_{i}\delta^{c}{}_{j}+g^{bc}\delta^{a}{}_{i}\delta^{d}{}_{j}-g^{ac}\delta^{b}{}_{i}
\delta^{d}{}_{j}-g^{bd}\delta^{a}{}_{i}\delta^{c}{}_{j}\right) \epsilon^{ij}.
\]

The commutation rules expressed in terms of $\epsilon_{ab}$, $\epsilon^{a}{}_{b}$, $\epsilon_{a}{}^{b}$ are immediately obtained from the above ones by the appropriate $g$-lowering of indices. The basic linear mappings $E_{a}{}^{b}$, $\epsilon^{ab}$ are built of the basic vectors $e_{a}$ in $V$. However to avoid a manifold of symbols, we do not use the more precise notation like $E[e]$, $\epsilon[e]$. Obviously, the matrix elements of $E_{a}{}^{b}$, $\epsilon^{ab}$ with respect to the basis $e$ are respectively given by 
\[
\left(E_{a}{}^{b}\right)^{i}{}_{j}=\delta_{a}{}^{i}\delta^{b}{}_{j}, \quad \left(\epsilon^{ab}\right)^{i}{}_{j}=g^{ai}\delta^{b}{}_{j}-g^{bi}\delta^{a}{}_{j}.
\]

If there is no danger of confusion, we use the same symbols for linear mappings $E_{a}{}^{b}$, $\epsilon^{ab}$, and their matrices $\left[\left(E_{a}{}^{b}\right)^{i}{}_{j}\right]$, $\left[ \left(\epsilon^{ab}\right)^{i}{}_{j}\right]$.

Let us note some important low-dimensional examples, relevant for physical applications.

If $n=2$ ("Flatland" \cite{1a}), then of course (using the mentioned identification),
\begin{eqnarray*}
E_{1}{}^{1}= \left[\begin{array}{cc}
1 & 0 \\
0 & 0
\end{array}\right], \
E_{1}{}^{2}= \left[\begin{array}{cc}
0 & 1 \\
0 & 0
\end{array}\right], \
E_{2}{}^{1}= \left[\begin{array}{cc}
0 & 0 \\
1 & 0
\end{array}\right], \
E_{2}{}^{2}= \left[\begin{array}{cc}
0 & 0 \\
0 & 1
\end{array}\right].
\end{eqnarray*}

If $g$ is positive definite and we use an orthonormal basis $e$, i.e.,
\begin{eqnarray*}
[g_{ab}]=
\left[\begin{array}{cc}
1 & 0 \\
0 & 1
\end{array}\right],
\end{eqnarray*}
then
\begin{eqnarray*}
\epsilon^{11}= \left[\begin{array}{cc}
0 & 0 \\
0 & 0
\end{array}\right], \ 
\epsilon^{12}= \left[\begin{array}{cc}
0 & 1 \\
-1 & 0
\end{array}\right], \
\epsilon^{21}= \left[\begin{array}{cc}
0 & -1 \\
1 & 0
\end{array}\right], \
\epsilon^{22}= \left[\begin{array}{cc}
0 & 0 \\
0 & 0
\end{array}\right].
\end{eqnarray*}

In this redundant system there is only one linearly independent element; we may take
\begin{eqnarray*}
\epsilon=\epsilon^{21}
= \left[\begin{array}{cc}
0 & -1 \\
1 & 0
\end{array}\right].
\end{eqnarray*}
If $n=2$ and $g$ is normal-hyperbolic,
\begin{eqnarray*}
[g_{ab}]=\left[\begin{array}{cc}
1 & 0 \\
0 & -1
\end{array}\right],
\end{eqnarray*}
then of course
\begin{eqnarray*}
\epsilon^{00}= \left[\begin{array}{cc}
0 & 0 \\
0 & 0
\end{array}\right], \
\epsilon^{01}= \left[\begin{array}{cc}
0 & 1 \\
1 & 0
\end{array}\right], \
\epsilon^{10}= \left[\begin{array}{cc}
0 & -1 \\
-1 & 0
\end{array}\right], \
\epsilon^{11}= \left[\begin{array}{cc}
0 & 0 \\
0 & 0
\end{array}\right],
\end{eqnarray*}
where the "relativistic" conventions for labels is used, $\mu=0, 1$. Obviously, the non-redundant basis is one-element one; usually one chooses
\[
\epsilon=\epsilon^{01}.
\]
This is a generator of the planar Lorentz transformations.

If $n=3$ and $g$ is positive definite (signature ($+++$)) then we have the well-known expressions (assuming that one uses the basis $e$ in which $g_{ab}=\delta_{ab}$):
\[
\epsilon^{23}=-\epsilon^{32}=\left[\begin{array}{ccc}
  0 & 0 & 0\\
  0 & 0 & 1\\
  0 & -1 & 0\\
\end{array}\right], \quad \epsilon^{31}=-\epsilon^{13}=\left[\begin{array}{ccc}
  0 & 0 & -1\\
  0 & 0 & 0\\
  1 & 0 & 0\\
\end{array}\right], 
\]
\[
 \epsilon^{12}=-\epsilon^{21}=\left[\begin{array}{ccc}
  0 & 1 & 0\\
  -1 & 0 & 0\\
  0 & 0 & 0\\
\end{array}\right].
\]
If we change the signature for the negatively definite one, $(---)$, i.e., if $g_{ab}=-\delta_{ab}$ in the underlying $e$-basis, then the signs of all $\epsilon^{ab}$ will be reversed.

The peculiarity of dimension $n=3$ is that one can use the dual one-label basic elements,
\[
\epsilon_{a}=-\frac{1}{2}\varepsilon_{abc}\epsilon^{bc}, \quad \epsilon^{ab}=-\varepsilon^{abc}\epsilon_{c},
\]
where $\varepsilon_{abc}$ is the totally antisymmetric Ricci symbol with the convention $\varepsilon_{123}=1$, $\varepsilon^{123}=1$. The same representation will be used for canonical coordinates,
\[
\omega^{a}=-\frac{1}{2}\varepsilon^{abc}\omega_{bc}, \quad \omega_{ab}=-\varepsilon_{abc}\omega^{c}.
\]
Then we have
\[
\omega^{a}\epsilon_{a}=\frac{1}{2}\omega_{ab}\epsilon^{ab}
\]
and the non-redundant canonical coordinates of the first kind, $\omega^{a}$, coincide with the components of the rotation vector.

Obviously, we have
\begin{eqnarray*}
\epsilon_{1}=\left[\begin{array}{ccc}
  0 & 0 & 0\\
  0 & 0 & -1\\
  0 & 1 & 0\\
\end{array}\right], \quad \epsilon_{2}=\left[\begin{array}{ccc}
  0 & 0 & 1\\
  0 & 0 & 0\\
  -1 & 0 & 0\\
\end{array}\right], \quad
 \epsilon_{3}=\left[\begin{array}{ccc}
  0 & -1 & 0\\
  1 & 0 & 0\\
  0 & 0 & 0\\
\end{array}\right]
\end{eqnarray*}
and the commutation relations have the standard form
\[
[\epsilon_{a}, \epsilon_{b}]=\varepsilon_{ab}{}^{c}\epsilon_{c};
\]
the "cosmetic" shift of indices is meant in the Kronecker-delta-sense, because in our coordinates $g_{ab}=\delta_{ab}$.

If $n=3$ and $g$ is normal-hyperbolic, i.e., pseudo-Euclidean with signature, e.g., $(+--)$, we again use the "relativistic" label convention, $\mu=0, 1, 2$, $a=1, 2$. Then, the non-vanishing elements of the redundant basis are given by
\[
\epsilon^{12}=-\epsilon^{21}=\left[\begin{array}{ccc}
  0 & 0 & 0\\
  0 & 0 & -1\\
  0 & 1 & 0\\
\end{array}\right], \quad \epsilon^{01}=-\epsilon^{10}=\left[\begin{array}{ccc}
  0 & 1 & 0\\
  1 & 0 & 0\\
  0 & 0 & 0\\
\end{array}\right], 
\]
\[
 \epsilon^{02}=-\epsilon^{20}=\left[\begin{array}{ccc}
  0 & 0 & 1\\
  0 & 0 & 0\\
 1 & 0 & 0\\
\end{array}\right].
\]
Denoting traditionally:
\[
\epsilon^{12}=M, \quad \epsilon^{01}=N_{1}, \quad \epsilon^{02}=N_{2},
\]
we have the Lorentz $(+--)$-commutation rules
\[
[N_{1}, N_{2}]=-M, \quad [M, N_{1}]=N_{2}, \quad [M, N_{2}]=-N_{1}.
\]

Using the reversed signature $(-++)$ of $g$ we would change the signs of all generators and commutation rules.

If $n=4$ and $g$ is positive definite (signature $(++++)$), then, using the "relativistic" labels $\mu=0, 1, 2, 3$ for basic vectors (and assuming $g_{\mu\nu}=\delta_{\mu\nu}$), we obtain 
\[
\epsilon^{01}=-\epsilon^{10}=\left[\begin{array}{cccc}
  0 & 1 & 0 & 0\\
  -1 & 0 & 0 & 0\\
  0 & 0 & 0 & 0\\
 0 & 0 & 0 & 0\\
\end{array}\right], \quad \epsilon^{02}=-\epsilon^{20}=\left[\begin{array}{cccc}
  0 & 0 & 1 & 0\\
  0 & 0 & 0 & 0\\
  -1 & 0 & 0 & 0\\
 0 & 0 & 0 & 0\\
\end{array}\right], 
\]
\[
 \epsilon^{03}=-\epsilon^{30}=\left[\begin{array}{cccc}
  0 & 0 & 0 & 1\\
  0 & 0 & 0 & 0\\
 0 & 0 & 0 & 0\\
 -1 & 0 & 0 & 0\\
\end{array}\right],
\]
\begin{eqnarray}\label{eq429}
\end{eqnarray}
\[
\epsilon^{23}=-\epsilon^{32}=\left[\begin{array}{cccc}
  0 & 0 & 0 & 0\\
  0 & 0 & 0 & 0\\
  0 & 0 & 0 & 1\\
 0 & 0 & -1 & 0\\
\end{array}\right], \quad \epsilon^{31}=-\epsilon^{13}=\left[\begin{array}{cccc}
  0 & 0 & 0 & 0\\
  0 & 0 & 0 & -1\\
  0 & 0 & 0 & 0\\
 0 & 1 & 0 & 0\\
\end{array}\right], 
\]
\[
 \epsilon^{12}=-\epsilon^{21}=\left[\begin{array}{cccc}
  0 & 0 & 0 & 0\\
  0 & 0 & 1 & 0\\
 0 & -1 & 0 & 0\\
 0 & 0 & 0 & 0\\
\end{array}\right].
\]
Denoting
\[
M_{1}=\epsilon^{32}, \quad M_{2}=\epsilon^{13}, \quad M_{3}=\epsilon^{21}, \quad N_{1}=\epsilon^{01}, \quad N_{2}=\epsilon^{02}, \quad N_{3}=\epsilon^{03},
\] 
one expresses the basic commutation rules as follows:
\[
[M_{i}, M_{j}]=\varepsilon_{ij}{}^{k}M_{k}, \quad [M_{i}, N_{j}]=\varepsilon_{ij}{}^{k}N_{k}, \quad [N_{i}, N_{j}]=\varepsilon_{ij}{}^{k}M_{k}.
\]

We mention that the Lie algebra of ${\rm SO}(V, g)$, ${\rm SO}(4, \mathbb{R})$ is not semisimple; this is the 
exceptional property of dimension $n=4$. It may be identified with the Cartesian product ${\rm SO}(3, \mathbb{R})' \times {\rm SO}(3, \mathbb{R})'$ of the Lie algebra of the three-dimensional rotation group. This is seen when we introduce the following combinations of the basic generators:
\[
X_{i}=\frac{1}{2}(M_{i}+N_{i}), \quad Y_{i}=\frac{1}{2}(M_{i}-N_{i}).
\]
Indeed, it is easy to see that 
\[
[X_{i}, X_{j}]=\varepsilon_{ij}{}^{k}X_{k}, \quad [Y_{i}, Y_{j}]=\varepsilon_{ij}{}^{k}Y_{k}, \quad [X_{i}, Y_{j}]=0.
\]

Obviously, in the commutation rules above, the shift of indices in the Ricci symbol is meant in the "cosmetic" sense of the Kronecker delta. We use it only to be formally correct with the rules of summation convention.

Warning: The Lie algebra splits, ${\rm SO}(4, \mathbb{R})'\simeq {\rm SO}(3, \mathbb{R})' \times {\rm SO}(3, \mathbb{R})'$, but there is no global identification on the level of groups; i.e., ${\rm SO}(4, \mathbb{R})$ is NOT the Cartesian product ${\rm SO}(3, \mathbb{R}) \times {\rm SO}(3, \mathbb{R})$.

If instead of the positive signature $(++++)$ we use the negative one $(----)$, then the signs of all basis generators and their commutators become inverted. 

If $n=4$ and $g$ is normal-hyperbolic with the convention $(+---)$, then the basic generators, i.e., basic elements of the Lie algebra of the restricted Lorentz group ${\rm SO}(V, g)^{\uparrow}\simeq {\rm SO}(1, 3)^{\uparrow}$ (the connected component of unity in the total Lorentz group ${\rm O}(V, g)\simeq {\rm O}(1, 3)$) have the following form:
\[
\epsilon^{01}=-\epsilon^{10}=\left[\begin{array}{cccc}
  0 & 1 & 0 & 0\\
  1 & 0 & 0 & 0\\
  0 & 0 & 0 & 0\\
 0 & 0 & 0 & 0\\
\end{array}\right], \quad \epsilon^{02}=-\epsilon^{20}=\left[\begin{array}{cccc}
  0 & 0 & 1 & 0\\
  0 & 0 & 0 & 0\\
  1 & 0 & 0 & 0\\
 0 & 0 & 0 & 0\\
\end{array}\right], 
\]
\[
 \epsilon^{03}=-\epsilon^{30}=\left[\begin{array}{cccc}
  0 & 0 & 0 & 1\\
  0 & 0 & 0 & 0\\
 0 & 0 & 0 & 0\\
 1 & 0 & 0 & 0\\
\end{array}\right],
\]
\[
\epsilon^{23}=-\epsilon^{32}=\left[\begin{array}{cccc}
  0 & 0 & 0 & 0\\
  0 & 0 & 0 & 0\\
  0 & 0 & 0 & -1\\
 0 & 0 & 1 & 0\\
\end{array}\right], \quad \epsilon^{31}=-\epsilon^{13}=\left[\begin{array}{cccc}
  0 & 0 & 0 & 0\\
  0 & 0 & 0 & 1\\
  0 & 0 & 0 & 0\\
 0 & -1 & 0 & 0\\
\end{array}\right], 
\]
\[
 \epsilon^{12}=-\epsilon^{21}=\left[\begin{array}{cccc}
  0 & 0 & 0 & 0\\
  0 & 0 & -1 & 0\\
 0 & 1 & 0 & 0\\
 0 & 0 & 0 & 0\\
\end{array}\right].
\]
Traditionally \cite{50} one uses the following symbols for the generators of boosts and rotations:
\[
M_{1}=\epsilon^{23}, \quad M_{2}=\epsilon^{31}, \quad M_{3}=\epsilon^{12}, \quad N_{1}=\epsilon^{01}, \quad N_{2}=\epsilon^{02}, \quad N_{3}=\epsilon^{03}.
\]
The corresponding commutation rules have the following standard form:
\[
[M_{i}, M_{j}]=-\varepsilon_{ij}{}^{k}M_{k}, \quad [M_{i}, N_{j}]=-\varepsilon_{ij}{}^{k}N_{k}, \quad [N_{i}, N_{j}]=-\varepsilon_{ij}{}^{k}M_{k},
\]
where, as usual, the indices of the Ricci symbol $\varepsilon_{ijk}$ are shifted in the "cosmetic" sense of the Kronecker delta. 

As usual, taking the reversed signature convention $(----)$ we change the signs of all basic generators and commutation rules.

We mention a fact used in the description of representations of ${\rm SO}(1, 3)^{\uparrow}$ and its covering ${\rm SL}(2, \mathbb{C})$. The idea consists in using complexification and introducing the following combinations of basic generators:
\[
X_{a}=\frac{1}{2}(M_{a}+iN_{a}), \quad Y_{a}=\frac{1}{2i}(M_{a}-iN_{a}).
\]
Their commutation rules have the form
\[
[X_{a}, X_{b}]=\varepsilon_{ab}{}^{c}X_{c}, \quad [Y_{a}, Y_{b}]=\varepsilon_{ab}{}^{c}Y_{c}, \quad [X_{a}, Y_{b}]=0,
\]
just like in ${\rm SO}(4, \mathbb{R})$. However, there is an important difference. After complexification we work in the $\mathbb{C}$-six-dimensional, i.e., $\mathbb{R}$-$12$-dimensional, group ${\rm O}(4, \mathbb{C})$. Its elements may be parametrized by six complex canonical coordinates of the first kind $z^{a}$, $w^{a}$, $a=1, 2, 3$,
\[
L(z, w)=\exp (z^{a}X_{a}+w^{a}Y_{a}).
\]

Taking $w^{a}=\overline{z^{a}}$ we obtain the subgroup isomorphic with ${\rm SO}(1, 3)^{\uparrow}$,
\[
L[k, x]=\exp (k^{a}M_{a}, \chi^{a}N_{a}),
\]
\[
k^{a}=\frac{1}{2}(z^{a}+\overline{z^{a}}) \in \mathbb{R}, \quad \chi^{a}=\frac{i}{2}(z^{a}-\overline{z^{a}}).
\] 

In spite of the compact ${\rm SO}(3, \mathbb{R})' \times {\rm SO}(3, \mathbb{R})'$ commutation rules for linear mappings $X_{a}$, $Y_{a}$, the above exponents generate the noncompact Lorentz group because of imaginary terms of complex coordinates $z^{a}$, $\overline{z^{a}}$.

Concerning the formulas above, e.g., (\ref{eq398})-(\ref{eq403}), it must be stressed that the $L$ given by (\ref{eq402}) acts on the vector $x=x^{c}e_{c}$ according to the usual matrix rule:
\[
(Lx)^{a}=L^{a}{}_{b}x^{b}, \quad x=x^{c}e_{c}.
\] 
Similarly, the corresponding affine, i.e., inhomogeneous linear transformations of the vector $x \in V$ are given coordinate-wise by 
\[
((L, l)(x))^{a}=L^{a}{}_{b}x^{b}+l^{a},
\]
where
\[
L^{a}{}_{b}=\langle e^{a}, Le_{b}\rangle, \quad l^{a}=\langle e^{a}, l\rangle.
\]
We mention that all these formulas are suited to field theory or quantum mechanics rather than to classical phase-space mechanical studies. The point is that in classical mechanics of discrete and continuous systems one often uses the distinction between material points and their spatial location, i.e., between physical space (or space-time) $M$ and the material space $N$. For example, configurations of continuous bodies are described by sufficiently smooth mappings from $N$ to $M$. The simplest model is the one in which both $N, M$ are considered as sufficiently smooth $n$-dimensional manifolds. Obviously, physically $n=3$, but it also convenient to take a general $n$, or different dimensions $n$, $m$ in $N$ and $M$. In the most important special case $m=n$, and for both $M, N$, being topologically $\mathbb{R}^{n}$, one can describe configurations of a continuum as is done in field theory, i.e., using only one set $M$ and representing instantaneous configurations by diffeomorphisms of $M$ onto $M$. But in mechanics this is rather artificial and must be effected by using as transformation groups the two different versions of apparently the same group, i.e., left- and right- acting natural translations of the corresponding diffeomorphism group onto itself. Therefore, in continuum mechanics, one uses two manifolds: the material space $N$, i.e., the set of material points, and the physical space $M$. In principle there is no reason to assume that they are globally diffeomorphic and even that they have the same dimension. For example, they may have quite different topologies. Quite often, $N$ is a compact manifold with a boundary with a non-trivial geometry, but $M$ is considered to be $\mathbb{R}^{n}$. However, here we do not enter into such problems and consider usually $M$ and $N$ as logically different manifolds, with both topologically equivalent to $\mathbb{R}^{n}$. The configuration space of a structureless continuum may be identified with ${\rm Diff}(N, M)\cong {\rm Diff}(\mathbb{R}^{n})$. The groups  ${\rm Diff} N={\rm Diff}(N, N)$,  ${\rm Diff} M={\rm Diff}(M, M)$ act on this set on the right and on the left as follows:
\[
{\rm Diff}(N, M) \ni \varphi \mapsto A \circ \varphi \circ B \in {\rm Diff} (N, M),
\]
\begin{equation}\label{eq443}
\end{equation}
\[
A \in {\rm Diff} M, \quad B \in {\rm Diff} N.
\]

They are isomorphic but different transformation groups. Let us also stress that if we took instead of $N, M$ diffeomorphic with $\mathbb{R}^{n}$ some two completely different sets $N, M$ and instead ${\rm Diff}(N, M)$ some other set of mappings of $N$ onto $M$, then even this isomorphism would be lost. Furthermore, the corresponding groups of transformations of $N$ onto itself and $M$ onto itself would be always mutually commuting groups acting respectively on the right and on the left in an appropriate set of mappings from $N$ onto $M$. An extreme, amorphous situation is to think about $M$ and $N$ as completely different abstract sets and in place of ${\rm Diff} (N, M)$, ${\rm Diff} N$, ${\rm Diff} M$, we take the sets of all injections of $N$ into $M$, and all bijections of $N$ and $M$ onto themselves. For example, $N$ and $M$ might be differential manifolds of different dimensions. If $N$ and $M$ are differential manifolds, then their coordinates $a^{k}$, $x^{i}$ are respectively interpreted as Lagrange and Euler coordinates.

In finite-dimensional mechanical models, e.g., in the theory of discretized continua, one takes some finite-dimensional manifolds of mappings from $N$ to $M$. The simplest possible models are based on affine geometry. So, from now on we declare that $N$ and $M$ are affine spaces, i.e., that we deal with triplets $(N, U, \rightarrow)$, $(M, V, \rightarrow)$. Here $N$ and $M$are the underlying point sets, respectively the set of material points and the set of their spatial positions. Linear spaces $U, V$ of the same dimension $n$ are, respectively, translation spaces, i.e., spaces of vectors, in $N$ and $M$. The arrow symbol, for simplicity the same in two spaces, denotes the vector unifying a pair of points, $\overrightarrow{AB}\in U$, $\overrightarrow{x, y}\in V$ for $A, B\in N$ and $x, y\in M$. This operation satisfies all basic axioms, i.e., the triangle rule, e.g.,
\[
\overrightarrow{xy}+\overrightarrow{yz}+\overrightarrow{zx}=0
\]
and the double rule, i.e., that for some, and then for every $y\in M$, the mapping
\[
M\ni x \mapsto \overrightarrow{yx}\in V,
\]
is a bijection of $M$ onto $V$. And, obviously, the same holds in $N$. The Euclidean metric concepts are introduced by fixing some metric tensors, $\eta\in U^{*}\otimes U^{*}$, $g\in V^{*}\otimes V^{*}$.

Then, instead the infinite-dimensional manifolds of all diffeomorphisms of $N$ onto $M$, one can use the finite-dimensional manifold ${\rm Aff I}(N, M)$ of all affine isomorphisms of $N$ onto $M$. This is the configuration space of the affinely-rigid body \cite{AM2}-\cite{m6}, \cite{JJS}-\cite{JJSc}. One may also consider the usual, i.e., metrically-rigid body. The configuration space is then restricted to constraints ${\rm E}(N, \eta; M, g)$, i.e., to the manifold of isometries of $N, \eta$, $M, g$. In finite-dimensional manifolds, i.e., configuration spaces,  ${\rm Aff I}(N, M)$, ${\rm E}(N, \eta; M, g)$ of the affinely-rigid and the metrically-rigid body, we take respectively the Lie groups ${\rm GAff}(M) \times {\rm GAff}(N)$ and ${\rm E}(M, g)\times {\rm E}(N, \eta)$ (or their special groups of isometries ${\rm SE}(M, g) \times {\rm SE}(N, \eta))$ as the natural groups of transformations. If $a^{K}$, $y^{i}$ are affine coordinates in $N, M$, then affine mappings from $N$ to $M$ are described as:
\begin{equation}\label{eq444}
y^{i}=\varphi^{i}{}_{K}a^{K}+x^{i},
\end{equation}
where $\varphi^{i}{}_{K}$, $x^{i}$ are independent of $a^{K}$ and depend only on time. Here $x^{i}$ is the global position of the body in space and $\varphi^{i}{}_{K}$ are internal variables of the relative motion. One usually, but not always, assumes that ${\rm det}[\varphi^{i}{}_{A}]\neq 0$. Also, one usually assumes certain constraints imposed on $\varphi^{i}{}_{K}$; for example in the case of rigid motion the constraints have the form
\[
g_{ij}\varphi^{i}{}_{A}\varphi^{j}{}_{B}=\eta_{AB}.
\]

Consider first an affine model of motion in $n$-dimensional affine space $(M, V, g)$. We are dealing with two basic affine spaces $M, N$, the physical space of positions and the material space of particles. The configuration space of an affine body is given by the manifold of affine isomorphisms of $N$ onto $M$. The configuration $\varphi \in {\rm Aff I}(N, M)$ is to be meant in such a way that the $a$-th material point $a \in N$ is located at the geometric position $y=\varphi(a)$; i.e., (\ref{eq444}) holds for them. Motion is described by some relatively smooth time-dependence of the configuration $\varphi$, i.e., by the time-dependence of $(x, \varphi)$. Generalized velocities are given by the systems of
\[
(v^{i}, V^{i}{}_{A})=\left( \frac{dx^{i}}{dt}, \frac{d}{dt}\varphi^{i}{}_{A}\right).
\]
From the tensor point of view, one deals here with the doubled objects like 
\[
(v, V) \in V \times {\rm L}(U, V)\simeq V \times (V \otimes U^{*})
\]
with indices partly in $V$, partly in $U$. The natural question arises as to the possibility of using only one kind of indices, i.e., ones in $V$ or ones in $U$.

PROPOSITION 31. {\it From the point of view of $(M, V)$ they are quantities:
\[
(v^{i}, \Omega^{i}{}_{j})= \left( \frac{dx^{i}}{dt}, \frac{d\varphi^{i}{}_{A}}{dt}\varphi^{-1A}{}_{j}\right);
\]
their co-moving representation is
\[
(\hat{v}^{A}, \hat{\Omega}^{A}{}_{B})= \left(\varphi^{-1A}{}_{i} \frac{dx^{i}}{dt}, \varphi^{-1A}{}_{i}\frac{d\varphi^{i}{}_{A}}{dt}\right);
\]
i.e., equivalently},
\[
(\hat{v}^{A}, \hat{\Omega}^{A}{}_{B})= \left(\varphi^{-1A}{}_{i} v^{i}, \varphi^{-1A}{}_{i}\Omega^{i}{}_{j} \varphi^{j}{}_{B}\right).
\]
This transformation rule between $(v^{i}, \Omega^{i}{}_{j})$, $(\hat{v}^{A}, \hat{\Omega}^{A}{}_{B})$ explains their names: current and co-moving ones. And their own transformation rules of quantities $(v^{i}, V^{i}{}_{A})$, $(v^{i}, \Omega^{i}{}_{j})$, $(\hat{v}^{A}, \hat{\Omega}^{A}{}_{B})$ under ${\rm GL}(V)\times {\rm GL}(U)$ read respectively as follows for $(A, B)\in {\rm GL}(V)\times {\rm GL}(U)$:
\[
(v^{i}, V^{i}{}_{A}) \mapsto (A^{i}{}_{j}v^{j}, A^{i}{}_{j}V^{j}{}_{K}B^{K}{}_{A}),
\]
\[
(v^{i}, \Omega^{i}{}_{j}) \mapsto (A^{i}{}_{j}, A^{i}{}_{j}\Omega^{j}{}_{k}B^{-1k}{}_{j}),
\]
\[
(\hat{v}^{M}, \hat{\Omega}^{K}{}_{L}) \mapsto (B^{-1M}{}_{P}\hat{v}^{P}, B^{-1K}{}_{P} \hat{\Omega}^{P}{}_{R} \hat{B}^{R}{}_{L}).
\]
Sometimes crazy mixtures like $(v^{i}, \hat{\Omega}^{A}{}_{B})$, $(\hat{v}^{A}, \Omega^{i}{}_{j})$ are used.

The canonical momenta conjugate to $(v^{i}, V^{i}{}_{A})$ are elements of the linear space $V^{*(n+1)}=V^{*}\times (V^{*})^{n}$.
They are analytically given by quantities
\[
(P_{i}, P^{A}{}_{i})
\]
which are dual to velocities $(v^{i}, V^{i}{}_{A})$ in the sense of the pairing:
\[
\langle(p_{i}, P^{A}{}_{i}), (v^{j}, V^{j}{}_{A})\rangle=p_{i}v^{i}+P^{A}{}_{i}V^{i}{}_{A},
\]
where the summation convention is meant over all indices. Replacing $(v^{i}, V^{i}{}_{A})$ by affine velocities $(v^{i}, \Omega^{i}{}_{j})$, $(\hat{v}^{A}, \hat{\Omega}^{A}{}_{B})$ we automatically replace the systems $(p, P)$ by $(p, \Sigma)$ or $(\hat{p}, \hat{\Sigma})$, where $p \in V^{*}$, $\Sigma \in {\rm L}(V)=V \otimes V^{*}$, $\hat{p} \in U^{*}$, $\hat{\Sigma}\in {\rm L}(U)=U \otimes U^{*}$. The systems are connected to $(p_{i}, P^{A}{}_{i})$ dually:
\[
\Sigma^{i}{}_{j}=\varphi^{i}{}_{A}P^{A}{}_{j}, \quad \hat{\Sigma}^{A}{}_{B}=P^{A}{}_{i}\varphi^{i}{}_{B}, \quad \hat{p}_{B}=p_{i}\varphi^{i}{}_{B}.
\]

Therefore, the following conditions hold:
\[
p_{i}v^{i}=\hat{p}_{A}\hat{v}^{A}
\]
\[
P^{A}{}_{i}V^{i}{}_{A}=\Sigma^{i}{}_{j}\Omega^{j}{}_{i}=\hat{\Sigma}^{A}{}_{B}\hat{\Omega}^{B}{}_{A}.
\]

The quantities $p_{i}$, $\Sigma^{i}{}_{j}$; $\hat{p}_{A}$, $\hat{\Sigma}^{A}{}_{B}$ are defined also at the phase space points where $\varphi^{i}{}_{A}$ is degenerate. They are evidently related to the Hamiltonian generators of the groups ${\rm GAff} M$, ${\rm GAff} N$, i.e., of $A, B$ operating in the sense of (\ref{eq443}) on the configuration space ${\rm GAff}(N, M)$, i.e., on the manifold of affine mappings of the material space $N$ onto $M$, the physical space of positions. More precisely, they are Hamiltonian generators of the corresponding extended point transformations acting in the cotangent bundle over ${\rm GAff} (N, M)$. In spite of their (non-canonical) isomorphism, the groups ${\rm GAff} M$, ${\rm GAff} N$ are different transformations groups of ${\rm Aff I}(N, M)$. In principle, such a situation occurs only in mechanics and it does not occur in field theory, including in quantum-mechanical problems. Let us discuss this question briefly in some details.

Let matter be distributed in $N$ with a distribution corresponding to some constant measure $\mu$. The total mass is given by
\[
m=\int_{N}d\mu(a).
\]
Lagrangian coordinates in $N$, $a^{K}$, will be chosen so that their origin coincides with the center of mass,
\[
\int a^{K}d\mu(a)=0.
\]
In the special case of a discrete or continuous affine body, the inertia is described by two parameters, the mass $m$ and the symmetric, constant inertial tensor $J$, namely
\[
J^{KL}=\int a^{K}a^{L}d\mu(a).
\]
In general there exists the total hierarchy of such objects; however, in this special case the higher-order inertial multipoles do not participate in affine motion. 

PROPOSITION 32. {\it The usual kinetic energy is then calculated after substituting affine constraints to the general expression for the unconstrained formula as} \cite{AM2}-\cite{m6}, \cite{JJS}-\cite{JJSc}:
\begin{equation}\label{eq459}
T=T_{tr}+T_{int}=\frac{m}{2}g_{ij}\frac{dx^{i}}{dt}\frac{dx^{j}}{dt}+\frac{1}{2}g_{ij}\frac{d\varphi^{i}{}_{A}}{dt}
\frac{d\varphi^{j}{}_{B}}{dt}J^{AB}.
\end{equation}
For the classical Lagrangians of the form $L=T-V(x, \varphi)$ (no generalized velocity-dependent potentials, e.g., magnetic ones), performing the Legendre transformations 
\[
p_{i}=mg_{ij}\frac{dx^{j}}{dt}, \quad P^{A}{}_{i}=g_{ij}\frac{d\varphi^{j}{}_{B}}{dt}J^{BA},
\]
one obtains the following expression for the geodetic Hamiltonian:
\begin{equation}\label{eq461}
\mathcal{T}=\mathcal{T}_{tr}+\mathcal{T}_{int}=\frac{1}{2m}g^{ij}p_{i}p_{j}+\frac{1}{2}J^{-1}{}_{AB}P^{A}{}_{i}P^{B}{}_{j}g^{ij}.
\end{equation}
We have then,
\[
H=\mathcal{T}+\mathcal{V}(x, \varphi)
\]
for the non-geodetic case.
When the center of mass in fixed in the material space $N$, and we fix also for technical reasons the origin of coordinates in the physical space $M$, then (\ref{eq459}) may be also written in the following equivalent forms:
\[
T=T_{tr}+T_{int}=\frac{m}{2}g_{ij}v^{i}v^{j}+\frac{1}{2}g_{ij}\Omega^{i}{}_{k}\Omega^{j}{}_{l}J[\varphi]^{kl}=
\]
\begin{equation}\label{eq463}
=\frac{m}{2}G_{KL}\hat{v}^{K}\hat{v}^{L}+\frac{1}{2}G_{KL}\hat{\Omega}^{K}{}_{A}\hat{\Omega}^{L}{}_{B}J^{AB}.
\end{equation}
Similarly, the corresponding geodetic expressions in Hamiltonian variables are given by
\[
\mathcal{T}=\mathcal{T}_{tr}+\mathcal{T}_{int}=\frac{1}{2m}g^{ij}p_{i}p_{j}+\frac{1}{2}J[\varphi]^{-1}{}_{ij}
\Sigma^{i}{}_{k}\Sigma^{j}{}_{l}g^{kl}=
\]
\begin{equation}\label{eq464}
=\frac{1}{2m}G^{-1KL}p_{K}p_{L}+\frac{1}{2}J^{-1}{}_{AB}\hat{\Sigma}^{A}{}_{K}\hat{\Sigma}^{B}{}_{L}G[\varphi]^{-1KL}
\end{equation}
where $G[\varphi]$, briefly $G$, denotes the Green deformation tensor and $J[\varphi]$ denotes the spatial tensor of inertia
\begin{equation}\label{eq465}
G[\varphi]=\varphi^{*}\cdot \eta, \quad G[\varphi]_{AB}=g_{ij}\varphi^{i}{}_{A}\varphi^{j}{}_{B}, \quad J[\varphi]^{ij}=\varphi^{i}{}_{A}\varphi^{j}{}_{B}J^{AB}.
\end{equation}
One introduces similarly the Cauchy deformation tensor $C[\varphi]$, namely
\[
C[\varphi]=\varphi^{-1*}\cdot \eta, \quad C[\varphi]_{ij}=\eta_{AB}\varphi^{-1A}{}_{i}\varphi^{-1B}{}_{j}.
\]
In formulas (\ref{eq463}) and (\ref{eq464}) the kinetic energy is expressed by the momentum mappings of ${\rm GAff} N$, ${\rm GAff} M$ as functions on the phase space with values in the Lie algebras of those groups. More precisely, $(\hat{p}_{A}, \hat{\Sigma}^{A}{}_{B})$ is the momentum mapping of ${\rm GAff} N$, and $(p_{i}, \Sigma^{i}{}_{j})$ is the momentum of ${\rm GAff} M$. Even more precisely, when the center of mass is fixed, then $\hat{\Sigma}^{A}{}_{B}$ describes the momentum mapping of ${\rm GL}(U)$ acting as
\[
(x, \varphi) \mapsto (x, \varphi B), \quad B \in {\rm GL}(U).
\]
When the origin in $M$ is somehow fixed by convention, then $(p_{i}, \Sigma^{j}{}_{i})$ are used to construct the momentum mapping of GAff N, e.g., as an affine moment with respect to the origin $o \in M$:
\[
I(o)^{i}{}_{j}=\Lambda(o)^{i}{}_{j}+\Sigma^{i}{}_{j}=x^{i}p_{j}+\Sigma^{i}{}_{j}.
\]
Obviously, $\Lambda(o)^{i}{}_{j}=x^{i}p_{j}$ is the orbital affine momentum with respect to $o \in M$, and $\Sigma^{i}{}_{j}$ is the affine spin.

The two pairs of metric tensors $(G, \eta)$, $(C, g)$ respectively in $U, V$ give rise in a generic non-degenerate case to two pairs of orthonormal frames in those spaces. Let us denote them by $(\cdots, R_{a}, \cdots)$, $(\cdots, L_{a}, \cdots)$. 
If, as it is the case in Euclidean spaces, $\eta$, $g$ are positive metrics in $U$, $g$, then we have
\[
\eta(R_{a}, R_{b})=\delta_{ab}, \quad g(L_{a},L_{b})=\delta_{ab}.
\]
Let us denote the corresponding $\eta$, $g$-eigenvalues of $G$, $C^{-1}$ with respect to $R_{a}$, $L_{a}$ by
\begin{equation}\label{eq470}
\lambda_{a}=(Q^{a})^{2}={\rm exp}(2q^{a}), \quad a=1, \cdots, n.
\end{equation}
Then the following eigenequations hold:
\[
\hat{G}R_{a}=\lambda_{a}R_{a}, \quad \hat{C}L_{a}=\lambda_{a}{}^{-1}L_{a},
\]
where $\hat{G}$, $\hat{C}$, are mixed tensors
\[
\hat{G}^{A}{}_{B}=\eta^{AK}G_{KB}, \quad \hat{C}^{i}{}_{j}=g^{ik}C_{kj}
\]
built out of $G$, $C$ with the help of $\eta$, $g$. Thus, in the positive case they differ trivially from $G$, $C$. If we identify the linear frames $(\cdots, L_{a}, \cdots)$, $(\cdots, R_{a}, \cdots)$ with linear mappings from $\mathbb{R}^{n}$ to $V$, $U$, and the dual co-frames $L^{-1}=(\cdots, L^{a}, \cdots)$, $R^{-1}=(\cdots, R^{a}, \cdots)$ with isomorphisms of $V$, $U$ onto $\mathbb{R}^{n}$, then we have for any $\varphi \in {\rm LI}(U, V)$, the obvious representation \cite{m4}, \cite{AM}, \cite{m2}
\begin{equation}\label{eq472}
\varphi=LDR^{-1}
\end{equation}
where $D:\mathbb{R}^{n} \rightarrow \mathbb{R}^{n}$ is the diagonal matrix with the nontrivial (diagonal) elements (\ref{eq470}). This two-polar decomposition is non-unique. Its non-uniqueness in the non-degenerate case is finite and corresponds to the simultaneous multiplication of $L$ and $R$ on the right by orthogonal matrices having only $\pm 1$ in any 
row and any column; any such multiplication is accompanied by the corresponding permutation of diagonal elements of $D$ so that that (\ref{eq472}) remains unchanged. The elements of $\varphi$ for which the spectrum (the set of deformation invariants) is degenerate possess an infinite-dimensional realization (\ref{eq472}). Nevertheless, this singularity is not particularly embarrassing. 

The bases $(\cdots, L_{a}, \cdots)$, $(\cdots, R_{a}, \cdots)$ are orthonormal, and the configuration space of the affine body splits into two configurations of rigid bodies in $V$ and $U$ and into the system of $n$ deformation invariants (\ref{eq470}). This concerns the internal configuration spaces. In addition there is of course the $M$-space of the center of mass positions.

Another often used representation is the usual polar decomposition of ${\rm LI}(U, V)$, or rather two equivalent forms of it. Unlike the two-polar representation, it is unique, i.e., one-valued. namely, given the bases $(\cdots, L_{a}[\varphi], \cdots)$, $(\cdots, R_{a}[\varphi], \cdots)$ characterizing $\varphi \in {\rm LI}(U, V)$,then there exists only one orthogonal mapping $U[\varphi] \in {\rm O}(U, \eta; V, g)$, i.e., such one that 
\begin{equation}\label{eq473}
\eta_{AB}=g_{ij}U[\varphi]^{i}{}_{A}U[\varphi]^{j}{}_{B}
\end{equation}
which maps the bases into each other,
\[
U[\varphi]R_{a}[\varphi]=L_{a}[\varphi], \quad a=1, \cdots, n.
\]
Then there exist symmetric and positive automorphisms $A[\varphi] \in {\rm GL}(U)$, $B[\varphi] \in {\rm GL}(V)$ such that 
\[
\varphi=U[\varphi]A[\varphi]=B[\varphi]A[\varphi].
\]
The symmetry and the positive definiteness are understood in the sense of metrics $\eta \in U^{*}\otimes U^{*}$, $g \in V^{*}\otimes V^{*}$,
\[
\eta(A[\varphi]u, v)=\eta(u, A[\varphi]v), \quad g(B[\varphi]w, z)=g(w, B[\varphi]z),
\]
where
\[
\eta(A[\varphi]u, u)>0, \quad g(B[\varphi]w, w)>0
\]
for any non-vanishing vector arguments.

$B[\varphi]$ and $A[\varphi]$ are $U[\varphi]$-related:
\[
B[\varphi]=U[\varphi]A[\varphi]U[\varphi]^{-1}.
\]
The special case of the rigid body consists in putting $\varphi=U[\varphi]\in {\rm LI}(U, \eta;V, g)$, or, in the two-polar case by taking $D={\rm Id}_{n}$ and then in glueing the matrix $[L^{a}[\varphi]R_{b}[\varphi]]$ into a single orthogonal mapping from $U$ onto $V$.

Let us give the basic Poisson brackets between our physical quantities. Thus 
\[
\{\Sigma^{i}{}_{j}, \Sigma^{k}{}_{l}\}=\delta^{i}{}_{l} \Sigma^{k}{}_{j}-\delta^{k}{}_{j} \Sigma^{i}{}_{l}
\]
and the same brackets for quantities $\Lambda^{i}{}_{j}$, $I^{i}{}_{j}$, i.e., for the orbital and total affine momenta. For the co-moving representants of those quantities we have the reversed-sign rules, 
\[
\{\hat{\Sigma}^{A}{}_{B}, \hat{\Sigma}^{C}{}_{D}\}=\delta^{C}{}_{B} \hat{\Sigma}^{A}{}_{D}-\delta^{A}{}_{D} \hat{\Sigma}^{C}{}_{B},
\]
and
\[
\{\Sigma^{i}{}_{j}, \hat{\Sigma}^{A}{}_{B}\}=0,
\]
\[
\{\hat{\Sigma}^{A}{}_{B}, \hat{p}_{C}\}=\delta^{A}{}_{C}\hat{p}_{B},
\]
\[
\{I^{i}{}_{j},p_{k}\}=\{\Lambda^{i}{}_{j},p_{k}\}=\delta^{i}{}_{k} p_{j}.
\]
For any function $F$ depending only on the configuration variables $x^{i}$, $\varphi^{j}{}_{A}$, we have 
\[
\{\Sigma^{i}{}_{j}, F\}=-\varphi^{i}{}_{A}\frac{\partial F}{\partial \varphi^{j}{}_{A}},
\]
\begin{equation}\label{eq482}
\{\Lambda^{i}{}_{j}, F\}=-x^{i}\frac{\partial F}{\partial x^{j}},
\end{equation}
\[
\{\hat{\Sigma}^{A}{}_{B}, F\}=-\varphi^{i}{}_{B}\frac{\partial F}{\partial \varphi^{i}{}_{A}}.
\]
These are just the technically useful formulas following from the group structure constants and the general definition of Poisson brackets in any symplectic manifold. 

Notice that the two-polar decomposition identifies the affine body with a pair of the usual metrically rigid bodies and the system of $n$ deformation invariants modulo the identifications mentioned. One of the rigid bodies rotates in the physical space, the other in the material space. They have both their angular velocities of rotation, i.e., skew-symmetric matrices $\chi$, $\hat{\chi}$ for the rigid body in space, and $\vartheta$, $\hat{\vartheta}$ in the material space. Again one deals here with two representations of motion: in $V$ and $U$ their angular velocities are $\chi, \vartheta$, and their co-moving representations in $\mathbb{R}^{n}$ are denoted by $\hat{\chi}$, $\hat{\vartheta}$. There are their conjugate spin angular momenta $\rho$, $\hat{\rho}$ (conjugate respectively to $\chi$, $\hat{\chi}$) and $\tau$, $\hat{\tau}$  (conjugate respectively to $\vartheta$, $\hat{\vartheta}$). Instead of canonical momenta $P_{i}$ conjugate to the deformation invariants $Q^{i}$, i.e., to diagonal elements of the matrix $D$, one may use the $p_{i}$ as conjugate momenta of $q^{i}={\rm ln}Q^{i}$. Then 
\[
P_{i}={\exp}(-q^{i})p_{i}=\frac{1}{Q^{i}}p_{i}.
\]
Notice that $\rho^{i}{}_{j}$ and $\tau^{i}{}_{j}$ coincide with the usual spin $S^{i}{}_{j}$ and the negative of vorticity  $V^{A}{}_{B}$, and thus with the $g$-skew-symmetric part of $\Sigma^{i}{}_{j}$ and the $\eta$-skew-symmetric part of $\hat{\Sigma}^{A}{}_{B}$,
\[
S^{i}{}_{j}=\Sigma^{i}{}_{j}-g^{ik}g_{jl}\Sigma^{l}{}_{k}, \quad V^{A}{}_{B}
=\hat{\Sigma}^{A}{}_{B}-\eta^{AC}\eta_{BD}\hat{\Sigma}^{D}{}_{C}.
\]
The $S^{i}{}_{j}$ and the $V^{A}{}_{B}$ are Hamiltonian generators, i.e., momentum mappings of rotations in $(V, g)$, $(U, \eta)$ - Euclidean spaces,
\[
\varphi \rightarrow A\varphi, \quad \varphi \mapsto \varphi B^{-1}, \quad A \in {\rm SO}(V, g), \quad B \in {\rm SO}(U, \eta). 
\]
The generators $\hat{\rho}^{a}{}_{b}$, $\hat{\tau}^{a}{}_{b}$ of "right transformations" in $L$, $R$ - variables,
\[
L \rightarrow LA, \quad R \rightarrow RB, \quad A, B \in {\rm SO}(n, \mathbb{R})
\]
are related to  $\rho^{i}{}_{j}$ and $\tau^{A}{}_{B}$ in the usual way:
\[
\rho=\hat{\rho}^{a}{}_{b}L_{a}\otimes L^{b}, \quad \tau=\hat{\tau }^{a}{}_{b}R_{a}\otimes R^{b}.
\]
We may use, depending on our purpose, any of the following two systems of canonical variables,
\[
(q, p; L, R; \rho, \tau), \quad (q, p; L, R; \hat{\rho}, \hat{\tau}).
\]
Being generators of orthogonal groups, these quantities satisfy the obvious Poisson rules:
\[
\{\hat{\rho}_{ab}, \hat{\rho}_{cd}\}=-g_{bd}\hat{\rho}_{ac}+g_{bc}\hat{\rho}_{ad}+g_{ad}\hat{\rho}_{bc}-g_{ac}\hat{\rho}_{bd},
\]
\[
\{\hat{\tau}_{ab}, \hat{\tau}_{cd}\}=-g_{bd}\hat{\tau}_{ac}+g_{bc}\hat{\tau}_{ad}+g_{ad}\hat{\tau}_{bc}-g_{ac}\hat{\tau}_{bd},
\]
\[
\{\hat{\rho}_{ab}, \hat{\tau}_{cd}\}=0.
\]
Their "spatial" generators in $V, U$, $S^{i}{}_{j}$, $\hat{V}^{A}{}_{B}$ satisfy Poisson brackets with reversed signs on the right-hand sides.
\[
\{q^{i}, p_{j}\}=\delta^{i}{}_{j},
\]
\[
\{q^{i}, \hat{\rho}_{ab}\}=\{q^{i}, \hat{\tau}_{ab}\}=\{q^{i}, S^{i}{}_{j}\}=\{q^{i}, \hat{V}^{A}{}_{B}\}=0,
\]
\[
\{p_{i}, \hat{\rho}_{ab}\}=\{p_{i}, \hat{\tau}_{ab}\}=\{p_{i}, S^{i}{}_{j}\}=\{p_{i}, \hat{V}^{A}{}_{B}\}=0.
\]
Now, $q^{i}$ and $p_{j}$ also Poisson-commute with the $L$- and $R$- configuration variables. The non-vanishing and important Poisson brackets are those among the quantities $\hat{\rho}_{ab}$, $\hat{\tau}_{ab}$, $S^{i}{}_{j}$, $\hat{V}^{A}{}_{B}$ and the configuration variables of $L$- and $R$- gyroscopes. But they have the standard geometric structure; i.e., they are given by the action of left and right translation generators on the corresponding configuration functions similar to (\ref{eq482}).

It is important to stress that all functions of the variables $(q^{i}, p_{j}, \hat{\rho}_{ab}, \hat{\tau}_{ab})$, or respectively of $(q^{i}, p_{j}, S^{i}{}_{j}, V^{A}{}_{B}$, form Poisson algebras under Poisson brackets. Incidentally, it is convenient to introduce new variables, namely 
\[
M_{ab}=-\hat{\rho}_{ab}-\hat{\tau}_{ab}, \quad N_{ab}=\hat{\rho}_{ab}-\hat{\tau}_{ab}.
\]
These variables enable one to perform practically important partial diagonalization of the kinetic energy. 

Furthermore, there is an important class of models with which one operates smoothly using the Poisson manifold of variables $(q^{i}, p_{j}, M_{ab}, N_{ab})$, forgetting in a sense about the origin of Poisson structures from the cotangent bundle over $M \times {\rm LI}(U, V)$. Moreover, it is the structure of this Poisson space that suggests some interesting models of the Poisson-Hamilton dynamical systems, expressing at the same time the particular meaning of the left- and right- acting group translations on the configuration space.

\section{Lattice aspects of the phase-space description of affine dynamics}

We now review some of the possible dynamical models of the previous section. We have seen that in the special case of an affinely constrained extended system of material points, one simply derives expressions (\ref{eq459}), (\ref{eq461}) for the kinetic energy, that is, the quadratic form of generalized velocities $(dx^{i}/dt, d/dt \ \varphi^{i}{}_{A})$, with constant coefficients. It is interesting to rewrite these expression using non-holonomic velocities, with configuration-dependent coefficients, (\ref{eq463}), (\ref{eq464}), (\ref{eq465}). These formulas are just another way of writing (\ref{eq459}), (\ref{eq461}). In the mechanics of the usual, i.e., metrically rigid, body this expression in terms of non-holonomic velocities would be identical with the usual expression of kinetic energy and would have an advantage of being a quadratic form with constant coefficients. Now it is an expression in terms of non-holonomic velocities, but with variable coefficients.

This suggests the following questions: May we treat (seriously) the expression in terms of non-holonomic coefficients but replacing its coefficients by constant ones? And by which ones? And how will we motivate this? The point is that it is only the mechanics of the usual collective systems with relatively small non-collective motions, where the usual algorithm of constraints and the usual d'Alembert principle are useful. By the d'Alembert principle we mean one where the reaction forces responsible for maintaining constraints are orthogonal to the constraint surface in the sense of the usual metric tensor of the physical space. This is an obvious restriction.

Consider, for example, the droplet model of nuclei based on the idea of affine vibrations of the droplet of a nuclear fluid. In no classical model underlying quantization, anything like the above form of d'Alembert's principle be seriously used. But it is clear that kinematics is based on the model of affine motion of "something." And this "something" may have rather unusual origins, based, e.g., on some field-theoretical model of affine motion of the resulting nuclear fluid. It is natural to expect that it is rather a symmetry principle that underlies both kinematics and dynamics of such affine vibrations. There are also some direct indications for this. And besides, in various models of condensed matter, in defect theory, in dynamics of fullerens and in some two-dimensional carbon physics, it may happen that instead of the usual physical metric, some effective dependence on physical phenomena may be necessary. This suggests to us considering seriously non-holonomic models with constant coefficients.
The simplest situation would be, e.g., to replace in (\ref{eq463}), (\ref{eq464}) the Green deformation tensor $G_{KL}$ by the constant material metric $\eta_{KL}$, i.e., to postulate something like 
\begin{equation}\label{eq492}
T=\frac{m}{2}\eta_{KL}\hat{v}^{K}\hat{v}^{L}+\frac{1}{2}\eta_{KL}\hat{\Omega}^{K}{}_{A}\hat{\Omega}^{L}{}_{B}J^{AB},
\end{equation}
or, after Legendre transformation,
\[
T=\frac{1}{2m}\eta^{KL}\hat{p}_{K}\hat{p}_{L}+\frac{1}{2}J^{-1}{}_{KL}\hat{\Sigma}^{K}{}_{A}\hat{\Sigma}^{L}{}_{B}\eta^{AB}.
\]
Happiness is achieved when $J^{KL}=\eta^{KL}I$, where $I$ denotes the constant scalar of internal inertia. Then,  (\ref{eq472}), (\ref{eq473}) become respectively:
\[
T=\frac{m}{2}\eta_{KL}\hat{v}^{K}\hat{v}^{L}+\frac{I}{2}\eta_{KL}\hat{\Omega}^{K}{}_{A}\hat{\Omega}^{L}{}_{B}\eta^{AB},
\]
\[
T=\frac{1}{2m}\eta^{KL}\hat{p}_{K}\hat{p}_{L}+\frac{1}{2I}\eta_{KL}\hat{\Sigma}^{K}{}_{A}
\hat{\Sigma}^{L}{}_{B}\eta^{AB}.
\]
Notice that this model of kinetic energy, just like (\ref{eq492}) in general, is invariant under the entire affine group ${\rm GAff}(M)$ acting in $M$, and through the left regular translations, acting also on ${\rm AffI}(N, M)$, i.e., on the manifold of affine isomorphisms of $N$ onto $M$.

The idea of affine invariance of kinetic energy looks rather attractive. And this model suggests our more general search. What would be the most general models of kinetic energy of affine bodies showing also affine invariance in the physical and material space, and perhaps in both of them \cite{44}, \cite{45}? 

PROPOSITION 33. {\it It is easy to answer that among all models quadratic in velocities and splitting into translational and internal parts, the most natural models invariant under ${\rm GAff}(M)$, ${\rm GAff}(N)$ have respectively the structures:
\begin{equation}\label{x}
T=T_{tr}+T_{int}=\frac{m}{2}\eta_{KL}v^{K}v^{L}+\frac{1}{2}\mathcal{L}^{B}{}_{A}{}^{D}{}_{C}\hat{\Omega}^{A}{}_{B}
\hat{\Omega}^{C}{}_{D},
\end{equation}
\begin{equation}\label{y}
T=T_{tr}+T_{int}=\frac{m}{2}g_{ij}v^{i}v^{j}+\frac{1}{2}\mathcal{R}^{j}{}_{i}{}^{l}{}_{k}\Omega^{i}{}_{j}
\Omega^{k}{}_{l},
\end{equation}
where $m$, $\mathcal{L}^{A}{}_{B}{}^{C}{}_{D}$, $\mathcal{R}^{i}{}_{j}{}^{k}{}_{l}$ are constant inertial quantities, translational and internal ones}. 

There is no model affinely invariant simultaneously in space and in matter. After all, the total affine group in not semisimple and contains a non-central normal divisor, namely the translation group. Translational kinetic energy may be affinely-invariant only in $M$ or in $N$, but not in both of them. Unlike this, there exist internal kinetic energies, i.e., metric tensors on ${\rm LI}(U, V)$, invariant both under ${\rm GL}(V)$ and ${\rm GL}(U)$. They are given by 
\begin{equation}\label{z}
T_{int}=\frac{A}{2}{\rm Tr}(\Omega^{2})+\frac{B}{2}{\rm Tr}(\Omega^{2})=\frac{A}{2}{\rm Tr}(\hat{\Omega}^{2})+\frac{B}{2}{\rm Tr}(\hat{\Omega}^{2}).
\end{equation}
The second term (the one multiplied by $B$) is merely a correction, because the corresponding metric tensor is evidently degenerate. The main term is the first one. Translational kinetic energy may be either invariant under ${\rm GAff}(M)\times {\rm E}(N, \eta)$ or under ${\rm E}(M, g)\times {\rm GAff}(N)$. Therefore, the largest admissible groups of motion of the total kinetic energy are as well ${\rm GAff}(M)\times {\rm E}(N, \eta)$ and ${\rm E}(M, g)\times {\rm GAff}(N)$. This also fixes our attention on models of the internal kinetic energy invariant under these groups. 

PROPOSITION 34. {\it Such models are given respectively by
\begin{equation}\label{eq497}
T_{int}=\frac{I}{2}{\rm Tr}(\hat{\Omega}^{T\eta}\hat{\Omega})+\frac{A}{2}{\rm Tr}(\hat{\Omega}^{2})+\frac{B}{2}({\rm Tr}\hat{\Omega})^{2},
\end{equation}
and 
\begin{equation}\label{eq498}
T_{int}=\frac{I}{2}{\rm Tr}(\Omega^{Tg}\Omega)+\frac{A}{2}{\rm Tr}(\Omega^{2})+\frac{B}{2}({\rm Tr}\Omega)^{2},
\end{equation}
where $\hat{\Omega}^{T\eta}$, $\Omega^{Tg}$ are respectively $\eta$-transposition in $U$ and $g$-transposition in $V$ of tensors} $\hat{\Omega}$, $\Omega$,
\[
(\hat{\Omega}^{T\eta})^{A}{}_{B}=\eta^{AC}\eta_{BD}\hat{\Omega}^{D}{}_{C}, \quad (\Omega^{Tg})^{i}{}_{j}=g^{ik}g_{jl}\Omega^{l}{}_{k}.
\]
It must be stressed that transpositions are related respectively to the metric tensors $\eta$, $g$, because $\hat{\Omega}$, $\Omega$ are mixed tensors. The second and third terms in (\ref{eq497}) and (\ref{eq498}) are respectively equal to each other. The corresponding Hamiltonian expressions have the form:
\begin{equation}\label{eq500}
\mathcal{T}_{int}=\frac{1}{2I'}{\rm Tr}(\hat{\Sigma}^{T\eta}\hat{\Sigma})+\frac{1}{2A'}{\rm Tr}(\hat{\Sigma}^{2})+\frac{1}{2B'}({\rm Tr}\hat{\Sigma})^{2},
\end{equation}
\begin{equation}\label{eq501}
\mathcal{T}_{int}=\frac{1}{2I'}{\rm Tr}(\Sigma^{Tg}\Sigma)+\frac{1}{2A'}{\rm Tr}(\Sigma^{2})+\frac{1}{2B'}({\rm Tr}\Sigma)^{2},
\end{equation}
with the new constants $I'$, $A'$, $B'$ following from the Legendre transformation and having the form:
\[
I'=(I^{2}-A^{2})/I, \quad A'=(A^{2}-I^{2})/A, B'=-(I+A)(I+A+nB)/B.
\]
That the only difference between (\ref{eq497}) and (\ref{eq498}) is in the first term; so, it is between (\ref{eq500}) and (\ref{eq501}).

The expressions (\ref{eq500}), (\ref{eq501}) may be rewritten in the following forms:
\begin{equation}\label{eq502}
\mathcal{T}_{int}=\frac{1}{2a}{\rm Tr}(\hat{\Sigma})^{2}+\frac{1}{2b}({\rm Tr}\hat{\Sigma})^{2}-\frac{1}{4c}{\rm Tr}(V^{2}),
\end{equation}
\begin{equation}\label{eq503}
\mathcal{T}_{int}=\frac{1}{2a}{\rm Tr}(\Sigma)^{2}+\frac{1}{2b}({\rm Tr}\Sigma)^{2}-\frac{1}{4c}{\rm Tr}(S^{2}),
\end{equation}
where the constants $a, b, c$ are given by
\begin{equation}\label{eq504}
a=I+A, \quad b=-(I+A)(I+A+nB)/B, \quad c=(I^{2}-A^{2})/I.
\end{equation}
If the term responsible for the breaking of the metric of the two sided affine symmetry vanishes, then the inverse of (\ref{eq504}) become:
\[
I=0: \quad 1/a=1/A, \quad 1/b=-B/A(A+nB), \quad 1/c=0.
\]
If in addition the correction term $B=0$, then, we have
\[
a=A, \quad 1/b=1/c=0.
\]

The last terms in (\ref{eq502}), (\ref{eq503}) are proportional to the squared magnitudes of spin and vorticity respectively, because the following holds:
\[
-\frac{1}{4c}{\rm Tr}(V^{2})=\frac{1}{2c}||V||^{2}, \quad -\frac{1}{4c}{\rm Tr}(S^{2})=\frac{1}{2c}||S||^{2}.
\]

In the special case of rigid body motion, when $A=0$, $B=0$, these two expressions become equal to each other and exactly equal to the kinetic energy of the metrically rigid body.

PROPOSITION 35. {\it Introducing the quantities $q^{a}$, $p_{a}$, $M_{ab}$, $N_{ab}$ into the main part of} (\ref{eq497})/(\ref{eq498}) {\it or} (\ref{eq502})/(\ref{eq503}), {\it i.e., to the part controlled by the parameter $a$, one obtains:
\begin{equation}\label{eq508}
\mathcal{T}=\frac{1}{2a}\sum_{a}p_{a}^{2}+\frac{1}{32a}\sum_{a,b}\frac{(M_{ab})^{2}}{\sinh^{2}\frac{q^{a}-q^{b}}{2}}-
\frac{1}{32a}\sum_{a,b}\frac{(N_{ab})^{2}}{\cosh^{2}\frac{q^{a}-q^{b}}{2}}.
\end{equation} 
We discuss here only this term of} (\ref{eq502})/(\ref{eq503}) {\it because it has all necessary properties, and the other terms introduce only corrections which are not very essential from the structural point of view}.

This is a very instructive element of the power of the phase-space and Poisson-manifold methods. And quite independently such dynamical models seem to be attractive, e.g., from the point of view of nuclear and other applications.

Let us look at (\ref{eq508}). In $n$-dimensional space there are $n$ deformation invariants, $q^{a}$, $a=1, \cdots, n$ in logarithmic scale. And there are $n(n-1)/2$ \ $L$-rotations and $n(n-1)/2$ \ $R$-rotations, combined into the same number of parameters $M_{ab}$, $N_{ab}$, on the level of phase space functions. There is a beautiful picture, namely, $(M_{ab})^{2}$, $(N_{ab})^{2}$ are respectively squares of these angular momenta; in (\ref{eq508}) they play the roles of repulsive and attractive strengths between deformation invariants as fictitious material points. Of course, in spatial problems there are only three deformation invariants and two in planar problems, but there is no reason to restrict our imagination. One can consider a system of $n$ material points on a straight line, interacting via the singular-repulsive and attractive springs $M_{ab}$, $N_{ab}$. This is a kind of lattice, the hyperbolic Sutherland lattice but with an attraction admitted. This is particularly striking in the special case $n=2$ \cite{1a} when $M_{ab}$, $N_{ab}$ are constants of motion. In geodetic, non-potential models, there is the dissociation threshold $|N_{12}|=|M_{12}|$. For the values of $|N_{12}|$  larger than $|M_{12}|$, attraction does prevail at large distances $|q_{2}-q_{1}|$. In a neighbourhood of small distances, when $q_{2}-q_{1} \rightarrow 0$, the attractive part is negative and finite, but the repulsive $M$-contribution tends to positive infinity. But if $|N_{12}|<|M_{12}|$, then the repulsion prevails all over, and we deal with a scattering situation. In this way essentially nonlinear elastic vibrations are possible even without any potential energy. In principle, they may be calculated in terms of the matrix exponential function of a matrix. The same holds in higher dimension; although then $M_{ab}$, $N_{ab}$ are non-constant. 

Then, solutions in terms of exponential mappings of the Lie algebra have the following form:
\[
\varphi(t)={\exp}(Et)\varphi_{0}=\varphi_{0}{\exp}(\hat{E}t),
\]
where 
\[
\hat{E}=\varphi_{0}^{-1}E\varphi_{0}, \quad E=\varphi_{0}\hat{E}\varphi_{0}^{-1}.
\]
The dependence  on $t$ of deformation invariants $q^{i}$ and variables $M_{ij}$, $N_{ij}$ (equivalently $\hat{\rho}_{ij}$, $\hat{\tau}_{ij}$) may be in principle obtained from these formulas, however it is also a non-automatic task. The dependence of $L_{a}$, $R_{a}$ on time is quite complicated a matter. It may be in principle obtained from the time-dependence of $q^{i}$, $\hat{\rho}_{ij}$, $\hat{\tau}_{ij}$ on $t$ by solving the non-autonomous differential equations:
\[
\frac{dL_{a}}{dt}=L_{b}\hat{\rho}^{b}{}_{a}, \quad \frac{dR_{a}}{dt}=R_{b}\hat{\tau}^{b}{}_{a}.
\]
But the main problem is to find the time dependence of deformation invariants $q^{i}(t)$, and for this purpose the complicated task of solving these equations is not necessary. In practice, it is only $|q_{a}-q_{b}|$ that performs the elastic vibrations, whereas the dilatation parameter
\[
q=\frac{1}{n}(q^{1}+q^{2}+\cdots +q^{n})
\]
is either constant or performs uniform motion with constant velocity. To prevent this undesired phenomenon, we can stabilize the motion of $q$ using some simple dilatational potential depending only on this variable and restricting its motion. This may be done, because ${\rm GL}(n, \mathbb{R})$ is the Cartesian product of the isochoric, i.e., volume-preserving group ${\rm SL}(n, \mathbb{R})$ and the one-dimensional dilatational group. Various potential wells or attractive harmonic oscillators are good models for vibrations of $q$ in a bounded domain \cite{AM}. There are also other lattice models of a similar kind, based on the phase-space and Poisson manifold geometry.

The above lattice was a hyperbolic Sutherland lattice with attraction. Replacing ${\rm GL}(n, \mathbb{R})$ by its other classical form $U(n)$ one obtains the usual Sutherland lattice with attraction. Namely, in the two-polar decomposition of configurations we replace the real diagonal matrix $D$ by a diagonal unimodular complex matrix with diagonal elements 
\[
Q^{a}={\exp}(iq^{a}).
\]

PROPOSITION 36. {\it For the kinetic internal energy we obtain the following expression:
\[
\mathcal{T}_{int}=\frac{1}{2a}\sum_{a}p_{a}^{2}+\frac{1}{32a}\sum_{i,j}\frac{(M_{ab})^{2}}{\sin^{2}\frac{q^{a}-q^{b}}{2}}+
\frac{1}{32a}\sum_{i,j}\frac{(N_{ab})^{2}}{\cos^{2}\frac{q^{a}-q^{b}}{2}}.
\] 
The plus sign of the $N$-term does not mean that the attraction is absent, because with the circular topology it is difficult to distinguish between repulsion and attraction}.

To finish these analogies, let us quote the corresponding expression for the usual model of an isotropic affine body, 
(\ref{eq459})/(\ref{eq461}) with isotropic inertia $J^{AB}=I\eta^{AB}$. 

PROPOSITION 37. {\it Then one obtains in terms of the two-polar decomposition:
\[
\mathcal{T}_{int}=\frac{1}{2I}\sum_{a}P_{a}^{2}+\frac{1}{8I}\sum_{a,b}\frac{(M_{ab})^{2}}{(Q^{a}-Q^{b})^{2}}+
\frac{1}{8I}\sum_{a,b}\frac{(N_{ab})^{2}}{(Q^{a}+Q^{b})^{2}}.
\]
It is seen that without affine invariance this modified Calogero model is practically useless, admitting only repulsion of deformation invariants}.

The partial separation of variables and the lattice structure appear only when the phase-space language is used. Obviously, we have reviewed above only the kinetic energy terms satisfying some interesting, perhaps even fascinating symmetry demands. And it turns out that even the purely geodetic models with Lagrangians just equal to those kinetic energies may describe the stable elastic vibrations. In any case it is so when we restrict ourselves to the $\rm{SL}$-invariant models, e.g., restricting the dilatational vibrations by some appropriately chosen one-dimensional potentials. But of course some more general potentials, first of all ones depending only on deformation invariants, $V(q^{1}, \cdots, q^{n})$ are admitted. Obviously, the class of such realistic models is rather very special, but it is always so when we wish to calculate something analytically, or at least reducing ourselves to solutions given by the known special functions of mathematical physics. And systems based on symmetries under Lie groups are very promising from this point of view. In any case, it is quite admissible to restrict ourselves to the very special classes of models like (\ref{eq502}), (\ref{x}) or (\ref{eq503}),  (\ref{y}) with additionally admitted potential terms like $V(q^{1}, \cdots, q^{n})$. The question remains however, concerning physical applicability of affine models. It was mentioned that they may be so useful in elastic problems like Maupertuis principle in theoretical mechanics. But there are also more direct applications. For example, models of the type (\ref{eq503}),  (\ref{y}) may be interpreted as a finite-dimensional discretization of the Arnold group-theoretic approach to the dynamics of incompressible ideal fluids (obviously when $\rm{Tr}\Sigma=\rm{Tr}\hat{\Sigma}=0$. Besides, the both kinds of models (\ref{eq502}), (\ref{x}) and (\ref{eq503}), (\ref{y}) are expected to be useful when dealing with some special objects like soap bubbles, atomic nuclei or neutron stars. Obviously, when trying to describe the last two subjects, one should use the quantized version of the theory. It is interesting that in the last two kinds of applications one should use the operator-based quantized version of the theory. Let us observe that in addition to the spectra of (\ref{z}) there appears in (\ref{eq502}) the term $\hbar^{2}/2c \ v(v+1)$, and in (\ref{eq503}) the term $\hbar^{2}/2c \ s(s+1)$. Here $v, s$ are quantum numbers which are non-negative integers, or rather half-integers when instead the $\rm{SL}(3, \mathbb{R})$-groups their non-linear coverings  $\overline{\rm{SL}(3, \mathbb{R})}$ are used. Obviously, we mean here the "physical" special case $n=3$. We do not feel astonished by the term $\hbar^{2}/2c \ s(s+1)$, because $\hbar^{2} s(s+1)$ is the eigenvalue of the quantum operator of the squared spin. But what is the meaning of the term $\hbar^{2}/2c \ v(v+1)$? The eigenvalue of the squared isospin of the nuclei? But if so, why the full linear combination of $\hbar^{2}/2c \ s(s+1)$, $\hbar^{2}/2c \ v(v+1)$ does not occur? To admit them one should modify the expression for the kinetic energy. Namely, it must contain all apriori possible metrically (orthogonally)-invariant terms and the two-side affinely invariant term (\ref{z}). Therefore, the kinetic energy should be given as:
\begin{eqnarray*}
T&=&\frac{1}{2}\left( m_{1}G_{AB}+m_{2}\eta_{AB}\right)\hat{v}^{A}\hat{v}^{B}\\
&+&\frac{1}{2}\left(I_{1}G_{KL}G^{MN}+I_{2}\eta_{KL}\eta^{MN}+I_{3}G_{KL}\eta^{MN}+I_{4}\eta_{KL}G^{MN}\right)\hat{\Omega}^{K}{}_{M}
\hat{\Omega}^{L}{}_{N}\\
&+&\frac{A}{2}\hat{\Omega}^{I}{}_{J}\hat{\Omega}^{J}{}_{I}+
\frac{B}{2}\hat{\Omega}^{I}{}_{I}\hat{\Omega}^{J}{}_{J}.
\end{eqnarray*}
Alternatively, this may be written as follows:
\begin{eqnarray*}
T&=&\frac{1}{2}\left( m_{1}g_{ij}+m_{2}C_{ij}\right)v^{i}v^{j}\\
&+&\frac{1}{2}\left(I_{1}g_{kl}g^{mn}+I_{2}C_{kl}C^{mn}+I_{3}g_{kl}C^{mn}+I_{4}C_{kl}g^{mn}\right)\Omega^{k}{}_{m}
\Omega^{l}{}_{n}\\
&+&\frac{A}{2}\Omega^{i}{}_{j}\Omega^{j}{}_{i}+
\frac{B}{2}\Omega^{i}{}_{i}\Omega^{j}{}_{j}.
\end{eqnarray*}

Both expressions for $T$ are mutually equal. And obviously, we can also use non-geodetic Lagrangians:
\[
L=T-V(q^{1}, \cdots, q^{n}).
\]
It justifies our patiency to transform the above expressions to the form similar to (\ref{eq502}), (\ref{eq503}). Nevertheless, it is possible and probably it will result in appearing the terms like a linear combination of $\rm{Tr}(V^{2})$, $\rm{Tr}(S^{2})$.

\section{Groups and phase spaces}

Let us make a detour at this point. We will follow the dictum of Eugene Wigner which we paraphrase: {\it physics resides in the symmetry group of the system}. In this setting, first one has a classical physical system. Then it may be placed in a phase space setting. This phase space has a dynamical or kinematic group $G$ of symplectic transformations; $G$ is a locally compact Lie group. So one obtains the Lie algebra $\mathfrak{g}$ from $G$. In the previous part the Lie algebra of $G$ was denoted by $G'$. For some reasons it is convenient to deviate from this notation in this section and to denote the Lie algebra by $\mathfrak{g}$. Suppose that $\mathfrak{g}$ is finite $(n)$ dimensional. Since $\mathfrak{g}$ is a linear space, choose a basis $\{X_{j}\}$ for $\mathfrak{g}$, and then since $\mathfrak{g}$ is also an algebra,
\begin{equation*}
[X_{i}, X_{j}]=\sum^{n}_{k=1}c^{k}{}_{ij}X_{k}, \quad c^{k}{}_{ij} \in \mathbb{R}.
\end{equation*}
The $c^{k}{}_{ij}$ are structure constants for $\mathfrak{g}$.
Let us define $\mathfrak{g}^{\wedge n}$ by $\mathfrak{g}=\mathfrak{g}^{1}$, $\mathfrak{g}\wedge\mathfrak{g}=\mathfrak{g}^{\wedge 2}$, $\mathfrak{g}^{\wedge n}=\mathfrak{g}\wedge\mathfrak{g}^{\wedge (n-1)}$. Thus, we obtain the chain complex $\wedge^{\ast}[\mathfrak{g}]$,
\begin{equation*}
\begin{array}{c}
 \\
0
\end{array}
\begin{array}{c}
 \\
\leftarrow\cdots
\end{array}
\begin{array}{c}
d_{n-1} \\
\leftarrow
\end{array}
\begin{array}{c}
 \\
\mathfrak{g}^{\wedge (n-1)}
\end{array}
\begin{array}{c}
d_{n} \\
\leftarrow
\end{array}
\begin{array}{c}
 \\
\mathfrak{g}^{\wedge n}
\end{array}
\begin{array}{c}
d_{n+1} \\
\leftarrow
\end{array}
\begin{array}{c}
 \\
\cdots
\end{array}
\end{equation*}
with
\[
d_{n}\left(X_{1}\wedge\wedge X_{n}\right)= \sum_{1\leq i < j\leq n}(-1)^{j}X_{1}\wedge...\wedge X_{i-1}\wedge[X_{i},X_{j}]\wedge X_{i+1}... \wedge\widehat{X}_{j}\wedge...\wedge X_{n},
\]
$X_{i}\in \mathfrak{g}$, and extended linearly. 

Let us define $\mathfrak{g}^{*}$ as the dual to the Lie algebra $\mathfrak{g}$. Define the dual basis $\{dX^{j}\}$ for $\mathfrak{g}^{*}$ by
\begin{equation*}
dX^{i}(X_{j})=\delta^{i}{}_{j}.
\end{equation*}

There is another form for all this. Now you may define the graded coboundary operator $\delta$ by
\begin{eqnarray*}
\delta_{0} f(p)&=&\sum^{n}_{i=1}\frac{\partial f}{\partial X^{j}}(p)\left(
dX^{j}\right)_{p}\quad {\rm for}\quad f\in C^{\infty}(M),\\ \nonumber
\delta_{1} \sum^{n}_{i=1}f_{i}(p)\left(dX^{i}\right)_{p}&=&\sum^{n}_{i,j=1}
\frac{\partial f_{i}(p)}{\partial X^{j}}\left(dX^{j}\right)_{p}\wedge \left(dX^{i}\right)_{p}+\sum^{n}_{i=1}f_{i}(p)\delta_{1}(dX^{i})_{p}, 
\end{eqnarray*}
\begin{equation*}
 f_{i}\in C^{\infty}(M),
\end{equation*}

and on higher order forms $X$, $Y\in \mathfrak{g}^{*}$ by
\begin{equation*}
\delta(X\wedge Y)=(\delta X)\wedge Y+(-1)^{q}X\wedge \delta Y,
\end{equation*}
where $X$ is a $q$-form, i.e., $X \in (\mathfrak{g}^{*})^{\wedge q}$. Then we obtain
\begin{equation*}
\begin{array}{c}
 \\
\cdots
\end{array}
\begin{array}{c}
\delta_{n-2} \\
\rightarrow
\end{array}
\begin{array}{c}
 \\
(\mathfrak{g}^{*})^{\wedge (n-1)}
\end{array}
\begin{array}{c}
\delta_{n-1} \\
\rightarrow
\end{array}
\begin{array}{c}
 \\
(\mathfrak{g}^{*})^{\wedge n}
\end{array}
\begin{array}{c}
\delta_{n} \\
\rightarrow
\end{array}
\begin{array}{c}
 \\
\cdots
\end{array}
\end{equation*}
and 
\begin{eqnarray*}
Z^{n}(\mathfrak{g})&=&{\rm Ker}\delta_{n}\equiv\ {\rm the\ space\ of\ cocycles},\\ 
B^{n}(\mathfrak{g})&=&{\rm image}\ \delta_{n-1}\equiv\ {\rm the\ space\ of\ coboundaries},\\ 
H^{n}(\mathfrak{g})&=&Z^{n}(\mathfrak{g})/B^{n}(\mathfrak{g}).
\end{eqnarray*}

We have that $\delta_{n-1}\delta_{n}=0$ for all $n\in \mathbb{N}$ or $\delta^{2}=0$. 

There is a result of V. Guillemin and S. Sternberg \cite{20} that says that all the symplectic spaces on which $G$ (or $\mathfrak{g}$) acts symplectically are isomorphic to the form given as follows:
\begin{enumerate}
\item Take $\omega$ from (one orbit of) $Z^{2}(\mathfrak{g})$.
\item Define $\mathfrak{h}_{\omega}=\left\{g\in\mathfrak{g}:\omega(g\wedge\cdot)=0\right\}$; $\mathfrak{h}_{\omega}$ is a Lie subgroup of $\mathfrak{g}$.
\item Form $H_{\omega}$ by exponentiating $\mathfrak{h}_{\omega}$. 
\item If $H_{\omega}$ is a closed subgroup of $G$, then $G/H_{\omega}$ is a symplectic space.
\item To obtain the general form of symplectic space, initially take
\[
\bigcup^{\cdot}_{{\rm some\ of\ the\ } \omega{\rm 's\ }}G/H_{\omega}.
\]
\end{enumerate}
For the usual classical mechanics, in which the action of $G$ is transitive, one takes the disjoint union over a single orbit of a single $\omega$ \cite{6a}. The proof of this fact is embedded in the proof provided by Guillemin and Sternberg. 
Now suppose $G/H_{\omega}$ is given and has dimension $2n$. 

The computation of all this is aided by the following which follows from $\delta^{2}=0$:
\begin{equation*}
\delta(dX^{k})=-\frac{1}{2}\sum_{1\leq i,j\leq n}c^{k}{}_{ij}dX^{i}\wedge dX^{j}.
\end{equation*}
Thus the coboundary operator is in fact determined by the structure of the Lie group in a direct way.

A particularly convenient form for $q$-forms for Lie groups is obtained by first taking the $G$-left invariant vector fields $X_{j}$ and then obtaining the $G$-left invariant  $q$-forms as a special case of the following:
\begin{equation*}
[g^{*}dX](Y)=(dX)(g_{*}Y)
\end{equation*}
for $dX$ a $q$-form and $Y$ an element of $\mathfrak{g}^{\wedge q}$.

The question of whether there are central extensions of $G$ is partially answered by the answer to the question "Is $H^{(1)}(\mathfrak{g})=H^{(2)}(\mathfrak{g})=\{0\}$ (no central extensions) or not?" It also has to do with whether or not the group has representations other than the coadjoint representations. For this, in the case where $\mathfrak{g}$ is semisimple, then it can be shown that $H^{(1)}(\mathfrak{g})=H^{(2)}(\mathfrak{g})=\{0\}$. This is the case for example of the Lorentz group on the Minkowski space $\mathfrak{M}^{4}$. The Euclidean group in three or more dimensions has  $H^{(1)}(\mathfrak{g})=H^{(2)}(\mathfrak{g})=\{0\}$ but is not semisimple. But the Galilei group has  ${\rm dim}(H^{(1)}(\mathfrak{g}))={\rm dim}(H^{(2)}(\mathfrak{g}))=1$ with the one degree of freedom being parametrized by mass $m$. In the cases of $\mathfrak{g}$ for which $=H^{(2)}(\mathfrak{g})=\{0\}$, then one has that every $\omega \in Z^{2}(\mathfrak{g})$ is of the form $\delta \omega_{1}$ where $\omega_{1}$ is some one-form. Thus we may characterize any two-form in $Z^{2}(\mathfrak{g})$ by the one-form from which it derives. Hence, every representation of $G$ is a coadjoint representation. Such a characterization is often in terms of the mass and/or spin. In cases such as the Galilei group, the coadjoint orbit representation corresponds to choosing the mass $m$ equal to zero, and only the representations with $m>0$ correspond to what we think of as the "proper" representations. See \cite{16a} for representations of the Galilei group.

Then a left-invariant measure, $\mu$, on $G/H_{\omega}$ is given as follows \cite{20}. We have a map $\rho:G \rightarrow G/H_{\omega}$. Then a symplectic form on $G/H_{\omega}$ is just $\overline{\omega}=\rho^{*}\omega$, and 
\[
\mu=\overline{\omega}^{\wedge n}.
\]
We will abbreviate this and just say 
\begin{equation}\label{ap}
\mu=\rho^{*}\omega^{\wedge n}.
\end{equation}

It may seem curious to have more than one $\omega$ from (an orbit of) $Z^{2}(\mathfrak{g})$ and thus more than one phase space for $G$. But this is not new. In a similar fashion, given a kinematical group for a particle of one mass/spin, you may generate the representations for all particles of any mass/spin. Here, we find that given one phase space on which $G$ acts, we may generate all phase spaces on which $G$ acts according to the number of orbits in $Z^{2}(\mathfrak{g})$.

We look more deeply at the at the factoring of $G$ into $G/H_{\omega}$. Let 
\begin{equation*}
\sigma: G/H_{\omega} \rightarrow G
\end{equation*}
be a Borel section of the group. Since every $x \in G/H_{\omega}$ is of the form $x=\sigma(x)H_{\omega}$, we have
\begin{equation*}
\sigma: x \in G/H_{\omega} \mapsto \sigma(x) \in G.
\end{equation*}
Then
\begin{equation*}
g\circ (\sigma(x)H_{\omega})=gx=\sigma(gx)H_{\omega}.
\end{equation*}
Hence, there is a unique $h(g, x) \in H_{\omega}$ such that 
\begin{equation*}
g\circ \sigma(x) \circ h(g, x)=\sigma(gx).
\end{equation*}
From the associative property of $G$, we obtain the generalized cocycle condition
\begin{equation}\label{ap2}
h(g_{1} \circ g_{2}, x)=h(g_{2}, x)h(g_{1}, g_{2}x).
\end{equation}

For the purposes of obtaining a (projective) representation of $G$ on $G/H_{\omega}$, we define
\begin{equation*}
\alpha: H_{\omega} \rightarrow \{c \in \mathbb{C} \ \ | \  || c ||=1\}.
\end{equation*}
Then we may perform Mackey's induced representation starting from this $\alpha$. This also gives the treatment of spin eventually.

Next, let us define a left action of $g \in G$ on $x=\sigma(x)H_{\omega}$ by
\begin{equation*}
L(g)x=g \circ (\sigma(x)H_{\omega})\equiv (g \circ \sigma(x))H_{\omega}.
\end{equation*}
This $L$ defines a representation of $G$. 

There is also a right action of $G$ on the disjoint union
\begin{equation}\label{eq38a}
\bigcup^{\cdot}_{ \omega ' \ \in \ {\rm orbit \ of \ }\omega}G/H_{\omega '}.
\end{equation}
Let $x$ now be an element of (\ref{eq38a}). Then 
\begin{eqnarray}\label{eq38b}
R(g^{-1})x&=&\bigcup^{\cdot}_{ \omega ' \ \in \ {\rm orbit \ of \ }\omega}\sigma(x)H_{\omega '}g^{-1} \nonumber \\
&=&\bigcup^{\cdot}_{ \omega ' \ \in \ {\rm orbit \ of \ }\omega}\sigma(x) \circ g^{-1}(gH_{\omega '}g^{-1}) \nonumber \\
&=& \bigcup^{\cdot}_{ \omega '' \ \in \ {\rm orbit \ of \ }\omega}\sigma(x) \circ g^{-1}H_{\omega ''}  \nonumber \\
& \in &\bigcup^{\cdot}_{ \omega '' \ \in \ {\rm orbit \ of \ }\omega}G/H_{\omega ''}       
\end{eqnarray}
where
\[
\omega ''=L^{*}(g)R^{*}(g^{-1})\omega '.
\]
This insures that we have a space 
\[
\bigcup^{\cdot}_{ \omega '' \ \in \ {\rm orbit \ of \ }\omega}G/H_{\omega ''}
\]
with a left action of $G$ on it in spite of it starting out as a right action of $G$ on it! This $R$ defines a representation of $G$. We may simplify this since the left and right translations commute with each other, and  
for left-invariant vector fields we have
\begin{equation*}
[L^{*}(g)\omega](X, Y)=\omega(L_{*}(g)X, L_{*}(g)Y)=\omega(X, Y).
\end{equation*}
Hence, we have 
\begin{equation*}
\omega ''=R^{*}(g^{-1})\omega '.
\end{equation*} 
Moreover, we have
\begin{eqnarray*}
L_{*}(g)R_{*}(g^{-1})\mathfrak{h}_{\omega}&=&R_{*}(g^{-1})\mathfrak{h}_{\omega} \nonumber \\
&=&\{R_{*}(g^{-1})X \in TG \ | \ \omega(R_{*}(g^{-1})X, Y)=0, \forall \ Y  \in TG \}\nonumber \\
&=&\{R_{*}(g^{-1})X \in TG \ | \ \omega(R_{*}(g^{-1})X, R_{*}(g^{-1})Y) =0, \forall \ Y \in TG \} \nonumber \\
&=& \{X \in TG \ | \ [R^{*}(g^{-1})\omega](X, Y) =0, \forall \ Y \in TG \} \nonumber \\
&=& \mathfrak{h}_{R^{*}(g^{-1})\omega}.  
\end{eqnarray*}

Thus, for $x=\sigma(x)H_{\omega} \in G/ H_{\omega}$, we may write
\begin{equation*}
R(g^{-1})x=\sigma(x)\circ g H_{R^{*}(g^{-1})\omega}.
\end{equation*} 
Similarly, for $x=\sigma(x)H_{\omega} \in G/ H_{\omega}$; so, $R(g^{-1})x=\sigma(x)\circ g H_{R^{*}(g^{-1})\omega}$, and
\[
R(g^{-1}{}_{1})R(g^{-1})x=[\sigma(x)\circ g] \circ g_{1}H_{R^{*}(g^{-1}{}_{1})R^{*}(g^{-1})\omega}
\]
or
\begin{equation*}
R(\{g \circ g_{1} \}^{-1})x=[\sigma(x) \circ \{g \circ g_{1} \}]H_{R^{*}(\{g \circ g_{1} \}^{-1})\omega}.
\end{equation*}
Hence, $g \mapsto R(g^{-1})$ defines again a representation of $G$. Furthermore, writing
\[
\{x_{g} \ | \ g \in G \} \in \bigcup^{'}_{g \in G}G/H_{R^{*}(g^{-1})\omega},
\]
or
\[
\{x_{g} \ | \ g \in G \} \in \bigcup^{'}_{g \in G}\sigma_{g}(x_{g})H_{R^{*}(g^{-1})\omega},
\]
we may write
\[
\sigma_{g}(x_{g})=\sigma_{e}(x)\circ g \equiv \sigma(x)\circ g
\]
where we have defined $\sigma_{e}\equiv \sigma$ as the section $\sigma: G/ H_{\omega} \rightarrow G$. This has the added advantage that $\sigma_{g}$ is a continuous function of $g \in G$. Hence, we have that for $\omega \in \mathbb{Z}^{2}(\mathfrak{g})$ fixed,
\begin{equation}\label{ap5}
\{x_{g} \ | \ g \in G \}= \bigcup^{'}_{g \in G}\sigma(x)\circ gH_{R^{*}(g^{-1})\omega}.
\end{equation}

Now, only in the case in which $H_{\omega}$ is normal in $G$ is $R$ an action on $G/H_{\omega}$ for a single $\omega$. This is the case for the Heisenberg group, but in almost all other groups, $H_{\omega}$ is not normal in $G$.

Consequently, we see that to obtain a right representation of $G$, we must take representations over an entire orbit of $G/H_{\omega}$ or of a union of such spaces. If it is a representation that is transitive, it is of the first type.

We might also frame this in terms of the right cosets and derive the analogies to $L(g)$ and $R(g)$. Realization of the homogeneous space in terms of the left or right cosets $G/H$ and $H \backslash G$ is primarily a matter of convention. We choose to work on $G/H$ exclosively to have the group acting on the left on $G/H$ as in "Let the group $G$ act on the left of manifold $X$ in the following" which is a common assumption. There are also some physical reasons for choosing a particular one of $G/H$ or $H \backslash G$.

\section{Quantum Mechanics and Classical Mechanics on Hilbert Spaces}

We now present a form of "quantum mechanics on phase space" that is a
generalization of the standard quantum mechanics to functions on an
arbitrary phase space. The reason we do this here is that this is outside
the repertoire of most physicists to date, and is very powerful. We mean by
quantum mechanics on phase space a formalism that is \textit{not} the
formalism provided by Wigner, Weyl, Moyal, and Ville; it is rather the
formalism related to the classical phase space in the form $G/H_{\omega }$\
where $G$ is a Lie group, etc. as we discussed above. We shall
first work on the Hilbert space $L^{2}{}_{\mu }(G/H_{\omega })$, and define
what operators are there as the image of the classical operators. In this
fashion, we shall obtain a meaning for the classical momentum, etc., but in $%
L^{2}{}_{\mu }(G/H_{\omega })$. Then we shall take every one of the Hilbert
spaces, $\mathcal{H}$, that occur in the standard quantum mechanics and that
are in fact irreducible representation spaces for $G$, and intertwine $%
\mathcal{H}$ with $L^{2}{}_{\mu }(G/H_{\omega })$. This will entail having a
vector $\eta \in \mathcal{H}$\ (with its coherent states) that somehow
describes the way that we shall measure every other vector $\psi \in 
\mathcal{H}$. Then we take any of the classical operators in
$L^{2}{}_{\mu }(G/H_{\omega })$ and drag them down to $\mathcal{H}$ by using the
intertwining operator, thereby obtaining what may be taken to be the quantum
counterpart to the momentum, etc. Using this form of quantization, we obtain
quantization without any obstructions. By this, we obtain a
language in quantum mechanics that is a valid reflection of the
corresponding classical situation. We will also obtain a formalism for
mapping from the classical setting to the quantum setting in which we
preserve any upper or lower bounds on the spectrum of the various operators,
whether classical or quantum mechanical. For example, we do not have the
problems with the energy of an electron having negative eigenvalues for an
energy that was classically positive. Many of the problems of quantum
mechanics thus disappear when we treat quantum mechanics on phase space,
just as in the case (section $7$) of singularities disappearing if one
reformulated the problem of the asymptotics of propagator properties for
Huyghen's principle in phase space.

There is another reason to transfer to $L^{2}{}_{\mu }(G/H_{\omega })$, or to
be more precise to $\cup _{g\in G}^{\bullet }L^{2}{}_{\mu }\newline(G/  (g^{-1}\circ
H_{\omega }\circ g))$. There one may transfer all the properties of
classical mechanics to quantum mechanics, just as the procedure of Koopman 
\cite{6a}, \cite{6b}, \cite{Ko} did for the formalism of the Heisenberg group. The
only difference between quantum mechanics and classical mechanics is how one
represents $G$ on $\cup _{g\in G}^{\bullet }L^{2}{}_{\mu }(G/(g^{-1}\circ
H_{\omega }\circ g))$; quantum mechanics has $G$ represented on the left
with $L(G)$,\ and classical mechanics on the right with $R(G)$. Thus, there
is no taking a limit as the physical constant $\hslash \rightarrow
0 $, and moreover, one may implement both classical mechanics and quantum
mechanics on the same space.

Hence, in this section, we will discuss several topics including:
\begin{itemize}
\item [$ 1)$]  that classical mechanics and quantum mechanics may be put on the same
Hilbert space which is, roughly speaking, $L^{2}{}_{\mu }(G/H_{\omega }),$

\item [$ 2)$] the physical interpretation of measurement in this scheme,

\item [$ 3)$] informational completeness in this scheme,

\item [$ 4)$] an application to solid state mechanics,

\item [$ 5)$] the major value of quantum mechanics on phase space.
\end{itemize}

\subsection{The Hilbert spaces}

In this subsection we will begin with the phase spaces derived in general in
section $1$ from a locally compact Lie group, $G$, with a finite
dimensional Lie algebra, $\mathfrak{g}$. As before, we compute $Z^{2}(%
\mathfrak{g});$ pick $\omega \in Z^{2}(\mathfrak{g});$ form $\mathfrak{h}%
_{\omega }$, $H_{\omega }$; and then form $G/H_{\omega }$ supposing $%
H_{\omega }$ is a closed subgroup of $G$. Finally we work on $\cup _{g\in
G}^{\bullet }G/H_{R_{g}^{\ast }\omega }$ or simply on $G/H_{\omega }$.
(Recall that $g^{-1}H_{\omega }g=H_{R_{g}^{\ast }\omega }$.) Included in
this formalism are the cases in which $H_{\omega }$\ is normal in $G$ so we
may work in a single $G/H_{\omega }$. We also work with the cases in which $%
\omega \in B^{2}(\mathfrak{g})$\ so we may use the coadjoint orbit
representation if we choose. But these are only special cases of the
formalism; the formalism is valid for all elements $\omega \in Z^{2}(%
\mathfrak{g})$\ such that $H_{\omega }$\ is a closed subgroup of $G$ -
period.

\subsubsection{$L_{\protect\mu }^{2}(G/H_{\protect\omega })$ as a
left-representation space of $G$}

We have the separable Hilbert space $L^{2}{}_{\mu }(G/H_{\omega })$\ where $%
\mu $\ is the left $G$-invariant measure coming from $\omega $\ as in (\ref{ap}). $%
L^{2}{}_{\mu }(G/H_{\omega })$\ has a natural left-regular representation $%
V_{L}$\ of $G$\ given by%
\begin{eqnarray*}
\lbrack V_{L}(g)\Psi ](\boldsymbol{x}) &=&\Psi (L(g^{-1})\boldsymbol{x}%
)=\Psi (g^{-1}\boldsymbol{x}),  \notag \\
\Psi &\in &L^{2}{}_{\mu }(G/H_{\omega }),\text{ }g\in G,\text{ }\boldsymbol{x}%
\in G/H_{\omega }.
\end{eqnarray*}%
From the left-regular representation, we obtain the projective
representations $V^{\alpha }{}_{L}$\ of $G$\ on $\Psi \in L^{2}{}_{\mu }(G/H_{\omega })$\ given by%
\begin{equation*}
\lbrack V^{\alpha }{}_{L}(g)\Psi ](\boldsymbol{x})=\alpha (h(g^{-1},%
\boldsymbol{x}))\Psi (L(g^{-1})\boldsymbol{x}),  \notag
\end{equation*}%
with $h$ satisfying the generalized cocycle condition, (\ref{ap2}), and $\alpha $ a
one dimensional representation of $H_{\omega }$. The fact that $%
V^{\alpha }{}_{L}$\ is a representation of $G$ follows from the generalized
cocycle condition. The operators $V^{\alpha }{}_{L}(g)$\ are unitary on $%
L^{2}{}_{\mu }(G/H_{\omega })$ \cite{Sch}.

This Hilbert space, however, is not a Hilbert space of a quantum mechanical
particle as it is not an irreducible representation space of $G$. To obtain
an irreducible representation, we first note that the eigenvalues of the
Casimir invariants of $G$ are also the parameters of $Z^{2}(\mathfrak{g})$.
We will choose them to be the same.

For example, if we take the group%
\begin{equation*}
G=(\text{phase group}\times \mathbb{R}_{{\rm configuration}}^{3}\times \mathbb{R}%
_{{\rm momentum}}^{3})\rtimes {\rm SO}(3)
\end{equation*}%
which is the Heisenberg group with the rotations added, and a basis for $%
\mathfrak{g}$ is $\{\Theta ,Q_{j},P_{j},J_{j}\}$. Then $Z^{2}(\mathfrak{g})$
includes $m\sum_{j=1}^{3}P_{j}^{\ast }\wedge Q_{j}^{\ast }+SJ_{1}^{\ast
}\wedge J_{2}^{\ast }$ with $m=$ the mass and $S=$ the $z$-component of
spin. Now%
\begin{equation*}
\mathfrak{h}_{\omega }=\{\text{the phases and }\beta J_{3}\text{, }\beta \in 
\mathbb{R}\}.
\end{equation*}%
From this, we obtain 
\begin{equation*}
H_{\omega }=\{(\text{the phase group,}\boldsymbol{0},\boldsymbol{0})\rtimes
e^{\beta J_{3}}\mid \beta \in \mathbb{R}\}.
\end{equation*}%
Thus, $H_{\omega }$\ is a closed subgroup of $G$; so, $G/H_{\omega }$\ is a
phase space. In fact, $G/H_{\omega }$\ is topologically just $(\mathbb{R}%
_{{\rm configuration}}^{3}\times \mathbb{R}_{{\rm momentum}}^{3})\rtimes \lbrack
{\rm SO}(3)]_{J_{3}-{\rm rotations}}$.

We may take 
\begin{eqnarray*}
\alpha (e^{i\theta },0,0,e^{\beta J_{3}}) &=&e^{i\lambda _{1}\theta
}e^{i\lambda _{2}\beta } \\
\text{for any fixed }\lambda _{1},\lambda _{2} &\in &\mathbb{R}  \notag
\end{eqnarray*}%
to obtain a representation $V^{\alpha }{}_{L}$ of $G$.

For a general $G$, there are a multitude of operators on $L^{2}{}_{\mu }(G/H_{\omega })$ 
of course. Since $G/H_{\omega }$\ is a classical
phase space, we take the classical observables to be 
\begin{equation*}
\mathcal{F}=\{f:G/H_{\omega }\rightarrow \mathbb{R}\mid f\text{ is }\mu 
\text{-measurable\}.}
\end{equation*}%
We take the operators $A(f)$\ on $\Psi \in L^{2}{}_{\mu }(G/H_{\omega })$\ to
be given by%
\begin{eqnarray*}
\lbrack A(f)\Psi ](\boldsymbol{x}) &=&f(\boldsymbol{x})\Psi (\boldsymbol{x})
\notag \\
x &\in &G/H_{\omega },\text{ }f\in \mathcal{F}.
\end{eqnarray*}%
In this fashion, we represent every classical observable, $f$, in $L^{2}{}_{\mu }(G/H_{\omega })$
 by a multiplication operator, $A(f)$.

Being operators of multiplication, the set of all $A(f)$\ commute among
themselves. Furthermore, the $A(f)$\ satisfy covariance:%
\begin{eqnarray*}
V^{\alpha }{}_{L}(g)A(f)V^{\alpha }{}_{L}(g)^{-1} &=&A(_{g}f),  \notag \\
(_{g}f)(\boldsymbol{x}) &=&f(L(g^{-1})\boldsymbol{x}).
\end{eqnarray*}

We next investigate the connection between (quantum) mechanics on $L^{2}{}_{\mu }(G/H_{\omega })$
and ordinary classical mechanics on $G/H_{\omega }$.

First of all, we define a projection operator $P_{\Psi }$\ on $L^{2}{}_{\mu }(G/H_{\omega })$ as usual:%
\begin{eqnarray*}
P_{\Psi }\Phi &=&\text{ }<\Psi ,\Phi >\Psi ,  \notag \\
\Psi ,\Phi &\in &L^{2}{}_{\mu }(G/H_{\omega }),\text{ }\parallel \Psi
\parallel =1.
\end{eqnarray*}%
Then any density operator, $\rho $, on $L^{2}{}_{\mu }(G/H_{\omega })$\ may be
written in terms of mixing coefficients as%
\begin{equation}\label{ap3}
\rho =\sum_{j}r_{j}P_{\Psi _{j}}
\end{equation}%
for some orthonormal basis $\{\Psi _{j}\}$\ of $L^{2}{}_{\mu }(G/H_{\omega })$%
, and $r_{j}\geq 0$, $\sum_{j}r_{j}=1$. Furthermore, we have for all $\Phi
,\Theta \in L^{2}{}_{\mu }(G/H_{\omega })$, that%
\begin{equation*}
<\Phi ,\rho \Theta >=\int{}_{G/H_{\omega }}d\mu (%
\boldsymbol{y}){}_{G/H_{\omega }}d\mu (\boldsymbol{x})\overline{%
\Phi (\boldsymbol{y})}\rho (\boldsymbol{y},\boldsymbol{x})\Theta (%
\boldsymbol{x})
\end{equation*}%
where we have defined the kernel $\rho (\boldsymbol{y},\boldsymbol{x})$
almost everywhere by%
\begin{equation}\label{ap4}
\rho (\boldsymbol{y},\boldsymbol{x})=\sum_{j}r_{j}\Psi _{j}(\boldsymbol{y})%
\overline{\Psi _{j}(\boldsymbol{x})}.
\end{equation}

Next we take the connection between states, observables, and probability to
be as is usual in quantum theory:%
\begin{equation*}
\text{The expected value of }A(f)\text{ when measured in state }\rho \text{
is\ }{\rm Tr}(A(f)\rho ).
\end{equation*}%
Consequently,%
\begin{equation*}
{\rm Tr}(A(f)\rho )=\int{}_{G/H_{\omega }}d\mu (\boldsymbol{x})f(\boldsymbol{%
x})\rho (\boldsymbol{x},\boldsymbol{x}).
\end{equation*}%
Comparing this with the classical expectation%
\begin{equation*}
{\rm Exp}(f;\rho _{cm})=\int{}_{G/H_{\omega }}d\mu (\boldsymbol{x})f(%
\boldsymbol{x})\rho _{cm}(\boldsymbol{x}),
\end{equation*}%
we see that we obtain equality iff we make the following identification:%
\begin{equation}\label{ap7}
\rho _{cm}(\boldsymbol{x})=\rho (\boldsymbol{x},\boldsymbol{x})\text{ for }%
a.e.\boldsymbol{x}\in G/H_{\omega }.
\end{equation}%
But since $\rho _{cm}$\ is Kolmogorov probability density, we have to check
that $\rho (\boldsymbol{x},\boldsymbol{x})$\ is also. But this follows from
(\ref{ap3}) and (\ref{ap4}).

We also may make the definition

DEFINITION 4. {\it $\rho $ and $\rho^{\prime}$ are classically equivalent (denoted $\rho
\approx \rho^{\prime }$) iff ${\rm Tr}(A(f)\rho )={\rm Tr}(A(f)\rho^{\prime })$ for
all $f\in \mathcal{F}.$}

Thus, $\rho \approx \rho ^{\prime }$\ iff $\rho (\boldsymbol{x},\boldsymbol{x%
})=\rho ^{\prime }(\boldsymbol{x},\boldsymbol{x})$ for almost every $\boldsymbol{x}%
\in G/H_{\omega }.$ Of course, the diagonal elements of the kernel of $\rho $%
\ do not determine $\rho $\ completely as a density operator on $L^{2}{}_{\mu }(G/H_{\omega })$.

\subsubsection{$\cup _{g\in G}^{\bullet }L^{2}{}_{R^{\ast }(g)\protect\mu %
}(G/H_{R^{\ast }(g)\protect\omega })$ as a right-representation space of 
$G$}

As usual, let $\sigma :G/H_{\omega }\rightarrow G$\ be a Borel
cross-section. As an element of $\cup _{g\in G}^{\bullet }\sigma (%
\boldsymbol{x})\circ gH_{R^{\ast }(g^{-1})\omega }$, we must take $%
\boldsymbol{x}$ as $\{\boldsymbol{x}_{g}\mid g\in G\}$. Let us take the
definition of $\{\boldsymbol{x}_{g}\mid g\in G\}$ directly from equation
(\ref{ap5}), i.e., 
\begin{equation*}
\{\boldsymbol{x}_{g}\mid g\in G\}=\cup _{g\in G}^{\bullet }\sigma (%
\boldsymbol{x})\circ gH_{R^{\ast }(g^{-1})\omega }.
\end{equation*}%
Now we write 
\begin{equation*}
\Psi =\{\Psi _{g}\in G/H_{R^{\ast }(g^{-1})\omega }\mid g\in G\}\in \cup
_{g\in G}^{\bullet }L^{2}{}_{R^{\ast }(g^{-1})\mu }(G/H_{R^{\ast
}(g^{-1})\omega })
\end{equation*}%
with%
\begin{equation*}
\Psi _{g}(\boldsymbol{x}_{g})=\Psi _{g}(\sigma (\boldsymbol{x})\circ
gH_{R^{\ast }(g^{-1})\omega }).
\end{equation*}%
We shall abbreviate $L^{2}{}_{R^{\ast }(g^{-1})\mu }$\ with just $L^{2}{}_{\mu }$%
\ with the caveat that we should take $R^{\ast }(g^{-1})\mu $\ rather than $%
\mu $ as the measure.\ Hence

\begin{equation*}
\text{"}\Psi \in \cup _{g\in G}^{\bullet }L^{2}{}_{\mu }(G/H_{R^{\ast
}(g^{-1})\omega }).\text{"}
\end{equation*}

On this space, we define%
\begin{equation*}
<\Psi ,\Phi >\text{ }=\text{ }\{<\Psi _{g},\Phi _{g}>\mid g\in G\},
\end{equation*}%
where the spaces $L^{2}{}_{\mu }(G/H_{R^{\ast }(g)\omega })$\ are related by
unitary representations of $G$. Hence, we obtain 
\begin{equation*}
<\Psi ,\Phi >\text{ }=\text{ }\{<\Psi _{e},\Phi _{e}>\}.
\end{equation*}

Now%
\begin{equation*}
\lbrack A(f)\Psi _{g}](\boldsymbol{x})=f(\boldsymbol{x})\Psi _{g}(%
\boldsymbol{x}),\text{ }f\in \mathcal{F}.
\end{equation*}%
This definition is also consistent with the previous definition of $A(f)$.
Moreover, it makes the definition of $A(\mathsf{H})$\ uniform for $\mathsf{H}
$ any Hamiltonian.

On $\cup _{g\in G}^{\bullet }L^{2}{}_{\mu }(G/H_{R^{\ast }(g)\omega })$\ we
define the right-regular representation $V_{R}$\ by%
\begin{eqnarray*}
V_{R}(g_{1}) &:&\Psi _{g}\rightarrow \Psi _{g\circ g_{1}},  \notag \\
\lbrack V_{R}(g_{1})\Psi _{g}](\boldsymbol{x}) &=&\Psi _{g\circ
g_{1}}(\{\sigma (\boldsymbol{x})\circ g\}\circ g_{1}H_{R^{\ast
}(g_{1}^{-1})R^{\ast }(g^{-1})\omega })  \notag \\
&=&\Psi _{g\circ g_{1}}(\sigma (\boldsymbol{x})\circ \{g\circ
g_{1}\}H_{R^{\ast }(\{g\circ g_{1}\}^{-1})\omega })  \notag \\
&\in &L^{2}{}_{\mu }(G/H_{R^{\ast }(\{g\circ g_{1}\}^{-1})\omega }).
\end{eqnarray*}%
Define $V^{\alpha }{}_{R}$\ by 
\begin{equation*}
\lbrack V^{\alpha }{}_{R}(g_{1})\Psi _{g}](\boldsymbol{x}_{g})=\alpha
(h(g_{1}^{-1},\boldsymbol{x}_{g}))[V_{R}(g_{1})\Psi _{g\circ g_{1}}](%
\boldsymbol{x}_{g\circ g_{1}}).
\end{equation*}%
We have that $V^{\alpha }{}_{R}(g_{1})$\ is a unitary operator on $\cup _{g\in
G}^{\bullet }L^{2}{}_{\mu }(G/H_{R^{\ast }(g^{-1})\omega })$. Again $\cup
_{g\in G}^{\bullet }L^{2}{}_{\mu }\newline (G/H_{R^{\ast }(g^{-1})\omega })$\ is not a
Hilbert space of a single quantum mechanical particle as it is not
irreducible.

Also, the $A(f)$\ are covariant under\ the action of the $V^{\alpha }{}_{R}$:%
\begin{eqnarray*}
V^{\alpha }{}_{R}(g_{1})A(f)V^{\alpha }{}_{R}(g_{1})^{-1} &=&A(f_{g_{1}}), 
\notag \\
f_{g_{_{1}}}(\sigma (\boldsymbol{x})gH_{R^{\ast }(g^{-1})\omega })
&=&f(\sigma (\boldsymbol{x})\circ \{g\circ g_{1}\}H_{R^{\ast }(\{g\circ
g_{1}\}^{-1})\omega }).
\end{eqnarray*}

We may make definitions similar to the definitions of $P_{\Psi }$, $\rho
=\{\rho _{g}\mid g\in G\}$, the kernel of $\rho $, ${\rm Tr}(\rho A(f))$\ as the
probability of measuring $A(f)$\ in state $\rho $, and get the equivalence
relation $\approx $\ in a manner similar to that before, the difference
being that we have to consider them on the orbit of the spaces $L^{2}{}_{\mu
}(G/H_{\omega })$.

It was to be expected that these definitions are necessarily complicated by
the fact that we are working in the disjoint union over an orbit of $%
G/H_{\omega }$. But we gain that%
\begin{equation}\label{ap6}
\text{for }X,Y\in \mathfrak{g}_{e}\text{ with }[X,Y]=0\text{ on }G/H_{\omega}
\end{equation}%
then%
\begin{equation*}
\lbrack (R_{g})_{\ast }X,(R_{g})_{\ast }Y]=0\text{ on }G/H_{(R_{g})^{\ast
}\omega }\text{ }\forall g\in G.
\end{equation*}%
This is necessary for us to be able to implement classical mechanics in $%
L^{2}{}_{\mu }(G/H_{\omega })$. To do this, we only have to satisfy (\ref{ap6}).

For example, we next treat a relativistic case, derive a phase space for it,
and then compute the left and right representations on the appropriate $%
L^{2}{}_{\mu }(G/H_{\omega })$\ or the disjoint orbit of this space.

We have the covering group for the Poincar\'{e} group as 
\begin{equation*}
G=\mathbb{R}^{4}\rtimes {\rm SL}(2,\mathbb{C})
\end{equation*}%
equipped with the Minkowski metric on $\mathbb{R}^{4}$, and for the ${\rm SL}(2,%
\mathbb{C})$ part, we have the Cayley representation of $\mathbb{R}^{4}$\ to
the complex $2\times 2$ matrices:%
\begin{equation*}
(t,x,y,z)\mapsto t\sigma _{0}+x\sigma _{1}+y\sigma _{2}+z\sigma _{3}
\end{equation*}%
where $\sigma _{0}$\ is the identity and $\sigma _{1},\sigma _{2},\sigma
_{3} $\ is the Pauli algebra.

A basis for the Lie algebra corresponding to the time and space
translations, the boosts, and the rotations is $\{Q_{0},Q_{l},P_{l},J_{l}%
\mid k=1,2,3\}.$ The commutation relations for this basis are%
\begin{eqnarray*}
\lbrack J_{1},J_{2}] &=&J_{3},\text{ }[P_{1},P_{2}]=J_{3},\text{ }%
[J_{1},P_{2}]=P_{3}, \\
\lbrack J_{1},Q_{2}] &=&-Q_{3},\text{ }[P_{j},Q_{j}]=-Q_{0},\text{ }%
[Q_{j},Q_{0}]=0, \\
\lbrack J_{l},Q_{0}] &=&0,\text{ }[P_{j},Q_{0}]=Q_{j},\text{ }%
[Q_{j},Q_{k}]=0,
\end{eqnarray*}%
and cyclically. Choose 
\begin{equation*}
\omega =a\delta J_{3}+b\delta P_{3}+c\delta Q_{3}+d\delta Q_{0}
\end{equation*}%
with $a\neq 0,b\neq 0,c\neq 0,d\neq 0$, and you have that your phase space
is just $G/H_{\omega }$. Here $H_{\omega }$\ is not normal in $G$. One may
write any $g\in G$ in the form 
\begin{equation*}
g=\exp (\boldsymbol{u}\cdot \mathfrak{h}_{\omega })\exp (\boldsymbol{v}\cdot
(\mathfrak{g}\backslash \mathfrak{h}_{\omega }))
\end{equation*}%
where we have abbreviated $\boldsymbol{u}\cdot \mathfrak{h}_{\omega }$\
meaning that $\mathfrak{h}_{\omega }$\ has a basis chosen from the basis for 
$\mathfrak{g}$, etc. and $\boldsymbol{u}_{j},\boldsymbol{v}_{j}\in \mathbb{R}
$. Then one has that 
\begin{equation*}
(G/H_{\omega })\circ \exp (\boldsymbol{u}\cdot \mathfrak{h}_{\omega
})=G/H_{\omega }.
\end{equation*}%
So, from the right, the action of $G$ reduces, and $\exp (\boldsymbol{u}%
\cdot \mathfrak{h}_{\omega })$ acts like $\boldsymbol{1}$. Said another way, 
$(\boldsymbol{u}\cdot \mathfrak{h}_{\omega })$\ acts like $\boldsymbol{0}$.
The action from the left has all the properties of the group.

Now for the action on the right on $\cup _{g\in G}^{\bullet }L^{2}{}_{\mu
}(G/H_{R^{\ast }(g^{-1})\omega }).$ Each of the spaces $L^{2}{}_{\mu
} (G/ \newline H_{R^{\ast }(g^{-1})\omega })$\ may be realized as a flat space;
only, the right action of $G$ on each carries it to another. If the action
is by an element in $\exp (\boldsymbol{u}\cdot \mathfrak{h}_{\omega })$, one
obtains of course that $L^{2}{}_{\mu }(G/H_{R^{\ast }(g^{-1})\omega })$\ maps
to itself. But the action on the left is another story. Here it will do to
choose to work in $G/H_{\omega }$\ and then in $L^{2}{}_{\mu }(G/H_{\omega }).$
This is stable under the left action of the group.

\subsubsection{The connection of $L_{\protect\mu }^{2}(G/H_{\protect\omega %
}) $\ with the usual irreducible representation spaces of quantum mechanics}

We investigate the connection between $L^{2}{}_{\mu }(G/H_{\omega })$\ and $%
\mathcal{H}$, the irreducible Hilbert space for a quantum particle of mass $m
$ and spin $S$ or a particle of mass zero and helicity $S$. Here $G$ is one
of the Galilei or Poincar\'{e} groups. We note that $\omega $\ is chosen
from $Z^{2}(\mathfrak{g})$\ by choosing the parameters to be $m$ and $S$ as
discussed in section $1$.

Let $U$ be a left-regular unitary irreducible representation of $G$ on $%
\mathcal{H}$; i.e., $\mathcal{H}$ is a space of functions over some space $%
\boldsymbol{X}$ with an operation of $G$ on the left. Thus, for $\psi \in 
\mathcal{H}$, $g\in G$, and $\boldsymbol{x\in X}$, $[U(g)\psi ](\boldsymbol{x%
})=\psi (L(g^{-1})\boldsymbol{x})=\psi (g^{-1}\boldsymbol{x}).$

Next, pick an $\eta \in \mathcal{H}$\ such that $\parallel \eta \parallel
=1. $ Define%
\begin{equation*}
\lbrack W^{\eta }\psi ](\boldsymbol{x})=\text{ }<U(\sigma (\boldsymbol{x}%
))\eta ,\psi >\text{, }\psi \in \mathcal{H},\text{ }\boldsymbol{x}\in
G/H_{\omega }.
\end{equation*}%
Note that it doesn't matter what $\mathcal{H}$ is, be it a collection of
functions over $p$'s, over $q$'s, over $p$'s and $q$'s, $\cdot \cdot \cdot $
that may have the component functions describing the spin, etc. The
functions $[W^{\eta }\psi ](\boldsymbol{x})$ describe a mapping from $(\psi ,%
\boldsymbol{x})$\ to $\mathbb{C}$, independent of the nature of $\mathcal{H}$%
. \ Also, note that\ we have used capitol Greek letters for elements of $%
L^{2}{}_{\mu }(G/H_{\omega })$\ and lower case Greek letters for elements of $%
\mathcal{H}$.

It turns out that we may choose the $\eta $\ such that $W^{\eta }$\ is an
isometry onto a closed subspace of $L^{2}{}_{\mu }(G/H_{\omega })$. This
property is that "$\eta $\ is $\alpha $-admissible" \cite{Sch}:

 DEFINITION 5. {\it Let $G$, $\omega $, $H_{\omega }$, $\mu $, $\sigma $, $\mathcal{H}$, $U$ be
as before. Then $\eta \in \mathcal{H}$\ is admissible with respect to $%
\sigma (G/H_{\omega })$\ iff $\eta \neq 0$\ and $\int_{G/H_{\omega
}}|<U(\sigma (\boldsymbol{x}))\eta ,\eta >^{2}d\mu (\boldsymbol{x})<\infty .$
Furthermore, $\eta $\ is $\alpha $-admissible iff it is admissible with
respect to $\sigma (G/H_{\omega })$\ and there exists a mapping $\alpha
:H_{\omega }\rightarrow \mathbb{C}$\ such that $U(h)\eta =\alpha (h)\eta $\
for all $h\in H_{\omega }$}.

Then, we have the canonical projection%
\begin{equation*}
P^{\eta }:L^{2}{}_{\mu }(G/H_{\omega })\rightarrow W^{\eta }\mathcal{H}\subset
L^{2}{}_{\mu }(G/H_{\omega }).
\end{equation*}%
Furthermore, the $W^{\eta }$\ intertwine the $U$'s and $V^{\alpha }$'s: 
\begin{equation*}
V^{\alpha }(g)W^{\eta }=W^{\eta }U(g);
\end{equation*}%
i.e.,%
\begin{equation*}
\lbrack V^{\alpha }(g)W^{\eta }\varphi ](\boldsymbol{x})=[W^{\eta
}U(g)\varphi ](\boldsymbol{x})
\end{equation*}%
as can be seen by direct computation.

We may also obtain the orthogonality theorem with the $\alpha $-admissibilty
condition on $\eta $, of which we give just an abbreviated version below 
\cite{Sch}:

THEOREM 1. {\it Let $G$, $\omega $, $H_{\omega }$, $\mu $, $\sigma $, $\mathcal{H}$, $U$ be
as before and let $\eta $\ satisfy the $\alpha $-admissibilty condition.
Then, there is a positive, self-adjoint invertible operator $C$ such that
for all} $\psi ,\varphi \in \mathcal{H}$,
\begin{eqnarray}
\int_{G/H_{\omega }} &<&\varphi ,U(\sigma (\boldsymbol{x}))\eta ><U(\sigma (%
\boldsymbol{x}))\eta ,\psi >d\mu (\boldsymbol{x})  \notag \\
&=&\text{ }\parallel C\eta \parallel ^{2}<\varphi ,\psi >.
\end{eqnarray}

We may renormalize this by $d\mu (\boldsymbol{x})\rightarrow ||C\eta
||^{-2}d\mu (\boldsymbol{x})$.

This theorem has several special cases of interest. For example, if $G$ is
compact, $H_{\omega }=\{e\}$, and $\sigma (\boldsymbol{x})=\boldsymbol{x}\in
G,$ then $C$ is a multiple of the identity; i.e., we obtain the standard
result in that case.

Also, because of the $\alpha $-admissability of $\eta $, we have for any $%
g\in G$,$\ $then there$\ $exists $\boldsymbol{x}\in G/H_{\omega }$ and $h\in
H_{\omega }$ such that $g=\sigma (\boldsymbol{x})h.$ Then $U(g)\eta
=U(\sigma (\boldsymbol{x})h)\eta =\alpha (h)U(\sigma (\boldsymbol{x}))\eta .$
Hence, we are expanding any $\psi $\ in the coherent state basis $\{U(\sigma
(\boldsymbol{x}))\eta \}$ when we apply this theorem.

Next, define the projections $T^{\eta }(\boldsymbol{x})$\ on any $\varphi
\in \mathcal{H}$ by 
\begin{equation*}
T^{\eta }(\boldsymbol{x})\varphi =\text{ }<U(\sigma (\boldsymbol{x}))\eta
,\varphi >U(\sigma (\boldsymbol{x}))\eta ,\text{ }\varphi \in \mathcal{H},
\end{equation*}%
or%
\begin{equation*}
T^{\eta }(\boldsymbol{x})=\text{ }\mid U(\sigma (\boldsymbol{x}))\eta
><U(\sigma (\boldsymbol{x}))\eta \mid .
\end{equation*}

We may then recast the theorem in the form%
\begin{equation*}
\int_{G/H_{\omega }}<\varphi ,T^{\eta }(\boldsymbol{x})\psi >d\mu (%
\boldsymbol{x})=\text{ }\parallel C\eta \parallel ^{2}<\varphi ,\psi >.
\end{equation*}

Next, define the operators $A^{\eta }(f)$\ by 
\begin{align}
A^{\eta }(f)& :\mathcal{H\rightarrow H},  \notag \\
A^{\eta }(f)& \equiv \lbrack W^{\eta }]^{-1}P^{\eta }A(f)W^{\eta }.
\end{align}%
From this definition, we automatically have that the $A^{\eta }(f)$\ are
covariant under the action of $U$ on $\mathcal{H}$. We may think of the
operators $A^{\eta }(f)$\ as the quantization of the classical statistical
observables $f$; i.e., $f$ is a real valued Borel function.

We add the following note on quantization. These operators form a
generalization of the stochastic quantization of Prugove\v{c}ki\ \cite{Prugo}
and \cite{Prugo2}. They also constitute a case of "prime quantization" \cite[%
pp. 459-465]{Ali} for each choice of $\eta $. Furthermore, it also makes
contact with coherent state quantization \cite[pp. 465-473]{Ali}. But it is
quite different from other forms of quantization as here we have a
quantization of every classical observable on $G/H_{\omega }$. The other
forms of quantization have obstructions to quantizing various of the
classical observables on $G$, and/or require the limit as $\hbar \rightarrow
0$\ as one of the inputs to obtain the commutation of the classical
observables \cite{Berezin} \cite{Klauder}.

In particular, we have shown in general \cite[pp. 516-517]{Sch} that when we
choose $f$ to be a linear function on $G/H_{\omega }$, then $A^{\eta }(f)$\
is the operator on $\mathcal{H}$ corresponding to the same function in the
Lie algebra of $G$. If we write $\boldsymbol{x}=(\boldsymbol{q},\boldsymbol{p%
},\boldsymbol{s})$, where $\boldsymbol{q}$ is the usual configuration
variable, $\boldsymbol{p}$ is the usual momentum variable, and $\boldsymbol{s%
}$ includes any other variables we may have such as\ the spin variable, we
have for $f(\boldsymbol{p},\boldsymbol{q},\boldsymbol{s})=q_{j}$, then $%
A^{\eta }(f)=\boldsymbol{Q}_{j}$, and for $f(\boldsymbol{p},\boldsymbol{q},%
\boldsymbol{s})=p_{j}$, then $A^{\eta }(f)=\boldsymbol{P}_{j}$. Here the $%
\boldsymbol{Q}_{j}$ and $\boldsymbol{P}_{j}$ are the operators for the $j$th
component of the position and momentum.

We have also computed \cite[pp. 517-518]{Sch} that in the case of the
Heisenberg group and choosing $\eta $\ to be a harmonic oscillator function, 
$A^{\eta }(f)$ corresponds to antinormal ordering of $f$ as a function of
the operators in $\mathcal{H}$. We have that in this example, the Husimi
transform of $\psi \in \mathcal{H}$, $\parallel \psi \parallel =1$,\ is just 
${\rm Tr}(T^{\eta }(\boldsymbol{x})P_{\psi })$ \cite{Sch3}. Thus we have these
examples that show that this form of quantization produces the correct
quantization.

By computing $<\varphi, A^{\eta }(f)\psi >$\ for arbitrary $\varphi $, $\psi
\in \mathcal{H}$, we also obtain%
\begin{equation*}
A^{\eta }(f)=\int_{G/H_{\omega }}f(\boldsymbol{x})T^{\eta }(\boldsymbol{x}%
)d\mu (\boldsymbol{x}).
\end{equation*}

From this, we see that if $f$ is positive, then $A^{\eta }(f)$\ is positive.
Thus, we have a form of quantization in which there is no difficulty, for
example, with the Hamiltonian which is positive going to something in $%
\mathcal{H}$ which is not always positive. Moreover, we have a positive
operator valued measure (P.O.V.M.) when we do the following: Let%
\begin{equation*}
\mathcal{F}_{1}=\{f\in \mathcal{F}\mid f\text{ is real-valued and }0\leq f(%
\boldsymbol{x})\leq 1\text{ for a.e.}\ \boldsymbol{x}\text{\}.}
\end{equation*}%
Then we prove \cite[pp. 30, 367]{Sch} that
\begin{equation*}
\{A^{\eta }(f)\mid f\in \mathcal{F}_{1}\}\text{ is a normalized P.O.V.M.}
\end{equation*}%
which contains no non-trivial projections. Moreover, it contains \cite{Schf}.

\subsection{A physical interpretation of $\protect\eta $}

We obtain an interpretation of the $\eta $\ in $A^{\eta }(f)$\ as follows:
For $\rho $\ a state on the space $\mathcal{H}$, we may form the Hilbert
space expectation functional ${\rm Tr}(A^{\eta }(f)\rho ).$ We have $\rho =\sum
r_{n}P_{\phi _{n}}$ for $\{\phi _{n}\}$\ some orthonormal basis for $%
\mathcal{H}$. After expanding the $A^{\eta }(f)$ we obtain 
\begin{eqnarray}\label{ap8}
{\rm Tr}(A^{\eta }(f)\rho ) &=&\int_{G/H_{\omega }}f(\boldsymbol{x}){\rm Tr}(T^{\eta }(%
\boldsymbol{x})\rho )d\mu (\boldsymbol{x})  \notag \\
&=&\sum_{n}r_{n}\int_{G/H_{\omega }}f(\boldsymbol{x})\mid <U(\sigma (%
\boldsymbol{x}))\eta ,\phi _{n}>\mid ^{2}d\mu (\boldsymbol{x}),
\end{eqnarray}%
which is an integral and sum over $f$ of the transition probability of $%
U(\sigma (\boldsymbol{x}))\eta $\ and $\phi _{n}$ and which is physically interpretable.
We are taking the coherent state $U(\sigma(x))\eta$ as describing the instrument by which
we measure \cite{Sch4}. We may consider%
\begin{equation*}
\rho \mapsto {\rm Tr}(T^{\eta }(\boldsymbol{x})\rho )
\end{equation*}%
as a connection between the density operators, $\rho $, on $\mathcal{H}$\
and the standard classical statistical states on $G/H_{\omega }$. In analogy
with (\ref{ap7}), we define $\rho _{{\rm classical}}$\ by 
\begin{equation*}
\rho _{{\rm classical}}(\boldsymbol{x})={\rm Tr}(\rho T^{\eta }(\boldsymbol{x})),
\end{equation*}%
and then we may rewrite (\ref{ap8}) in the form%
\begin{equation*}
\int_{G/H_{\omega }}f(\boldsymbol{x})\rho _{{\rm classical}}(\boldsymbol{x})d\mu (%
\boldsymbol{x})={\rm Tr}(A^{\eta }(f)\rho ).
\end{equation*}%
In this way, we explicitly make the connection between classical statistical
expectation values on $G/H_{\omega }$ and quantum mechanical expectation
values on $L^{2}{}_{\mu }(G/H_{\omega })$. Furthermore, the joint distribution
function, $\rho _{{\rm classical}}$ in the various variables in $\boldsymbol{x}$
is measurable, never negative and integrates to 1 over $G/H_{\omega }$.
Hence $\rho _{{\rm classical}}$ is a Kolmogorov probability density on $%
G/H_{\omega }$.

\subsection{Informational completeness of the $A^{\protect\eta }(f)$}

We turn to the question of what happens with other self-adjoint operators on 
$L^{2}{}_{\mu }(G/ \newline H_{\omega })$. For this, we introduce the following:

DEFINITION 6. {\it The set, $\mathcal{O}$, of self-adjoint operators on $\mathcal{H}$ is
informationally complete if and only if, for $\rho ,\rho ^{\prime }\in $
states in} $\mathcal{H}$,
\begin{align}
{\rm Tr}(A\rho )& ={\rm Tr}(A\rho ^{\prime })\text{ for all }A\in \mathcal{O}  \notag \\
& \Longleftrightarrow \rho =\rho ^{\prime }.
\end{align}

We now define $\mathcal{A}_{1}$ as the set %
\[
\{A^{\eta }(f),\text{ }A^{\eta }(f)A^{\eta }(h) \ \mid \ 
f,h \in \{\text{real-valued, measurable functions}\},
\]
\begin{equation}
f,h \ \in \ L^{1}{}_{\mu }(G/H_{\omega })\cap L^{\infty }{}_{\mu }(G/H_{\omega
})\}.
\end{equation}
(In $\mathcal{A}_{1}$ we mean that we break the product $A^{\eta }(f)A^{\eta
}(h)$ into the self-adjoint real and imaginary parts.) Then using this definition 
for $\mathcal{A}_{1}$, we obtain \cite{FS} that 
\begin{equation}
\mathcal{A}_{1}\text{ is informationally complete in }\mathcal{H}\text{ if
and only if }<U(\sigma (\boldsymbol{x}))\eta ,\eta >\text{ }\neq 0.
\end{equation}%
We may simply say "the set of the \ $A^{\eta }(f)A^{\eta }(h)$ is
informationally \ complete" \ since $A^{\eta } \newline (\chi _{G/ H})=1$. Then by a
theorem of P. Busch \cite{Busch}, the set $\mathcal{A}_{1}$ is dense in the
bounded operators on $\mathcal{H}$ in a topology given in terms of the $%
{\rm Tr}(A\rho )$, $A\in \mathcal{O}$. Consequently, the set $\{{\rm Tr}(\rho A^{\eta
}(f)A^{\eta }(h))\}$\ determines a unique $\rho $.

There is a difference between the condition for informational completeness
and the condition for $\approx $\ defined on $L^{2}{}_{\mu }(G/H_{\omega })$
in Definition $1$. This condition for informational completeness has as a
result, that $\eta $\ is spread everywhere in $G/H_{\omega }$. This means
that the set of $A^{\eta }(f)$\ are maximally non-local operators.

Moreover, we have a mapping from $\Delta $\ to $A^{\eta }(\chi _{\Delta
}),\chi _{\Delta }$ the characteristic function for the Borel set $\Delta $\
of $G/H_{\omega }.$ From this we obtain again the result that the set of $%
{\rm Tr}(A^{\eta }(\chi _{\Delta })\rho )$\ is a Kolmogorov probability for all
states $\rho $. However, because of the projection, $P^{\eta }$, appearing
in the definition of $A^{\eta }(f)$, we no longer have a necessarily
one-to-one mapping in $f\rightarrow A^{\eta }(f)$. ${\rm Tr}(\rho T^{\eta }(%
\boldsymbol{x}))$ gives an effect valued measure in $\mathcal{H}$,
none-the-less \cite{BelBug}.

There is an important corollary to this:

COROLLARY 1. {\it Let $\mathcal{H}$ and $A^{\eta }(f)$\ be as before. By the informational
completeness of $\mathcal{A}_{1}$, the values $\{{\rm Tr}(A^{\eta }(f)A^{\eta
}(h)\rho )\}$\ for any state $\rho $\ are sufficient to uniquely
characterize $\rho $. But $\mathcal{A}_{1}$ has no non-trivial projections
in it}. The proof of this is in \cite{FS}.

This has additional consequences which are fundamental because of the
following:

\begin{itemize}
 \item[$1)$] The projections on non-trivial closed subspaces of $\mathcal{H}$ are
excluded from $\mathcal{A}_{1}$; on the basis of the values ${\rm Tr}(A^{\eta
}(f)A^{\eta }(h)\rho )$, it would be impossible to conclude\ that $\rho $\
was supported in any closed subspace of $\mathcal{H}$. This is the property
of measurement being strictly non-local, but for which $\mathcal{A}_{1}$ is
informationally complete.

\item[$2)$] $\eta $\ is a wave function for a particle describing the experimental
apparatus used to measure $\rho $\ in a truly quantum mechanical
measurement. Thus the scheme we have devised makes no use of Gleason's
Theorem \cite{Gleason}, or any collapse of $\rho $.

\item[$3)$] The fact that non-trivial projections on intervals in the phase space
are nowhere to be found in $\mathcal{A}_{1}$ makes the interpretation of the
phase space experimentally problematical. Only the contextual (the $\eta $)
interpretation of the phase space is available to us by means of the average
values of the variables of $(\boldsymbol{p},\boldsymbol{q},\boldsymbol{s})$\
in $\eta $.\ With this contextual interpretation, there is a resolution of
Bloch's paradox that there is an apparent dependence on the particular frame
in which, relativistically, a collapse is to occur \cite{Sch4}.

\item[$4)$] Assuming that $\eta $\ leads to informational completeness, then we have
that for almost any bounded operator in $\mathcal{H}$, we may write it as
(limits of) the integral of $f(\boldsymbol{x},\boldsymbol{y})T^{\eta }(%
\boldsymbol{x})T^{\eta }(\boldsymbol{y}).$ However, the projections $T^{\eta
}(\boldsymbol{x}),T^{\eta }(\boldsymbol{y})$ do not commute; so, we have a
generalization of the spectral operators and the spectral representation of
the ordinary quantum mechanics to non-commuting operators \cite{Sch}.
\end{itemize}

\subsection{The connection of $\cup _{g\in G}^{\bullet }L^{2}{}_{\protect\mu %
}(G/H_{R^{\ast }(g^{-1})\protect\omega })$\ with the classical
representation spaces of classical mechanics}

A general transitive\ classical representation space of classical
(statistical) mechanics is given by either $G/H_{\omega }$\ or $\cup _{g\in
G}^{\bullet }G/H_{R^{\ast }(g^{-1})\omega }$, as we have discussed before.
Suppose we have certain generators $X$ and $Y$ of the symmetry group of the
classical situation that have a Poisson bracket $\{X,Y\}$ that is
non-vanishing. Remember that we may specify the values of the density
matrices in terms of the $X$ and $Y$ and other generators which seem to
commute. What do they map to (say $X\mapsto X^{\prime }$) in the $L^{2}$
sense. This is apparently a very trickey question if the $X^{\prime }$ and $%
Y^{\prime }$ have a commutator that is a non-trivial operator, or so it
would seem. But we have the following.

For example, suppose we take the operators for rotations, $iL_{x}$\ and $%
iL_{y},$ compute $iL_{x}^{\prime }$\ and $iL_{y}^{\prime }$ after another
rotation\ and compute their commutator which is $[iL_{x}^{\prime
},iL_{y}^{\prime }]=iL_{z}^{\prime }.$ How can we get the operators to
commute when they obviously do not? But, we have omitted any reference to the
space on which they are to commute! Suppose we take just $G={\rm SO}(3)$, and $%
\mathfrak{g}={\rm so}(3)$ with generators $\{iL_{x},iL_{y},iL_{z}\}$ satisfying
the usual Lie algebra relations. Compute $Z^{2}(\mathfrak{g})$\ to obtain 
\begin{equation*}
Z^{2}(\mathfrak{g})=\{aL_{x}^{\ast }\wedge L_{y}^{\ast }+bL_{y}^{\ast
}\wedge L_{z}^{\ast }+cL_{z}^{\ast }\wedge L_{x}^{\ast }\mid a,b,c\in 
\mathbb{R}\}.
\end{equation*}%
We will choose 
\begin{equation*}
\omega =L_{x}^{\ast }\wedge L_{y}^{\ast }.
\end{equation*}%
Then 
\begin{equation*}
\mathfrak{h}_{\omega }=\{aiL_{z}\mid a\in \mathbb{R}\},
\end{equation*}%
and 
\begin{equation*}
H_{\omega }=\{e^{iaL_{z}}\mid a\in \mathbb{R}\}.
\end{equation*}%
Since $H_{\omega }$ is closed as a subspace of ${\rm SO}(3)$, ${\rm SO}(3)/H_{\omega }$
is a phase space. On ${\rm SO}(3)/H_{\omega }$, everything in $H_{\omega }$\ just
acts like the identity. Consequently $\mathfrak{h}_{\omega }$\ acts just
like the $0$-operator. Thus the commutation relations are $[iL_{x},iL_{y}]=0$%
\ on ${\rm SO}(3)/H_{\omega }$. When we promote this to the right operation of $%
{\rm SO}(3)$ on $L^{2}{}_{\mu }({\rm SO}(3)/H_{\omega })$, we again obtain the same result
for the commutation relations. The only point on which we should be careful
is that we should map $L^{2}{}_{\mu }({\rm SO}(3)/H_{\omega })$ to $L^{2}{}_{\mu
}({\rm SO}(3)/H_{R^{\ast }(g^{-1})\omega })$\ when using the right
representation of ${\rm SO}(3)$. On ${\rm SO}(3)/H_{R^{\ast }(g^{-1})\omega }$\ the
commutation relations are no longer $[iL_{x},iL_{y}]=0$\ but are instead $%
[giL_{x}g^{-1},giL_{y}g^{-1}]=0$ with $giL_{x}g^{-1}$\ the abbreviation for
the action of the angular momentum\ on the functions $f(g\circ
e^{iaL_{x}^{\prime }}\circ g^{-1})$ for $f\in C^{\infty }({\rm SO}(3))$, and
similarly for $iL_{x}^{\prime }$ and $iL_{y}^{\prime }$ on $L^{2}{}_{\mu
}({\rm SO}(3)/H_{R^{\ast }(g^{-1})\omega })$.\ 

The above is another example like the example in the Poincar\'{e} group in
$14.1$. It is generalizable. This is precisely what also happens for
any Lie group, $G$, when operating from the right, except we have to replace 
${\rm SO}(3)$\ with $G$. By $\int_{G/H_{\omega }}d\mu (g)$ we are effectively
integrating over the tangent space of $L^{2}{}_{\mu }(G/H_{\omega })$, which
is flat! Hence there are always coordinates for which the commutators are
zero on $G/H_{\omega }$; thus there are coordinates for which the
commutators are zero on $g\circ G/H_{\omega }\circ g^{-1}=G/H_{R^{\ast
}(g^{-1})\omega }$. We conclude: We may choose $H_{\omega }$ so that all
commutators on $\mathfrak{g}_{e}$ (Lie algebra relations) are effectively
zero on $L^{2}{}_{\mu }(G/H_{\omega })$, and then rotate, boost, and translate
to obtain the same relations for the elements of $\mathfrak{g}_{g}$.

The right-representation spaces are the ones usually taken to represent
classical mechanics on a Hilbert space in general, and the
left-representation spaces are the ones usually taken to represent quantum
mechanics. The reason for this is that, with the right-representations, we
just have to translate all our equations $[\cdot ,\cdot ]=0$\ with the
group, while for the left-representations we have a problem with the
equations $[\cdot ,\cdot ]=0$\ when we translate with the group elements.

\subsection{An application to solid state physics}

We have applications to the Heisenberg group, or the Galilei group, or the
Poincar\'{e} group which we could give for $G/H_{\omega }$\ topologically
isomorphic to $R^{2n}\ltimes spin$ $group$. But we would prefer to give an
application of the Heisenberg group which displays the full force of phase
space methods, namely to treat a finite crystal lattice. We take the
following from \cite{G.Ali}:

We choose the Heisenberg group as the group of phase shifts by $\lambda $,
translations of $\mathbb{R}^{3}$ by $\boldsymbol{q}$, and boosts of $\mathbb{%
R}^{3}$ by $\boldsymbol{p}$. Specifically,%
\begin{equation*}
\mathcal{W}=\{(\lambda ,\boldsymbol{q},\boldsymbol{p})\mid \boldsymbol{q},%
\boldsymbol{p}\in \mathbb{R}^{3},\text{ }\lambda \in \mathbb{R}/2\pi \}
\end{equation*}%
with the multiplication law%
\begin{equation*}
(\lambda ^{\prime },\boldsymbol{q}^{\prime },\boldsymbol{p}^{\prime })\circ
(\lambda ,\boldsymbol{q},\boldsymbol{p})=(\lambda ^{\prime }+\lambda +(%
\boldsymbol{q}^{\prime }\cdot \boldsymbol{p}-\boldsymbol{q}\cdot \boldsymbol{%
p}^{\prime })/2,\boldsymbol{q}^{\prime }+\boldsymbol{q},\boldsymbol{p}%
^{\prime }+\boldsymbol{p}).
\end{equation*}%
The center of $\mathcal{W}$ is the set%
\begin{equation*}
Z(\mathcal{W})=\{(\lambda ,\boldsymbol{0},\boldsymbol{0})\mid \lambda \in 
\mathbb{R}/2\pi \}.
\end{equation*}

A basis of generators of $\mathfrak{g}$\ for $\mathcal{W}$ are $\{i%
\boldsymbol{1},i\boldsymbol{Q}_{j},i\boldsymbol{P}_{k}\mid j,k=1,2,3\}$\
satisfying%
\begin{equation*}
\lbrack i\boldsymbol{Q}_{j},i\boldsymbol{P}_{k}]=\delta _{j,k}i\boldsymbol{1}%
,
\end{equation*}%
the other commutators being zero. (We may write the basis with the imaginary 
$i$ subsumed into the $Q$'s, $P$'s, and $i\boldsymbol{1}=\mathbb{J}$, but we
will assume that we may represent the group in a complex Hilbert space so we
may as well write the basis for $\mathfrak{g}$ as is.)\ Using%
\begin{equation*}
\boldsymbol{p}\cdot \boldsymbol{P}=\sum_{j}\boldsymbol{p}_{j}\boldsymbol{P}%
_{j},\text{ }\boldsymbol{p}_{j}\in \mathbb{R},
\end{equation*}%
and%
\begin{equation*}
\boldsymbol{q}\cdot \boldsymbol{Q}=\sum_{j}\boldsymbol{q}_{j}\boldsymbol{Q}%
_{j},\text{ }\boldsymbol{q}_{j}\in \mathbb{R},
\end{equation*}%
we obtain, for a general element $X$ of $\mathfrak{g}$, 
\begin{equation*}
X=i(\lambda \boldsymbol{1}+\boldsymbol{q}\cdot \boldsymbol{Q}+\boldsymbol{p}%
\cdot \boldsymbol{P}).
\end{equation*}%
Then for $X^{\prime }$\ defined similarly, 
\begin{equation*}
\lbrack X,X^{\prime }]=-(\boldsymbol{q}^{\prime }\cdot \boldsymbol{p}-%
\boldsymbol{q}\cdot \boldsymbol{p}^{\prime })i\boldsymbol{1.}
\end{equation*}%
Also, we have%
\begin{equation*}
(\lambda ,\boldsymbol{q},\boldsymbol{p})\equiv \exp \{i(\lambda \boldsymbol{1%
}+\boldsymbol{q}\cdot \boldsymbol{Q}+\boldsymbol{p}\cdot \boldsymbol{P})\}.
\end{equation*}

The dual basis for $\mathfrak{g}$ is given by $\{-i\boldsymbol{1}^{\ast },-i%
\boldsymbol{Q}_{j}^{\ast },-i\boldsymbol{P}_{k}^{\ast }\mid j,k=1,2,3\}$\
such that%
\begin{eqnarray*}
\boldsymbol{1}^{\ast }(\boldsymbol{1}) &=&1,\text{ }\boldsymbol{1}^{\ast }(%
\boldsymbol{Q}_{k})=0,\text{ }\boldsymbol{1}^{\ast }(\boldsymbol{P}_{k})=0, 
\notag \\
\boldsymbol{Q}_{j}^{\ast }(\boldsymbol{1}) &=&0,\text{ }\boldsymbol{Q}%
_{j}^{\ast }(\boldsymbol{Q}_{k})=\delta _{jk},\text{ }\boldsymbol{Q}%
_{j}^{\ast }(\boldsymbol{P}_{k})=0,  \notag \\
\boldsymbol{P}_{j}^{\ast }(\boldsymbol{1}) &=&0,\text{ }\boldsymbol{P}%
_{j}^{\ast }(\boldsymbol{Q}_{k})=0,\text{ }\boldsymbol{P}_{j}^{\ast }(%
\boldsymbol{P}_{k})=\delta _{jk}.
\end{eqnarray*}

From these equations we obtain the coboundary operator, $\delta $ \cite%
[pp. 344-350]{Sch}:%
\begin{equation*}
\delta (\boldsymbol{1}^{\ast })=\sum_{j}\boldsymbol{P}_{j}^{\ast }\wedge 
\boldsymbol{Q}_{j}^{\ast },\text{ }\delta (\boldsymbol{Q}_{j}^{\ast })=0,%
\text{ }\delta (\boldsymbol{P}_{j}^{\ast })=0,
\end{equation*}%
and so%
\begin{equation*}
\delta (\boldsymbol{Q}_{j}^{\ast }\wedge \boldsymbol{Q}_{k}^{\ast })=\delta (%
\boldsymbol{P}_{j}^{\ast }\wedge \boldsymbol{Q}_{k}^{\ast })=\delta (%
\boldsymbol{P}_{j}^{\ast }\wedge \boldsymbol{P}_{k}^{\ast })=\delta (%
\boldsymbol{1}^{\ast }\wedge \boldsymbol{1}^{\ast })=0.
\end{equation*}%
Thus the coboundary of a general two-form $\omega _{2}$\ is of the form%
\begin{eqnarray*}
\delta \omega _{2} &=&\delta \left( \sum_{j}(\alpha _{j}\boldsymbol{1}^{\ast
}\wedge \boldsymbol{Q}_{j}^{\ast }+\beta _{j}\boldsymbol{1}^{\ast }\wedge 
\boldsymbol{P}_{j}^{\ast })\right)  \notag \\
&=&\delta (\boldsymbol{1}^{\ast })\wedge \sum_{j}(\alpha _{j}\boldsymbol{Q}%
_{j}^{\ast }+\beta _{j}\boldsymbol{P}_{j}^{\ast })
\end{eqnarray*}%
for $\alpha _{j},\beta _{j}\in \mathbb{R}$; so, $Z^{2}(\mathfrak{g})\equiv
{\rm Kernel}(\delta |_{2-{\rm forms}})=\{\omega _{2}$ not involving $\boldsymbol{1}%
^{\ast }\wedge \sum (\alpha _{j}\boldsymbol{Q}_{j}^{\ast }+\beta _{j}%
\boldsymbol{P}_{j}^{\ast })\}$, or%
\begin{equation*}
Z^{2}(\mathfrak{g})=\left\{ \omega _{2}=\sum_{j,k}(\alpha _{jk}\boldsymbol{P}%
_{j}^{\ast }\wedge \boldsymbol{P}_{k}^{\ast }+\beta _{jk}\boldsymbol{Q}%
_{j}^{\ast }\wedge \boldsymbol{P}_{k}^{\ast }+\gamma _{jk}\boldsymbol{Q}%
_{j}^{\ast }\wedge \boldsymbol{Q}_{k}^{\ast })\right\}
\end{equation*}%
for $\alpha _{jk},\beta _{jk},\gamma _{jk}\in \mathbb{R}$.

We may choose (almost) any $\omega $\ from $Z^{2}(\mathfrak{g})$, obtain $%
\mathfrak{h}_{\omega }$\ and $H_{\omega }$\ from it, and then form the phase
space $G/H_{\omega }$ if $H_{\omega }$\ is closed. To obtain the usual phase
space, which we shall not use here except as an illustration, there is one
particular choice for $\omega $\ that will give it, namely%
\begin{equation*}
\omega _{0}=\sum_{j}\boldsymbol{Q}_{j}^{\ast }\wedge \boldsymbol{P}%
_{j}^{\ast }.
\end{equation*}%
Then%
\begin{equation*}
\mathfrak{h}_{\omega _{0}}=\{i\lambda \boldsymbol{1}\mid \lambda \in \mathbb{%
R}\}
\end{equation*}%
from which we obtain%
\begin{equation*}
H_{\omega _{0}}=\{e^{i\lambda \boldsymbol{1}}\mid \lambda \in \mathbb{R}\}.
\end{equation*}%
But this is closed; so, $\mathcal{W}/H_{\omega _{0}}$\ is a phase space. $%
H_{\omega _{0}}$ is also a normal subgroup; so, $\mathcal{W}/H_{\omega _{0}}$%
\ is even a group, and there is little difference in the action of $\mathcal{%
W}$\ on $\mathcal{W}/H_{\omega _{0}}$ from the left or right. $\mathcal{W}%
/H_{\omega _{0}}$ just is the usual space of $\boldsymbol{q}$'s and $%
\boldsymbol{p}$'s.

We now search for another phase space, one more suited to the problem of a
crystal. For a single cell crystal, let us define the basic lattice $%
\mathcal{L}^{\prime }$ of the physical system of the crystal as the set of
linear combinations with integer coefficients of the non-colinear "primitive
vectors" $\{\boldsymbol{a}_{1},\boldsymbol{a}_{2},\boldsymbol{a}_{3}\}$\
with coefficients in $\mathbb{Z}$. Then the crystal lattice is defined as
the set of linear combinations with integer coefficients of the primitive
vectors. (There are several ways to pick the primitive vectors for a given
lattice; pick one.) This lattice is invariant under translation by the
primitive vectors, inversions in a finite class of planes,\ (and by a
certain finite set of rotations which we will not consider here). Set 
\begin{equation*}
\nu =2\pi (\boldsymbol{a}_{1}\cdot (\boldsymbol{a}_{2}\times \boldsymbol{a}%
_{3}))^{-1}
\end{equation*}%
and 
\begin{equation*}
\boldsymbol{b}_{1}=\nu \boldsymbol{a}_{2}\times \boldsymbol{a}_{3},\text{ }%
\boldsymbol{b}_{2}=\nu \boldsymbol{a}_{3}\times \boldsymbol{a}_{1},\text{ }%
\boldsymbol{b}_{3}=\nu \boldsymbol{a}_{1}\times \boldsymbol{a}_{2}.
\end{equation*}%
The vectors that are linear combinations of the $\{\boldsymbol{b}_{1},%
\boldsymbol{b}_{2},\boldsymbol{b}_{3}\}$\ with coefficients in $\mathbb{Z}$\
form a lattice $\mathcal{L}^{\prime \prime }$, called\ the Heisenberg
lattice. Furthermore, we have 
\begin{equation*}
\boldsymbol{a}_{j}\cdot \boldsymbol{b}_{k}=2\pi \delta _{jk}.
\end{equation*}%
Thus, the lattice $\mathcal{L}^{\prime \prime }$\ is also known as the reciprocal
lattice to $\mathcal{L}^{\prime }$. We will set%
\begin{equation*}
\mathcal{L}=\mathcal{L}^{\prime }\times \mathcal{L}^{\prime \prime }.
\end{equation*}%
We will use $\boldsymbol{a}\in \mathcal{L}_{+}^{\prime }$\ to denote that $%
\boldsymbol{a}$ has a "first" coordinate that is positive and $\boldsymbol{a}%
\in \mathcal{L}^{\prime }$.

We now have the notation necessary to define a different phase space for the
Heisenberg group, one which we shall use for the crystal. Define%
\begin{eqnarray*}
\omega &=&\sum_{\boldsymbol{a}\in \mathcal{L}_{+}^{\prime },\boldsymbol{b}%
\in \mathcal{L}^{"}\text{ }}(\boldsymbol{a}\cdot \boldsymbol{Q}^{\ast }+%
\boldsymbol{b}\cdot \boldsymbol{P}^{\ast })\wedge \left( -\boldsymbol{a}%
\cdot \boldsymbol{Q}^{\ast }+\boldsymbol{b}\cdot \boldsymbol{P}^{\ast
}\right)  \notag \\
&=&\sum_{\boldsymbol{a}\in \mathcal{L}_{+}^{\prime },\boldsymbol{b}\in 
\mathcal{L}^{"}\text{ }}2\boldsymbol{a}\cdot \boldsymbol{Q}^{\ast }\wedge 
\boldsymbol{b}\cdot \boldsymbol{P}^{\ast }.
\end{eqnarray*}%
Then, for $\boldsymbol{a}^{\prime }\in \mathcal{L}^{\prime }$ and $%
\boldsymbol{b}^{\prime }\in \mathcal{L}^{\prime \prime },$ 
\begin{equation*}
\lbrack \boldsymbol{a}\cdot \boldsymbol{Q}^{\ast }+\boldsymbol{b}\cdot 
\boldsymbol{P}^{\ast }](\boldsymbol{a}^{\prime }\boldsymbol{P}+\boldsymbol{b}%
^{\prime }\boldsymbol{Q})=\boldsymbol{b}\cdot \boldsymbol{a}^{\prime }+%
\boldsymbol{a}\cdot \boldsymbol{b}^{\prime }\in 2\pi \mathbb{Z}.
\end{equation*}%
But for $n\in \mathbb{Z}$, $\exp \{2\pi ni\boldsymbol{1}\}=\exp \{0i%
\boldsymbol{1}\}=1$. Thus,\ we obtain%
\begin{equation*}
\mathfrak{h}_{\omega }=\{i\lambda \boldsymbol{1+ia\cdot P}+i\boldsymbol{%
b\cdot Q}\mid \lambda \in \mathbb{R},(\boldsymbol{a},\boldsymbol{b})\in 
\mathcal{L}\}
\end{equation*}%
and%
\begin{equation*}
H_{\omega }=\{e^{i(\lambda \boldsymbol{1+a}\cdot \boldsymbol{P}+\boldsymbol{b%
}\cdot \boldsymbol{Q)}}\mid \lambda \in \mathbb{R}\},(\boldsymbol{a},%
\boldsymbol{b})\in \mathcal{L}\}.
\end{equation*}%
Moreover, $H_{\omega }$\ is closed as a subgroup of $\mathcal{W}$. Hence $%
\mathcal{W}/H_{\omega }$\ is a phase space. It is the phase space for
particles confined to the crystal.

Furthermore, from the group relations, $H_{\omega }$\ is normal in $\mathcal{%
W}$. Hence $\mathcal{W}/H_{\omega }$\ is a group. $H_{\omega }$\ is also
commutative. These two properties have as consequences the fact that $%
\mathcal{W}/H_{\omega }$\ is a very special phase space as we shall see, and
one that is not necessary for a general phase space.

To obtain a unitary representation of $\mathcal{W}$ on $\mathcal{W}%
/H_{\omega }$, we follow the procedure of Mackey \cite{Mack}. First obtain a
unitary representation, $\Lambda $, of $H_{\omega }$: Since $H_{\omega }$\
is a commutative group, this is simply%
\begin{equation*}
\Lambda (e^{i(\lambda \boldsymbol{1+a\cdot P}+\boldsymbol{b\cdot Q)}})=\exp
\left\{ i\left( \lambda \alpha +\sum_{j}(\boldsymbol{a}_{j}\cdot \boldsymbol{%
\beta }_{j}+\boldsymbol{b}_{j}\cdot \boldsymbol{\gamma }_{j})\right) \right\}
\end{equation*}%
where $\alpha \in \mathbb{R},$ and $\boldsymbol{\beta }_{j},\boldsymbol{%
\gamma }_{j}\in \mathbb{R}^{3}$ for some choice of $\alpha $, $\beta _{j}$,
and $\gamma _{j}$. The $\Lambda $\ act on $S^{1}$. The choices $\alpha =0,%
\boldsymbol{\beta }_{j}\in \mathcal{L}^{\prime \prime },\boldsymbol{\gamma }%
_{j}\in \mathcal{L}^{\prime }$\ are most convenient. Then induce a
representation, $V$, for all of $\mathcal{W}$.

One obtains a representation 
\begin{equation*}
\mathcal{W}=\{(\lambda ,\boldsymbol{a},\boldsymbol{b})=e^{i(\lambda 
\boldsymbol{1+}\widetilde{\boldsymbol{a}}\cdot \boldsymbol{P}+\widetilde{%
\boldsymbol{b}}\cdot \boldsymbol{Q)}}\mid \lambda \in \mathbb{R},\text{ and }%
\boldsymbol{a},\boldsymbol{b}\in \mathbb{R}^{3}\}.
\end{equation*}%
with 
\begin{equation*}
\widetilde{\boldsymbol{a}}=\boldsymbol{a}{\rm mod}(\mathcal{L}^{\prime }), \quad
\widetilde{\boldsymbol{b}}=\boldsymbol{b}{\rm mod}(\mathcal{L}^{\prime
\prime }).
\end{equation*}%
Thus one obtains a unitary representation $\mathcal{W}$ on a compact set,
practically speaking! Here, we also have the direct interpretation of the $%
\widetilde{\boldsymbol{p}}\cdot \boldsymbol{P}$\ and $\widetilde{\boldsymbol{%
q}}\cdot \boldsymbol{Q}$\ being the translations and boosts. This is quite
different from the usual (mis)interpretation \cite{Ash}, \cite{Slater}.

Next we take the left-regular (quantum mechanical) representations of $%
\mathcal{W}$ on square integrable functions, $\psi $, in $L^{2}(\mathcal{W}%
/H_{\omega })$, or the right-regular (classical mechanical) representations
on the disjoint union over an orbit of these. One may easily check that the $%
V_{L}$\ and $V_{R}$ \textit{are} representations of $\mathcal{W}$. In
particular, there is no difficulty in showing that a particular $\psi $\ is
square integrable or not, it being an integral over a compact set. This
again is much different from the false claim that in $\mathbb{R}^{3}$, the
periodic analogs of $\psi $\ and $\mathcal{F}\psi $, $\mathcal{F}$\ the
Fourier transform,\ are square integrable over all of $\mathbb{R}^{3}.$

We see that it has been very profitable to express the results in terms of
the Heisenberg lattice, etc. as this has allowed us to express the
restrictions on both the position and momentum vectors at the same time.

We also remark that we could have made the factoring by any lattice that has
primitive vectors of the form $(n_{1}\boldsymbol{\alpha }_{1},n_{2}%
\boldsymbol{\alpha }_{2},n_{3}\boldsymbol{\alpha }_{3}),n_{j}\in Z_{>0}.$
This would take care of the case of a finite crystal, and not just a single
crystal cell.

Finally, we mention that we could have started with the Galilei group or the
Poincar\'{e} group and perform similar computations and tricks. See \cite
{G.Ali} for this.

\subsection{The value of working in quantum mechanics on phase space}

We have seen that there are many points in which working in one of the
configuration or momentum spaces we have many apparent paradoxes, while
working in the phase space formalism we have apparently circumvented these
paradoxes. We mention the facts that we do not have any obstructions to
quantizing any classical observable, there is no difficulty with the
ordering problem as that is just a property of the $\eta $\ that we choose,
there is no difficulty with quantizing a positive operator and obtaining
something which has positive and negative values as in the energy of the
electron, there is no difficulty with collapse of the wave function as we
have just the transition probabilities of going from one wave function to
another in measurement, there is no problem with talking about being "at $%
\boldsymbol{x}$" as this means "transitioning to a wave function having
expectation values at $\boldsymbol{x}$," there is no difficulty with making
classical mechanical theories in the Hilbert space setting, and there is no
difficulty with making a theory of particles embedded in a finite crystal
with the appropriate momentum constraints. Neither is there any difficulty
with the Wigner quasi-distribution as that is not a theory based on the
phase space as we have defined it and which we have discussed in \cite{Sch3}. 
We add that there is no necessity to "correct" the position operator as in
zitterbewegung of the electron, and the spurious derivation of the
anti-electron which came about because the electron had negative energy
states in the ordinary theory. We also could have obtained a theory of
particles of zero mass and various helicities by the same methods we have
employed here \cite{Bro}. We could have derived a field theory for the
quantum mechanical particles based on phase space, as we have the Hilbert
spaces decomposable into reproducing kernel Hilbert spaces with all the properties
of the wave functions to which we transist; so, we have the anti-particles
none-the-less \cite{Sch6}. However, these latter subjects take us away from
the point of this paper, which is that we can make a great deal of progress
simply by considering quantum mechanics on phase space.

\bigskip
\noindent \textbf{Acknowledgements}

The two of us (J. J. S\l awianowski, A. Martens) in preparing this publication were supported by the 
Research Grant No 501 049 540 of the National Scientific Center in Poland (NCN). We are grateful for this support.


\begin{thebibliography}{99}
\bibitem{1a} E.A. Abbott: {\em Flatland: A Romance of Many Dimensions}, Seely and Co., London 1884.

\bibitem{1} R. Abraham, J.E. Marsden: {\em Foundations of Mechanics} (second edition), The Benjamin and Cummings Publishing Company, Inc. London, Amsterdam, don Mills, Ontario, Sydney, Tokyo 1978.

\bibitem{Ali} S.T. Ali, M. Engli\v{s}: {\em Rev. Mod. Phys.} {\bf 17} (2005), 391-490.

\bibitem{G.Ali} G. Ali, R. Beneduci, G. Mascali, F.E. Schroeck, Jr., J.J. S\l awianowski: {\em Int. J. Theor. Phys.} {\bf 52} doi: 10.1007/s10773-013-1912-9, 2013.

\bibitem{Prugo2} S.T. Ali, E. Prugove\v{c}ki: {\em Acta Appl. Math.} {\bf 6} (1986), 1-18, 19-45, 47-62.

\bibitem{2} V.I. Arnold: {\em Mathematical Methods of Classical Mechanics\/}, Springer Graduate Texts in Mechanics {\bf 60}, Springer Verlag, New York 1978.

\bibitem{3} V.I. Arnold, A. Avez: {\em Ergodic Problems of Classical Mechanics,\/} Benjamin-Cummings, Reading-Massachussets.

\bibitem{Ash} N.W. Ashcroft, N.D. Mermin: {\em Solid State Physics\/}, Harcourt College Pub. 1976.

\bibitem{4} L. Ballentine: {\em Quantum Mechanics: A Modern Development}, World Scientific Publishing Co. Ltd., Singapore-New Jersey-London-Hong-Kong 1998.

\bibitem{5} V. Bargmann: {\em On Unitary Ray Representations of Continuous Groups}, Ann. Math. {\bf 59} (1954), 1-46.

\bibitem{6} A.O. Barut, R. R\c{a}czka: {\em Theory of Group Representations and Applications}, 2nd ed., World Scientific Publishing Co., Singapore 1986. 

\bibitem{BelBug} E.C. Beltrametti, S. Bugajski: {\em J. Phys. A.: Math. Gen.} {\bf 28} (1995), 3329-3343.

\bibitem{6a} R. Beneduci, J. Brook, R. Curran, F.E. Schroeck, Jr.: {\em Int. J. Theor. Phys.}  {\bf 50} (2011), 3682-3696.

\bibitem{6b} R. Beneduci, J. Brook, R. Curran, F.E. Schroeck, Jr.: {\em Int. J. Theor. Phys.}  {\bf 50} (2011), 3697-3723.

\bibitem{Berezin} F.A. Berezin: {\em Commun. Math. Phys.} {\bf 40} (1975), 153-174.

\bibitem{7} P.G. Bergmann: {\em Phys. Rev.} {\bf 144} (1966), 1078-1080.

\bibitem{8} P.G. Bergmann, I. Goldberg: {\em Phys. Rev.} {\bf 98} (1955), 531-538. 

\bibitem{B} L. Boltzmann: {\em Uber die Beziehung zwischen dem zweiten Haupsatz der mechanischen W\"{a}rmtheorie und der Wahrscheinlichkeitsrechnung\/}, Wiener Academische Sitz., 1876, publ. 1977; to be found in R.H. Ellis,  {\em Entropy, Large Deviations and Statistical Mechanics\/} (1985), 373-435,

\bibitem{9} M. Born: {\em Vorlesungen \"{u}ber Atommechanik\/}, Springer, Berlin 1925.

\bibitem{10} M. Born, E. Wolf: {\em Principles of Optics \/}, Pergman Press, London 1964.

\bibitem{Bro} J.A. Brooke, F.E. Schroeck, Jr.: {\em J. Math. Phys.} {\bf 37} (1996), 5958-5986.

\bibitem{Busch} P. Busch: {\em Int. J. Theor. Phys.} {\bf 30} (1991), 1217-1227.

\bibitem{11} C. Caratheodory:  {\em Variationsrechnung und Particlle Differentialgleichungen Erster Ordung\/}, B.G. Teubner Leipzig 1956.

\bibitem{Dirac_wiezy} P.A.M. Dirac: {\em Canad. J. Math.} {\bf 2} (1950), 129. 

\bibitem{13} P.A.M. Dirac:  {\em Proc. Royal Soc. London} {\bf A246} (1958), 326-332.

\bibitem{14} P.A.M. Dirac:  {\em Proc. Royal Soc. London} {\bf A246} (1958), 333.

\bibitem{15} P.A.M. Dirac: {\em Proc. R.I.A.} {\bf 63A} 49-59.

\bibitem{18} C.M. Edvards, J.T. Levis: {\em Comm. Math. Phys.} {\bf 13}, 2 (1969), 119.

\bibitem{16} A. Erd\'{e}yi: {\em  Asymptotic Expansions\/}, Dover, New York  1956.

\bibitem{16a} B.R. Fischer: {\em On the Geometric Quantization of Symplectic Lie Group Actions\/}, dissertation, Florida Atlantic University 1995.

\bibitem{17} N. Fr\"{o}man, P.O. Fr\"{o}man: {\em JWKB Approximation\/}, North-Holland Publishing Company, Amsterdam 1965.

\bibitem{G} J.W. Gibbs: {\em Elementary Principles in Statistical Mechanics\/}, Yale Univ. Press 1902.

\bibitem{Gleason} A.M. Gleason: {\em J. Math. and Mech.} {\bf 6} (1957), 885-893.

\bibitem{GuiSte} V. Guillemin, S. Sternberg: {\em Geometric Asymptotics\/}, American Mathematical Society, Providence, Rhode Island 1977.

\bibitem{20} V. Guillemin, S. Sternberg: {\em Symplectic Techniques in Physics\/}, Cambridge University Press, Cambridge 1984.

\bibitem{21} K.E. Hellwig, B. Wegner: {\em Mathematik und Theoretische Physik. Ein Integrierter Grundkurs f\"{u}r Physiker und Mathematiker\/}, Walter de Gruyter, Berlin-New York I-1992, II-1993.

\bibitem{22} K. Huang, {\em Statistical Mechanics\/}, John Wiley \& Sons, Inc., New York, London 1963.

\bibitem{23} R.L. Hudson: {\em Rep. on Math. Phys.} {\bf 10} (1974), 766-789.

\bibitem{24} M.V. Karasev (Editor): {\em Quantum Algebras and Poisson Geometry in Mathematical Physics\/}, Providence, Rhode Island, American Mathematical Society, 2005. American Mathematical Society Translations. Series 2. {\bf 216} Advances in the Mathematical Sciences-57 (Formerly-Advances in Sovjet Mathematics.)

\bibitem{KarMas} M.V. Karasev, V.P. Maslov: {\em Nonlinear Poisson Brackets. Geometry and Quantization\/}, Nauka, Moscow 1991 (and references therein) (in Russian).

\bibitem{KH63} K. Huang: {\em Statistical mechanics}, John Wiley \& Sons, Inc., New York-London 1963.

\bibitem{26} A.A. Kirillov: {\em Elements of the Theory of Representations\/}, Springer-Verlag, New York 1976.

\bibitem{27} A.A. Kirillov: {\em Lectures on the Orbit Method\/}, Graduate Studies in Mathematics {\bf 64}, American Mathematical Society, Providence, Rhode Island 2004. 

\bibitem{Klauder} J.R. Klauder: {\em Beyond Conventional Quantization\/}, Academic Press, New York 2000.

\bibitem{28} K. Kobayashi, S. Nomizu: {\em Foundation of Differential Geometry\/}, Interscience, New York 1963.

\bibitem{29} A.N. Kolmogorov: {\em Proc. 1954 Intern. Congr. Math.}, North-Holland, Amsterdam 1, cf. also Appendix in (Abraham-Marsden) (1957), 315-333.

\bibitem{30} B. Kostant: {\em Lecture Notes in Mathematics} {\bf 170} 87-207, Springer, New York 1970.

\bibitem{Ko} B.O. Koopman:  {\em Proc. Natl. Acad. Sci.} {\bf 7} (1931), 315-318.

\bibitem{LandL} L.D. Landau, E.M. Lifshitz: {\em Course of Theoretical Physics, Vol. III, Quantum Mechanics}, Pergaman Press, London 1958.

\bibitem{32} L. D. Landau, E.M. Lifshitz: {\em Statistical Physics, Vol. V, Part I}, Pergamon Press, London 1958.

\bibitem{33} L.H. Loomis: {\em An Introduction to Abstract Harmonic Analysis}, D. Van Nostrand Company, Inc., Princeton-New Jersey-Toronto-London-New York 1953.

\bibitem{Mack} G.W. Mackey: {\em Ann. Math.} {\bf 58} (1953), 101-139.

\bibitem{34} G.W. Mackey:  {\em The Mathematical Foundation of Quantum Mechanics\/}, W.
A. Benjamin, Inc., New York, Amsterdam 1963.

\bibitem{35} B. Malgrange:  {\em Ideals of Differentiable Functions\/}, Oxford University Press, Oxford 1966.

\bibitem{36} J.E. Marsden, T. Ratiu:  {\em Introduction to Mechanics and Symmetry\/}, Springer, New York 1994.

\bibitem{37} J.E. Marsden, T. Ratiu:  {\em Introduction to Mechanics and Symmetry. A Basic Exposition of Classical Mechanical Systems (second ed.)\/}, Springer, New York 1999.

\bibitem{AM2}
A. Martens: {\em Rep. Math. Phys.} {\bf 49} (2002), 295-303.

\bibitem{m3}
A. Martens: {\em Rep. Math. Phys.} {\bf 51} (2003), 287-295.

\bibitem{m4}
A. Martens: {\em J.\ of Nonlinear Math.\ Phys.} {\bf 11}, Supplement (2004), 145-150.

\bibitem{m5}
A. Martens: {\em J.\ of Nonlinear Math.\ Phys.} {\bf 11}, Supplement (2004), 151-156.

\bibitem{AM} A. Martens: {\em Rep. Math. Phys.} {\bf 62}, 2 (2008), 145-155.

\bibitem{m2} A. Martens, J.J. S\l awianowski: {\em Acta\ Phys.\ Pol. B\/} {\bf 41} (2010), 1847-1880.

\bibitem{m6}
A. Martens: {\em Rep. Math. Phys.} {\bf 71}, 3 (2013), 381-398 .

\bibitem{38} K. Maurin: {\em General Eigenfunctions Expansions and Unitary Representations of Topological Groups}, PWN-Polish Scientific Publishers, Warsaw 1968.

\bibitem{39} K. Maurin:  {\em Methods of Hilbert Spaces}, PWN-Polish Scientific Publishers, Warsaw 1972.

\bibitem{40} A. Messiah:  {\em Quantum Mechanics}, North-Holland Publishing Co., Amsterdam 1965.

\bibitem{41} J.E. Moyal:  {\em Proc. Cambridge Phil. Soc.} {\bf 45} (1949), 99.

\bibitem{42} J.v. Neumann:  {\em Mathematische Grundlagen der Quantenmechanik}, Springer, Berlin 1932.

\bibitem{43} L. Pontryagin:  {\em Topological Groups}, Princeton University Press, Princeton, New Jersey 1956.

\bibitem{Prugo} E. Prugove\v{c}ki: {\em Stochastic Quantum Mechanics and Quantum Spacetime\/}, D. Reidel, Dordrecht-Boston 1984.

\bibitem{44} T. Ratiu:  {\em Euler-Poisson Equations on Lie Algebras}, Thesis, Berkley 1980.

\bibitem{45} T. Ratiu:  {\em Am. J. Math.} {\bf 104} (1982), 409-448.

\bibitem{46} M.E. Rose,  {\em Elementary Theory of Angular Momentum}, Dover Publications 1965.

\bibitem{47} A. Rubinowicz,  {\em Quantum Mechanics}, Amsterdam, New York (etc), Elsevier Pub. Co, Warsaw, PWN-Polish Scientific Publishers 1968.

\bibitem{48} W. Rudin:  {\em Fourier Analysis on Groups}, Interscience Publishers, New York-London 1962.

\bibitem{Sch6} F.E. Schroeck, Jr.:  {\em Rep. on Math. Phys.} {\bf 26} (1988), 197-210.

\bibitem{Sch} F.E. Schroeck, Jr.:  {\em Quantum mechanics on Phase Space}, Kluwer Academic Publishers, Dordrecht, the Netherlands 1996.

\bibitem{Schf} F.E. Schroeck, Jr.: {\em Int. J. Theor. Phys.} {\bf 44} (2005), 2091-2100 (Theorem 7).

\bibitem{Sch4} F.E. Schroeck, Jr.: {\em J. Phys. A: Math. Theor.} {\bf 42} doi:10.1088/1751-8113/42/15/155301, 2009.

\bibitem{Sch3} F.E. Schroeck, Jr.: {\em J. Phys. A: Math. Theor.} {\bf 45}  doi: 10.1088/1751-8113/45/6/065303, 2012.

\bibitem{FS} F.E. Schroeck, Jr.:  {\em The C* Axioms and the Phase Space Formalism of Quantum Mechanics\/}, in Contributions in Mathematical Physics, a Tribute to Gerard G. Emch, S.T. Ali and K.B. Sinha,
eds., Hindustan Book Agency, New Delhi, 2007, pp. 197-212; and Int. J. Theor. Phys. {\bf 47} (2008), 175-184.

\bibitem{50} S.S. Schweber:  {\em An Introduction to Relativistic Quantum Field Theory}, Harper \& Row, Pubs., Inc., New York, 1961 or 1962.

\bibitem{Sh} C.E.Shannon: {\em Bell System Tech. J.} {\bf 27} (1948), 379-423, 623-656.

\bibitem{SiWo} D.J. Simms, N.M.J. Woodhouse: {\em Lecture Notes in Physics} {\bf 53} Springer, Berlin-Heidelberg-New York 1976.

\bibitem{Slater} J.C. Slater: {\em Symmetry and Energy Bands in
Crystals\/}, Dover Publs. Inc., N. Y., (1972), 151-154.

\bibitem{JJS} J.J. S\l awianowski: {\em Geometry of Phase Spaces},  John Wiley \& Sons, Chichester, New York, Brisbane, Toronto, Singapore, PWN-Polish Scientific Publishers, Warszawa 1991. 

\bibitem{JJSa} J.J. S\l awianowski, V. Kovalchuk, A. Martens, B. Go\l ubowska, E.E. Ro\.{z}ko: {\em Math. Meth. Appl. Sci.} {\bf 34} (2011), 1512-1540. 

\bibitem{JJS1} J.J. S\l awianowski, V. Kovalchuk, A. Martens, B. Go\l ubowska, E.E. Ro\.{z}ko: {\em Journal of Geometry and Symmetry in Physics} {\bf 21} (2011), 61-94.

\bibitem{JJS2} J.J. S\l awianowski, V. Kovalchuk, A. Martens, B. Go\l ubowska, E.E. Ro\.{z}ko: {\em Journal of Geometry and Symmetry in Physics} {\bf 22} (2011), 67-94.

\bibitem{JJS3} J.J. S\l awianowski, V. Kovalchuk, A. Martens, B. Go\l ubowska, E.E. Ro\.{z}ko: {\em Journal of Geometry and Symmetry in Physics} {\bf 23} (2011), 59-95. 

\bibitem{JJSc} J.J. S\l awianowski, V. Kovalchuk, A. Martens, B. Go\l ubowska, E.E. Ro\.{z}ko: {\em Discrete and Continuous Dynamical Systems-Series B} {\bf 17}, 2 (2012), 699-733. 

\bibitem{Synge} J.L. Synge: {\em Classical Dynamics}, Springer Verlag, Berlin 1960.

\bibitem{53} J.L. Synge: {\em Geometrical Mechanics and de Broglie Waves}, Cambridge University Press, Cambridge 1954.

\bibitem{54} J.L. Synge: {\em Phys. Rev.} {\bf 89} (1953), 467.

\bibitem{Sn} J. \'{S}niatycki: {\em Geometric Quantization and Quantum Mechanics}, Applied Mathematical Sciences Series of Springer-Verlag, New York 1980.

\bibitem{55} W.M. Tulczyjew: {\em Unpublished Lectures and Private Communications}, Warsaw 1964-1968.

\bibitem{56} J.H. Van Vleck: {\em Proc. Nat. Acad. Sci.} {\bf 14} (1928), 178-188.

\bibitem{56a} A. Wehrl: {\em Rev. Mod. Phys.} {\bf 50} (1978), 221-260.

\bibitem{57} H. Weyl: {\em Z. Phys.} {\bf 46} 1, 1928.

\bibitem{58} H. Weyl: {\em The Theory of Groups and Quantum Mechanics}, Dover, New York 1931.

\bibitem{59} H. Weyl: {\em Symmetry}, Princeton University Press, Princeton, New Jersey 1952.

\bibitem{60} E.P. Wigner: {\em Gruppentheorie und Ihre Anwendungen auf die Quantenmechanik der Atomspektren}, Vieweg Verlag, Braunschweig, 1931. English Translation by J. Griffin, Group Theory and its Applications to the Quantum Mechanics of Atomic Spectra, Academic Press, New York 1959.

\bibitem{61} S. Woronowicz: {\em On a Theorem of Mackey, Stone and von Neumann}, Studia Mathematica {\bf 24} (1964) 101-105. 

\bibitem{62} D.N. Zubarev: {\em Nonequilibrium Statistical Thermodynamics}, Plenum, New York 1974.


\end{thebibliography}
\end{document}